\documentclass[prx, reprint, amsmath,amssymb,showpacs,floatfix,longbibliography, aps, twocolumn, superscriptaddress,nofootinbib]{revtex4-1}
\usepackage{times} 
\usepackage{graphicx}
\usepackage{dcolumn}
\usepackage{bm}
\usepackage{color}
\usepackage{xcolor}
\usepackage{hyperref}
\usepackage{enumitem}
\usepackage{cancel}
\hypersetup{colorlinks=true,citecolor=blue,linkcolor=blue, urlcolor=blue}
\hypersetup{linktocpage}
\newcolumntype{M}[1]{>{\centering\arraybackslash}m{#1}}
\newcolumntype{N}{@{}m{0pt}@{}}
\usepackage{environ}

\bibpunct{[}{]}{,}{n}{}{}

\usepackage{amsfonts,amssymb,amsmath}
\usepackage[T1]{fontenc}
\usepackage{tikz}
\usetikzlibrary{arrows,snakes,backgrounds}

\newcommand{\ZZ}{{\mathbb Z}}
\newcommand{\RR}{{\mathbb R}}

\newcommand{\GG}{{\mathbb G}}

\newcommand{\ba}{\boldsymbol{a}}

\newcommand{\bv}{\boldsymbol{v}}
\newcommand{\be}{\boldsymbol{e}}
\newcommand{\bface}{\boldsymbol{f}}
\newcommand{\bt}{\boldsymbol{t}}

\newcommand{\bsig}{\boldsymbol{\Sigma}}
\newcommand{\corho}{\overline{\boldsymbol{\rho}}_{\scriptscriptstyle{\{g_v\}}}}
\newcommand{\conu}{{\overline{\boldsymbol{\nu}}}_{\scriptscriptstyle{\{g_v\}}}}

\newcommand{\crho}{{\boldsymbol{\rho}}_{\scriptscriptstyle{\{g_v\}}}}
\newcommand{\cnu}{{{\boldsymbol{\nu}}}_{\scriptscriptstyle{\{g_v\}}}}

\newcommand{\calpha}{\boldsymbol{\alpha}_{\scriptscriptstyle{\{g_v\}}}}
\newcommand{\coalpha}{\overline{\boldsymbol{\alpha}}_{\scriptscriptstyle{\{g_v\}}}}
\newcommand{\coeps}{\boldsymbol{\epsilon}_{\scriptscriptstyle{\{g_v\}}}}
\newcommand{\kappah}{\widetilde{{\boldsymbol{\rho}}}^{\, h}_{\scriptscriptstyle{\{g_v\}}}}
\newcommand{\Uspt}{\mathcal{U}_1}
\newcommand{\Ugspt}{\mathcal{U}_2}
\newcommand{\onepara}{H_0}

\newcommand{\onetriv}{\Psi_0}

\newcommand{\Us}{\mathcal{U}_s}
\newcommand{\Uf}{\mathcal{U}_f}
\newcommand{\rhohath}{{\hat{\overline{\boldsymbol{\rho}}}^h}}

\newcommand{\rhohat}{{\hat{\overline{\boldsymbol{\rho}}}}}
\newcommand{\nuhat}{{\hat{\overline{\boldsymbol{\nu}}}}}
\newcommand{\nhat}{{\hat{\overline{\boldsymbol{n}}}}}
\newcommand{\etahat}{{\hat{\overline{\boldsymbol{\eta}}}}}
\newcommand{\betahat}{{\hat{\overline{\boldsymbol{\beta}}}}}

\newcommand{\bdryv}{{w}}

\NewEnviron{eqs}{%
\begin{equation}\begin{split}
    \BODY
\end{split}\end{equation}
}

\newcommand{\bra}[1]{\langle {#1} |}
\newcommand{\ket}[1]{ | {#1} \rangle}
\newcommand{\lr}[1]{ \langle {#1} \rangle}

\newcommand{\ra}{\rightarrow}
\newcommand{\eps}{\epsilon}

\DeclareMathOperator{\coker}{coker}
\DeclareMathOperator{\Ima}{Im}

\begin{document}

\title{Disentangling supercohomology symmetry-protected topological phases in three spatial dimensions}

\author{Yu-An Chen}
\email[E-mail: ]{yuanchen718@gmail.com}
\affiliation {California Institute of Technology, Pasadena, CA 91125, USA}

\author{Tyler D. Ellison}
\email[E-mail: ]{ellisont@uw.edu}
\affiliation {University of Washington, Seattle, WA 98195, USA}

\author{Nathanan Tantivasadakarn}
\email[E-mail: ]{ntantivasadakarn@g.harvard.edu}
\affiliation {Harvard University, Cambridge, MA 02138, USA}

\date{\today}

\begin{abstract}

\noindent We build exactly solvable lattice Hamiltonians for fermionic symmetry-protected topological (SPT) phases in ($3+1$)D classified by group supercohomology. A central benefit of our construction is that it produces an explicit finite-depth quantum circuit (FDQC) that prepares the ground state from an unentangled symmetric state. The FDQC allows us to clearly demonstrate the characteristic properties of supercohomology phases - namely, symmetry fractionalization on fermion parity flux loops -- predicted by continuum formulations. By composing the corresponding FDQCs, we also recover the stacking relations of supercohomology phases. Furthermore, we derive topologically ordered gapped boundaries for the supercohomology models by extending the protecting symmetries, analogous to the construction of topologically ordered boundaries for bosonic SPT phases.
% in Ref.~\cite{WWW19}. 
Our approach relies heavily on dualities that relate certain bosonic $2$-group SPT phases with supercohomology SPT phases. We develop physical motivation for the dualities in terms of explicit lattice prescriptions for gauging a $1$-form symmetry and for condensing emergent fermions. We also comment on generalizations to supercohomology phases in higher dimensions and to fermionic SPT phases outside of the supercohomology framework.

\end{abstract}
\maketitle

\tableofcontents

\section{Introduction}

% Symmetry-protected topological (SPT) phases of matter are classified by quantized invariants that capture a characteristic response
% % Symmetry-protected topological (SPT) phases of matter can be classified by enumerating the consistent quantized responses
% %of a system
% to probing with symmetry defects \cite{LG12,K14,KTTW15,W15,X18,GJ19,CTW17,WLG17,WL15,W16,ZWWG19,EN15,C15,FV17}. Correspondingly, the classification of fermionic SPT (fSPT) phases is notably distinct from the classification of bosonic SPT (bSPT) phases -- due to the inherent fermion parity symmetry obeyed by fSPT phases. In particular, in three spatial dimensions and with unitary internal symmetries, the bSPT phases are believed to be classified by group cohomology, while the fSPT phases are only \textit{partially} classified by the more rich, group supercohomology theory \cite{EN15,CGLW13,GW14,GJ19,WG18}. 

Symmetry-protected topological (SPT) phases of matter are classified by quantized invariants that capture a characteristic response
% Symmetry-protected topological (SPT) phases of matter can be classified by enumerating the consistent quantized responses
%of a system
to probing with symmetry defects \cite{LG12,K14,KTTW15,W15,X18,GJ19,CTW17,WLG17,WL15,W16,ZWWG19,EN15,C15,FV17}. SPT phases built from fermionic degrees of freedom (d.o.f.) must conserve fermion parity, and the associated fermion parity symmetry defects can be used to probe the system. Consequently, the classification of fermionic SPT (fSPT) phases, where the constituent d.o.f. may be fermions, is notably distinct from the classification of bosonic SPT (bSPT) phases, composed of only bosonic d.o.f.. In particular, in three spatial dimensions and with unitary internal symmetries, the bSPT phases are believed to be classified by group cohomology, while the fSPT phases are only \textit{partially} classified by the more rich, group supercohomology theory \cite{EN15,CGLW13,GW14,GJ19,WG18}. 

The algebraic data of group cohomology can be used to construct an exactly solvable model belonging to a bSPT phase \cite{CGLW13}. The celebrated group cohomology models yield a transparent connection between the quantized invariant - namely, a group cocycle $\nu$ - and a lattice Hamiltonian. Another feature of these models is that a finite-depth quantum circuit (FDQC) \cite{CGW10} that prepares the ground state from a tensor product state can be written expressly in terms of $\nu$. Further, in Ref.~\cite{WWW19}, it was shown that the group cohomology data could be used to identify symmetric topologically ordered gapped boundaries for the group cohomology models, by enlarging the protecting symmetry group on the boundary. 

In this work, we construct exactly solvable models for (${3+1}$)D fSPT phases directly from the group supercohomology data that characterizes the phases. The resulting supercohomology models describe fSPT phases protected by finite unitary internal symmetries of the form $G_f = G \times \ZZ_2^f$, where $\ZZ_2^f$ denotes the fermion parity symmetry. The supercohomology data can be written as a certain pair of $G$-dependent functions $(\rho, \nu)$, where heuristically, $\nu$ corresponds to the data that characterizes the bSPT phases, while $\rho$ captures a response that is intrinsic to fSPT phases \cite{C15,FV17}. With this, our principal contributions can be stated as follows. For any choice of supercohomology data $(\rho,\nu)$ characterizing a $(3+1)$D fSPT phase: 

\begin{enumerate}[label=(\roman*)]
    \item We construct a representative fSPT Hamiltonian with mutually commuting un-frustrated terms and verify that the quantized responses of the model correspond to the data $(\rho,\nu)$ - by explicitly computing the $G$-symmetry fractionalization on fermion parity fluxes.
    \item We identify a FDQC that prepares the ground state from a symmetric product state and determine the stacking rules for supercohomology phases from the composition of the FDQCs. 
    \item We use an extension of the symmetry to build symmetric topologically ordered gapped boundaries for the supercohomology model.
\end{enumerate}

\begin{figure*}[htb]
\centering
\includegraphics[width=.8\textwidth, trim={0 405 0 405},clip]{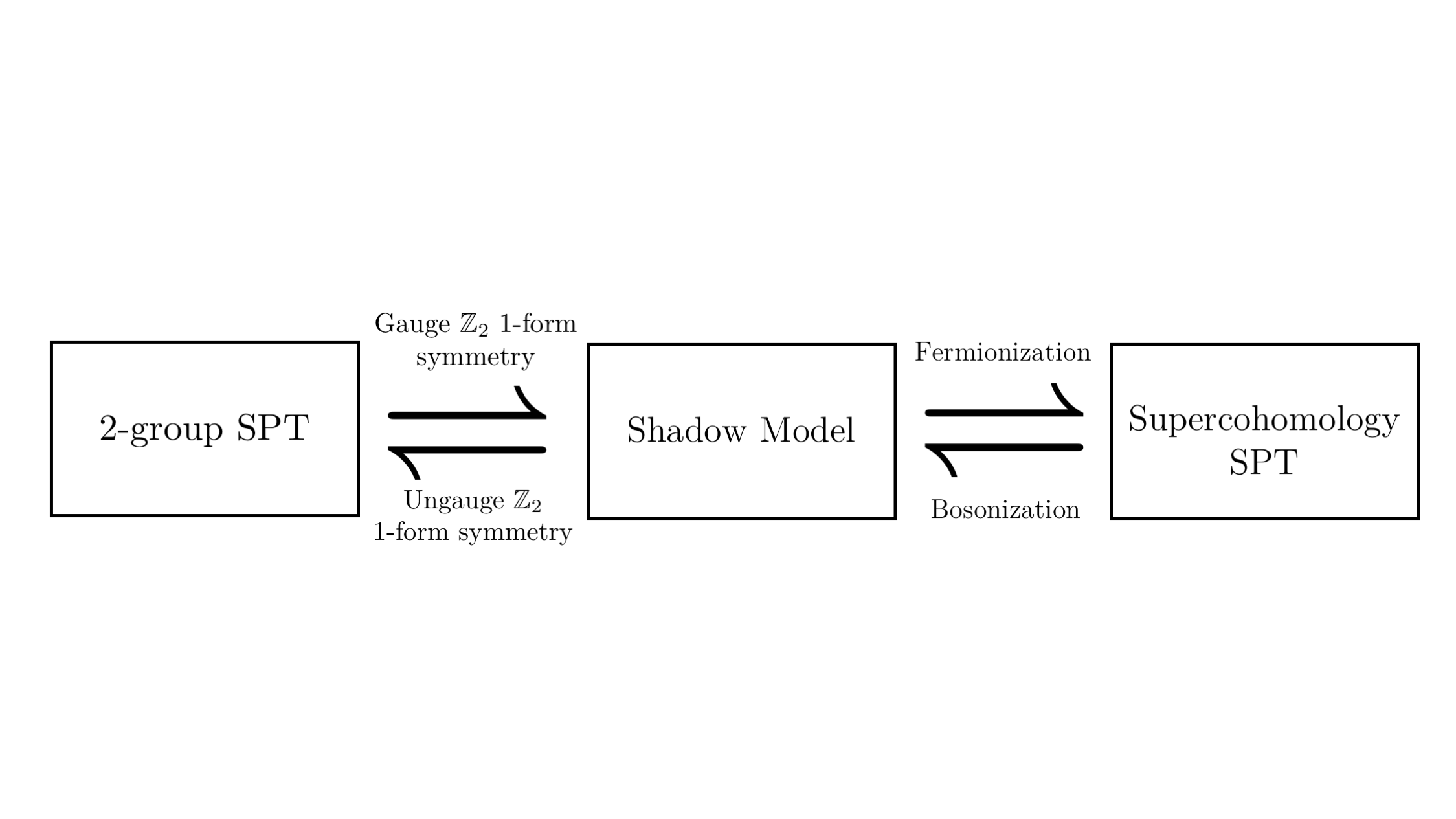}
\caption{To construct a $G_f=G \times \ZZ_2^f$ supercohomology SPT model we start with a model for a particular $2$-group SPT phase determined by the supercohomology data $(\rho,\nu)$. Next, we gauge the $\ZZ_2$ $1$-form symmetry of the $2$-group to build the shadow model. We then condense the fermion in the shadow model, or apply the fermionization duality, to obtain a model for the supercohomology SPT phase corresponding to $(\rho,\nu)$. }
\label{fig:Gf_outline}
\end{figure*}

Our strategy is largely motivated by the spacetime formulation in Ref.~\cite{KT17}, wherein fSPT phases are related to particular 2-group bSPT phases through a process of bosonization. More specifically, our construction can be broken down into the three succinct steps outlined below, and shown schematically in Fig.~\ref{fig:Gf_outline}. 

\begin{enumerate}[label=(\arabic*)]
    \item Starting with a choice of supercohomology data $(\rho, \nu)$, we first build an auxiliary bSPT model with a $2$-group symmetry. The $2$-group symmetry contains a $\ZZ_2$ $1$-form symmetry as a subgroup. 
    \item We gauge the $\ZZ_2$ $1$-form subgroup by minimally coupling the model to a $2$-form gauge field. This produces a symmetry-enriched $\ZZ_2$ gauge theory with an emergent fermion, referred to as the `shadow model' \cite{BGK17}.
    \item Finally, we pair the emergent fermion with a physical fermion and condense the composite excitation: the $\ZZ_2$ gauge theory is dual to a fermionic theory \cite{CK18}. The result is a model for a fSPT phase characterized by the supercohomology data $(\rho,\nu)$. 
\end{enumerate}

\vspace{.2mm}
\noindent \begin{center}\emph{Relation to previous work:}\end{center}
\vspace{.2mm}

Our work can be viewed as a generalization of the exactly solvable supercohomology models in $(2+1)$D, described in Refs.~\cite{EF19,TV18,BGK17}. Similarly, the first step is to construct an auxiliary bSPT model from the supercohomology data. In the case discussed here, however, the auxiliary bSPT phase is protected by a higher-form symmetry. 

In the pioneering work of Ref.~\cite{GW14}, representative wave functions for supercohomology phases were identified by studying re-triangulation invariant non-linear $\sigma$ models on a discrete spacetime manifold. Later, Ref.~\cite{WG18} provided a comprehensive classification of fSPT phases on a spatial lattice, by solving for consistent domain wall decorations.
% spatial renormalization schemes via re-triangulations of the manifold. 
Our work builds on these results by constructing an explicit parent Hamiltonian for their fixed point wave functions along with FDQCs that prepare the ground states from a product state. Within our framework, we are also able to demonstrate that the supercohomology models indeed exhibit the universal responses to symmetry fluxes captured by the supercohomology data $(\rho,\nu)$.  

Our strategy for constructing the supercohomology models mirrors the methods employed at the level of spacetime partition functions in Refs.~\cite{KT17,LZW18,KOT19}. In particular, Ref.~\cite{LZW18} studies supercohomology phases by constructing a Lagrangian for the associated shadow model. However, we go beyond studying the shadow model and explicitly implement the fermionization duality to establish supercohomology data as quantized invariants of lattice Hamiltonians. In recent work, Ref.~\cite{KOT19} constructed gapped boundaries for spacetime models of supercohomology phases using a symmetry extension (see also Ref.~\cite{GOPWW20}).
We employ a similar symmetry extension to construct the supercohomology Hamiltonians on a manifold with boundary.

We note that many of the models constructed in this text describe intrinsically interacting fSPT phases \cite{WLG17,CTW17}. That is, there are neither interacting bosonic counterparts nor free-fermionic representations of the phases. Hence, in particular, our work falls outside of the scope of Refs.~\cite{SRFL08,K09,RSFL10}.

\vspace{.2mm}
\noindent \begin{center}\emph{Structure of the paper:}\end{center}
\vspace{.2mm}

In Section~\ref{sec: supercohomology models}, we define the quantized invariants of supercohomology phases and present our supercohomology models in terms of the associated data. Subsequently, we describe the derivation of the bulk supercohomology models in Sections~\ref{sec: bulk Z2f} and \ref{sec:bulkG}. In Section~\ref{sec: bulk Z2f}, we give an example of our construction, for the case where the protecting symmetry is simply $G_f = \ZZ_2^f$. We use the opportunity to introduce the notation of cohomology on a manifold $M$, which is used throughout the text. Furthermore, in Section~\ref{sec ttc} and Section~\ref{sec: atomic insulator}, we detail a lattice prescription for gauging a $1$-form symmetry and condensing an emergent fermion, respectively. Section~\ref{sec:bulkG} describes the construction of the supercohomology models more generally, where the protecting symmetry is $G_f = G \times \ZZ_2^f$. We show that the lattice Hamiltonians are indeed characterized by the supercohomology data and recover the additive group structure of supercohomology phases under the operation of stacking by composing the corresponding FDQCs in Section~\ref{sec: supercohomology model}.  Section~\ref{sec: sym extension boundary} presents the symmetry extension method for constructing symmetric gapped boundaries for the supercohomology models; we leave the detailed derivation to Appendix~\ref{app: super from 2gauge}. In Appendix~\ref{app:term}, we compile the notation used in the text. We discuss spin structure and the bosonization duality of Ref.~\cite{CK18} in Appendix~\ref{sec: review of boson-fermion duality}. In Appendices~\ref{sec: review of (2+1)D supercohology fSPT} and \ref{sec: 2-group extension}, we give a concise review the construction of supercohomology models in $(2+1)$D and $2$-group gauge theory, respectively. Lastly, Appendix~\ref{sec:example} gives an example of the symmetry extensions used to construct the models with a boundary. The remaining appendices provide the technical details and explicit calculations used in the derivation of our models.

\section{Supercohomology models} \label{sec: supercohomology models}

Before discussing the construction of the supercohomology models in Sections~\ref{sec: bulk Z2f} and \ref{sec:bulkG}, we give a concise description of the models themselves.  We begin with a definition for the supercohomology data $(\rho,\nu)$, and we assume familiarity with group cohomology. The group cohomology notation used here is summarized in Appendix~\ref{app:groupcoho}. In Section~\ref{sec: group cohomology models}, we then review the group cohomology models of Ref.~\cite{CGLW13}. We finish with Section~\ref{sec: supercohomology models 2}, where we define the more general supercohomology models and describe the stacking rules for the supercohomology phases, derived from the composition of FDQCs.

\subsection{Supercohomology data} \label{sec: supercohomology data2}

The supercohomology data gives quantized invariants for ($3+1$)D fSPT phases protected by a finite onsite\footnote{\unexpanded{An onsite representation of a $0$-form $G$ symmetry is a representation of $G$ in which, for all $g\in G$ the representation $V(g)$ is a tensor product of linear representations of $G$ on each site.}} unitary ${G_f = G \times \ZZ_2^f}$ symmetry. To streamline the discussion, we refer to Appendix~\ref{app:groupcoho} for a review of the notation from group cohomology. We freely use the notion of group cochains, the coboundary operator $\delta$, and the cup-$n$ products $\cup_n$ with $n \in \{0,1,2\}$ in the discussion below.

% The quantized invariants are most naturally described using the language of group cohomology. Therefore, to streamline the discussion, we refer to Appendix~\ref{app:groupcoho} for a review of the notation from group cohomology, including a definition of group cochains, the coboundary operator $\delta$, and the cup-$n$ product $\cup_n$ with $n \in \{0,1,2\}$.

For a finite group $G$, the supercohomology data is given by a pair of group cochains $(\rho,\nu)$ belonging to:
\begin{align}
    (\rho,\nu) \in C^3(G,\ZZ_2) \times C^4(G,\mathbb{R}/ \ZZ). 
\end{align}
Furthermore, $\rho$ and $\nu$ satisfy the relations \cite{GW14}:
\begin{align} \label{guweneqs}
    \delta \rho = 0, \quad \delta \nu = \frac{1}{2}\rho \cup_1 \rho.
\end{align}
Note that if $\rho=0$, then $\nu$ is a group cocycle ($\delta \nu =0$). In this case, the supercohomology data reduces to the data that characterizes bSPT phases within the group cohomology framework \cite{CGLW13}.

The supercohomology data is further organized into equivalence classes. Two sets of supercohomology data $(\rho,\nu)$ and $(\rho',\nu')$ are considered equivalent if there exists:
\begin{align}
    \beta \in C^2(G,\ZZ_2),\quad \eta \in C^3(G,\mathbb{R}/\ZZ),
\end{align}
such that:
\begin{eqs}
    \rho' &= \rho + \delta \beta \\
    \nu' &= \nu + \delta \eta + \frac{1}{2}\beta \cup \beta + \frac{1}{2}\beta \cup_{1} \delta \beta + \frac{1}{2}\rho \cup_2 \delta \beta.
\end{eqs}
In Section~\ref{sec: supercohomology model}, we give physical motivation for the equivalence relation and show that if $(\rho,\nu)$ and $(\rho',\nu')$ are equivalent, then the corresponding supercohomology models belong to the same fSPT phase.
% In Section~\ref{sec: supercohomology model}, we give physical motivation for the equivalence relation, by showing that $(\rho,\nu)$ and $(\rho',\nu')$ are equivalent if and only if the corresponding supercohomology models belong to the same fSPT phase.\footnote{We emphasize that we assume $G_f = G \times \ZZ_2^f$ for a unitary $G$.}

Throughout the text, we use the convention that group cochains are homogeneous. Therefore, in what follows, we take $\rho$ and $\nu$ to be functions:
\begin{eqs}
    \rho: G^4 &\to \ZZ_2 = \{ 0,1 \}, \\
    \nu:G^5 &\to \mathbb{R} / \ZZ = [0,1),
\end{eqs}
which are homogeneous, i.e., for any $h\in G$:
\begin{eqs} \label{rho homogeneous}
        \rho(g_0,g_1,g_2,g_3)&=\rho(hg_0,hg_1,hg_2,hg_3), \\ 
        \nu(g_0,g_1,g_2,g_3,g_4)&=\nu(hg_0,hg_1,hg_2,hg_3,hg_4).
\end{eqs}

\subsection{Review of group cohomology models} \label{sec: group cohomology models} 

When $\rho$ is zero, the supercohomology data is equivalent to the familiar group cohomology data, which characterizes ($3+1$)D bSPT phases with a finite onsite unitary $G$ symmetry. As a consequence, the corresponding group cohomology models are a special case of the supercohomology models. We build up to the supercohomology models in Section~\ref{sec: supercohomology models 2} by first reviewing the group cohomology models of Ref.~\cite{CGLW13}.

\begin{figure}[t]
\centering
\includegraphics[width=0.45\textwidth, trim={580 400 550 360},clip]{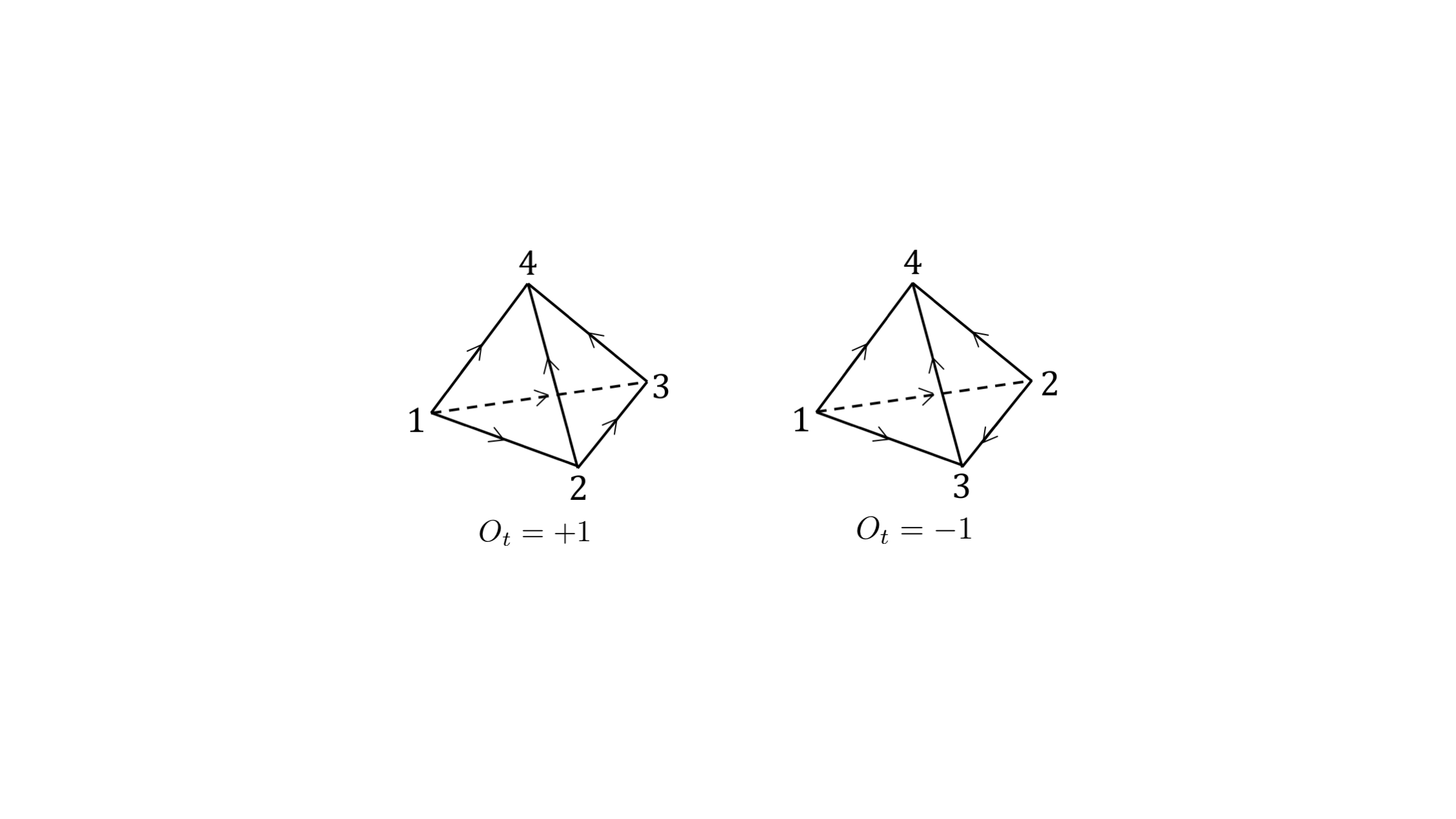}
\caption{A branching structure determines an ordering of the vertices of a tetrahedron. The vertices are ordered by the number of edges oriented towards the vertex. The branching structure also gives an orientation of each tetrahedron relative to the orientation of $M$. We use the convention that the tetrahedron pictured on the left is positively oriented ($O_t=+1$), and the tetrahedron to the right is negatively oriented ($O_t=-1$).}
\label{fig: branchingstructure}
\end{figure}

The group cohomology models are defined on an arbitrary triangulation of an orientable closed $3$-manifold $M$. The triangulation of $M$ gives a decomposition of $M$ into vertices, edges, faces, and tetrahedra. We further require that the triangulation is equipped with a branching structure -- an assignment of an orientation to each edge in such a way that there are no cycles around any of the faces. A branching structure yields both an ordering of the vertices of each tetrahedron as well as an orientation $O_t \in \{-1,+1\}$ of any tetrahedron $t$ relative to the orientation of $M$ (see Fig.~\ref{fig: branchingstructure}).

\begin{figure}[t]
\centering
\includegraphics[width=0.3\textwidth, trim={800 440 800 360},clip]{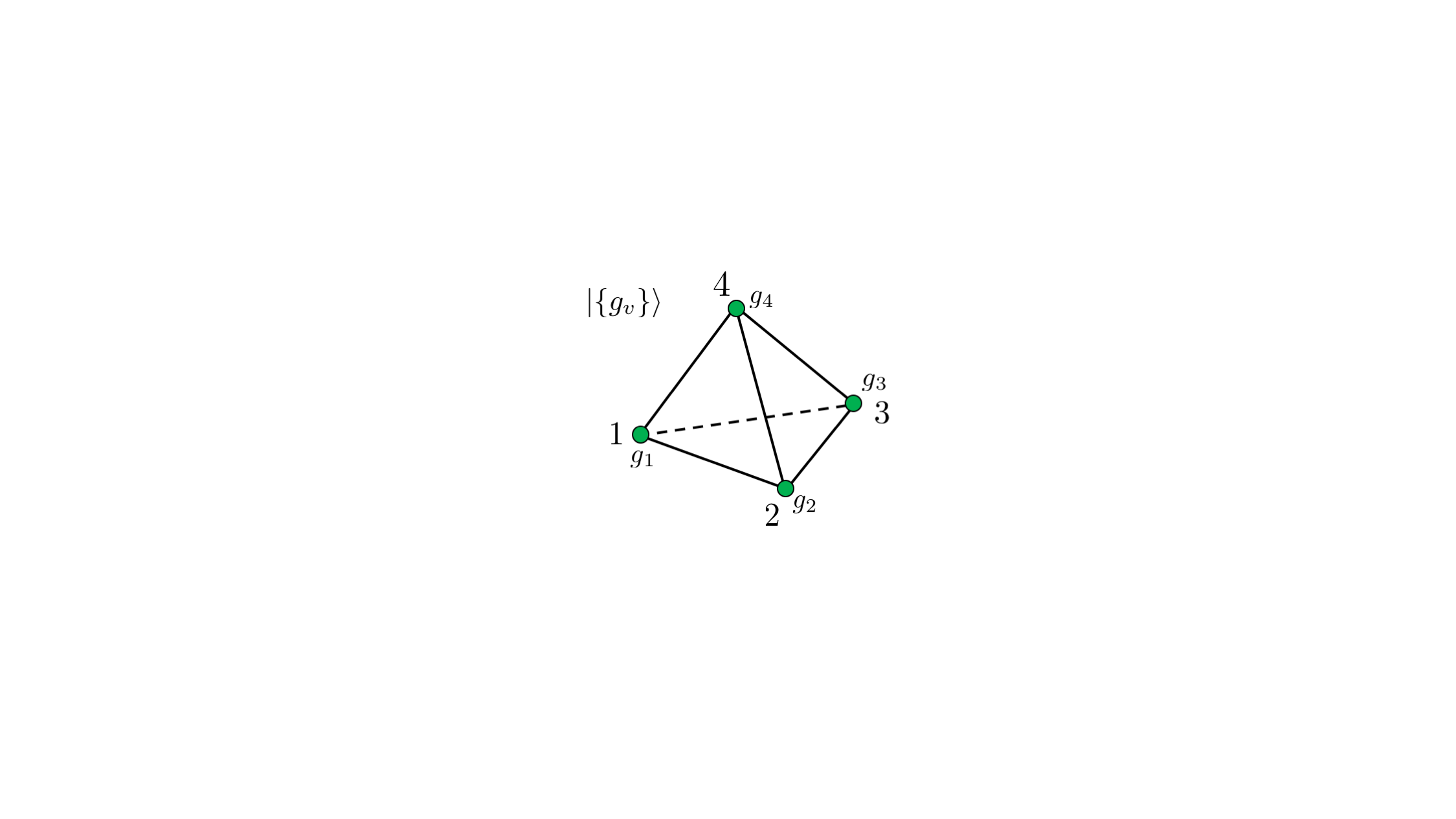}
\caption{The group cohomology models are defined on a Hilbert space consisting of a $G$ d.o.f. on every vertex. The configuration states $\ket{\{g_v\}}$ form a basis for the Hilbert space.}
\label{fig: groupcohodof}
\end{figure}

The Hilbert spaces for the group cohomology models are formed by placing a $G$ {d.o.f.} on each vertex of $M$ (Fig.~\ref{fig: groupcohodof}). A basis for the $|G|$ dimensional Hilbert space at vertex $v$ is given by states $\ket{g_v}$ labeled by elements of $G$. Furthermore, a basis for the full Hilbert space is given by product states of the form $\ket{\{g_v\}}$, in which, the state at vertex $v$ is $\ket{g_v}$. The $G$ symmetry is represented using the regular representation, i.e., for any $h\in G$, $h$ is represented by:
\begin{align}
    V(h) \equiv \sum_{\{g_v\}} \ket{\{hg_v\}} \bra{\{g_v\}}.
\end{align}

The group cohomology models can be built from a $G$-paramagnet Hamiltonian -- a Hamiltonian belonging to the trivial SPT phase. The $G$-paramagnet Hamiltonian is given by:
\begin{align} \label{gpara}
    H^G \equiv -\sum_v \mathcal{P}_v,
\end{align}
where the sum is over vertices in $M$, and $\mathcal{P}_v$ is a projector onto a symmetric state at the vertex $v$, i.e.:
\begin{align} \label{G projector}
    \mathcal{P}_v \equiv \frac{1}{|G|} \left(  \sum_{g_v} \ket{g_v} \right)\left( \sum_{g_v} \bra{g_v} \right).
\end{align}
The ground state $\ket{\Psi^G}$ of the $G$-paramagnet Hamiltonian is a tensor product of a symmetric state at each vertex. This can be written as an equal amplitude superposition over all $\{g_v\}$ configurations:
\begin{align}
    \ket{\Psi^G} \equiv \sum_{\{g_v\}}\ket{\{g_v\}}.
\end{align}
Here, as elsewhere in the text, we omit the normalization of the state for notational convenience.

We build the group cohomology model corresponding to the group cocycle $\nu$ by conjugating $H^G$ by a FDQC $\mathcal{U}_b$. $\mathcal{U}_b$ is defined in terms of the data $\nu$ as \cite{CGLW13}:
\begin{align}
    \mathcal{U}_b \equiv \sum_{\{g_v\}} \prod_{t=\langle 1234 \rangle} e^{2 \pi i O_t \nu(1,g_1,g_2,g_3,g_4)} \ket{\{g_v\}}\bra{\{g_v\}},
\end{align}
where the product is over all tetrahedra in $M$, the vertices specifying the tetrahedron $\langle 1234 \rangle$ are ordered according to the branching structure, and $1$, in the argument of $\nu$, denotes the identity in $G$. To simplify the notation, we introduce an operator $\nuhat\boldsymbol{(} \langle 1234 \rangle \boldsymbol{)}$, for each tetrahedron $\langle 1234 \rangle$ in $M$:
\begin{align}
    \nuhat \boldsymbol{(} \langle 1234 \rangle \boldsymbol{)} \equiv \sum_{\{g_v\}}\nu(1,g_1,g_2,g_3,g_4)\ket{\{g_v\}}\bra{\{g_v\}}.
\end{align}
With this, $\mathcal{U}_b$ can be written more compactly as:
\begin{align}
    \mathcal{U}_b = \prod_{t} e^{2 \pi i O_t \nuhat(t)}.
\end{align}
The Hamiltonian for the group cohomology model is then:
\begin{align}
    H_b \equiv \mathcal{U}_b H^G \mathcal{U}_b^\dagger,
\end{align}
with the unique ground state:
\begin{equation} \label{groupcoho gs} \begin{split} 
    \ket{\Psi_b} &\equiv \mathcal{U}_b \ket{\Psi^G}\\
    &=\sum_{\{g_v\}} \prod_{t=\langle 1234 \rangle} e^{2 \pi i O_t \nu(1,g_1,g_2,g_3,g_4)} \ket{\{g_v\}}.
    \end{split}
\end{equation}

The Hamiltonian $H_b$ indeed describes a bSPT phase. This is because the Hamiltonian is both symmetric and has a unique short-range entangled (SRE) ground state. The symmetry of the Hamiltonian follows from the fact that $\mathcal{U}_b$ is symmetric, which can be shown using the property $\delta \nu = 0$. The ground state is SRE, since it can be prepared from a product state by the FDQC $\mathcal{U}_b$.\footnote{\unexpanded{We note that, although ${\mathcal{U}_b}$ is symmetric, this does not imply that the state $\ket{\Psi_b}$ in Eq.~\eqref{groupcoho gs} belongs to the trivial SPT phase. This is only the case if ${\mathcal{U}_b}$ can additionally be expressed as a FDQC composed of symmetric local unitaries.}} Furthermore, the group cohomology models exhibit the characteristic responses encoded by $\nu$, as can be checked by introducing symmetry defects or gauging the $G$ symmetry and analyzing the properties of the symmetry fluxes \cite{EN15,WL15}.

\subsection{Definition of supercohomology models} \label{sec: supercohomology models 2}

We now generalize the discussion to supercohomology models, which describe fSPT phases. We leave the explicit derivation of the models from a choice of supercohomology data $(\rho,\nu)$ to Sections~\ref{sec: bulk Z2f} and \ref{sec:bulkG}.  Similarly to the group cohomology models, the supercohomology models are prepared from a Hamiltonian in a trivial SPT phase by conjugation with a FDQC. Further, the FDQCs for $(\rho,\nu)$ and $(\rho',\nu')$ can be used to recover the stacking laws for supercohomology phases, discussed at the end of this section. 

\begin{figure}[t]
\centering
\includegraphics[width=0.20\textwidth, trim={850 500 1000 300},clip]{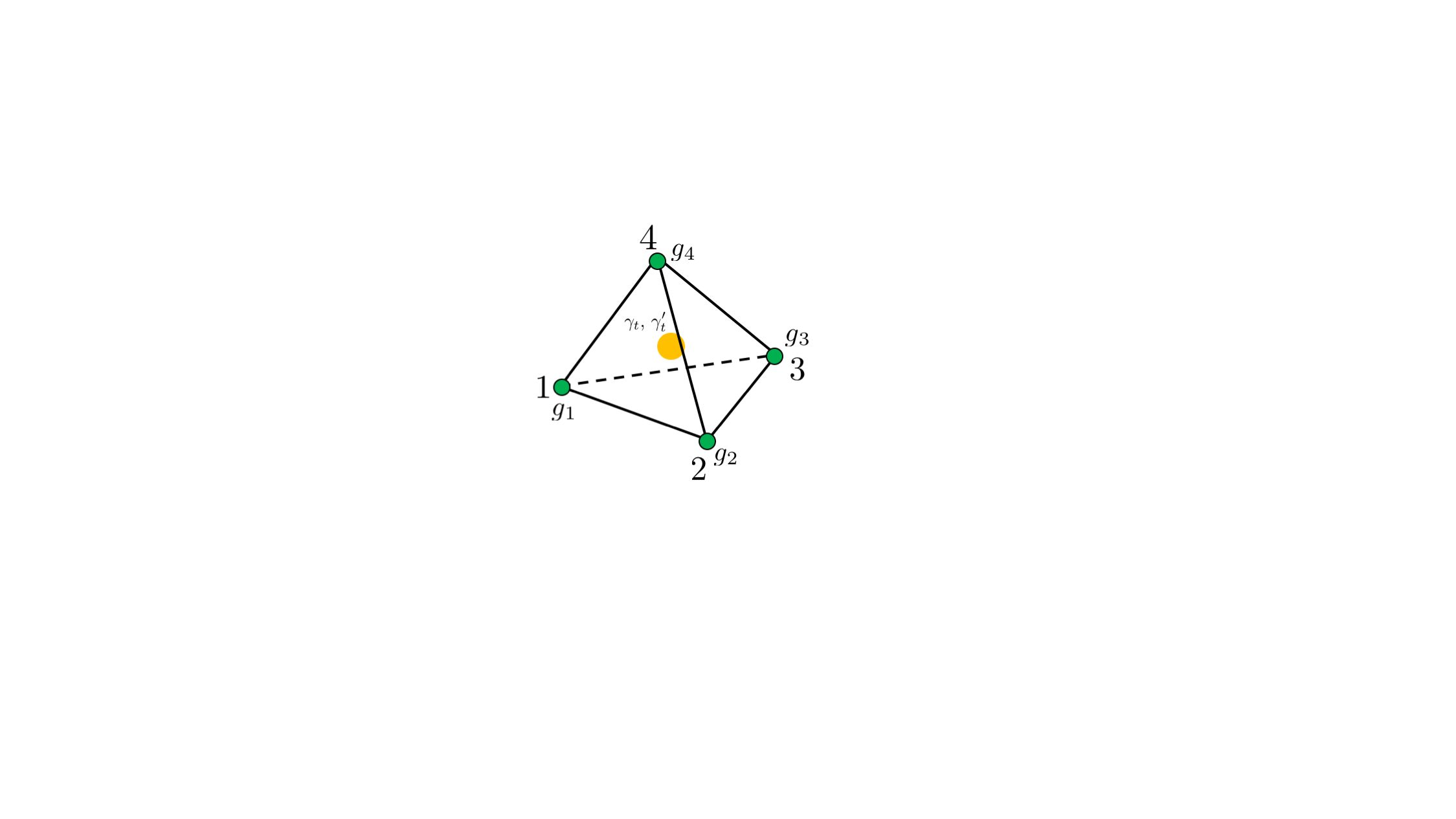}
\caption{Our model for a fSPT phase is defined on a Hilbert space with $G$ {d.o.f.} on the vertices of a triangulation of $M$ and a single spinless complex fermion at the center of each tetrahedron. }
\label{fig: Gai dof2}
\end{figure}

\begin{figure}[t]
\centering
\includegraphics[width=0.4\textwidth, trim={700 450 700 400},clip]{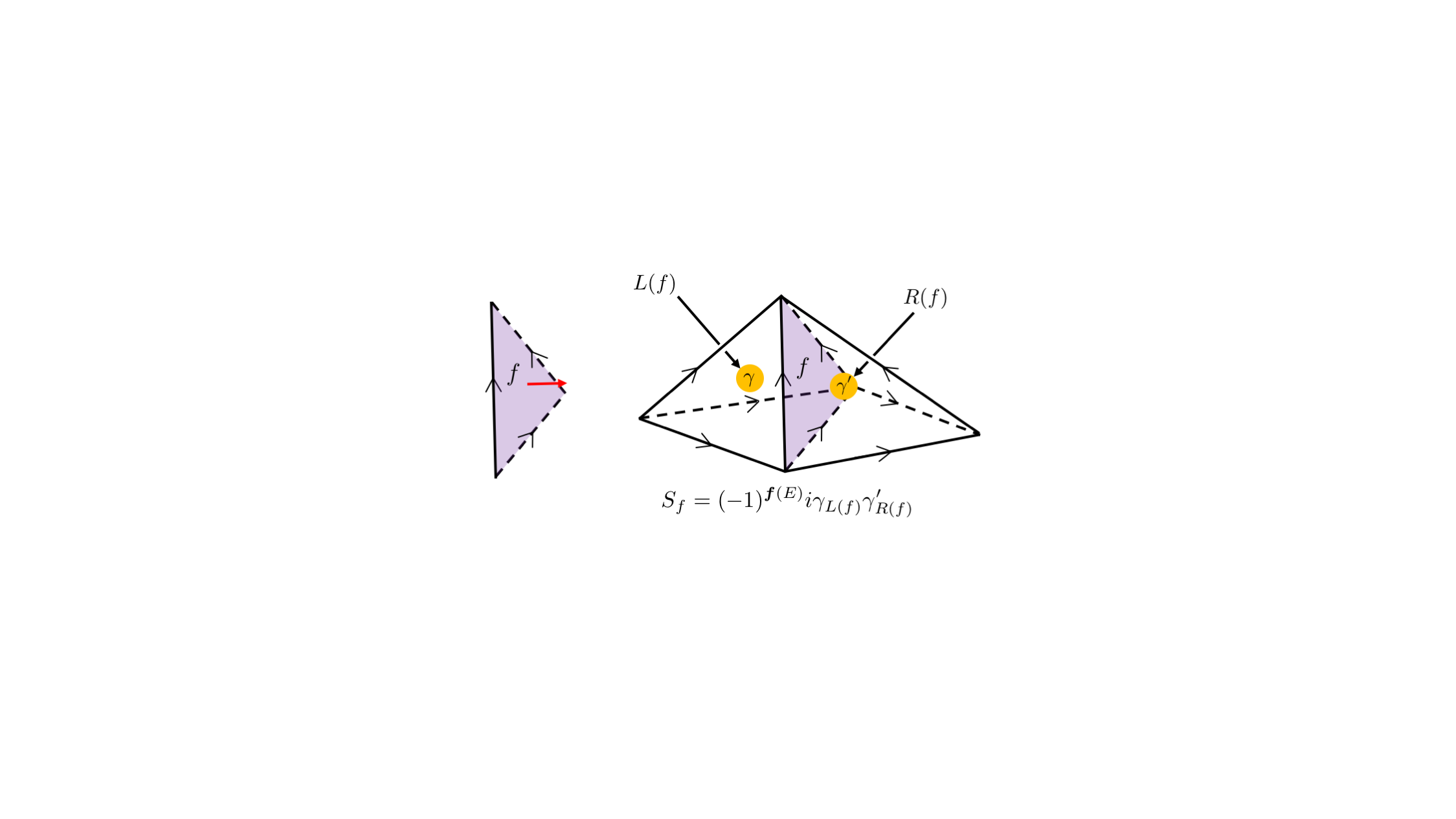}
\caption{The orientation of a face $f$ (red vector) is determined by the branching structure. The orientation of $f$ points out of the tetrahedron $L(f)$ and into the tetrahedron $R(f)$. The hopping operator $S_f$ acts with $\gamma$ on the complex fermion Hilbert space at $L(f)$ and $\gamma'$ on the site at $R(f)$.}
\label{fig:Sf def2}
\end{figure}

The supercohomology models are defined on a Hilbert space with $G$ {d.o.f.} at the vertices of $M$, as in the previous section, along with a fermionic d.o.f. at each tetrahedron.\footnote{We note that the manifold admits a spin structure. This is always true for orientable $3$-manifolds.}
Specifically, we place a single spinless complex fermion at the center of each tetrahedron and label the two Majorana operators at the tetrahedron $t$ by $\gamma_t$ and $\gamma'_t$ (see Fig.~\ref{fig: Gai dof2}). The fermion parity at $t$ is then given by:
\begin{align}
    P_t \equiv -i\gamma_t \gamma'_t,
\end{align}
and we also introduce a ``hopping'' operator $S_f$ that changes the fermion parity on either side of the face $f$: 
\begin{align} \label{hopping}
    S_f \equiv { (-1)^{\boldsymbol{f}(E)} }  i \gamma_{L(f)} \gamma'_{R(f)}.
\end{align}
Here, $L(f)$ and $R(f)$ are the tetrahedra neighboring $f$ such that the orientation of $f$ points {out of} the tetrahedron $L(f)$ and {into} the tetrahedron $R(f)$ (see Fig.~\ref{fig:Sf def2}). {$\boldsymbol{f}(E) \in \{0,1\}$ corresponds to a choice of spin-structure and is determined by the branching structure of the triangulation of $M$.} 
% {The $2$-chain $E \in C_2 (M,\ZZ_2)$ corresponds to a choice of spin-structure. The value $\boldsymbol{f} (E) = 0, 1$ depends on the branching structure of the triangulation $M$.} 
We refer to Appendix~\ref{sec: review of boson-fermion duality} and Ref.~\cite{C19-2} for the explicit form of $\boldsymbol{f}(E)$. 
% The hopping operators satisfy the commutation relations:
% \begin{align} \label{hopping commutation}
%     S_f S_{f'} = (-1)^{\delta_{R(f),R(f')}+\delta_{L(f),L(f')}} {S}_{f^\prime} {S}_f,
% \end{align}
% where $\delta_{R(f),R(f')}$ and $\delta_{L(f),L(f')}$ are Kronecker delta functions.
As before, the $G$ symmetry is represented with the regular representation, and here the global fermion parity symmetry is generated by $\prod_t P_t$.

The supercohomology models are built from a Hamiltonian belonging to a trivial fSPT phase  - namely, an atomic insulator with a decoupled $G$-paramagnet. The trivial fSPT Hamiltonian is explicitly:
\begin{align} \label{GAI H def2}
    H_\text{AI}^G \equiv -\sum_t P_t - \sum_v \mathcal{P}_v.
\end{align}
The ground state $\ket{\Psi_\text{AI}^G}$ of $H_\text{AI}^G$ is a product state with zero fermion occupancy at each tetrahedron and a symmetric state at each vertex.

Given a choice of supercohomology data $(\rho,\nu)$, we prepare the supercohomology model from $H_\text{AI}^G$ by conjugation with the FDQC $\Uf$:
% \begin{multline} \label{eq: uf def2}
%   \Uf \equiv \\ \prod_t e^{2 \pi i O_t \nuhat(t)} \xi_{\bar{\rho}}(M){\prod_{f}} S_f^{\rhohat(f)} \prod_{t=\langle 1234 \rangle} P_t^{\rhohat\boldsymbol{(} \langle 123 \rangle \boldsymbol{)}+\rhohat\boldsymbol{(} \langle 134 \rangle \boldsymbol{)}}. 
% \end{multline}
\begin{align} \label{eq: uf def2}
  \Uf \equiv \prod_t e^{2 \pi i O_t \nuhat(t)} \xi_{\bar{\rho}}(M){\prod_{f}} S_f^{\rhohat(f)} \prod_{t} P_t^{\int \rhohat \cup_2 \bt}. 
\end{align}
Let us unpack the notation used in the definition of $\Uf$. First of all, the $\nuhat(t)$ term is analogous to the FDQC $\mathcal{U}_b$ in Section~\ref{sec: group cohomology models}, tensored with the identity on the fermionic d.o.f..  Second, the product over hopping operators in Eq.~\eqref{eq: uf def2} depends on a choice of ordering for the faces $f \in M$, since not all hopping operators commute. However, $\xi_{\bar{\rho}}(M)$ is an order dependent sign that compensates for the choice of ordering. Therefore, in the end, the FDQC is independent of the choice of ordering for the faces in $M$. We give the explicit form of $\xi_{\bar{\rho}}(M)$ in Section~\ref{sec: supercohomology model}.\footnote{\unexpanded{We note that while $\xi_{\bar{\rho}}(M)$ depends on a global ordering of the faces in $M$, it can nonetheless be implemented by a FDQC. This can be seen from the derivation of $\xi_{\bar{\rho}}(M)$ in Appendix~\ref{app: fermionization shadow circuit}.}}  
$\rhohat(f)$ is the $\rho$-dependent operator:
\begin{align}
    \rhohat\boldsymbol{(}\langle 123 \rangle \boldsymbol{)} \equiv \sum_{\{g_v\}} \rho(1,g_1,g_2,g_3) \ket{\{g_v\}}\bra{\{g_v\}},
\end{align}
for an arbitrary face $\langle 123 \rangle$ and implicitly tensored with the identity on the fermionic d.o.f.. Finally, for a tetrahedron $t=\langle 1234\rangle$, $\int \rhohat \cup_2 \bt$ is shorthand for:
\begin{align}
    \int \rhohat \cup_2 \bt = \rhohat\boldsymbol{(} \langle 123 \rangle \boldsymbol{)}+\rhohat\boldsymbol{(} \langle 134 \rangle \boldsymbol{)}.
\end{align}
The exactly-solvable fermionic Hamiltonian produced by conjugating $H_\text{AI}^G$ by $\Uf$ is thus:
\begin{align} \label{eq: fSPT H2}
    H_f \equiv \Uf H_\text{AI}^G \Uf^\dagger,
\end{align}
which has the unique ground state $\ket{\Psi_f}$:
\begin{align} \label{supercohomology gs}
    \ket{\Psi_f} \equiv \Uf \ket{\Psi_\text{AI}^G}.
\end{align}

$H_f$ describes a system in a $G \times \ZZ_2^f$ fermionic SPT phase, because (i) $H_f$ is symmetric and (ii) it has a unique, SRE ground state [Eq.~\eqref{supercohomology gs}]. The Hamiltonian in Eq.~\eqref{eq: fSPT H2} is symmetric, since both $H_\text{AI}^G$ and $\Uf$ are invariant under the symmetry -- we argue that $\Uf$ is symmetric in Section~\ref{sec: supercohomology model}.\footnote{Although $\Uf$ is symmetric, it cannot be decomposed into a FDQC comprised of symmetric local unitaries.}  The ground state is unique and SRE, because $H_f$ is unitarily equivalent to a trivial fSPT Hamiltonian with a unique ground state and the unitary is a FDQC.

Most importantly, $H_f$ belongs to the fSPT phase characterized by the corresponding supercohomology data $(\rho,\nu)$. We show this in Section~\ref{sec: supercohomology model}, by gauging the fermion parity symmetry of $H_f$. This results in a $G$-symmetry-enriched $\ZZ_2$ gauge theory, where the $G$ symmetry fractionalizes on the fermion parity flux loops, as determined by $\rho$. The appropriate responses to $G$-symmetry defects follow from the bosonic, group cohomology case. 

We can gain intuition for the fSPT Hamiltonian by inserting the right hand side of Eq~\eqref{GAI H def2} into the expression for $H_f$:
\begin{align}
    H_f= -\sum_t \left( \Uf P_t \Uf^\dagger \right) -\sum_v \left( \Uf \mathcal{P}_v \Uf^\dagger \right).
\end{align}
By commuting the hopping operators of $\Uf$ past the parity operator $P_t$ and using that $\delta \rho =0$, the tetrahedron terms become: 
\begin{align} \label{parity term2}
    -\sum_t \left( \Uf P_t \Uf^\dagger \right)  = -\sum_{t=\langle 1234 \rangle} (-1)^{{\rho(g_1,g_2,g_3,g_4)}}P_t.
\end{align}
In the ground state, the fermion occupancy depends on the $\{ g_v \}$-configuration. For a $\{g_v\}$-configuration $\ket{\{g_v\}}$, it is energetically preferable for the fermion occupancy at the tetrahedron $\langle 1234 \rangle$ to be equal to $\rho(g_1,g_2,g_3,g_4)$. In this way, complex fermions are bound to junctions of symmetry domains. The vertex terms of $H_f$, on the other hand:
\begin{align}
    -\sum_v \left( \Uf \mathcal{P}_v \Uf^\dagger \right),
\end{align}
are more difficult to compute, in general. Heuristically, they fluctuate the $G$ {d.o.f.} and create, move, and annihilate fermions without affecting the tetrahedron terms in Eq.~\eqref{parity term2}. The ground state is thus a weighted superposition of $\{g_v\}$-configurations with the fermion occupancy at each tetrahedron $\langle 1234 \rangle$ equal to $\rho(g_1,g_2,g_3,g_4)$. This is in agreement with the fixed point wave functions in Ref.~\cite{WG18}.

% We can gain intuition for the fSPT ground state $\ket{\psi_f}$ by considering applying the terms in $\Uf$ to $\ket{\psi_\text{AI}^G}$ sequentially. First, the product over parity operators in Eq.~\eqref{eq: uf def2} leaves $\ket{\psi_\text{AI}^G}$ invariant, since $\ket{\psi_\text{AI}^G}$ has zero fermion occupancy at every tetrahedron. Although it is inessential for the preparation of the ground state, this term is important for ensuring that $\Uf$ is symmetric. Next, the product of hopping operators in $\Uf$ creates physical fermions according to the $G$-configurations. It can be checked using the group cocycle relation or each configuration state $\ket{\{g_v\}}$, the hopping operators create nonzero fermion occupancy at tetrahedra $\langle 1234 \rangle$ for which $\rho(g_1,g_2,g_3,g_4)=1$. Finally, similar to bosonic SPT states within the group cohomology classification \cite{CGLW13}, the $\nuhat$ term assigns a $G$-configuration dependent phase to each tetrahedron.

\begin{figure*}[!t]
\centering
\includegraphics[width=.8\textwidth, trim={0 405 0 405},clip]{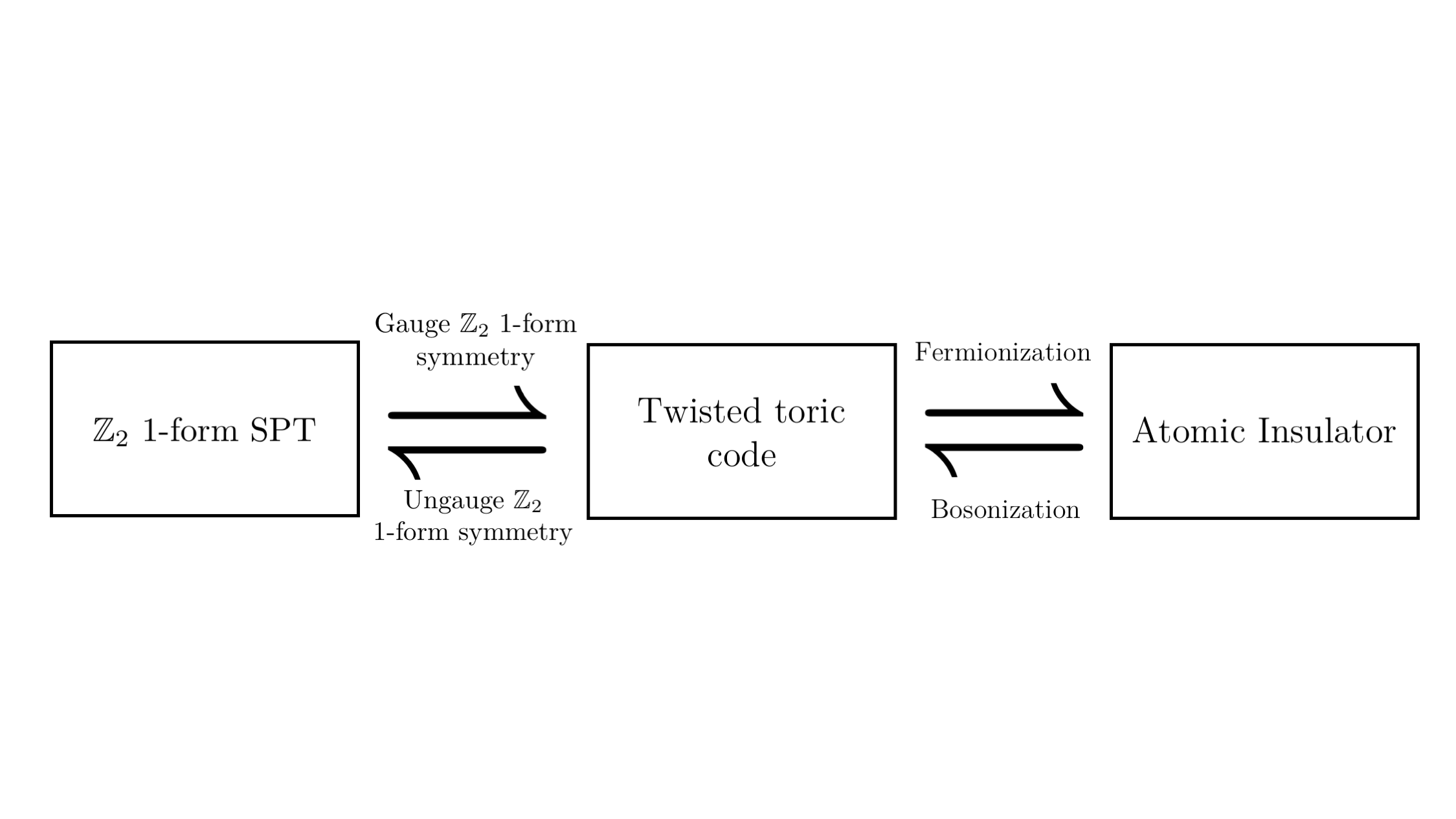}
\caption{In the case of $G_f=\ZZ_2^f$, the construction of a fermionic model starts with a model for a certain $\ZZ_2$ $1$-form SPT phase. We then gauge the $\ZZ_2$ $1$-form symmetry to obtain a twisted toric code. Lastly, we fermionize the twisted toric code, and the result is a model for an atomic insulator.}
\label{fig:z2f_outline}
\end{figure*}

\vspace{1.5mm}
\noindent \begin{center}\emph{Stacking rules for supercohomology phases:}\end{center}
\vspace{1.5mm}

% We now deduce the stacking rules for supercohomology phases from the algebraic properties of the FDQCs $\Uf$, presented above. We recall that two states can be stacked by taking the tensor product. Given two $G$-SPT states, this induces an operation $\boxtimes$ at the level of SPT phases. In Ref.~\cite{EF19}, it was argued that, if the two $G$-SPT states belong to the same Hilbert space (with the same representation of the symmetry), then the stacking of the states 
% In other words, for two $G$-SPT states 

Having defined the models, we can now deduce the stacking rules for supercohomology phases. We recall that two states can be stacked by taking the tensor product. Given two $G$-SPT states $\ket{\Psi_{\text{SPT}_1}}$ and $\ket{\Psi_{\text{SPT}_2}}$, the stacked state $\ket{\Psi_{\text{SPT}_1}}\otimes \ket{\Psi_{\text{SPT}_2}}$ also belongs to a $G$-SPT phase. Thus, the stacking of $G$-SPT states induces an operation $\boxtimes$ at the level of the SPT phases. 

In Ref.~\cite{EF19}, it was argued that the stacking operation $\boxtimes$ on SPT phases can be determined from the composition of FDQCs. To state the result from Ref.~\cite{EF19}, we define $\mathcal{U}_{\text{SPT}_1}$ and $\mathcal{U}_{\text{SPT}_2}$ to be symmetric FDQCs that prepare the $G$-SPT states $\ket{\Psi_{\text{SPT}_1}}$ and $\ket{\Psi_{\text{SPT}_2}}$, respectively, from a symmetric product state. According to Ref.~\cite{EF19}, if $\ket{\Psi_{\text{SPT}_1}}$ and $\ket{\Psi_{\text{SPT}_2}}$ belong to the same Hilbert space, then the composition of the FDQCs $\mathcal{U}_{\text{SPT}_1}$ and $\mathcal{U}_{\text{SPT}_2}$ prepares a state belonging to the same $G$-SPT phase as $\ket{\Psi_{\text{SPT}_1}}\otimes \ket{\Psi_{\text{SPT}_2}}$.

With this, we determine the group law under stacking for $(3+1)$D supercohomology phases by composing the FDQCs $\Uf$ defined in Eq.~\eqref{eq: uf def2}. Given two sets of supercohomology data $(\rho,\nu)$ and $(\rho',\nu')$ both characterizing fSPT phases with a $G \times \ZZ_2^f$ symmetry, we consider stacking the ground states of the corresponding supercohomology models, denoted by $\ket{\Psi_f^{\rho\nu}}$ and $\ket{\Psi_f^{\rho'\nu'}}$, respectively. The result from Ref.~\cite{EF19} tells us that the stacked state $\ket{\Psi_f^{\rho\nu}} \otimes \ket{\Psi_f^{\rho'\nu'}}$ belongs to the same phase as the state prepared by applying $\Uf^{\rho'\nu'}\Uf^{\rho\nu}$ to a symmetric product state. Here, $\Uf^{\rho\nu}$ and $\Uf^{\rho'\nu'}$ are the FDQCs from Eq.~\eqref{eq: uf def2} that prepare $\ket{\Psi_f^{\rho\nu}}$ and $\ket{\Psi_f^{\rho'\nu'}}$ from an unentangled symmetric state, respectively. In Appendix \ref{app:composition laws}, we show that the composition $\Uf^{\rho'\nu'}\Uf^{\rho\nu}$ is equivalent to a FDQC $\Uf^{\rho''\nu''}$ corresponding to a set of supercohomology data $(\rho'',\nu'')$:
\begin{align} \label{stacking by composition}
    (\rho'',\nu'')  = (\rho + \rho',\nu + \nu' + \frac{1}{2} \rho \cup_2 \rho').
\end{align}
Therefore, at the level of the supercohomology data, the stacking operation $\boxtimes$ is:
\begin{align}
    (\rho,\nu) \boxtimes (\rho',\nu') = (\rho + \rho',\nu + \nu' + \frac{1}{2} \rho \cup_2 \rho'),
\end{align}
in agreement with Ref.~\cite{GW14}.

\section{Bulk construction: $G_f=\ZZ_2^f$} \label{sec: bulk Z2f}

We begin by illustrating our construction of exactly-solvable models for fSPT phases in the simplest possible case -- for fSPT phases protected by only fermion parity symmetry $\ZZ_2^f$. While the resulting fSPT model is trivial (an atomic insulator), we nonetheless find this example instructive in demonstrating the general strategy. Moreover, we use this as an opportunity to introduce notation used throughout the paper.

To start, we describe a model for a certain bosonic SPT phase protected by a $\ZZ_2$ $1$-form symmetry in (3+1)D. The special property of this bosonic SPT is that, upon gauging the $\ZZ_2$ $1$-form symmetry, we obtain a $\ZZ_2$ gauge theory with an emergent fermion.  We refer to this $\ZZ_2$ gauge theory as the \textit{twisted} toric code. In the final step, we employ the fermionization duality of Ref.~\cite{CKR18} to map the twisted toric code to a model with a fundamental fermion. The construction is shown schematically in Fig.~\ref{fig:z2f_outline}, for the case of $G_f=\ZZ_2^f$. 

\subsection{$1$-form SPT and notation} \label{sec: 1formspt}

SPT phases protected by higher form symmetries, including $1$-form symmetries, were first introduced in Ref.~\cite{KT15}.  Subsequently, fixed point Hamiltonians for $1$-form SPT phases were described in detail in Refs.~\cite{Y16} and \cite{TW19}. We note that the Hamiltonian discussed in this section agrees with the model in Ref.~\cite{TW19} and is closely related to the $\ZZ_2 \times \ZZ_2$ $1$-form SPT model of Ref.~\cite{Y16}.\footnote{Specifically, our model is equivalent to the model in Ref.~\cite{Y16} on a four-colorable triangulation and upon restricting to the diagonal $\ZZ_2$ of the $\ZZ_2 \times \ZZ_2$ symmetry.}

% Higher form SPT phases were first introduced in Ref.~ and 

% Lattice models for higher form SPT phases were first studied in  Ref.~\cite{Y16}. More recently, fixed point models for $1$-form SPT phases have been described in detail in Refs.~\cite{TW19} and \cite{Delcamp19}. We note that the model discussed in this text agrees with the construction in Refs.~\cite{TW19} and \cite{Delcamp19} and is closely related to the $\ZZ_2 \times \ZZ_2$ $1$-form SPT model introduced in Ref.~\cite{Y16}.\footnote{Specifically, our model is equivalent to the model in Ref.~\cite{Y16} on a four-colorable triangulation and upon restricting to the diagonal $\ZZ_2$ of the $\ZZ_2 \times \ZZ_2$ symmetry.}

\begin{figure}[t]
\centering
\includegraphics[width=0.35\textwidth, trim={800 440 700 360},clip]{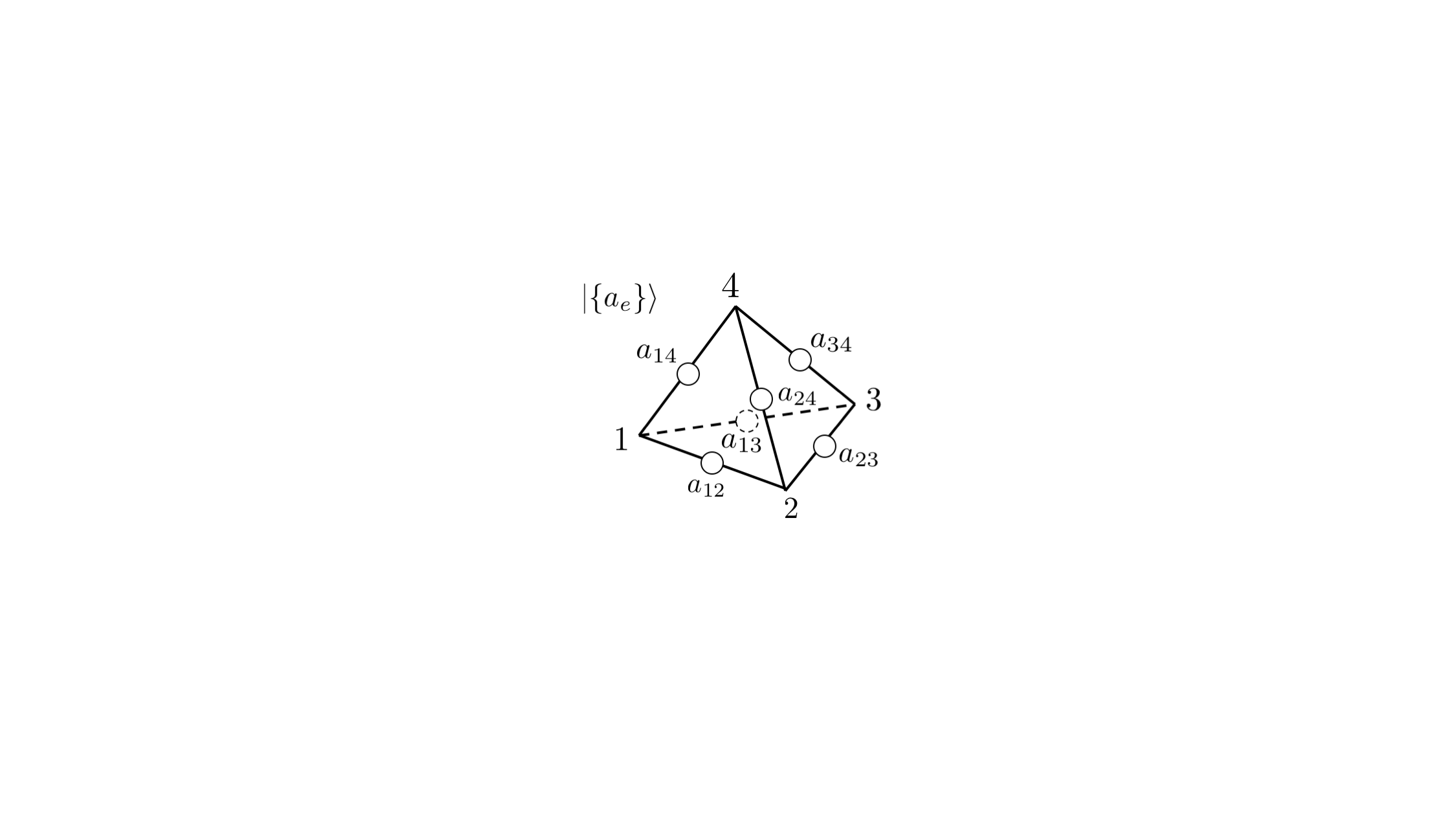}
\caption{The $1$-form SPT model is defined on a triangulation where each edge hosts a $\ZZ_2$ {d.o.f.} (represented by a circle). A state in the configuration basis is given by a value $a_e \in \{0,1\}$ chosen for each edge $e$. We have suppressed the branching structure for clarity.}
\label{fig: 1formdof}
\end{figure}

Our model for a nontrivial $\ZZ_2$ $1$-form SPT phase can be defined on an arbitrary triangulation of an oriented closed $3$-manifold $M$ equipped with a branching structure, as described in Section~\ref{sec: group cohomology models}. We define a Hilbert space on $M$ using the triangulation of the manifold -- at each edge of the triangulation, we place a single $\ZZ_2$ degree of freedom. Correspondingly, a basis for the Hilbert space at edge $e$ is given by states $\ket{a_e}$ with $a_e$ valued in $\{0,1\}$. The Pauli Z and Pauli X operators at each $e$ act as:
\begin{align}
    Z_e \ket{a_e}= (-1)^{a_e} \ket{a_e}, \quad X_e \ket{a_e}=\ket{a_e+1},
\end{align}
where addition is taken modulo $2$.
A basis for the total Hilbert space consists of states $\ket{\{a_e\}}$ labeled by configurations $\{a_e\}$ (Fig.~\ref{fig: 1formdof}). Here, the state $\ket{\{a_e\}}$ denotes a product state with the {d.o.f.} at edge $e$ in the state $\ket{a_e}$. 

% We define our model for a nontrivial $\ZZ_2$ $1$-form SPT phase on the following Hilbert space.  We consider an arbitrary triangulation of an oriented $3$-manifold $M$ equipped with a branching structure. We recall that a branching structure is an assignment of an orientation to each edge in such a way that there are no cycles around any plaquette. As such, a branching structure yields an ordering of the vertices of each tetrahedron as well as an orientation of the tetrahedron relative to the orientation of $M$ (see Fig.~\ref{fig: branchingstructure}). 
% At each edge of the triangulation, we place a single $\ZZ_2$ {d.o.f.}. Correspondingly, a basis for the Hilbert space at edge $e$ is given by states $\ket{a_e}$ with $a_e$ valued in $\{0,1\}$. The Pauli Z and Pauli X operators at $e$ then act as:
% \begin{align}
%     Z_e \ket{a_e}= (-1)^{a_e} \ket{a_e}, \quad X_e \ket{a_e}=\ket{a_e+1},
% \end{align}
% where addition is taken modulo $2$.
% A basis for the total Hilbert space $\mathcal{H}_1$ consists of states $\ket{\{a_e\}}$ labeled by configurations $\{a_e\}$ (Fig.~\ref{fig: 1formdof}). Here, the state $\ket{\{a_e\}}$ denotes a product state with the {d.o.f.} at edge $e$ in the state $\ket{a_e}$. 
% % We sometimes find it convenient to label the state as...introduce cohomology notation 

\begin{figure}[!t]
\centering
\includegraphics[width=0.27\textwidth, trim={900 440 880 500},clip]{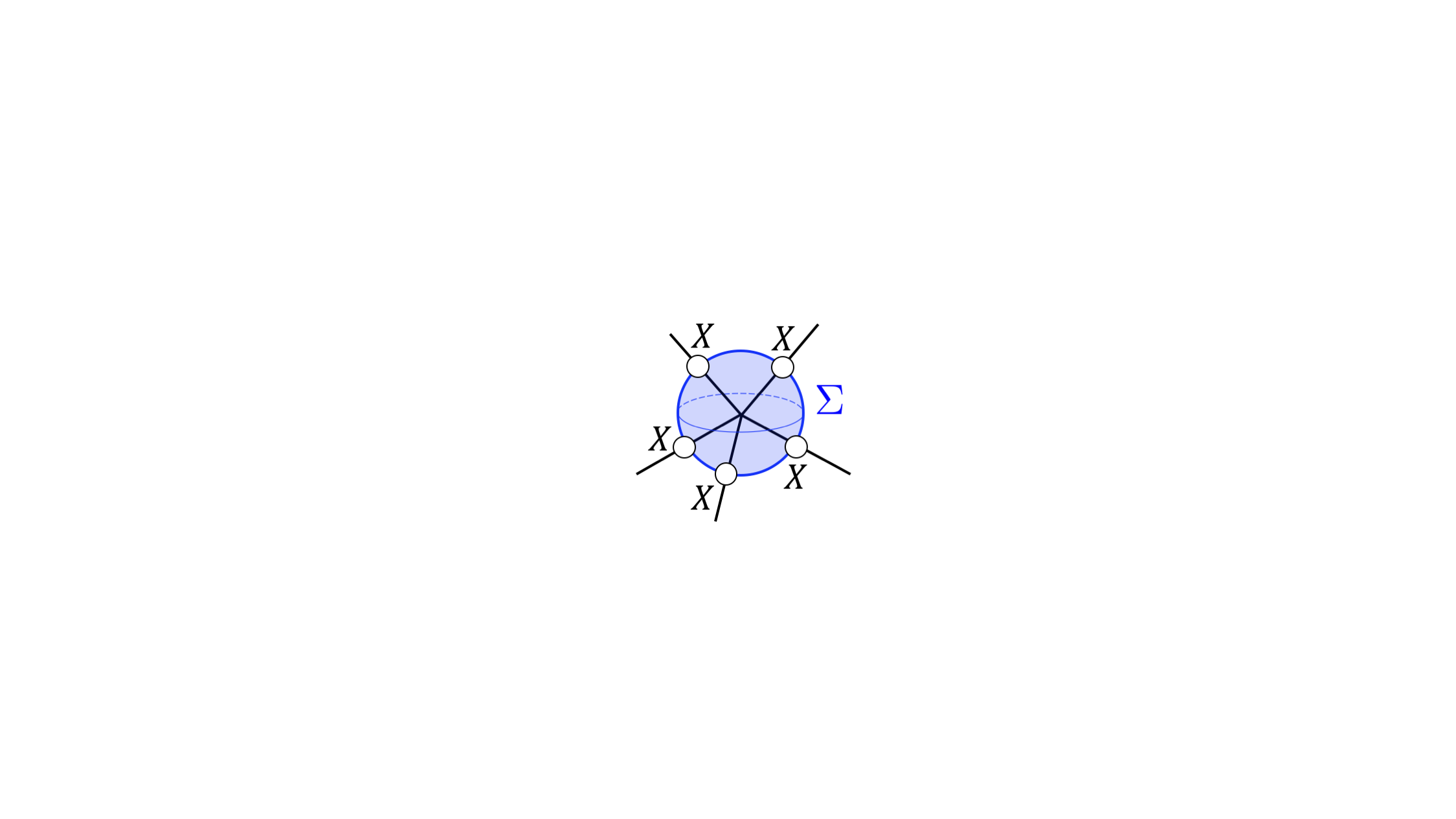}
\caption{The $1$-form symmetry operators act on closed surfaces in the dual lattice. Pauli X operators are applied to each edge intersected by the surface. The figure shows a surface $\Sigma$ (blue) that encloses a single vertex.}
\label{fig: 1formsym}
\end{figure}

The $\ZZ_2$ $1$-form symmetry acts on closed codimension-$1$ submanifolds of the dual lattice. In particular, we represent the symmetry action on a closed surface $\Sigma$ of the dual lattice with the operator:
% \textcolor{orange}{It might not be unreasonable to instead use $X_\Sigma$.}
\begin{align} \label{1formgenerator}
    A_\Sigma \equiv \prod_{e \perp \Sigma}X_e,
\end{align}
where the product is over edges intersected by the surface $\Sigma$ (see Fig.~\ref{fig: 1formsym}).

%We note that unlike in continuum descriptions, the representation is unfaithful. That is, the symmetry operators are not topological -- making a deformation to the surface $\Sigma$ on the dual lattice corresponds to a different symmetry operator. \Yuan{I don't quite understand this paragraph. I thought the $1$-form symmetry here is topological, i.e., the deformation of $\Sigma$ not affecting $A_\Sigma$.} \textcolor{orange}{What I mean to say here is that on a lattice it is not topological, deforming $\Sigma$ gives a different operator: $A_\Sigma \neq A_{\Sigma'}$. Maybe this is a matter of interpretation. In any event, this doesn't really need to be stated or discussed here. We can remove it. I can work on removing it later. Wenjie and Xiao-Gang refer to the lattice representation as a higher symmetry, and the topological, continuum representation as the higher-form symmetry. I think this is confusing.}

We construct our model for the nontrivial SPT phase starting with a Hamiltonian for a $\ZZ_2$ $1$-form paramagnet.
The Hamiltonian for the $1$-form paramagnet is given by:
\begin{align} \label{1formparamagnet}
    H_0 = - \sum_e X_e. 
\end{align}
$\onepara$ is certainly symmetric, as it commutes with $A_\Sigma$ for every surface $\Sigma$ of the dual lattice. Further, the unique ground state of $\onepara$ is a product state with the $+1$ eigenstate of $X_e$ at each edge $e$. This state can be expressed in the configuration basis as:
\begin{align}\label{Psi0}
    \ket{\onetriv} \equiv \sum_{\{a_e\}}\ket{\{a_e\}},
\end{align}
with the sum over all configurations $\{a_e\}$. 

% The model for the nontrivial $\ZZ_2$ $1$-form SPT phase is constructed from the $1$-from paramagnet in Eq.~\eqref{1formparamagnet} by conjugation with the diagonal (in the configuration basis) unitary:
% \begin{align} \label{1formcircuit}
%     {\Uspt} \equiv \prod_{t=\langle 1234 \rangle} CZ_{(\langle{12}\rangle,\langle{23}\rangle)}CZ_{(\langle{12}\rangle,\langle{34}\rangle)} CZ_{(\langle{12}\rangle,\langle{24}\rangle)}.
% \end{align}
% In equation Eq.~\eqref{1formcircuit}, the product is over tetrahedra $\langle 1234 \rangle$, and the ordering of the vertices is given by the branching structure. $CZ_{(e',e'')}$ is a control Z operator, defined by:
% \begin{align}
%     CZ_{(e',e'')} \ket{\{a_e\}} = (-1)^{a_{e'}a_{e''}}\ket{\{a_e\}}.
% \end{align}
% Explicitly, the Hamiltonian of the nontrivial $\ZZ_2$ $1$-form SPT model is:
% \begin{align}
%     {\Uspt} \onepara {\Uspt^\dagger},
% \end{align}
% and the ground state is:
% \begin{align}
%     {\Uspt} \ket{\onetriv} &= \sum_{\{a_e\}} {\Uspt} \ket{\{a_e\}} \\ \nonumber
%     &= \sum_{\{a_e\}} (-1)^{a_{12}(a_{23}+a_{34}+a_{24})} \ket{\{a_e\}}.
% \end{align}

Now, our model for the nontrivial $\ZZ_2$ $1$-form SPT phase is built from the $1$-form paramagnet in Eq.~\eqref{1formparamagnet} by conjugation with the FDQC:
\begin{align} \label{1formcircuit}
    {\Uspt} \equiv \sum_{\{a_e\}} \prod_{t=\langle 1234 \rangle} (-1)^{a_{12}(a_{23}+a_{34}+a_{24})} \ket{\{a_e\}}\bra{\{a_e\}}.
\end{align}
% In the formula for ${\Uspt}$, the product is over tetrahedra $\langle 1234 \rangle$ with the vertices ordered according to the branching structure. 
Specifically, the Hamiltonian of the nontrivial $\ZZ_2$ $1$-form SPT model is:
\begin{align} \label{1formsptH}
    H_{1} &\equiv {\Uspt} \onepara {\Uspt^\dagger} =-\sum_e {\Uspt} X_e {\Uspt^\dagger}.
\end{align}
Indeed, $H_1$ describes a \textit{nontrivial} $\ZZ_2$ $1$-form SPT phase. In the next section, we show this by gauging the $1$-form symmetry. The $1$-form paramagnet $\onepara$ is mapped to a $\ZZ_2$ gauge theory with an emergent boson - the usual $3$D toric code, while $H_1$ is mapped to a $\ZZ_2$ gauge theory with an emergent {fermion} - a ``twisted toric code''.

For now, we note that the model in Eq.~\eqref{1formsptH} is symmetric and exactly solvable. In Appendix \ref{App: 1form}, we show that ${\Uspt}$ is symmetric under the $\ZZ_2$ $1$-form symmetry. Consequently, $H_1$ is also symmetric. Furthermore, the model is exactly solvable, since by construction, the terms in $H_1$ are mutually commuting and unfrustrated. The unique ground state is then expressly:
\begin{align} \label{1formsptwf}
    \ket{\Psi_1} &\equiv {\Uspt} \ket{\onetriv} \\ \nonumber
    &= \sum_{\{a_e\}} \prod_{t=\langle 1234 \rangle} (-1)^{a_{12}(a_{23}+a_{34}+a_{24})} \ket{\{a_e\}}.
\end{align}
% We remark that, in analogy with the group cohomology models of Ref.~\cite{CGLW13}, the amplitude of each configuration is given by a product of phases assigned to each tetrahedron.
\begin{figure}[t]
\centering
\includegraphics[width=0.37\textwidth, trim={700 300 830 325},clip]{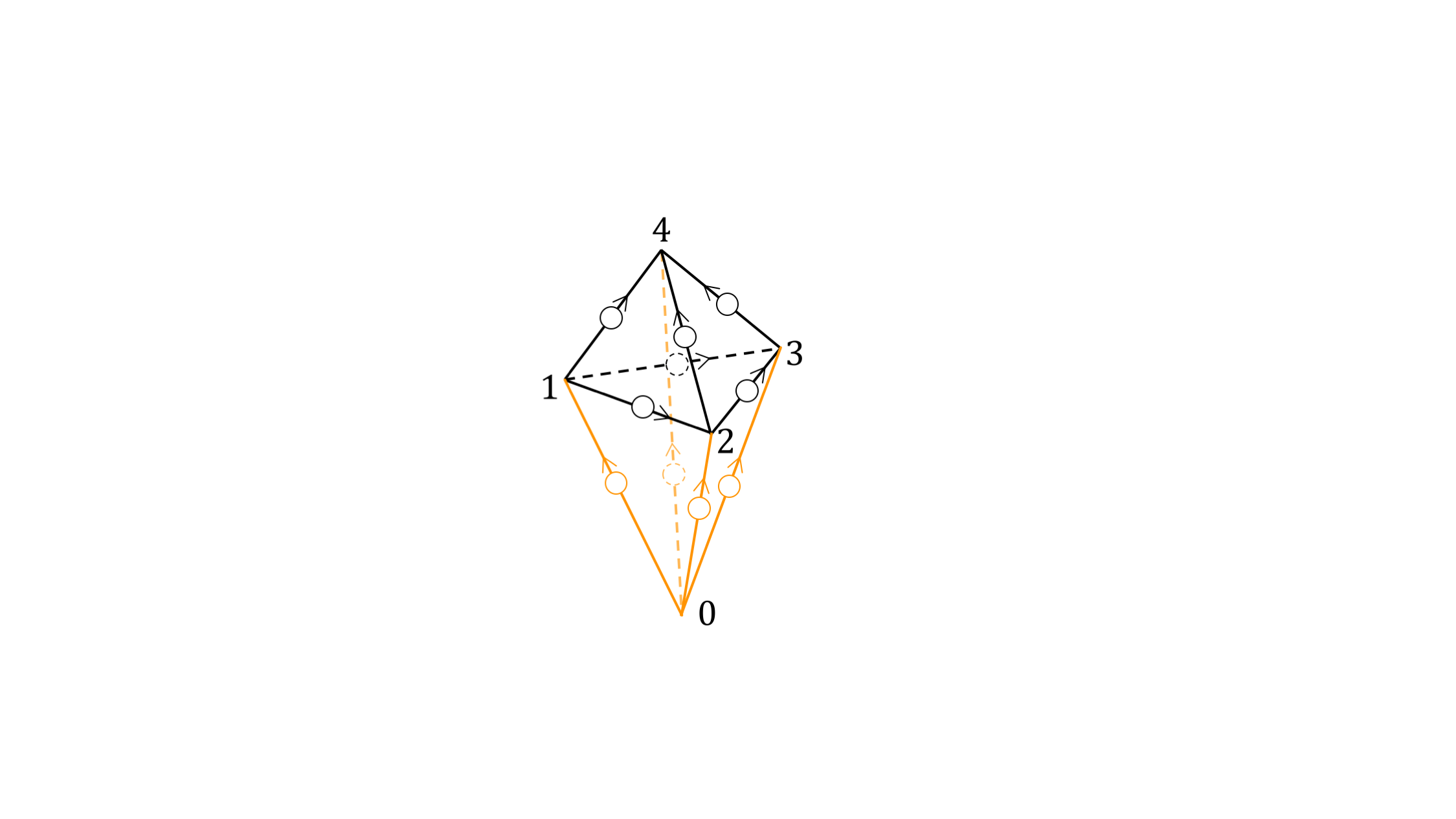}
\caption{The cone of $M$ is constructed by connecting the vertices of $M$  to an additional spacetime point labeled by $0$. A tetrahedron of the spatial manifold $M$ is shown in black. The time-like edges (orange) are oriented away from the $0$ vertex. The $1$-form SPT partition function is defined on $CM$ with $\ZZ_2$ d.o.f. on the time-like edges (orange circles). Due to the re-triangulation invariance of the partition function, $a_{01}$, $a_{02}$, and $a_{03}$ can be set to $0$.
}
\label{fig: CM}
\end{figure}

To further motivate this model, we recount the spacetime construction of the $\ZZ_2$ $1$-form SPT phase in Ref.~\cite{TW19}. We consider a partition function for the SPT phase on the cone of $M$, denoted $CM$, which is the $(3+1)$D spacetime formed by connecting a single spacetime vertex to each vertex of the closed manifold $M$ as shown in Fig. \ref{fig: CM}. This produces a manifold with a boundary equal to $M$. We refer to the edges connected to the additional spacetime point as ``time-like'' edges and extend the branching structure so that the time-like edges have an orientation pointing away from the additional spacetime vertex. Then, the partition function for the $1$-form SPT model is \cite{TW19}:\footnote{Using notation introduced later in the text, the expression:
\unexpanded{
\begin{align}\nonumber
    (a_{01}+a_{12}+a_{02})(a_{23}+a_{34}+a_{24}),
\end{align} 
can be written as $\delta \ba_e \cup \delta \ba_e \boldsymbol{(}\langle 01234 \rangle \boldsymbol{)}$. $\delta \ba_e \cup \delta \ba_e$ is a nontrivial $\ZZ_2$ $1$-form cocycle that has been pulled back to $CM$.}} 
\begin{align}
    \mathcal Z_1=\sum_{\{a_e\}}\prod_{\langle 01234 \rangle}(-1)^{(a_{01}+a_{12}+a_{02})(a_{23}+a_{34}+a_{24})},
\end{align} 
with the product over spacetime $4$-simplices. The amplitude for a fixed configuration $\{a_e\}$:
\begin{align}
        \Psi_1(\{a_e\}) \equiv \prod_{\langle 01234 \rangle}(-1)^{(a_{01}+a_{12}+a_{02})(a_{23}+a_{34}+a_{24})},
\end{align}
is topological in the sense that it is invariant under re-triangulations of the spacetime manifold. 
% Due to the topological nature of the partition function, it is invariant under re-triangulations of the spacetime lattice.
Through re-triangulations, it can be seen that the values of $a_e$ on the time-like edges do not affect the amplitude. Therefore, we may set their value to $0$, for simplicity. As a result, the amplitude for a configuration $\{a_e\}$ on $M$ reduces to:
\begin{align}
    \Psi_1(\{a_e\}) \equiv \prod_{t=\langle 1234 \rangle} (-1)^{a_{12}(a_{23}+a_{34}+a_{24})},
\end{align}
where the product is over tetrahedra $t=\langle 1234 \rangle$ on the boundary of $CM$. This gives the amplitude $\Psi_1(\{a_e\})$ for a wave function on $M$ with the configuration $\{a_e\}$, as in Eq.~\eqref{1formsptwf}. We  remark that this construction parallels the approach for building $0$-form SPT Hamiltonians in Ref.~\cite{CGLW13}.

% \subsubsection{$\ZZ_2$ cohomology notation}

\vspace{1.5mm}
\noindent \begin{center}\emph{$\ZZ_2$ cohomology on $M$}\end{center}
\vspace{1.5mm}

At this point, we find it convenient to introduce the language of $\ZZ_2$ cohomology on $M$. The cohomology notation allows for compact expressions and, in our opinion, more transparent calculations. Here, we only describe the necessary ingredients, and we leave a more thorough summary to Appendix~\ref{app: terminology}. In the process of introducing concepts from $\ZZ_2$ cohomology on $M$, we re-express the $1$-form SPT model using the corresponding notation. In particular, we aim to write $H_1$ in an explicit form.

To begin, we define a $p$-cochain as a linear, $\ZZ_2$-valued function of $p$-simplices in $M$. For example, we can consider the $1$-cochain $\be$ defined by:
\begin{align} \label{ecochain}
    \be(e') = \begin{cases} 
      1 & e'=e \\
      0 & \text{otherwise}. 
   \end{cases}
\end{align} 
In words, $\be$ evaluates to $1$ on the edge $e$ and $0$ on all other edges.
More general $1$-cochains can be formed from linear combinations of $1$-cochains of the form in Eq.~\eqref{ecochain}.
Specifically, for each configuration $\{a_e\}$ we can define a corresponding $1$-cochain $\ba_e$ as:
\begin{align} \label{boldadef}
    \ba_e \equiv \sum_{\be} a_e \be.
\end{align}
Evaluating $\ba_e$ on an edge $e'$ gives:
\begin{align}
    \ba_e(e') = \sum_{\be} a_e \be(e') = a_{e'}.
\end{align}

Given the correspondence between $1$-cochains and configurations $\{a_e\}$ in Eq.~\eqref{boldadef}, we can label a configuration state $\ket{\{a_e\}}$ by the $1$-cochain $\ba_e$:
\begin{align} \label{mapto1cochainstate}
    \ket{\{a_e\}} \ra \ket{\ba_e}.
\end{align}
In this notation, a Pauli Z operator at edge $e'$ acts on the state $\ket{\ba_e}$ as:
\begin{align}
    Z_{e'} \ket{\ba_e} = (-1)^{\ba_e(e')} \ket{\ba_e},
\end{align}
and an $X_{e'}$ operator acts as:
\begin{align}
    X_{e'} \ket{\ba_e} = \ket{\ba_e + \be'}.
\end{align}
Moreover, the action of the $1$-form symmetry operator $A_\Sigma$ on a configuration state is:
\begin{align} \label{1formsym on basis}
    A_\Sigma \ket{\ba_e} = \prod_{e \perp \Sigma} X_e \ket{\ba_e} = \ket{\ba_e + \sum_{e \perp \Sigma} \be} = \ket{\ba_e + \bsig},
\end{align}
where we have defined the $1$-cochain $\bsig$ as:
\begin{align} \label{Sigmacochain}
    \bsig \equiv \sum_{e \perp \Sigma} \be.
\end{align}

Next, we introduce the coboundary operator $\delta$, which is a linear map taking $p$-cochains to $(p+1)$-cochains. Specifically, it maps a $p$-cochain $\boldsymbol{c}$ to the $(p+1)$-cochain $\delta \boldsymbol{c}$ for which:
\begin{align} \label{coboundarydef}
    \delta \boldsymbol{c}(s) = \boldsymbol{c}(\partial s),
\end{align}
where $s$ is any $(p+1)$-simplex and $\partial s$ denotes a formal sum of $p$-simplices in the boundary of $s$. Simply put, the $(p+1)$-cochain $\delta \boldsymbol{c}$ is evaluated on a $(p+1)$-simplex $s$ by evaluating $\boldsymbol{c}$ on the boundary components of $s$.
The coboundary of $\ba_e$, for example, is a $2$-cochain satisfying:
\begin{align}
    \delta \ba_e \boldsymbol{(}\langle 123 \rangle\boldsymbol{)} &= \ba_e\boldsymbol{(}\langle {12} \rangle + \langle {23} \rangle + \langle {13}\rangle \boldsymbol{)} \\ \nonumber
     &= \ba_e\boldsymbol{(}\langle {12} \rangle\boldsymbol{)} + \ba_e\boldsymbol{(}\langle {23} \rangle\boldsymbol{)} + \ba_e\boldsymbol{(}\langle {13}\rangle \boldsymbol{)} \\ \nonumber
    &=a_{12}+a_{23}+a_{13},
\end{align}
for a face $\langle 123 \rangle$.

If the coboundary of a $p$-cochain is $0$, we call the $p$-cochain closed. The $1$-cochain $\bsig$ in Eq.~\eqref{Sigmacochain} is closed, i.e.:
\begin{align}
    \delta \bsig = 0.
\end{align}
This is because $\Sigma$ is a closed surface of the dual lattice. As such, for any face $f$, the boundary of $f$ contains an even number of edges intersected by $\Sigma$. 

Further, we define the cup product $\cup$. The cup product maps a $p$-cochain $\boldsymbol{c}$ and a $q$-cochain $\boldsymbol{d}$ to a $(p+q)$-cochain $\boldsymbol{c}\cup\boldsymbol{d}$. Specifically, $\boldsymbol{c} \cup \boldsymbol{d}$ evaluated on a $(p+q)$-simplex ${\langle 0, \ldots , p+q \rangle }$ is:
\begin{align}
    \boldsymbol{c} \cup \boldsymbol{d} \boldsymbol{(}\langle 0, \ldots, p+q \rangle\boldsymbol{)} = \boldsymbol{c} \boldsymbol{(}\langle 0, \ldots, p \rangle \boldsymbol{)} \boldsymbol{d}\boldsymbol{(}\langle p,  \ldots, p+q\rangle\boldsymbol{)}.
\end{align}
$\boldsymbol{c}$ is evaluated on the $p$-simplex formed by the first $p+1$ vertices, while $\boldsymbol{d}$ is evaluated on $q$-simplex formed by the last $q+1$ vertices. 

A suggestive example of the cup product comes from considering $\ba_e \cup \delta \ba_e$. $\ba_e \cup \delta \ba_e$ evaluated on a tetrahedron $\langle 1234 \rangle$ gives:
\begin{align}
   \ba_e \cup \delta \ba_e \boldsymbol{(} \langle 1234 \rangle \boldsymbol{)} &= \ba_e \boldsymbol{(} \langle 12 \rangle \boldsymbol{)} \delta \ba_e\boldsymbol{(}\langle 234 \rangle \boldsymbol{)} \\ \nonumber
   &= a_{12}(a_{23}+a_{34}+a_{24}).
\end{align}
Referring to Eq.~\eqref{1formcircuit}, we see that the FDQC ${\Uspt}$ can be written as:
\begin{align}
    {\Uspt} &= \sum_{\ba_e} \prod_{t} (-1)^{\ba_e \cup \delta \ba_e (t)} \ket{\ba_e}\bra{\ba_e},
\end{align}
with the sum over all $1$-cochains. 

To simplify the notation, we use the shorthand:
\begin{align}
    \int_{N}\boldsymbol{c} \equiv \sum_{s \in N} \boldsymbol{c}(s),
\end{align}
where $\boldsymbol{c}$ is a $p$-cochain, $N$ is a $p$-dimensional manifold, and the sum is over $p$-simplices $s$ in $N$. Throughout the text, unless specified otherwise, it should be assumed that the integral is over the manifold $M$. In particular, we can make the replacement:
\begin{align}
    \int \ba_e \cup \delta \ba_e = \sum_{t \in M} \ba_e \cup \delta \ba_e (t).
\end{align} 
With this, the circuit $\Uspt$ is:
\begin{align}
    {\Uspt} = \sum_{\ba_e} (-1)^{ \int \ba_e \cup \delta \ba_e} \ket{\ba_e}\bra{\ba_e},
\end{align}
and the ground state in Eq.~\eqref{1formsptwf} is:
\begin{align} \label{1formsptstate}
    \ket{\Psi_1} = \sum_{\ba_e} (-1)^{\int \ba_e \cup \delta \ba_e} \ket{\ba_e}.
\end{align}

Lastly, we introduce the cup-$1$ product $\cup_1$. Although abstract, the cup-$1$ product allows for a convenient form of the Hamiltonian $H_1$ and is key to our analysis of the twisted toric code in the subsequent section. The cup-$1$ product takes a $p$-cochain $\boldsymbol{c}$ and a $q$-cochain $\boldsymbol{d}$ to a $(p+q-1)$-cochain $\boldsymbol{c} \cup_1 \boldsymbol{d}$.  Explicitly, $\boldsymbol{c} \cup_1 \boldsymbol{d}$ evaluated on a $(p+q-1)$-simplex $\langle 0, \ldots, p+q-1 \rangle$ is:
\begin{align} \label{defcup1}
    &\boldsymbol{c} \cup_1 \boldsymbol{d} \boldsymbol{(} \langle 0, \ldots, {p+q-1} \rangle \boldsymbol{)}= \\ \nonumber
    &\sum_{i=0}^{p-1}\boldsymbol{c}\boldsymbol{(} \langle 0, \ldots , i, {q+i}, \ldots ,{p+q-1} \rangle \boldsymbol{)} \boldsymbol{d} \boldsymbol{(} \langle i,...,{q+i} \rangle \boldsymbol{)}. 
\end{align}

A useful example, relevant to our expression for $H_1$, is the cup-$1$ product of $\bface$ and $\delta \be$. Here, $\bface$ is the $2$-cochain that evaluates to $1$ on the face $f$ and $0$ for all other faces:
\begin{align} 
    \bface(f') = \begin{cases} 
      1 & f'=f \\
      0 & \text{otherwise}. 
   \end{cases}
\end{align} 
The $3$-cochain $\bface \cup_1 \delta \be$ evaluated on a tetrahedron $\langle 1234 \rangle$ is [using Eq.~\eqref{defcup1}]:
\begin{align}
    \bface \cup_1 \delta \be \boldsymbol{(} \langle 1234 \rangle \boldsymbol{)} = \bface \boldsymbol{(} \langle 134 \rangle \boldsymbol{)} & \delta \be \boldsymbol{(} \langle 123 \rangle \boldsymbol{)} \\ \nonumber
    +&\bface \boldsymbol{(} \langle 124 \rangle \boldsymbol{)} \delta \be \boldsymbol{(} \langle 234 \rangle \boldsymbol{)}.
\end{align}

\begin{figure}[t]
\centering
\includegraphics[width=0.28\textwidth, trim={700 300 700 150},clip]{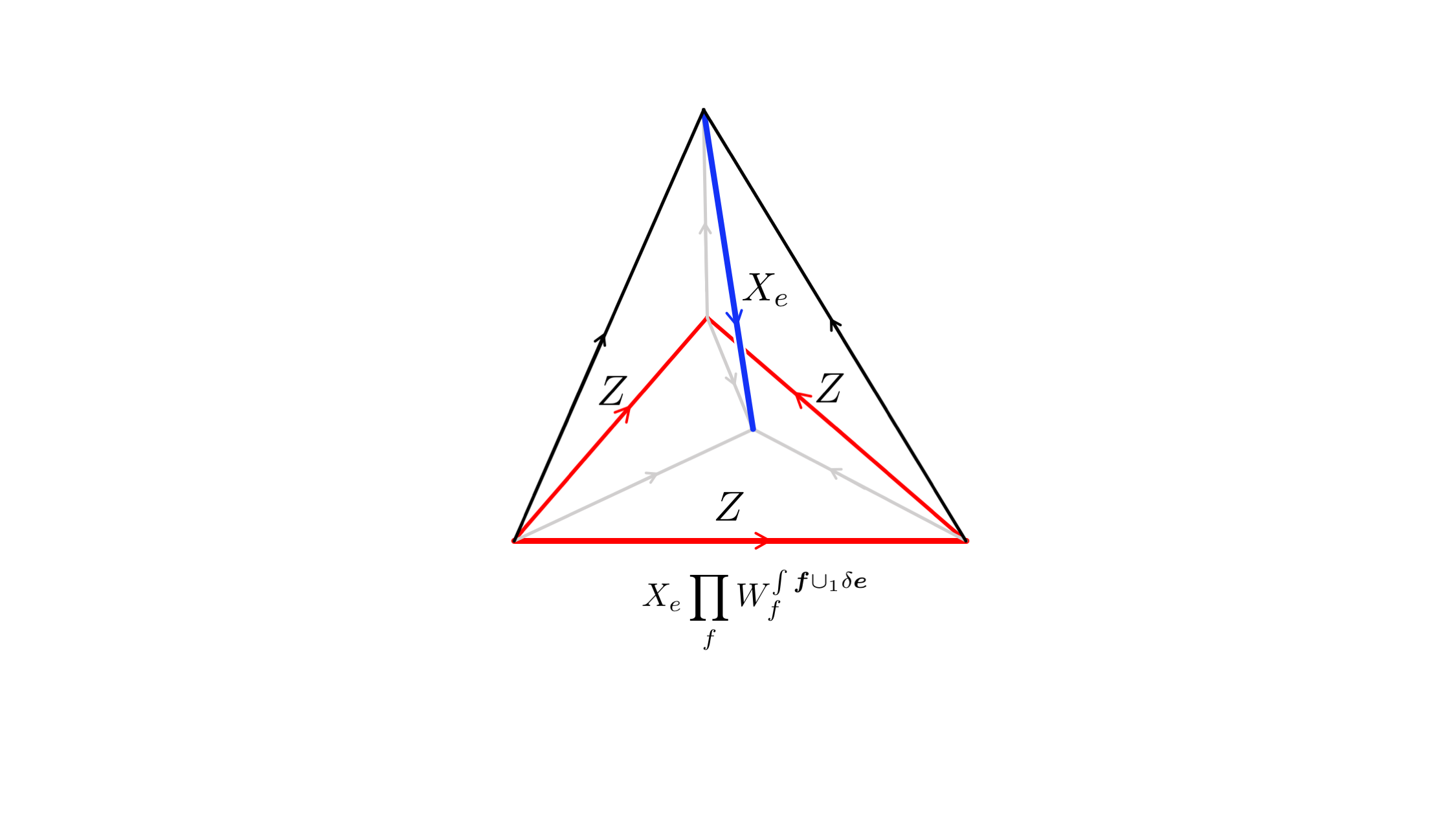}
\caption{Pictured above is an example of a term in $H_1$ associated to an edge $e$ with three tetrahedra meeting at $e$. The operator acts with $X_e$ on $e$ (blue) and with Pauli Z operators on edges nearby (shown in red). The placement of the Pauli Z operators depends on the branching structure through the exponent $\int \bface \cup_1 \delta \be$.}
\label{fig: 1formHterm}
\end{figure}

Now, with the notation from $\ZZ_2$ cohomology on $M$, we can express $H_1$ in Eq.~\eqref{1formsptH} in a compact form. In Appendix \ref{App: 1form}, we show that: 
\begin{align} \label{1formsptHsimplified}
    H_1 = -\sum_e \Big(X_e \prod_{f}  W_f^{\int \bface \cup_1 \delta \be } \Big).
\end{align}
Above, we have used $W_f$ to denote the product of $Z_e$ around the face $f$:
\begin{align} \label{Wfdef}
    W_f \equiv \prod_{e\subset f}Z_e.
\end{align}
While the product in Eq.~\eqref{1formsptHsimplified} is over all faces in $M$, the Hamiltonian is indeed local. This is because ${\bface \cup_1 \delta \be (t)=0}$, if the face $f$ and edge $e$ do not both belong to the tetrahedron $t$.
Heuristically, $X_e$ is ``dressed'' with loops of Pauli Z operators around certain faces near $e$ [Fig.~\ref{fig: 1formHterm}].

% We further show, in Appendix \ref{App: 1form}, that $H_1$ is symmetric under the $1$-form symmetry. We do so by showing that:
% \begin{align}
%   A_\Sigma {\Uspt} A_\Sigma = {\Uspt} ,
% \end{align}
% for any closed surface of the dual lattice $\Sigma$.
% Therefore, ${\Uspt}$ is symmetric under the $\ZZ_2$ $1$-form symmetry, and consequently, $H_1$ is also symmetric [see Eq.~\eqref{1formsptH}].  We note that, although ${\Uspt}$ is symmetric, this does not imply that the state $\ket{\Psi_1}$ in Eq.~\eqref{1formsptwf} belongs to the trivial SPT phase. This is only the case if ${\Uspt}$ can additionally be expressed as a FDQC composed of symmetric local unitaries. 

% Indeed, $H_1$ describes a \textit{nontrivial} $\ZZ_2$ $1$-form SPT phase. In the next section, we show this by gauging the the $1$-form symmetry. The $1$-form paramagnet $\onepara$ is mapped to a $\ZZ_2$ gauge theory with a bosonic gauge charge - the usual $3$D toric code, while $H_1$ is mapped to a $\ZZ_2$ gauge theory with a \textit{fermionic} gauge charge - a ``twisted toric code''. 

\subsection{Twisted toric code} \label{sec ttc}

The twisted toric code is constructed from $H_1$ in Eq.~\eqref{1formsptHsimplified} by gauging the $\ZZ_2$ $1$-form symmetry. In this section, we provide a physical description of the gauging procedure following the steps outlined in Ref.~\cite{LG12}. In Appendix \ref{app: 1form gauging quantum states}, we discuss the subtleties of the procedure and describe the process at the level of states, as in Ref.~\cite{Y16}. After gauging the $1$-form symmetry of $H_1$, we show that the resulting twisted toric code admits localized excitations with fermionic statistics.

% \subsubsection{Gauging a $1$-form symmetry}

The prescription for gauging a $\ZZ_2$ $0$-form symmetry in Ref.~\cite{LG12} naturally generalizes to gauging a $\ZZ_2$ $1$-form symmetry. In particular, the $1$-form symmetry is gauged according to the following steps.
\begin{enumerate}
    \item We introduce $\ZZ_2$ {d.o.f.} on faces corresponding to $2$-form gauge fields. We denote the Pauli Z and Pauli X operators at the face $f$ by $Z_f$ and $X_f$, respectively.
    % \item We impose a coupling constraint:
    % \begin{align}
    %     Z_f = W_f \equiv \prod_{e \subset f} Z_e,
    % \label{eq: KW gauge constraint}
    % \end{align}
    % where the product is over edges $e$ contained in the face $f$.
    % This constraint can be interpreted as coupling the background fields to the flux operators.
    % %This constraint can be interpreted as a $1$-form Gauss law.
    
    \item We impose a gauge constraint at each edge $e$:
    \begin{align}
        X_e \prod_{f \supset e}X_f=1,
    \end{align}
    where the product is over faces containing $e$. This constraint can be interpreted as a $1$-form Gauss law. We note that the operator $X_e \prod_{f \supset e}X_f$ defines a local action of the $1$-form symmetry. That is, a product of $X_e \prod_{f \supset e}X_f$ over edges intersected by a closed surface $\Sigma$ in the dual lattice yields the $1$-form symmetry operator $A_\Sigma$.
    
    \item To make coupling to the gauge field in the subsequent step unambiguous, we energetically enforce a ``no flux condition''. The point-like $1$-form gauge flux can be detected by the operator $W_t$, where $W_t$ is a product of $Z_f$ operators around a tetrahedron $t$:
    \begin{align}
        W_t \equiv \prod_{f \subset t} Z_f.
    \end{align}
    Therefore, the no flux condition is enforced by adding to the Hamiltonian the term:
    \begin{align} \label{nofluxcondition}
        - \sum_t W_t.
    \end{align}
    % with $J$ large compared to the other energy scales in the system.
    In addition, we conjugate each Hamiltonian term by a local projector onto the zero flux subspace in the vicinity of the term. That is, for a Hamiltonian term whose support\footnote{\unexpanded{The support of an operator is the set of sites on which the operator acts non-identically.}} is contained in the bounded region $R$, we conjugate by a projector: 
    \begin{align}
        \mathcal{P}_R^{\text{0-flux}}\equiv \prod_{t \in R} \frac{(1+W_t)}{2},
        \label{equ:P0flux_R}
    \end{align} 
    where the product is over tetrahedra in $R$.
    
    \item We then minimally couple the $\ZZ_2$ $1$-form symmetric model to the gauge fields, so as to make the model invariant under the gauge constraint. In particular, $W_f$ is coupled to the gauge field as:
    \begin{align}
        W_f \rightarrow W_fZ_f.
    \label{eq: minimal coupling}
    \end{align}
    
    % \item The operator $X_e$ is no longer an allowed operator since it violates the constraint \eqref{eq: KW gauge constraint}. Instead, we modify it as $X_e \prod_{f \supset e} X_f$. Since all of these operators commute, we can focus on their simultaneously diagonalized eigenspace:
    % \begin{equation}
    %     X_e \prod_{f \supset e} X_f = +1.
    % \label{eq: KW gauge constraint 2}
    % \end{equation}
    % We can assume they are all $+1$ since the sign can be fixed by some basis transformation.
    
    % \item
    % This gauging procedure \eqref{eq: KW gauge constraint} and \eqref{eq: KW gauge constraint 2} induces the operator duality summarized in TABLE. \ref{table: gauging 1-form}. Notice that the property $\prod_{f \subset t} W_f = 1$ and \eqref{eq: KW gauge constraint} requires the 1-form gauge symmetry for spins living on faces:
    % \begin{equation}
    %     \prod_{f \supset t} Z_f = 1,
    % \end{equation}
    % which is shown in the last line of TABLE. \ref{table: gauging 1-form}. On the other hand, $\prod_{e \supset v} (\prod_{f \supset e} X_f) = 1$ and \eqref{eq: KW gauge constraint 2} requires the 1-form symmetry for spins living on edges:
    % \begin{equation}
    %     \prod_{e \supset v} X_e = 1,
    % \end{equation}
    % which is included in the penultimate line of TABLE. \ref{table: gauging 1-form}. We refer to Appendix \ref{app: 1form gauging quantum states} for further details.
    
    \item We fix a gauge by mapping gauge invariant states to representative states in which the eigenvalue of $Z_e$ is $1$ at every edge $e$\footnote{\unexpanded{More precisely, we can form an over-complete basis for the gauge invariant Hilbert space by projecting configuration basis states to the gauge invariant subspace with the operator:
    \begin{align}
        \prod_e \left( 1 + X_e\prod_{f\supset e}X_f \right).
    \end{align}
    Each gauge invariant state in this (over-complete) basis is a superposition of configuration states and includes exactly one state for which the eigenvalue of $Z_e$ is $1$ at every edge. By gauge fixing, we mean that the over-complete basis states are mapped to the ``representative'' state with the eigenvalue of $Z_e$ equal to $1$ at each edge. The representative states form a basis for the gauge fixed Hilbert space.}}. This gauge fixed Hilbert space is equivalent to a Hilbert space with only the gauge field {d.o.f.} on the faces. The action of $X_e$ on gauge invariant states is replaced by $\prod_{f\supset e}X_f$ after fixing the gauge, and $W_fZ_f$ in Eq.~\eqref{eq: minimal coupling} becomes equivalent to $Z_f$ in the gauge fixed Hilbert space. Therefore, the gauge invariant operators $X_e$ and $W_fZ_f$ are mapped according to:
    \begin{align}
        X_e \rightarrow \prod_{f \supset e}X_f, \quad W_fZ_f \rightarrow Z_f.
    \end{align}
\end{enumerate}

\begin{table}
    \centering
    \normalsize
    {\renewcommand{\arraystretch}{1.5}
    \begin{tabular}{ | >{\centering\arraybackslash}m{4cm} | >{\centering\arraybackslash} m{4cm} |}
    \hline
    Model with  $\ZZ_2$            &  Model with dual $\ZZ_2$ \\ 
    $1$-form symmetry &  $1$-form symmetry  \\
    \hline
    $X_e$ & $\displaystyle \prod_{f \supset e} X_f$ \\ 
    \hline
    $\displaystyle W_f = \prod_{e \subset f} Z_e$ & $Z_f$  \\ 
    \hline
    $\displaystyle A_\Sigma=\prod_{e\perp \Sigma} X_e, \, \delta\boldsymbol{\Sigma}=0$ & 1  \\ 
    \hline
    1 & $\displaystyle M_\sigma = \prod_{f\subset \sigma} Z_f, \, \partial \sigma =0$ \\ 
    \hline
\end{tabular}}
\caption{In the process of gauging the $1$-form symmetry, the generators of local, $1$-form symmetric operators are mapped according to the duality above. The symmetry operators $A_\Sigma$ are mapped to the identity in the dual theory. The system on the right-hand side has a $\ZZ_2$ $1$-form symmetry, generated by membrane operators $M_\sigma$, where $\sigma$ is a closed $2$D surface on the direct lattice. }
\label{table: gauging 1-form}
\end{table}

% \begin{table}
%     \centering
%     \normalsize
%     {\renewcommand{\arraystretch}{1.5}
%     \begin{tabular}{ | >{\centering\arraybackslash}m{4cm} | >{\centering\arraybackslash} m{4cm} |}
%     \hline
%     Theories with              &  $2$-form  \\ 
%     $1$-form $\ZZ_2$ symmetry & $\ZZ_2$ gauge theory  \\
%     \hline
%     $X_e$ & $\prod_{f \supset e} X_f$ \\ 
%     \hline
%     $W_f = \prod_{e \subset f} Z_e$ & $Z_f$  \\ 
%     \hline
%     $\prod_{e\supset v} X_e$ & 1  \\ 
%     \hline
%     1 & $W_t = \prod_{f\subset t} Z_f$ \\ 
%     \hline
% \end{tabular}}
% \caption{caption}
% \label{table: gauging 1-form}
% \end{table}

We remark that, operationally, the gauging procedure is equivalent to a certain operator duality. In particular, the duality maps the $1$-form symmetric operators $X_e$ and $W_f$ according to:
\begin{align}
    X_e \rightarrow \prod_{f \supset e}X_f, \quad W_f \rightarrow Z_f.
\end{align}
We have summarized the corresponding operator duality in Table~\ref{table: gauging 1-form}, and we refer to Appendix \ref{app: 1form gauging quantum states} for further details.

\begin{figure}[t]
\centering
\includegraphics[width=0.23\textwidth, trim={800 440 840 360},clip]{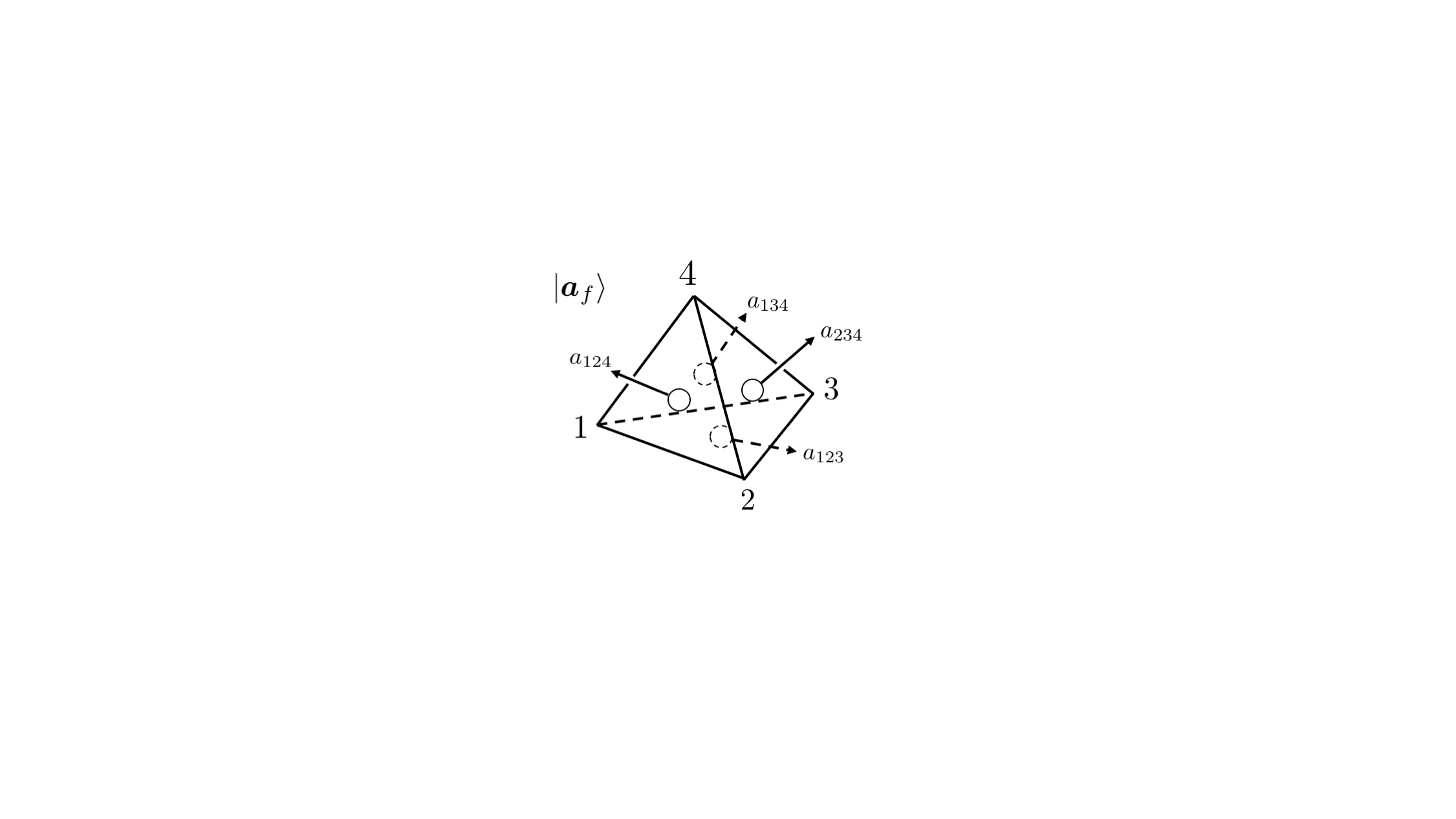}
\caption{We define the twisted toric code on a triangulation with $\ZZ_2$ {d.o.f.} at each face (represented by a circle). A configuration state is given by a value $a_f \in \{0,1\}$ chosen for each face $f$.}
\label{fig: H2configuration}
\end{figure}

As a result of applying steps 1-5 above to our model for the nontrivial $1$-form SPT phase, we obtain the twisted toric code. The twisted toric code is defined on a Hilbert space composed of $\ZZ_2$ {d.o.f.} attached to each face of the triangulation of $M$ (Fig.~\ref{fig: H2configuration}). Further, a basis for the Hilbert space is given by the product states $\ket{\{a_f\}}$, where the state at the face $f$ is $\ket{a_f}$ (with $a_f \in \{ 0,1 \}$). In analogy to Eq.~\eqref{mapto1cochainstate}, a configuration state $\ket{\{a_f\}}$ can be labeled by a $2$-cochain $\ba_f$:
\begin{align}
    \ba_f \equiv \sum_{f} a_f \bface.
\end{align}
With this notation, the Pauli Z and Pauli X operators acting on the face $f'$ can be written as:
\begin{align} \label{Paulionf}
    Z_{f'} \ket{\ba_f} = (-1)^{\ba_f(f')}\ket{\ba_f}, \quad
     X_{f'} \ket{\ba_f} = \ket{\ba_f+\bface'}.
\end{align}

\begin{figure}[t]
\centering
\includegraphics[width=0.28\textwidth, trim={700 260 700 150},clip]{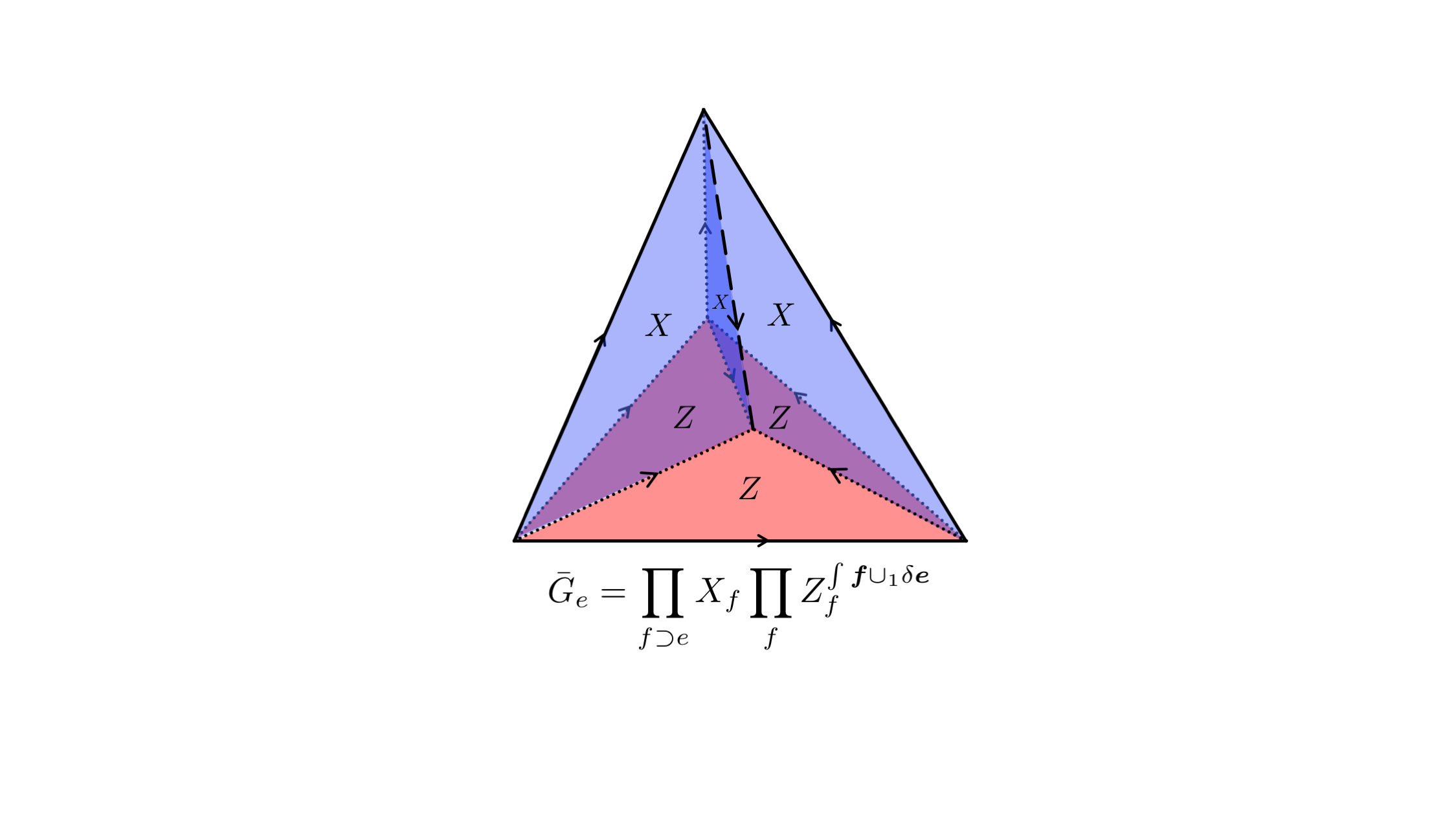}
\caption{An example of $\bar{G}_e$ for the edge $e$ (dashed line) is shown above. Pauli X operators act on the faces (shaded blue) adjoined at $e$ and Pauli Z operators (shaded red) act on nearby faces according to $\int \bface \cup_1 \delta \be$.}
\label{fig: barGe_example}
\end{figure}

By gauging the $1$-form symmetry 
% \Yuan{refer to Appendix \ref{app: 1form gauging quantum states} or simply describe the gauging here?} \textcolor{orange}{I referenced Appendix \ref{app: 1form gauging quantum states} at the beginning of the section. Here, we give the heuristic decription of the gauging procedure. I can reference the appendix again here, but the point is that the procedure above produces $H_\text{ttc}$. Update: okay, I referred to the appendix again to mention that there is an operator duality.} 
of $H_1$ [Eq.~\eqref{1formsptHsimplified}], we find the twisted toric code Hamiltonian:
% \footnote{It suffices to take 
%  \unexpanded{$J=1$} in Eq.~\eqref{nofluxcondition}.}
\begin{align}\label{ttcdef}
    H_\text{ttc} \equiv -\sum_{e}\bar{G}_e - \sum_t W_t,
\end{align}
where $\bar{G}_e$ is:
\begin{align} \label{Gedef}
    \bar{G}_e \equiv \prod_{f \supset e}X_f \prod_{f}  Z_f^{\int \bface \cup_1 \delta \be}.
\end{align}
An example of the term $\bar G_e$ is shown in Fig \ref{fig: barGe_example}. For simplicity, we have omitted the local projectors from step 3 of the gauging procedure. They do not affect the discussion in this section. The terms of $H_\text{ttc}$ are all mutually commuting, since the gauging procedure preserves the commutation relations. We show in Appendix \ref{app: ttc gs} that a ground state of the model is: 
\begin{align} \label{ttc gs}
    \ket{\Psi_\text{ttc}}\equiv \sum_{\ba_e} (-1)^{\int \ba_e \cup \delta \ba_e}\ket{\delta \ba_e}.
\end{align}
Note that while the expression for $\ket{\Psi_\text{ttc}}$ has a sum over $1$-cochains $\ba_e$, there are no {d.o.f.} on the edges. Rather, by summing over $1$-cochains $\ba_e$, $\ket{\Psi_\text{ttc}}$ is a ground state of $H_\text{ttc}$ with trivial holonomy. 

\vspace{1.5mm}
\noindent \begin{center}\emph{Excitations in the twisted toric code:}\end{center}
\vspace{1.5mm}

There are two types of excitations of the twisted toric code. The first, is a line-like $\ZZ_2$ $1$-form gauge charge corresponding to violations of the edge terms $\bar{G}_e$. A small loop of gauge charge around the face $f$ is created by acting with the face operator $Z_f$.  A larger loop of gauge charge can be created by acting with $Z_f$ on all faces contained in a 2D membrane $\sigma$ of the direct lattice:
% \textcolor{orange}{The notation here is potentially confusing, since in all other instances, $W_s$ has $Z$ operators acting on the boundary elements of the simplex $s$. Suggestions for notation? It will later be used to describe symmetry fractionalization on fermion parity flux loops.}
% \begin{align}
%     W_\sigma \equiv \prod_{f \subset \sigma}Z_f.
% \end{align}
\begin{align} \label{membrane operator}
   {M}_\sigma \equiv \prod_{f \subset \sigma}Z_f.
\end{align}
We think of the gauge charge as lying along the boundary of $\sigma$, since the membrane operator $M_\sigma$ anti-commutes with the edge terms $\bar{G}_e$ for which $e$ is in the boundary of $\sigma$. 

\begin{figure}[t]
\centering
\includegraphics[width=0.3\textwidth, trim={900 390 700 400},clip]{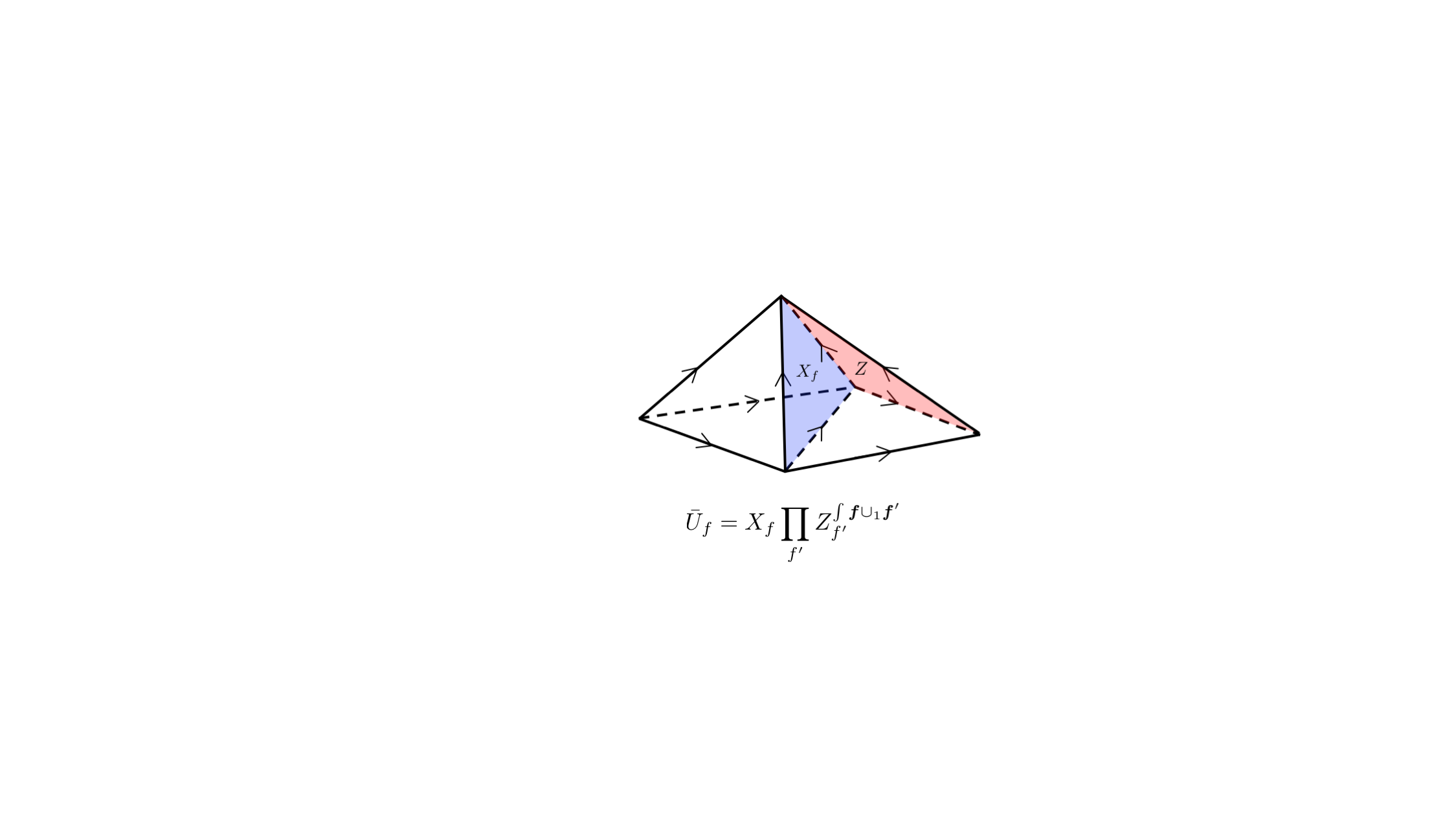}
\caption{The operator $\bar{U}_f$ applies a Pauli X operator at $f$ (blue) and Pauli $Z$ operators on certain faces of the neighboring tetrahedra (red). For a tetrahedron $t = \langle {1234} \rangle$, the cup-$1$ product of $\bface$ and $\bface'$ evaluates to $\bface \cup_1 \bface'(t)=\bface \boldsymbol{(}\langle 134 \rangle \boldsymbol{)}\bface' \boldsymbol{(}\langle 123 \rangle \boldsymbol{)} + \bface \boldsymbol{(}\langle 124 \rangle \boldsymbol{)} \bface' \boldsymbol{(}\langle 234 \rangle \boldsymbol{)}$. In the figure above, $f$ is the $\langle 124 \rangle$ face of the tetrahedron $\langle 1234 \rangle$ on the right. Thus, $\bar{U}_f$ applies a Pauli Z to $f'= \langle 234 \rangle$.}
\label{fig: ufexample}
\end{figure}

The second type of excitation of $H_\text{ttc}$ is a point-like $\ZZ_2$ $1$-form gauge flux corresponding to a violation of a $W_t$ term. A pair of gauge fluxes can be created at neighboring tetrahedra by the short string operator:\footnote{\unexpanded{To avoid confusion with the operator $U_f \equiv X_f \prod_{f^\prime} Z^{\int \bface' \cup_1 \bface}_{\bface^\prime}$ in Ref.~\cite{CK18}, we use $\bar{U}_f$.}}
\begin{align} \label{Ufdef}
    \bar{U}_f \equiv X_f \prod_{f^\prime} Z^{\int \bface \cup_1 \bface'}_{\bface^\prime},
\end{align}
pictured in Fig.~\ref{fig: ufexample}.
$\bar{U}_f$ anticommutes with the tetrahedron terms $W_t$ on either side of the face $f$. Thus, we interpret the gauge fluxes as living at the centers of tetrahedra. The Pauli Z operators in Eq.~\eqref{Ufdef} ensure that for any $f$ and any $e$, $\bar{U}_f$ commutes with $\bar{G}_e$. The short string operators satisfy the commutation relations \cite{CK18}:
\begin{align}  \label{Ufcommutation}
  \bar{U}_f \bar{U}_{f'} = (-1)^{\int (\bface^\prime \cup_1 \bface+\bface \cup_1 \bface^\prime)} \bar{U}_{f^\prime} \bar{U}_f,
\end{align}
for any faces $f$ and $f'$.

Longer string operators can be formed by composing the short string operators in Eq.~\eqref{Ufdef}. However, a simple product of $\bar{U}_f$ operators along a path $p$ in the dual lattice is ambiguous. This is because, given the commutation relations in Eq.~\eqref{Ufcommutation}, a product of $\bar{U}_f$ operators is generically order dependent. To remove the ambiguity, we define the following notation. For any set $F$ of faces, we define an order independent product of $\bar{U}_f$ by:
% \begin{align} \label{barproduct def}
%     \overline{\prod_{f \in F}}U_f \equiv \prod_{f \in F} \bigg( \prod_{f^\prime} Z^{\sum_t \bface' \cup_1 \bface(t)}_{\bface^\prime} \bigg) \prod_{f \in F}X_f.
% \end{align}
\begin{align} \label{barproduct def}
     \overline{\prod_{f \in F}}\bar{U}_f \equiv 
    \prod_{f \in F} \bigg( \prod_{f^\prime} Z^{\int \bface \cup_1 \bface'}_{\bface^\prime} \bigg) \prod_{f \in F}X_f.
\end{align}
Here, all of the Pauli Z operators from the definition of $\bar{U}_f$ appear to the left of the Pauli X operators.
We can then unambiguously define a gauge flux string operator $\mathcal{S}_p$ along a path $p$ in the dual lattice by:
\begin{align} \label{Emergent fermion string def}
   \mathcal{S}_p \equiv \overline{\prod_{f \in F_{\perp p}}} \bar{U}_f, 
\end{align}
where $F_{\perp p}$ denotes the set of faces intersected by $p$.

\begin{figure}[t]
\centering
\includegraphics[width=0.15\textwidth, trim={850 390 1000 200},clip]{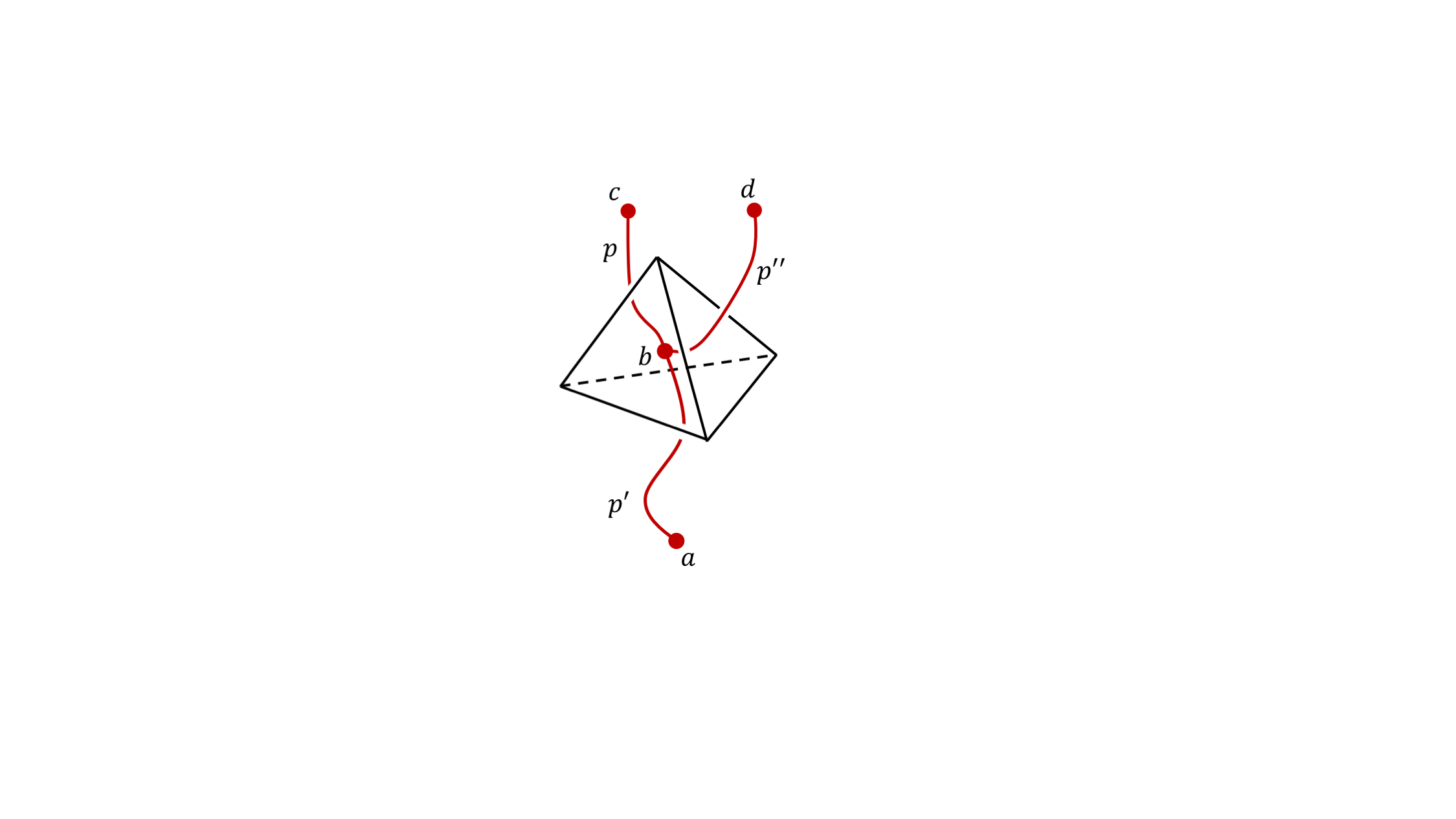}
\caption{The paths $p$, $p'$, and $p''$ share a single common endpoint $b$ inside of a tetrahedron $t$. Furthermore, the paths intersect distinct faces of $t$. Other details of the paths are unimportant in the computation of the statistics of the gauge fluxes.}
\label{fig: fluxstatistics}
\end{figure}

Remarkably, the gauge fluxes are emergent fermions. To see this, we use the methods developed in Refs.~\cite{LW03} and \cite{KL19} for computing the statistics of anyons from microscopic models. We consider three paths $p$, $p'$, and $p''$ along the dual lattice sharing a common endpoint, as in Fig.~\ref{fig: fluxstatistics}. The statistics of the gauge fluxes can be deduced by comparing the product $\mathcal{S}_p\mathcal{S}_{p'}\mathcal{S}_{p''}$ to $\mathcal{S}_{p''}\mathcal{S}_{p'}\mathcal{S}_p$. 

To gain intuition for this comparison, we imagine gauge fluxes at $a$ and $b$ in Fig.~\ref{fig: fluxstatistics}. In the first process $\mathcal{S}_p\mathcal{S}_{p'}\mathcal{S}_{p''}$, the gauge flux at $b$ is moved to $c$ and the gauge flux at $a$ is moved to $d$. Whereas, in the second process $\mathcal{S}_{p''}\mathcal{S}_{p'}\mathcal{S}_p$, the gauge flux at $b$ is moved to $d$ and the gauge flux at $a$ is moved to $c$. The final configurations of the gauge fluxes differ by an interchange of the position of the gauge fluxes. Consequently, the difference between $\mathcal{S}_p\mathcal{S}_{p'}\mathcal{S}_{p''}$ and $\mathcal{S}_{p''}\mathcal{S}_{p'}\mathcal{S}_p$ determines the statistics of the gauge fluxes.

It can be shown that the gauge flux string operator satisfies:
\begin{align} \label{statistics of gauge fluxes}
    \mathcal{S}_p\mathcal{S}_{p'}\mathcal{S}_{p''}=-\mathcal{S}_{p''}\mathcal{S}_{p'}\mathcal{S}_p.
\end{align}
This follows from an explicit computation using the commutation relations of the operators $\bar{U}_f$ in Eq.~\eqref{Ufcommutation}.\footnote{More precisely, this can be checked by explicitly computing the sign for each possible intersection at a tetrahedron. The sign for only four of the possible orientations of the tetrahedron needs to be verified, as the others follow from symmetries of the calculation.}
% :
% \begin{align}  \label{Ufcommutation}
%   U_f U_{f'} = (-1)^{\sum_t (\bface^\prime \cup_1 \bface+\bface \cup_1 \bface^\prime)(t)} {U}_{f^\prime} {U}_f,
% \end{align}
% for any faces $f$ and $f'$.
Therefore, the gauge fluxes are emergent fermions. We note that this is equivalent to saying that the twisted toric code has an anomalous $\ZZ_2$ $2$-form symmetry \cite{CK18,KT17}. The $2$-form symmetry, which, by definition, acts on closed codimension-$2$ subspaces, is generated by loops of the emergent fermion string operator. It is called anomalous, simply because the gauge fluxes have fermionic statistics.

Before leveraging our understanding of the twisted toric code to construct a model of physical fermions, we would like to point out that the emergent fermion string operator is not unique. In fact, we can define an alternative emergent fermion string operator built from the short segments: 
\begin{align}
    \tilde{U}_f\equiv \bar{U}_fW_{R(f)},
\end{align}
with $R(f)$ denoting the tetrahedron neighboring $f$ in the direction of the orientation of $f$ (see Fig.~\ref{fig:Sf def2}). The corresponding string operator along a path $p$ in the dual lattice is:
\begin{align} \label{tSstringop}
    \tilde{\mathcal{S}}_p \equiv  \overline{\prod_{f \in F_{\perp p}}} \bar{U}_f \prod_{f \in F_{\perp p}} W_{R(f)}.
\end{align}

An important observation moving forward is that $\bar{G}_e$ is equivalent to a small loop of $\tilde{\mathcal{S}}_p$ string around the edge $e$ (see Appendix~\ref{fermioncondensationHttc}):
\begin{align} \label{Ge tildeS}
    \bar{G}_e = \tilde{\mathcal{S}}_{p_e},
\end{align}
where the path $p_e$ intersects only the faces adjoined at $e$.
Therefore, the $\bar{G}_e$ operators are local generators of an anomalous $\ZZ_2$ $2$-form symmetry. Since $\bar{G}_e$ commutes with $H_\text{ttc}$, we see explicitly that the twisted toric code has an anomalous $2$-form symmetry.

\subsection{Atomic insulator} \label{sec: atomic insulator}

\begin{figure}[t]
\centering
\includegraphics[width=0.21\textwidth, trim={800 500 1000 300},clip]{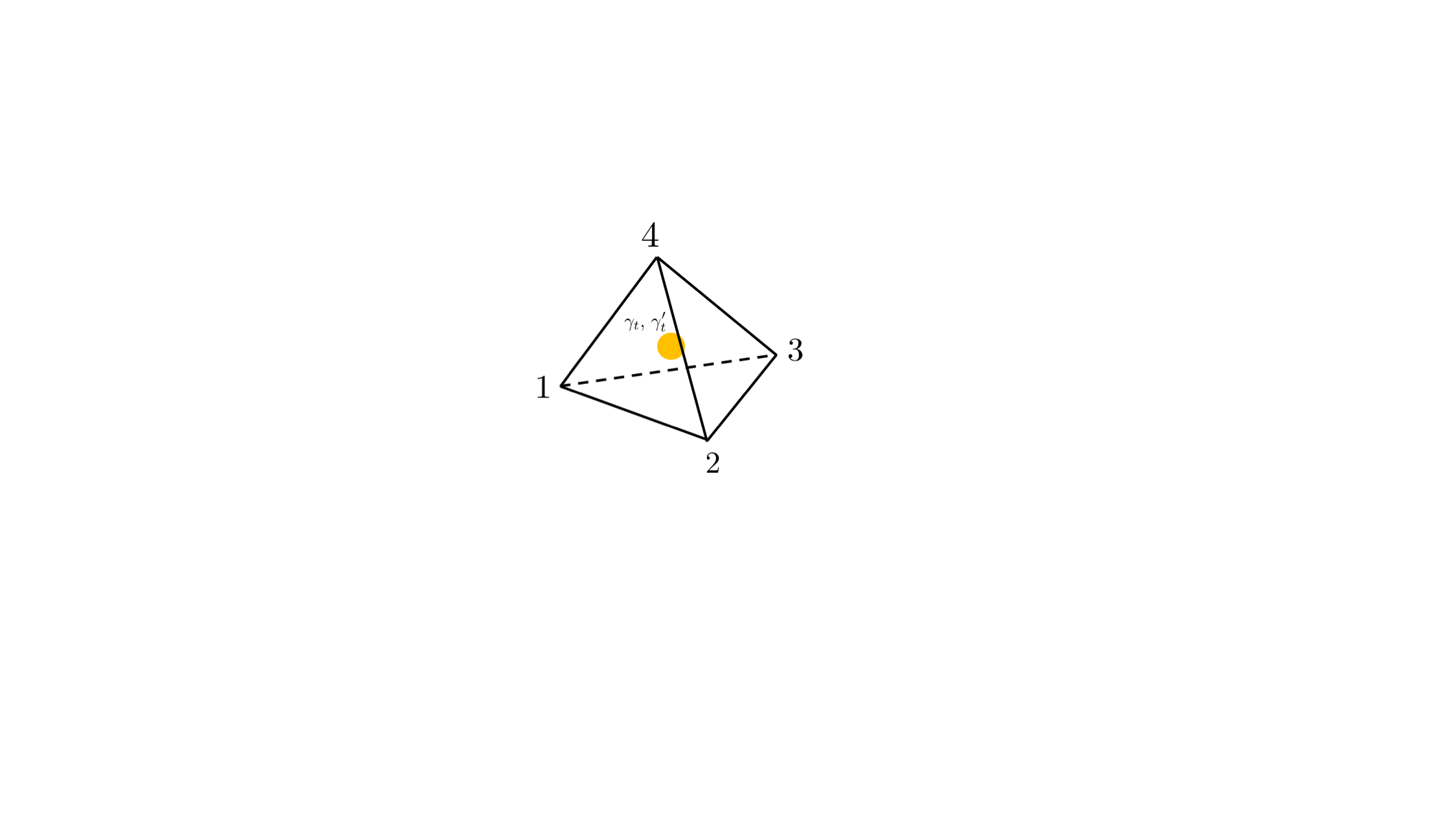}
\caption{The atomic insulator is defined on a Hilbert space with a single complex fermion {d.o.f.} (yellow circle) at each tetrahedron. The operator algebra at the tetrahedron $t=\langle 1234 \rangle$ is generated by the Majorana operators $\gamma_t$ and $\gamma'_t$.}
\label{fig:ai dof}
\end{figure}

The last step of our $G=\ZZ_2^f$ example is to convert the twisted toric code into a model with physical fermions. This can be accomplished by applying the $(3+1)$D fermionization duality introduced in Ref.~\cite{CK18}, reviewed in Appendix \ref{sec: review of boson-fermion duality}. In this section, we instead opt to describe the fermionization process in terms of fermion condensation \cite{Aasen17,EF19}. That is, we construct the fermionic model by pairing emergent fermions with physical fermions and condensing the composite bosonic excitations. The fermion condensation procedure, described below, can be interpreted as a generalization of Refs.~\cite{LG12} and \cite{SETstringnet} to gauging an anomalous $2$-form symmetry. Although an anomalous symmetry typically implies an obstruction to gauging the symmetry, we bypass the obstruction by employing fermionic {d.o.f.} for the gauge fields. 
% \footnote{Conventionally, an anomalous symmetry implies an obstruction to gauging the symmetry. We bypass the obstruction here by using fermionic {d.o.f.} for the gauge fields.}. \Nat{perhaps we should emphasize that the catch is that the gauge fields are themselves fermions. Otherwise people might not like the term "gauging an anomalous symmetry"} \textcolor{orange}{This is a good point, it sounds unnatural to say gauging an anomalous symmetry. I've added a footnote to clarify. We can include it in the main text instead, if you think it is worth it. We also comment on the fact that we are using fermionic {d.o.f.} in the procedure for gauging the symmetry.}

% \subsubsection{Fermion condensation}

Our prescription for fermion condensation starts by introducing a spinless complex fermion {d.o.f.} at the center of each tetrahedron (Fig.~\ref{fig:ai dof}). Thus, to prepare for the discussion of fermion condensation, we recall the notation for the operators on the fermionic Hilbert space, defined in Section~\ref{sec: supercohomology models 2}. The fermion parity operator at $t$ is:
\begin{align}
    P_t = -i\gamma_t \gamma'_t,
\end{align}
and the hopping operator $S_f$ across the face $f$ is: 
\begin{align}
    S_f = { (-1)^{\boldsymbol{f}(E)} }  i \gamma_{L(f)} \gamma'_{R(f)}.
\end{align}
$L(f)$ and $R(f)$ are defined below Eq.~\eqref{hopping}, and $E$ is a formal sum of $2$-simplices that amounts to a choice of spin-structure; see Appendix~\ref{sec: review of boson-fermion duality} for the explicit form of $E$.\footnote{Note that every orientable $3$-manifold admits a spin structure, so a choice of $E$ is guaranteed.}
% {\color{red} Notice that we have assumed that the spatial manifold is a spin manifold, where the spin structure $E$ exists.}
The hopping operators satisfy the commutation relations:
\begin{align} \label{hopping commutation}
    S_f S_{f'} = (-1)^{\int (\bface^\prime \cup_1 \bface+\bface \cup_1 \bface^\prime)} {S}_{f^\prime} {S}_f,
\end{align}
which we note matches the commutation relations of $\bar{U}_f$ in Eq.~\eqref{Ufcommutation}. Lastly, we define an alternative hopping operator $\tilde{S}_f$, instrumental for the fermion condensation prescription below:
\begin{align}
    \tilde{S}_f \equiv P_{L(f)} S_f .
\end{align}

\begin{figure}[t]
\centering
\includegraphics[width=.4\textwidth, trim={580 390 550 175},clip]{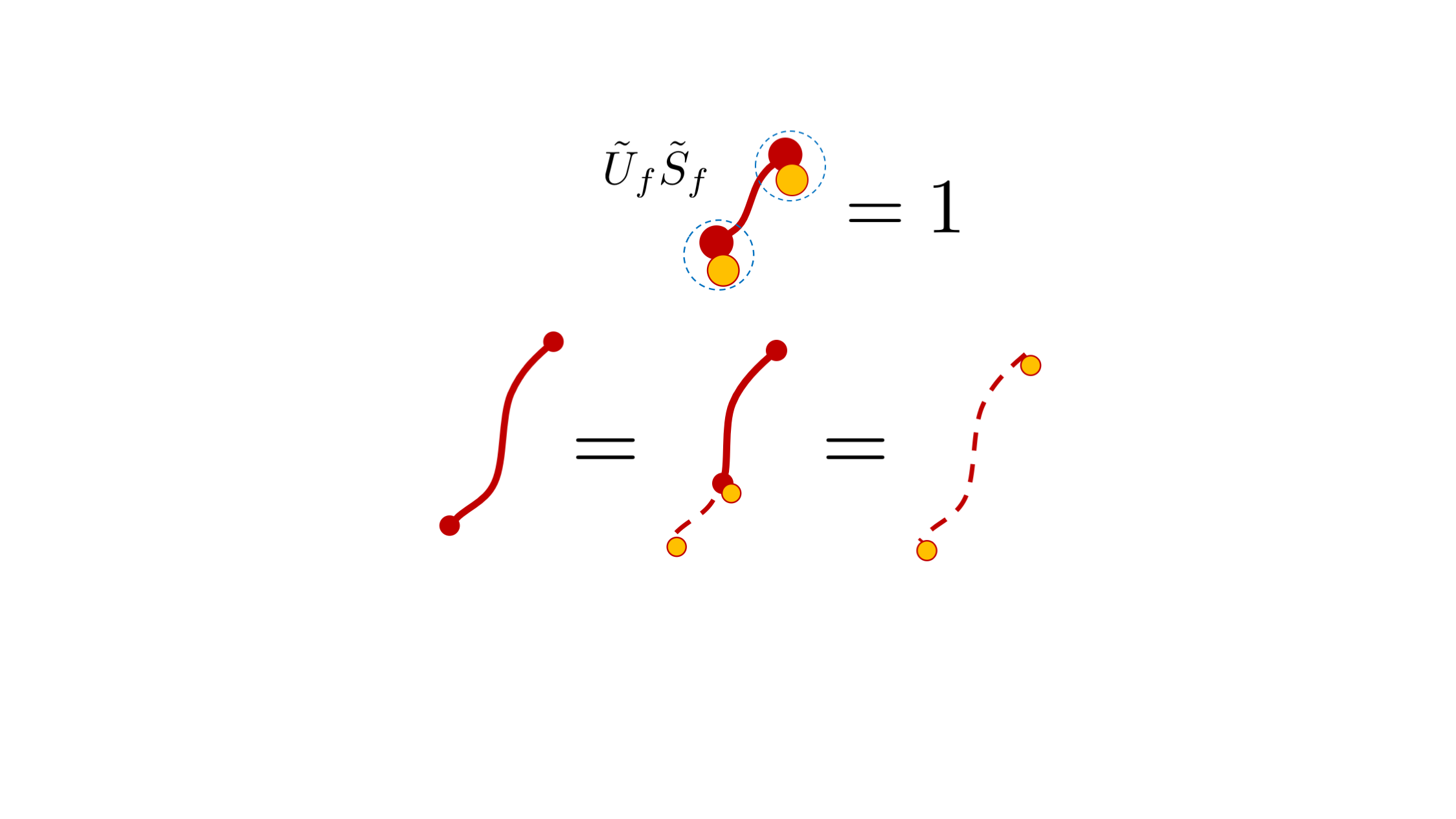}
\caption{To condense the emergent fermion, we impose the gauge constraint $\tilde{U}_f\tilde{S}_f=1$. Acting on the vacuum, $\tilde{U}_f$ creates a pair of emergent fermions (red) and $\tilde{S}_f$ creates a pair of physical fermions (yellow). The composite excitation (dashed blue circle) has bosonic statistics, so it may be condensed. Heuristically, emergent fermions can be replaced with physical fermions in the constrained Hilbert space.}
\label{fig:condenseconstraint}
\end{figure}

With notation for the fermionic {d.o.f.} defined, we can describe the fermion condensation procedure. Our procedure applies to any Hamiltonian with an emergent fermion created by the string operator $\tilde{\mathcal{S}}_p$ in Eq.~\eqref{tSstringop}. In other words, the fermion condensation procedure applies to any Hamiltonian with an anomalous $\ZZ_2$ $2$-form symmetry locally represented by the operators $\bar{G}_e$. Fermion condensation then proceeds as follows. 

\begin{enumerate}
    \item We introduce a spinless complex fermion {d.o.f.} at each tetrahedron. In terms of gauging the $2$-form symmetry, the anomalous nature requires that the ``$3$-form gauge fields'' are fermionic {d.o.f.}.
    
    % \item We couple the flux to the the fermions field:
    % \begin{align}
    %     P_t = W_t \equiv \prod_{f \subset t} Z_f,
    % \label{eq: JW gauge constraint}
    % \end{align}
    % where the product is over edges $f$ contained in the tetrahedron $t$.
    % This constraint can be interpreted as coupling the background on-site fermion parity to the flux operators.
    
    \item We impose a gauge constraint:
    \begin{align} \label{fermioncondensation}
        \tilde{U}_f \tilde{S}_f=1, 
    \end{align}
    for each face $f$. This constraint enforces a proliferation (or condensation) of composite excitations composed of an emergent fermion and a physical fermion. This is because $\tilde{U}_f$ is a short segment of emergent fermion string operator [Eq.~\eqref{tSstringop}], which creates emergent fermions at the tetrahedra on either side of $f$, while $\tilde{S}_f$ creates physical fermions at the corresponding tetrahedra (Fig.~\ref{fig:condenseconstraint}). Importantly, the constraints at different faces commute, due to the matching commutation relations of $\tilde{U}_f$ and $\tilde{S}_f$ (see Appendix~\ref{fermioncondensationHttc}). We note that the gauge constraint in Eq.~\eqref{fermioncondensation} is a local action of the $\ZZ_2$ anomalous $2$-form symmetry, in the sense that the product of $\tilde{U}_f\tilde{S}_f$ around an edge returns $\bar{G}_e$; this is guaranteed by the spin structure dependent sign in the definition of the hopping operator.
    % \footnote{\unexpanded{Similar to $U_f$, the commutation relations of $\tilde{U}_f$ are:
    % \begin{align}
    %     \tilde{U}_f \tilde{U}_{f'} = (-1)^{\sum_t (\bface^\prime \cup_1 \bface+\bface \cup_1 \bface^\prime)(t)} \tilde{U}_{f^\prime} \tilde{U}_f.
    % \end{align}
    % Using the commutation relations of the hopping operators in %Eq.~\eqref{hopping commutation}, we have:
    % \begin{align}
    %   \left( \tilde{U}_fS_f \right) \left( \tilde{U}_{f'}S_{f'} \right) = \left( \tilde{U}_{f^\prime}S_{f'} \right) \left( \tilde{U}_fS_{f}\right).
    % \end{align} }}

    % \item The operator $U_f$ is no longer an allowed operator since it violates the constraint \eqref{eq: JW gauge constraint}. Instead, we modify it as $U_f S_f$. Since all of these operators commute (the anomaly of $U_f$ and $S_f$ cancels out), we focus on their simultaneously diagonalized eigenspace:
    % \begin{equation}
    %     U_f S_f = +1.
    % \label{eq: JW gauge constraint 2}
    % \end{equation}
    % We can assume they are all $+1$ since the sign can be fixed by some basis transformation. The physical interpretation of \eqref{eq: JW gauge constraint 2} is that we condense the composite particles of the emergent fermion and the background fermion.
    
    \item To make the Hamiltonian gauge invariant, i.e., commute with the constraints in Eq.~\eqref{fermioncondensation}, we couple the Hamiltonian to the fermionic {d.o.f.}. Since the Hamiltonian commutes with $\bar{G}_e$, it can be expressed in terms of $\bar{U}_f$  and $W_t$ operators \cite{CK18}. We couple $\bar{U}_f$ operators and $W_t$ operators to the gauge fields as:
    \begin{align}
      \bar{U}_f \to \bar{U}_f P_{L(f)}, \quad  W_t \to W_t P_t.
    \label{eq: fermion minimal coupling}
    \end{align}
    To avoid possible ambiguity, we require that the coupling preserves the locality of the Hamiltonian.
    % \footnote{The potential ambiguity comes from the relation \unexpanded{$\prod_{t}W_t=1$. In any event, we map to fermionic states with even fermion parity, so the choice does not affect the eigenstates of the fermion condensed model.}} 
    Any local, gauge invariant operators can be expressed in terms of the operators $\bar{U}_fP_{L(f)}$ and $W_t P_t$.
    
    % \item
    % This gauging procedure \eqref{eq: JW gauge constraint} and \eqref{eq: JW gauge constraint 2} induces the operator duality summarized in TABLE. \ref{table: gauging fermion parity}. Notice that the property $\prod_{t} W_t = 1$ and \eqref{eq: JW gauge constraint} requires the total fermion parity symmetry:
    % \begin{equation}
    %     \prod_{t} P_t = 1,
    % \end{equation}
    % which is shown in the last line of TABLE. \ref{table: gauging fermion parity}. On the other hand, the identity (\cite{CK18,C19-2} and Appendix \ref{sec: review of boson-fermion duality}) \textcolor{orange}{I think we should still discuss the gauging section, but I think it would be nice to not have to introduce the identity below and Stiefel-Whitney classes in the main text. I think it can be avoided.}
    % \begin{equation}
    %      {\color{red} (-1)^{\boldsymbol e (w_2)} }
    %      S_{\delta \boldsymbol e} \prod_{t} P_t^{\int \boldsymbol e \cup_1 \boldsymbol t + \boldsymbol t \cup_1 \boldsymbol e} = 1
    % \end{equation}
    % {\color{red} (where $w_2 \in C_1(M,\ZZ_2)$ is the second Stiefel-Whitney class) }
    % and \eqref{eq: JW gauge constraint 2} requires the 2-form gauge constraints for spins living on faces:
    % \begin{equation}
    %     G_e \equiv \prod_{f \supset e}X_f   \prod_{f}  Z_f^{\sum_t \delta \be \cup_1 \bface (t)} = 1,
    % \end{equation}
    % which is included in the penultimate line of TABLE. \ref{table: gauging fermion parity}. We refer to Appendix \ref{sec: review of boson-fermion duality} for further details.

    \item We fix a gauge in which the eigenvalue of $Z_f$ is $1$ at each face $f$. The action of $\bar{U}_fP_{L(f)}$ on the constrained space is replaced by $S_f$ in the gauge fixed Hilbert space.\footnote{\unexpanded{This can be seen by multiplying $\bar{U}_fP_{L(f)}$ by $\tilde{U}_f\tilde{S}_f=1$. We obtain:
    \begin{eqs} \nonumber
        \bar{U}_f P_{L(f)} = \bar{U}_f P_{L(f)}\tilde{U}_f\tilde{S}_f
        =\bar{U}_f P_{L(f)} \bar{U}_f W_{R(f)} P_{L(f)} S_f = W_{R(f)}S_f.
    \end{eqs}
    $W_{R(f)}$ acts as the identity in the fixed gauge.}} Further, the gauge invariant operator $W_tP_t$ becomes $P_t$ after fixing the gauge. The generators of local, gauge invariant operators are thus mapped according to:
    \begin{align} \label{fermioncondensationduality}
        \bar{U}_fP_{L(f)} \rightarrow S_f, \quad W_t P_t \rightarrow P_t.
    \end{align}
    The mapping in Eq.~\eqref{fermioncondensationduality} produces a fermionic Hamiltonian defined on a Hilbert space with a single spinless complex fermion {d.o.f.} at each tetrahedron, as depicted in Fig.~\ref{fig:ai dof}.

    % We map gauge invariant operators in the constrained Hilbert space to operators in an unconstrained Hilbert space with only fermionic {d.o.f.}. In particular, we map:
    % \begin{align} \label{fermioncondensationduality}
    %     U_f \tilde{U}_f S_f \rightarrow S_f, \quad W_t P_t \rightarrow P_t,
    % \end{align}
    % where the relations satisfied by $U_f \tilde{U}_f S_f$ and $W_t P_t$ match the relations satisfied by $S_f$ and $P_t$.
    % The mapping in Eq.~\eqref{fermioncondensationduality} produces a fermionic Hamiltonian defined on a Hilbert space with a single spinless complex fermion {d.o.f.} at each tetrahedron, as depicted in Fig.~\ref{fig:ai dof}.
    
    %   {\color{red} Gauge fixing: we can use gauge transformation \eqref{fermioncondensation} to make $Z_f = 1 ~\forall f$. $W_t P_t$ in \eqref{eq: fermion minimal coupling} becomes simply $P_t$. The generators of gauge-invariant operator operators under the $2$-form gauge constraint $G_e=1$ is mapped to the fermionic theory with with fermionic {d.o.f.} on the tetrahedra:
    % \begin{equation}
    %     U_f \rightarrow S_f, \quad W_t \rightarrow P_t
    % \end{equation}
    % where the gauge constraint is $W_t \equiv \prod_{f\subset t} Z_f = 1$. The duality is summarized in TABLE. \ref{table: gauging fermion parity}.
    % }
    
\end{enumerate}

\begin{table}
    \centering
    \normalsize
    {\renewcommand{\arraystretch}{1.5}
    \begin{tabular}{ | >{\centering\arraybackslash}m{4cm} | >{\centering\arraybackslash} m{4cm} |}
    \hline
    Model with   &  Model with  \\ 
    emergent fermions           & even fermion parity  \\
    \hline
    $\bar{U}_f$ & $S_f$ \\ 
    \hline
    $W_t = \prod_{f \subset t} Z_f$ & $P_t$  \\ 
    \hline
    $\bar{G}_e$ & 1  \\ 
    \hline
    1 & $ \prod_{t} P_t$ \\ 
    \hline
\end{tabular}}
\caption{Fermion condensation implements an operator duality, wherein operators describing a model with an emergent fermion (commute with $\bar{G}_e$) are mapped to operators that act on a fermionic Hilbert space and have even fermion parity (commute with $\prod_{t}P_t$). For simplicity, we have only listed the local generators $\bar{G}_e$ of the anomalous $2$-form symmetry.}
\label{table: gauging fermion parity}
\end{table}

By condensing the emergent fermion in the twisted toric code, we obtain a model for an atomic insulator. More specifically, applying the fermion condensation procedure to $H_\text{ttc}$ yields the atomic insulator Hamiltonian (Appendix \ref{fermioncondensationHttc}):
\begin{align} \label{atomicinsulatorH}
    H_\text{AI} \equiv - \sum_t P_t.
\end{align}
This Hamiltonian has a unique ground state $\ket{\Psi_\text{AI}}$, a product state with zero fermion occupancy at each tetrahedron. Excitations are physical fermions, where $P_t$ has eigenvalue $-1$.
% \footnote{\unexpanded{The excitations corresponding to nontrivial fermion occupancy are, in other words, anomalous $2$-form gauge fluxes. Whereas fermion parity fluxes are equivalent to the anomalous $2$-form gauge charges.}}

The process of gauging a non-anomalous symmetry, such as the $1$-form symmetry in Section~\ref{sec ttc}, can be stated as an operator duality \cite{LG12}. Likewise, the fermion condensation procedure can be implemented by a mapping of operators. We summarize the corresponding duality in Table~\ref{table: gauging fermion parity} and provide more details in Appendix~\ref{sec: review of boson-fermion duality}. 
% Any Hamiltonian that commutes with $G_e$ can be expressed in terms of the operators $U_f$ and $W_t$ \cite{CK18}. 
Notably, the duality corresponding to fermion condensation maps:
\begin{align}
  \bar{U}_f \rightarrow S_f, \quad W_t \rightarrow P_t.
\end{align}
Combining Eqs.~\eqref{eq: fermion minimal coupling} and \eqref{fermioncondensationduality}, we see that the duality is functionally the same as the fermion condensation procedure outlined above.

\section{Bulk construction: $G_f=G \times \ZZ_2^f$} \label{sec:bulkG}

We now generalize the discussion of Section \ref{sec: bulk Z2f} to construct fSPT models protected by a $G_f=G \times \ZZ_2^f$ symmetry. In this case, we require a choice of supercohomology data $(\rho,\nu)$. Therefore, before outlining the construction of the fSPT models, we first review the supercohomology data $(\rho, \nu)$, and introduce corresponding cochains on the manifold $M$. Then, we use the supercohomology data to build a $2$-group SPT model, which, loosely speaking is protected by an interdependent $1$-form and $0$-form symmetry. Next, we gauge the $1$-form symmetry of the $2$-group to obtain the so-called shadow model -- a symmetry-enriched twisted toric code. The shadow model is such that fermion condensation produces a model for the fSPT phase corresponding to the supercohomology data $(\rho,\nu)$. The construction is shown schematically in Fig. \ref{fig:Gf_outline}.

\subsection{Supercohomology data on $M$} \label{sec: supercohomology data}

In Section~\ref{sec: supercohomology data2}, we introduced the supercohomology data $(\rho,\nu)$ as homogeneous functions:
\begin{align}
    \rho: G^4 \to \ZZ_2, \quad \nu:G^5 \to \mathbb{R} / \ZZ,
\end{align}
which, as group cochains, satisfy the relations:
\begin{align} \label{guweneqs2}
    \delta \rho = 0, \quad \delta \nu = \frac{1}{2} \rho \cup_1 \rho.
\end{align}
In what follows, we find it convenient to work with cochains on $M$ -- functions of simplices in the triangulation of $M$. Therefore, in this section, we describe how functions of $G$ variables,
% such as $\rho$ and $\nu$ 
can be pulled back to cochains on $M$.

We use the functions $\bar{\rho}$, $\bar{\rho}^h$, and $\bar{\nu}$ as examples for describing the pull back of functions to cochains on $M$. $\bar{\rho}$, $\bar{\rho}^h$, and $\bar{\nu}$ are defined by:
\begin{align}
    \bar{\rho}(g_1,g_2,g_3) &\equiv \rho(1,g_1,g_2,g_3), \\
    \bar{\rho}^{h}(g_1,g_2) &\equiv \rho(1,h^{-1},g_1,g_2), \\
    \bar{\nu}(g_1,g_2,g_3,g_4) &\equiv \nu(1,g_1,g_2,g_3,g_4),
\end{align}
with $1$ denoting the identity in $G$. The pull backs of these functions play an important role in the construction of the supercohomology models below. We note that the functions are not group cochains, since they fail to be homogeneous. 

To define cochains on $M$, corresponding to $\bar{\rho}$, $\bar{\rho}^h$, and $\bar{\nu}$, we assign an element of $G$ to each vertex in the triangulation of $M$. We refer to the set of $G$ labels $\{g_v\}$ as a $\{g_v\}$-configuration. With $G$ labels on the vertices of $M$, functions of $G^p$ can be pulled back to $p$-cochains on $M$. For each $\{g_v\}$-configuration, we define the cochains $\corho$, $\corho^h$, and $\conu$ on $M$ satisfying:
% \footnote{\unexpanded{We remark that $\bar{\rho}$, $\bar{\rho}^h$, and $\bar{\nu}$ are not group cochains, since they fail to be homogeneous. However, $\corho$, $\corho^h$, and $\conu$ are indeed cochains on $M$. }}
\begin{align} \label{bold bar rho}
    \corho\boldsymbol{(}\langle 123 \rangle\boldsymbol{)} &\equiv \bar{\rho}(g_1,g_2,g_3), \\
    \corho^h\boldsymbol{(}\langle 12 \rangle\boldsymbol{)} &\equiv \bar{\rho}^{h}(g_1, g_2), \\
    \conu\boldsymbol{(}\langle 1234 \rangle\boldsymbol{)} &\equiv \bar{\nu}(g_1,g_2,g_3,g_4),
\end{align}
for an arbitrary face $\langle 123 \rangle$, edge $\langle 12 \rangle$, and tetrahedron $\langle 1234 \rangle$.

The Hamiltonians discussed below are defined on triangulated manifolds with a $G$ {d.o.f.} at every vertex, such as in Figs.~\ref{fig: groupcohodof} and \ref{fig: Gai dof2}, and a set of basis states can be labeled by $\{g_v\}$-configurations. Hence, to simplify the notation in the construction of the supercohomology models, we introduce diagonal operators for each $\{g_v\}$-dependent cochain on $M$. The operator associated to the $\{g_v\}$-dependent $p$-cochain $\boldsymbol{c}_{\scriptscriptstyle{\{g_v\}}}$ and a $p$-simplex $s$ is defined as:
\begin{align}
    \hat{\boldsymbol{c}}(s) \equiv \sum_{\{g_v\}} \boldsymbol{c}_{\scriptscriptstyle{\{g_v\}}}(s)\ket{\{g_v\}}\bra{\{g_v\}}.
\end{align}
% We often omit the circumflex, i.e., $\hat{\boldsymbol{c}}(s) \to \boldsymbol{c}(s)$, to de-clutter the notation. The resulting operator should not be confused with a $\ZZ_2$ or $\mathbb{R}/\ZZ$ valued cochain on $M$. 
The operators associated to $\corho$, $\corho^h$, and $\conu$ are thus:
\begin{align}
    \rhohat(f) &\equiv \sum_{\{g_v\}}\corho(f)\ket{\{g_v\}}\bra{\{g_v\}} \\
    \rhohath(e) &\equiv \sum_{\{g_v\}} \corho^h(e) \ket{\{g_v\}}\bra{\{g_v\}}\\
    \nuhat(t) &\equiv \sum_{\{g_v\}} \conu(t) \ket{\{g_v\}}\bra{\{g_v\}}
\end{align}
for a choice of face $f$, edge $e$, and tetrahedron $t$.
Unless otherwise stated, it should be assumed that the operators are tensored with the identity on any other {d.o.f.} in the model.

The coboundary operator and cup products can naturally be extended to the operators at the cochain level. For example, the coboundary of $\rhohat$ is the operator:\footnote{\unexpanded{In fact, $\delta \rhohat$ is equivalent to the operator $\hat{\boldsymbol{\rho}}$, corresponding to the cochain $\boldsymbol{\rho}_{\scriptscriptstyle{\{g_v\}}}$ and the function $\rho$ in the natural way. This follows from the coboundary relation in Eq.~\eqref{guweneqs2}. We note that, although $\delta \corho$ is indeed equal to $\boldsymbol{\rho}_{\scriptscriptstyle{\{g_v\}}}$, this does not imply that $\rho$ is a group coboundary. The equality holds for cochains on $M$. Lacking homogeneity, $\bar{\rho}$ is not a group cochain.}}
%$\hat{\boldsymbol{\rho}}$ corresponding to the cochain $\boldsymbol{\rho}_{\scriptscriptstyle{\{g_v\}}}$ and the function $\rho$:
\begin{align}
    \delta \rhohat (t) = \sum_{\{g_v\}} \delta \corho (t)\ket{\{g_v\}}\bra{\{g_v\}}, 
    % \\ \nonumber
    % &= \sum_{\{g_v\}} \boldsymbol{\rho}_{\scriptscriptstyle{\{g_v\}}} (t)\ket{\{g_v\}}\bra{\{g_v\}},
\end{align}
where $t$ is an arbitrary tetrahedron. Similarly, the cup-$1$ product $\rhohat \cup_1 \bface(t)$ should be interpreted as:
\begin{align}
    \rhohat \cup_1 \bface = \sum_{\{g_v\}} \corho \cup_1 \bface (t) \ket{\{g_v\}}\bra{\{g_v\}}.
\end{align}
Using these operators we now build a Hamiltonian describing a certain $2$-group SPT phase.

% The next three sections are devoted to constructing a fSPT model whose responses to probing with symmetry defects correspond to a choice of supercohomology data $(\rho,\nu)$. To make the connection explicit, we express the fSPT Hamiltonian in terms of the operators $\rhohat(f)$ and $\nuhat(t)$. The first step is to use $\rho$ and $\nu$ to build a Hamiltonian describing a certain $2$-group SPT phase.

\subsection{$2$-group SPT} \label{sec: 2-groupspt}

In this section, we construct a model for a $2$-group SPT phase, given a set of supercohomology data $(\rho,\nu)$. The particular $2$-group symmetry is dependent upon $G$ and the group cochain $\rho$. For simplicity, we describe the relevant $2$-group symmetry in terms of its representation on a lattice. More details on $2$-groups including a formal definition can be found in Appendix \ref{sec: 2-group extension}. Our model for the $2$-group SPT phase is based on a Euclidean spacetime picture presented in Ref.~\cite{KT17}, and we elaborate on the connection to this perspective at the end of this section.

% The first step in our construction of a $G \times \ZZ_2^f$ fSPT model corresponding to the supercohomology data $(\rho,\nu)$ is to build a model for a certain $2$-group SPT phase. The particular $2$-group is determined by $G$ and the group cochain $\rho$. For simplicity, we describe the relevant $2$-group symmetry in terms of its representation on a lattice. However, we provide more details on $2$-groups including a formal definition in Appendix [APPENDIX]. Our construction is based on the Euclidean spacetime picture presented in Ref.~\cite{KT17}. We elaborate on the connection to the Euclidean spacetime perspective in Appendix [APPENDIX].

\begin{figure}[t]
\centering
\includegraphics[width=0.3\textwidth, trim={760 440 800 360},clip]{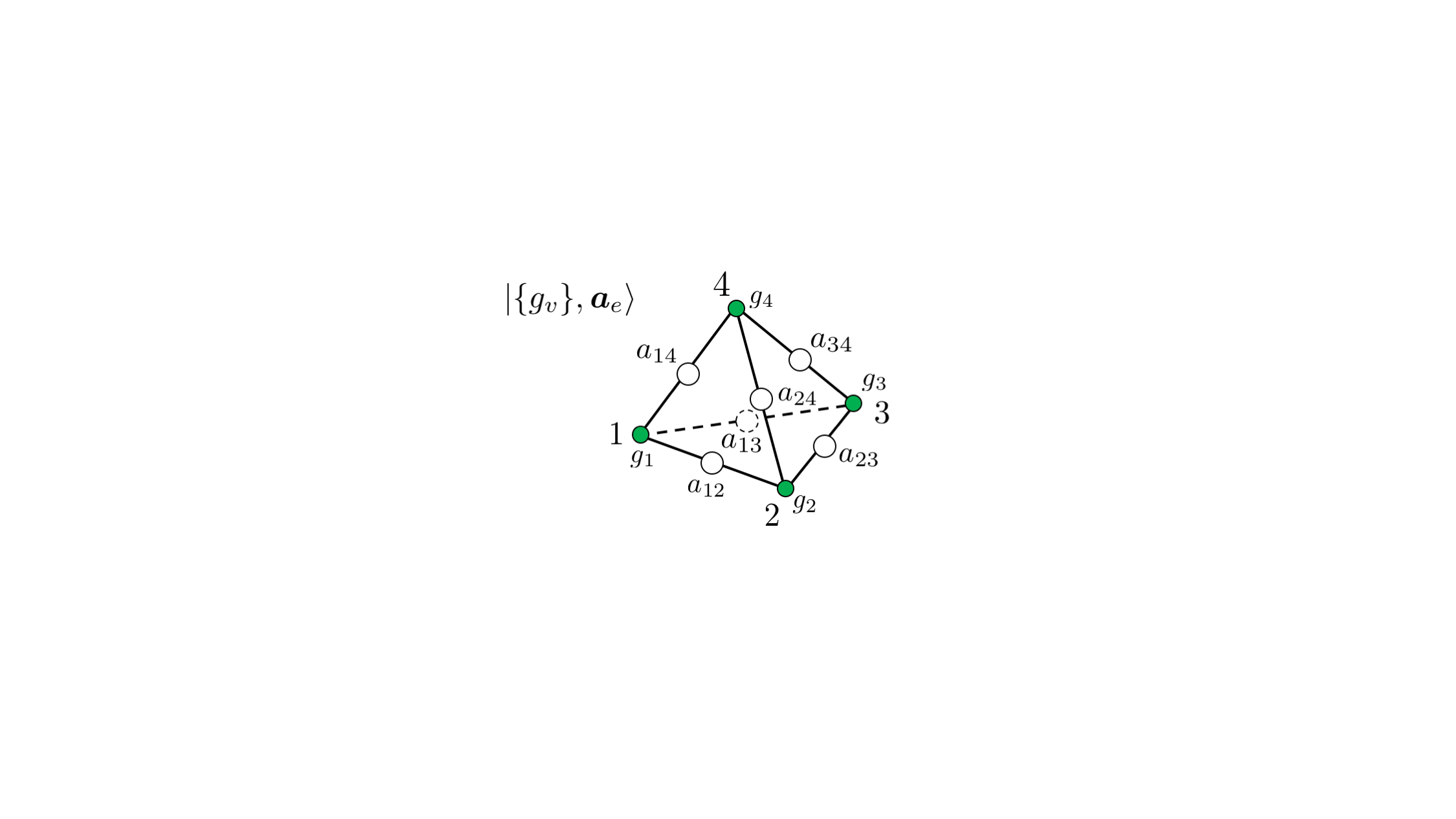}
\caption{The Hilbert space for the $2$-group SPT model has $G$ {d.o.f.} on vertices and $\ZZ_2$ {d.o.f.} on edges. We have labeled the states at the edges by the value of $\ba_e(e)=a_e$.}
\label{fig: G1formdof}
\end{figure}

The model for the $2$-group SPT phase is defined on a Hilbert space consisting of a $\ZZ_2$ {d.o.f.} at each edge and a $G$ {d.o.f.} at each vertex of a triangulation of $M$ (see Fig.~\ref{fig: G1formdof}). As described in the previous section, a basis for the $G$ {d.o.f.} is given by configuration states $\ket{\{g_v\}}$, and as in section \ref{sec: 1formspt}, a basis for the edge {d.o.f.} can be formed by states $\ket{\ba_e}$ labeled by $\ZZ_2$ $1$-cochains $\ba_e$. Thus, we use the collection of states of the form $\ket{\{ g_v \},\ba_e}$ as a basis for the total Hilbert space.

The $2$-group symmetry has both a $\ZZ_2$ $1$-form symmetry and a certain $0$-form symmetry parameterized by elements of $G$. Similar to Section \ref{sec: 1formspt}, we represent the $1$-form symmetry using the operators:
\begin{align} \label{2group 1formpart}
    A_\Sigma = \prod_{e \perp \Sigma}X_e,
\end{align}
where the product is over edges intersected by the closed surface $\Sigma$ on the dual lattice. 
The action of the $1$-form symmetry operator on a basis state is explicitly:
\begin{align}
    A_\Sigma \ket{\{g_v\},\ba_e} = \ket{\{g_v\},\ba_e+\bsig},
\end{align}
for a closed $1$-cochain $\bsig$ corresponding to $\Sigma$ [see Eq.~\eqref{1formsym on basis}]. 

We represent the $0$-form symmetry action associated to $h \in G$ as:
\begin{align} \label{2group 0formpart}
     V_\rho(h) \equiv V(h) \prod_e X_e^{\rhohath(e)}.
\end{align}
Here, $V(h)$ acts by (left) group multiplication of $h$ on the vertex {d.o.f.}:
\begin{align} 
    V(h)\equiv \sum_{\{g_v\},\ba_e}\ket{\{hg_v\},\ba_e}\bra{\{g_v\},\ba_e}.
\end{align}
Therefore, the action of the $0$-form symmetry operator $V_\rho(h)$ on a basis state is:
\begin{align} 
    V_\rho(h)\ket{\{g_v\},\ba_e} = \ket{\{hg_v\},\ba_e+\corho^h}.
\end{align}

We now construct a model for an SPT phase protected by the $2$-group symmetry generated by the operators in Eqs.~\eqref{2group 1formpart} and \eqref{2group 0formpart}. We start with the $1$-form paramagnet Hamiltonian $H_0$ in Section~\ref{sec: 1formspt} and the decoupled $G$ paramagnet $H^G$ from Section~\ref{sec: group cohomology models}:
\begin{align}
    H_0^G \equiv H_0 + H^G.
\end{align}
% The $G$ paramagnet Hamiltonian $H^G$ is given by a sum of projectors $\mathcal{P}_v$ onto a $G$ symmetric state at the vertex $v$, i.e.:
% \begin{align}
%     H^G \equiv - \sum_v \mathcal{P}_v,
% \end{align}
% with $\mathcal{P}_v$ defined as: 
% % \footnote{Again, we ignore normalization for convenience.}
% \begin{align}
%     \mathcal{P}_v \equiv \frac{1}{|G|} \left(  \sum_{g_v} \ket{g_v} \right)\left( \sum_{g_v} \bra{g_v} \right).
% \end{align}
The $2$-group SPT Hamiltonian is prepared from $H^G_0$ by conjugation with the FDQC $\Ugspt$, which is a composition of two FDQC:
\begin{align} \label{u2 def2}
    \Ugspt = \Ugspt'\Uspt.
\end{align}
% \begin{align} \label{u2 def}
%     \Ugspt = \prod_{t} \Big[ e^{2 \pi i O_t \left[\nuhat(t)+\frac{1}{2}\rhohat \cup_1 \rhohat(t)\right]}  \prod_{f \subset t} W_{f}^{\rhohat \cup_1 \bface(t)}\Big].
% \end{align}
$\Uspt$ is the FDQC from Section~\ref{sec: 1formspt} 
% (tensored with the identity on the $G$ {d.o.f.}) 
-- it prepares a $1$-form SPT model from a $1$-form paramagnet:
\begin{align}\label{u1 2group}
    \Uspt = \sum_{\{g_v\},\ba_e} (-1)^{\int \ba_e \cup \delta \ba_e}\ket{\{g_v\},\ba_e}\bra{\{g_v\},\ba_e}.
\end{align}
$\Ugspt'$ is a FDQC that couples the vertex {d.o.f.} to the edge {d.o.f.} and ensures that the model is $2$-group symmetric. Explicitly, $\Ugspt'$ is:
\begin{align} \label{u2 def}
    \Ugspt' \equiv \prod_{t} \Big[ e^{2 \pi i O_t \left[\nuhat(t)+\frac{1}{2}\rhohat \cup_1 \rhohat(t)\right]}  \prod_{f \subset t} W_{f}^{\rhohat \cup_1 \bface(t)}\Big].
\end{align}
Here, the $O_t$ is the orientation of each tetrahedron, defined in Fig. \ref{fig: branchingstructure}, the second product is over the faces in the boundary of $t$, and $W_f$ is given in Eq.~\eqref{Wfdef} .
% Here, the second product is over faces in the boundary of $t$, and $\rhohat \cup_1 \bface(t)$ is the operator:
% \begin{align}
%     \rhohat \cup_1 \bface(t) \equiv \sum_{\{g_v\},\ba_e} \corho \cup_1 \bface(t) \ket{\{g_v\},\ba_e}\bra{\{g_v\},\ba_e}.
% \end{align}
The $2$-group SPT Hamiltonian $H_2$ is thus:
\begin{align} \label{2group from 1form}
    H_2 \equiv \Ugspt H_0^G \Ugspt^\dagger.
\end{align}
In Appendix~\ref{app: 2group sym}, we prove that $H_2$ is indeed invariant under the $2$-group symmetry operators in Eqs.~\eqref{2group 1formpart} and \eqref{2group 0formpart}. 

% The ground state $\ket{\Psi_2}$ of $H_2$ is obtained from the ground state of $H_0^G$ by applying $\Ugspt$. This yields the state:
% \begin{align} \label{2group spt state}
%     \ket{\Psi_2} = \sum_{\{g_v\},\ba_e} \prod_t e^{2 \pi i O_t \coalpha(t)}\ket{\{g_v\}, \ba_e},
% \end{align}
% where $\coalpha$ is the $\mathbb{R}/\ZZ$ valued $3$-cochain on $M$ given by:
% \begin{multline}
%     \coalpha \equiv \conu + \frac{1}{2}\corho \cup_1 \corho + \frac{1}{2}\corho \cup_1 \delta \ba_e \\ + \frac{1}{2}\ba_e\cup \delta \ba_e.
% \end{multline}
% We note that the first three terms of $\coalpha$ are produced by $\Ugspt'$, while the last term is a result of the action by $\Uspt$.

% \begin{figure*}[htb]
% \centering
% \includegraphics[width=0.75\textwidth]{configuration.png}
% \caption{We have used the normalized condition for $\rho$, i.e.,  $\rho(1,1,g,h)=0$.}
% \label{fig: dPhi_1}
% \end{figure*}

The important property of this $2$-group SPT Hamiltonian is revealed after gauging the $1$-form symmetry of the total $2$-group symmetry. In the next section, we show that the remaining $0$-form $G$ symmetry ``fractionalizes'' on the loop-like $1$-form gauge charges according to the group cochain $\rho$. This property is key to characterizing the fSPT phase that results from this construction.

Before gauging the $1$-form symmetry, however, we motivate the $2$-group SPT model from the spacetime construction in Ref.~\cite{KT17}. The spacetime model is defined on $CM$, the cone of the closed $3$-manifold $M$. We orient the time-like edges, those connected to the additional spacetime point, towards the vertices of $M$ (Fig.~\ref{fig: CM}). We also place a $G$ {d.o.f.} on the additional spacetime point of $CM$ as well as a $\ZZ_2$ {d.o.f.} on each time-like edge. The partition function for the $2$-group SPT phase is taken to be:
\begin{align}
   \mathcal Z_2 \equiv \sum_{\{g_v\}, \ba_e} \prod_{\Delta_4} e^{2 \pi i O_{\Delta_4} \calpha(\Delta_4)},
\label{eq: Z2}
\end{align}
where the product is over $4$-simplices $\Delta_4$, and $\calpha$ is a certain $\{g_v\}$-dependent $\mathbb{R}/\ZZ$ valued cochain on $M$. In particular, $\calpha$ is defined in terms of the supercohomology data $(\rho,\nu)$ as:
\begin{align} \label{alpha def}
    \calpha \equiv \cnu+\frac{1}{2} \crho \cup_1 \coeps + \frac{1}{2} \coeps \cup \coeps.
\end{align} 
To define $\calpha$, we have introduced the cochains $\crho$, $\cnu$, and $\coeps$. $\crho$ is the $\ZZ_2$ $3$-cochain given by:
\begin{align}
    \crho \boldsymbol{(} \langle 0123 \rangle \boldsymbol{)} \equiv \rho(g_0,g_1,g_2,g_3),
\end{align}
for any $3$-simplex $\langle 0123 \rangle$, and $\cnu$ is the $\mathbb{R}/\ZZ$ valued $4$-cochain:
\begin{align}
    \cnu \boldsymbol{(} \langle 01234 \rangle \boldsymbol{)} \equiv \nu (g_0,g_1,g_2,g_3,g_4),
\end{align}
where $\langle 01234 \rangle$ is an arbitrary $4$-simplex. Lastly, $\coeps$ denotes the $2$-cochain:
\begin{align}
    \coeps \equiv \corho + \delta \ba_e.
\end{align}
{As elaborated on in Appendix~\ref{sec: 2-group extension}, $\calpha$ is the pullback of a 2-group cocycle \cite{KT17}:
\begin{equation}
    \alpha \equiv \nu + \frac{1}{2} \rho \cup_1 \eps + \frac{1}{2} \eps \cup \eps,
\label{eq: alpha_4 as 2-group cocycle}
\end{equation}
which acts on the 2-group classifying space.}
% {One important property of $\coeps$ is $\delta \coeps = \crho$, which implies that $\calpha$ is closed, i.e. $\delta \calpha = 0$. $\calpha$ is the pullback of the 2-group cocycle
% \begin{equation}
%     \alpha \equiv \nu + \frac{1}{2} \rho \cup_1 \eps + \frac{1}{2} \eps \cup \eps,
% \label{eq: alpha_4 as 2-group cocycle}
% \end{equation}
% which acts on the 2-group classifying space (defined in Appendix \ref{sec: 2-group extension}).}

%Evaluating $\alpha$ for the spacetime configuration $\ket{ d\Phi}$ defined in Fig. \ref{fig: CM_dPhi} gives the partition function $\mathcal Z_2$ in Eq. \eqref{eq: Z2}.}

To write down the SPT state, we consider the amplitude for a fixed configuration:
\begin{align}
    \Psi_2 \left(\{g_v\}, \ba_e \right) \equiv \prod_{\Delta_4} e^{2 \pi i O_{\Delta_4} \calpha(\Delta_4)}.
\end{align}
The amplitude is invariant under re-triangulations of $CM$, so  by a series of re-triangulations, we can remove both the dependence on the $G$ {d.o.f.} at the additional spacetime point and the dependence on  the $\ZZ_2$ {d.o.f.} at the time-like edges. Therefore, without affecting the amplitude, the additional {d.o.f.} can be set to the identity state. It can be checked that the resulting amplitude gives precisely the wave function for the ground state of the $2$-group SPT Hamiltonian $H_2$.

\subsection{Shadow model} \label{sec: shadow model}

The next step in our construction is to build the shadow model. We start by gauging the $\ZZ_2$ $1$-form symmetry of the $2$-group SPT Hamiltonian $H_2$. Then, we perform a change of basis to ensure that the remaining $0$-form symmetry forms an onsite representation of $G$. The result is the shadow model -- a $G$-symmetry-enriched twisted toric code, where the $G$ symmetry fractionalizes on the loop-like $1$-form gauge charges. We compute the symmetry fractionalization on the loop-like excitations explicitly, and show that the fractionalization is governed by the group cochain $\rho$ in the supercohomology data $(\rho,\nu)$. In the subsequent section, we condense the emergent fermion in the shadow model to complete the construction of the $G\times \ZZ_2^f$ fSPT model corresponding to the supercohomology data $(\rho,\nu)$.

\begin{figure}[t]
\centering
\includegraphics[width=0.3\textwidth, trim={760 440 800 360},clip]{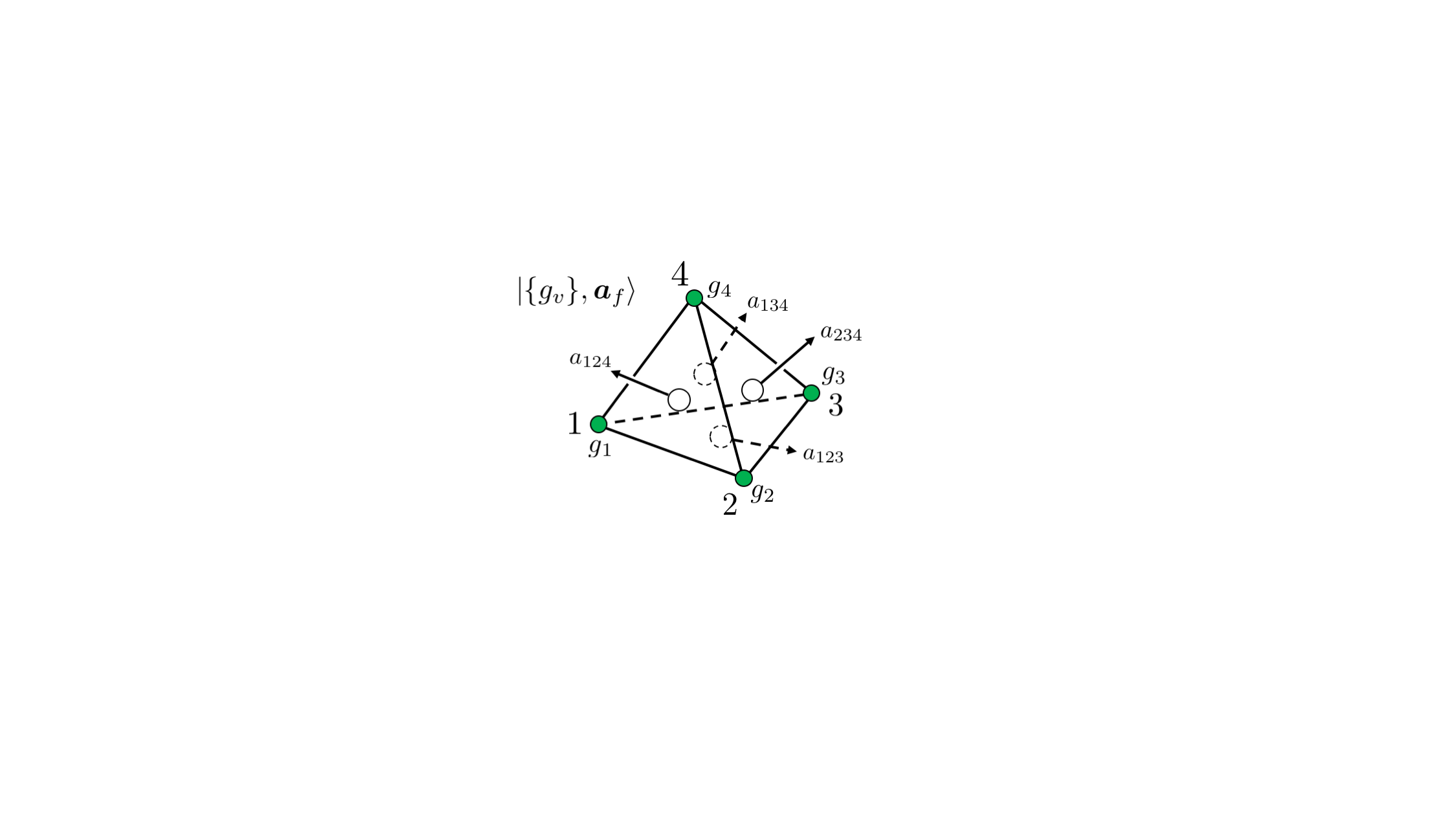}
\caption{The Hilbert space for the shadow model consists of $G$ {d.o.f.} on vertices and $\ZZ_2$ {d.o.f.} on faces. In the state $\ket{\{g_v\},\ba_f}$, the values of the $\ZZ_2$ {d.o.f.} at the faces are given by $\ba_f(f)=a_f$.}
\label{fig: GH2configuration}
\end{figure}

We gauge the $1$-form symmetry of $H_2$ following the discussion in Section~\ref{sec ttc} and Appendix~\ref{app: 1form gauging quantum states}. As described earlier in the text, the procedure for gauging the $1$-form symmetry is functionally equivalent to applying a duality that maps the $1$-form symmetric operators $X_e$ and $W_f$ according to:
\begin{align} \label{1form gauging duality2}
     X_e \rightarrow \prod_{f \supset e}X_f, \quad W_f \rightarrow Z_f.
\end{align}
For the $2$-group SPT model, the gauging procedure maps to a Hilbert space composed of $\ZZ_2$ {d.o.f.} on faces and $G$ {d.o.f.} on vertices (Fig.~\ref{fig: GH2configuration}). A basis for this Hilbert space is given by states of the form $\ket{\{g_v\},\ba_f}$, where we have used notation from Section \ref{sec ttc} to label a configuration of the face {d.o.f.} with a $2$-cochain $\ba_f$.

To apply the operator duality in Eq.~\eqref{1form gauging duality2} to $H_2$, we rewrite $H_2$ as:
\begin{align}
    H_2 = \Ugspt' \left(\Uspt H^G_0 \Uspt^\dagger \right) \Ugspt'^\dagger,
\end{align}
and map $\Uspt H^G_0 \Uspt^\dagger$ and $\Ugspt'$ independently. $\Uspt$ prepares the $1$-form SPT model in Section~\ref{sec: 1formspt} from $H^G_0$, so gauging the $1$-form symmetry of $\Uspt H^G_0 \Uspt^\dagger$ yields $H_\text{ttc}^G$, a twisted toric code with a decoupled $G$ paramagnet:
\begin{align}
    H_\text{ttc}^G \equiv H_\text{ttc} + H_G^0.
\end{align}
Importantly, we retain the local projectors $\mathcal{P}_R^\text{0-flux}$ in Eq. \eqref{equ:P0flux_R} from the gauging procedure (for appropriate choices of regions $R$). They are needed to ensure that the shadow model is $G$ symmetric.\footnote{\unexpanded{Without the projectors, the shadow model is only guaranteed to be $G$ symmetric up to factors of $W_t$.}}

We next apply the duality to $\Ugspt'$. Before doing so, however, we multiply $\Ugspt'$ by the identity:\footnote{\unexpanded{The operator $\prod_{f \subset t} W_f$ is identically $1$, since: 
    $\prod_{f \subset t} W_f=\prod_{f \subset t}\prod_{e \subset f}Z_e = \prod_{e \subset t}Z^2_e=1$,
 where the last product is over edges $e$ in $t$.}}
\begin{align} \label{Wf U2 identity}
    1 = \prod_t \Big( \prod_{f \subset t} W_f\Big)^{\int \rhohat \cup_2 \boldsymbol{t} }.
\end{align}
Here, we have used the cup-$2$ product $\cup_2$, defined in Appendix~\ref{app: terminology}. The $\cup_2$ product of $\rhohat$ and $\boldsymbol{t}$ evaluated on the tetrahedron $\langle 1234 \rangle$ is:
\begin{align}
    \rhohat \cup_2 \boldsymbol{t}\boldsymbol{(}\langle 1234 \rangle \boldsymbol{)} = \left[\rhohat\boldsymbol{(}\langle123\rangle\boldsymbol{)}+\rhohat\boldsymbol{(}\langle134\rangle\boldsymbol{)}\right]\boldsymbol{t}\boldsymbol{(}\langle 1234 \rangle \boldsymbol{)}.
\end{align}
% \begin{multline}
%     \Us' \equiv \sum_{\{g_v\},\ba_f}\prod_{t}  \Big[e^{2 \pi i \left(\conu+\frac{1}{2}\corho \cup_1 \corho\right)(t)} \\ \times \prod_{f\subset t}Z_{f}^{\corho \cup_1 \bface (t)}\Big]\ket{\{g_v\},\ba_f}\bra{\{g_v\},\ba_f}.
% \end{multline}
After multiplying by the operator in Eq.~\eqref{Wf U2 identity}, $\Ugspt'$ becomes:
\begin{multline}
    \Ugspt' \equiv \prod_{t}  \Big[e^{2 \pi i O_t \left[\nuhat(t)+\frac{1}{2}\rhohat \cup_1 \rhohat(t)\right]} \prod_{f\subset t}W_{f}^{\rhohat \cup_1 \bface (t)} \Big] \\ \times \prod_t\Big( \prod_{f \subset t} W_f\Big)^{\int \rhohat \cup_2 \boldsymbol{t} }.
\end{multline}
While the modification to $\Ugspt'$ has no affect on the ground state subspace of the shadow model, it is crucial to the symmetry of the FDQC in Section~\ref{sec: supercohomology model} that prepares the fSPT ground state (see Appendix~\ref{app: sym var Us} for details).
After the modification, $\Ugspt'$ maps to the FDQC $\Us'$:
\begin{multline}
    \Us' \equiv \prod_{t}  \Big[e^{2 \pi i O_t \left[\nuhat(t)+\frac{1}{2}\rhohat \cup_1 \rhohat(t)\right]} \prod_{f\subset t}Z_{f}^{\rhohat \cup_1 \bface (t) }\Big] \\ \times \prod_t W_t^{\int \rhohat \cup_2 \boldsymbol{t} }.
\end{multline}
The gauging procedure produces the Hamiltonian:
\begin{align} \label{preshadowH}
    H_s' \equiv \Us'H_\text{ttc}^G\Us'^\dagger.
\end{align}

$H_s'$ is invariant under a $0$-form $G$ symmetry given by applying the gauging duality to the $0$-form symmetry of the $2$-group [Eq.~\eqref{2group 0formpart}]. The $0$-form symmetry operator corresponding to $h\in G$ is mapped to:
\begin{align} \label{shadow 0form 1}
   V_\rho'(h) &\equiv V(h) \prod_{f=\langle 123 \rangle} X_f^{\rhohath\boldsymbol{(}\langle 23 \rangle\boldsymbol{)}+\rhohath\boldsymbol{(}\langle 13 \rangle\boldsymbol{)}+\rhohath\boldsymbol{(}\langle 12 \rangle\boldsymbol{)}} \\
   &= \sum_{\{g_v\},\ba_f}\ket{\{hg_v\},\ba_f+\delta \corho^h}\bra{\{g_v\},\ba_f}.
\end{align}
We notice that the resulting symmetry $V_\rho'(h)$ is not necessarily onsite. This is because the term $\delta \corho^h(f)$ in the second line of Eq.~\eqref{shadow 0form 1} depends on the $\{g_v\}$-configuration at the vertices of $f$, in general.

To obtain the shadow model, we make a local change of basis -- implemented by a FDQC, which makes the remaining $0$-form $G$ symmetry onsite. We implement the change of basis with the unitary operator:
%\footnote{By local, we mean that the basis transformation can be implemented by a FDQC.} change of basis, which guarantees that the $0$-form $G$ symmetry is onsite. We implement the change of basis with the unitary operator:
\begin{align}
   \mathcal{R} \equiv  \prod_f X_f^{\rhohat(f)}.
\label{eq: bulk basis transform}
\end{align}
Every state $\ket{\Psi}$ and operator $\mathcal{O}$ is transformed as:
\begin{equation} 
            \ket{\Psi} \rightarrow \mathcal{R} \ket{\Psi}, \quad
            \mathcal{O} \rightarrow \mathcal{R} \mathcal{O} \mathcal{R}^\dagger.
\label{shadow basis trans}
\end{equation}
Under the transformation above, $V_\rho'(h)$ becomes:
\begin{align} \label{symactionbasistrans}
    \sum_{\{g_v\},\ba_f}\ket{\{hg_v\},\ba_f+\overline{\boldsymbol{\rho}}_{\scriptscriptstyle{\{hg_v\}}}+\delta \corho^h} \bra{\{g_v\},\ba_f+\corho}.
\end{align}
It can be shown, using that $\rho$ is a cocycle [Eq.~\eqref{guweneqs2}], that $\corho$ and $\corho^h$ are related by:
\begin{align} \label{corhohrelation}
    \overline{\boldsymbol{\rho}}_{\scriptscriptstyle{\{hg_v\}}}+\delta \corho^h = \corho.
\end{align}
Thus, the operator in Eq.~\eqref{symactionbasistrans} is equivalent to the onsite symmetry action:
\begin{align} \label{symmetry after basis}
    V(h) \equiv \sum_{\{g_v\},\ba_f}\ket{\{hg_v\},\ba_f} \bra{\{g_v\},\ba_f}.
\end{align}

We apply the basis transformation $\mathcal{R}$ to $H'_s$ by conjugation. This produces the shadow model:
\begin{align} \label{shadowH1}
    H_s&\equiv \mathcal{R}  H_s' \mathcal{R}^\dagger = \mathcal{R} \left(\Us' H^G_\text{ttc} \Us' \right)\mathcal{R}^\dagger = \Us H_\text{ttc}^G \Us^\dagger,
\end{align}
% \begin{align} \label{shadowH1}
%     H_s &\equiv \mathcal{R}  H_s' \mathcal{R}^\dagger \\ \nonumber &= \mathcal{R} \left(\Us' H_\text{ttc} \Us' \right)\mathcal{R}^\dagger   \\ \nonumber &= \Us H_\text{ttc}^G \Us^\dagger,
% \end{align}
where, in the last line, we have introduced $\Us$:
\begin{align} \label{eq: Us original1}
    \Us \equiv \mathcal{R}\Us'.
\end{align}
The FDQC $\Us$ is explicitly:
\begin{multline} \label{eq: Us def}
    \Us = \prod_f X_f^{\rhohat(f)} \prod_{t} \Big[e^{2 \pi i O_t \left[\nuhat(t)+\frac{1}{2}\rhohat \cup_1 \rhohat(t)\right]} \prod_{f\subset t}Z_{f}^{\rhohat \cup_1 \bface (t)}\Big] \\ \times \prod_t W_t^{\int \rhohat \cup_2 \boldsymbol{t} }.
\end{multline}

The Hamiltonian $H_s$ describes a symmetry-enriched twisted toric code. This is because it is both $G$ symmetric and can be constructed from $H^G_\text{ttc}$ using the FDQC $\Us$. The symmetry of $H_s$ follows from the symmetry of the $2$-group SPT Hamiltonian $H_2$ (Appendix~\ref{app: 2group sym}).

Similar to the twisted toric code, $H_s$ admits loop-like excitations as well as point-like excitations with fermionic statistics. In the twisted toric code, the loop-like excitations are created at the boundary of a surface $\sigma$ using the membrane operator $M_\sigma$, defined in Eq.~\eqref{membrane operator}.
% :
% \begin{align}
%     M_\sigma = \prod_{f \subset \sigma}Z_f.
% \end{align}
Therefore, the loop-like excitations can be created in the shadow model with the operator:
\begin{align}
    M^s_\sigma \equiv \Us M_\sigma \Us^\dagger = \prod_{f \subset \sigma} \left[(-1)^{\rhohat(f)}Z_f\right],
\end{align}
where in the second equality we have commuted $\Us$ past the Pauli Z operators in $M_\sigma$. Likewise, emergent fermion string operators in the shadow model can be formed by conjugating the twisted toric code string operators $\mathcal{S}_p$ and $\tilde{\mathcal{S}}_p$ by $\Us$:
\begin{align}
    \mathcal{S}^s_p \equiv \Us \mathcal{S}_p \Us^\dagger, \quad \tilde{\mathcal{S}}^s_p \equiv \Us \tilde{\mathcal{S}}_p \Us^\dagger.
\end{align}

For certain choices of supercohomology data $(\rho,\nu)$, the corresponding shadow model describes a \textit{nontrivial}, symmetry-enriched twisted toric code. In particular, the group cochain $\rho$ determines the fractionalization (defined below) of the $G$ symmetry on the loop-like $1$-form gauge charges. This symmetry fractionalization partially characterizes the $G$-symmetry-enriched twisted toric code phase \cite{C15}.

\vspace{1.5mm}
\noindent \begin{center}\emph{Symmetry fractionalization on loop-like excitations:}\end{center}
\vspace{1.5mm}

In what follows, we compute the symmetry fractionalization on the loop-like excitations of the shadow model, explicitly. We do so by considering an excited state of $H_s$ obtained by applying $M^s_\sigma$ to a ground state $\ket{\Psi_s}$ of $H_s$.  The resulting state $M^s_\sigma \ket{\Psi_s}$ has a single loop of $1$-form gauge charge along $\partial \sigma$, the boundary of $\sigma$. We study the effect of the $G$-symmetry action on this state to determine the fractionalization of the symmetry on the loop-like excitation. 

To set up the computation and make the discussion more precise, we introduce an effective Hilbert space in the vicinity of the $1$-form gauge charge. We define the state $|\psi_\text{ttc},\{g_v\}_{\partial \sigma}\rangle$ to be the ground state of the twisted toric code Hamiltonian $H^G_\text{ttc}$ with fixed $G$ configuration $\{g_v\}_{\partial \sigma}$ at vertices $v$ contained in $\partial \sigma$. The set of states $\big\{\ket{\Psi_\text{ttc},\{g_v\}_{\partial \sigma}}\big\}$ spans a Hilbert space with dimension $|G|^N$, where $N$ is the number of vertices in $\partial \sigma$. The effective Hilbert space is then formed by the states $\big \{ \ket{\Psi^{\partial \sigma}_s,\{g_v\}_{\partial \sigma}} \big \}$, where $\ket{\Psi^{\partial \sigma}_s,\{g_v\}_{\partial \sigma}}$ is defined by:
\begin{align}
    \ket{\Psi^{\partial \sigma}_s,\{g_v\}_{\partial \sigma}}\equiv M^s_\sigma \Us \ket{\Psi_\text{ttc},\{g_v\}_{\partial \sigma}}.
\end{align}
We note that, in particular, the state $M^s_\sigma \ket{\Psi_s}$ belongs to the effective Hilbert space:
\begin{align}
    M^s_\sigma \ket{\Psi_s}=\sum_{\{g_v\}_{\partial \sigma}}\ket{\Psi^{\partial \sigma}_s,\{g_v\}_{\partial \sigma}}.
\end{align}
Heuristically, a state $\ket{\Psi^{\partial \sigma}_s,\{g_v\}_{\partial \sigma}}$ in the effective Hilbert space resembles the ground state $\ket{\Psi_s}$ far away from $\partial \sigma$.\footnote{\unexpanded{More concretely, any reduced density matrix of the state $\ket{\Psi^{\partial \sigma}_s,\{g_v\}_{\partial \sigma}}\bra{\Psi^{\partial \sigma}_s,\{g_v\}_{\partial \sigma}}$ will agree with the reduced density matrix of $\ket{\Psi_s}\bra{\Psi_s}$ on regions far from $\partial \sigma$.}} Since $\ket{\Psi_s}$ is symmetric, we expect the symmetry to act as the identity on $\ket{\Psi^{\partial \sigma}_s,\{g_v\}_{\partial \sigma}}$ away from $\partial \sigma$. However, the symmetry may act non-identically on the states in the effective Hilbert space, and we define the projection of the symmetry action to the effective Hilbert space to be the \textit{effective} symmetry action on a $1$-form gauge charge.  The fractionalization of the $G$-symmetry action is an obstruction to realizing the effective symmetry action onsite.
% can then be deduced from this effective symmetry action.

We next determine the effective symmetry action on the loop-like excitation by acting on an arbitrary state $\ket{\Psi_s^{\partial \sigma},\{g_v\}_{\partial \sigma}}$ in the effective Hilbert space with the symmetry operator $V(h)$:
\begin{align}\label{eff sym flux 1}
    V(h)\ket{\Psi_s^{\partial \sigma},\{g_v\}_{\partial \sigma}} = V(h)M^s_\sigma \Us \ket{\Psi_\text{ttc},\{g_v\}_{\partial \sigma}}.
\end{align}
The expression on the right hand side of Eq.~\eqref{eff sym flux 1} can be evaluated further by commuting $V(h)$ past $M^s_\sigma$ and $\Us$. Using the relation in Eq.~\eqref{corhohrelation}, we find:
\begin{align}
    V(h) M_\sigma^s = \Big[\prod_{e \subset \partial \sigma}(-1)^{\rhohat^{h^{-1}}(e)}\Big]M^s_\sigma V(h),
\end{align}
and as shown in Appendix~\ref{app: sym var Us}, $V(h)$ commutes with $\Us$ up to factors of $\bar{G}_e$:
\begin{align} \label{Us commutativity}
    V(h) \Us  =  \Us \Big[\prod_e \bar{G}_e^{\rhohat^{h^{-1}}(e)} \Big] V(h).
\end{align}
Therefore, the action of $V(h)$ on a state in the effective Hilbert space is:
\begin{align}
    V(h)\ket{\Psi_s^{\partial \sigma},\{g_v\}_{\partial \sigma}} = \prod_{e \subset \partial \sigma}(-1)^{\rhohat^{h^{-1}}(e)} \ket{\Psi_s^{\partial \sigma},\{hg_v\}_{\partial \sigma}},
\end{align}
where we have used that $\ket{\Psi_\text{ttc},\{hg_v\}_{\partial \sigma}}$ is a $+1$ eigenstate of $\bar{G}_e$. The effective symmetry action for $h \in G$ is explicitly:
\begin{multline}
    \mathcal{V}_{\partial \sigma}(h) \equiv \\  \sum_{\{g_v\}_{\partial \sigma}} \prod_{e \subset \partial \sigma}(-1)^{\rhohat^{h^{-1}}(e)}\ket{\Psi_s^{\partial \sigma},\{hg_v\}_{\partial \sigma}}\bra{\Psi_s^{\partial \sigma},\{g_v\}_{\partial \sigma}}.
\end{multline}

\begin{figure}[t]
\centering
\includegraphics[width=0.35\textwidth, trim={500 330 500 300},clip]{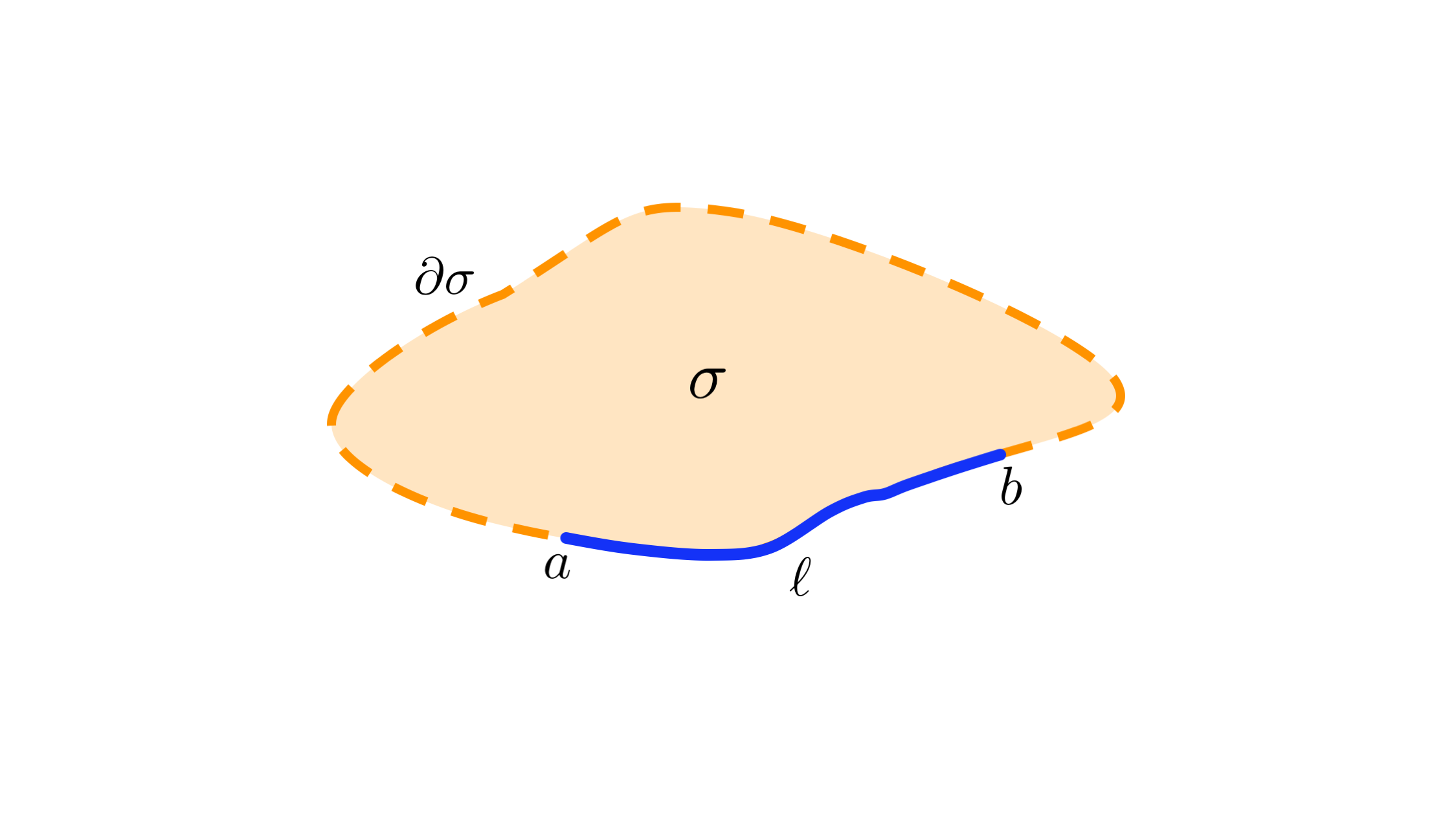}
\caption{A loop-like excitation is created at $\partial \sigma$ (dashed orange line) by applying the operator $M^s_\sigma$ on the surface $\sigma$ of the direct lattice. $\ell$ (thick blue line) is a connected submanifold of $\partial \sigma$ used to determine the fractionalization of the symmetry on the gauge charge. $a$ and $b$ label the end points of $\ell$.}
\label{fig:ell def}
\end{figure}

We identify the fractionalization of the symmetry, or an obstruction to an onsite representation, by considering the effective symmetry action restricted to a connected submanifold $\ell$ of $\partial \sigma$ (Fig.~\ref{fig:ell def}):\footnote{\unexpanded{There is some ambiguity in defining the restriction near the boundaries of $\ell$. In fact, the calculation is unaffected by the particular choice. For more details, we refer to Ref.~\cite{EN15}.}}
\begin{multline} \label{vupell def}
     \mathcal{V}_{\ell}(h) \equiv \\
     \sum_{\{g_v\}_{\ell}} \prod_{e \subset \ell}(-1)^{\rhohat^{h^{-1}}(e)}\ket{\Psi_s^{\partial \sigma},\{hg_v\}_{\ell}}\bra{\Psi_s^{\partial \sigma},\{g_v\}_{\ell}},
\end{multline}
with $\mathcal{V}_{\ell}(h)$ acting as the identity on sites in $\partial \sigma$ not contained in $\ell$. While $\mathcal{V}_{\partial \sigma}(h)$ satisfies the group law:
\begin{align}
    \mathcal{V}_{\partial \sigma}(h_1) \mathcal{V}_{\partial \sigma}(h_2)= \mathcal{V}_{\partial \sigma}(h_1h_2),
\end{align}
$\mathcal{V}_{\ell}(h)$ only satisfies the group law up to an operator $\Omega(h_1,h_2)$:
\begin{align}
    \mathcal{V}_{\ell}(h_1)\mathcal{V}_{\ell}(h_2)=\Omega(h_1,h_2)\mathcal{V}_{\ell}(h_1h_2).
\end{align}
For $\mathcal{V}_{\ell}(h)$ given in Eq.~\eqref{vupell def}, we find: 
% \begin{multline}
%     \Omega(h_1,h_2)= \\
%     \sum_{\{g_v\}_{\partial \ell}}\,\prod_{v \subset \partial \ell}(-1)^{\rhohat^{h_1,h_2}(v)} |\psi^{\partial \sigma}_s, \{g_v\}_{\partial \ell}\rangle \langle \psi^{\partial \sigma}_s, \{g_v\}_{\partial \ell}|,
% \end{multline}
\begin{multline}
    \Omega(h_1,h_2)= 
    \sum_{g_a,g_b}\Big[(-1)^{\rho(1,h_1,h_1h_2,g_a)+\rho(1,h_1,h_1h_2,g_b)} \\ \ket{\Psi^{\partial \sigma}_s, g_a, g_b} \bra{ \Psi^{\partial \sigma}_s, g_a, g_b}\Big],
\end{multline}
where we have labeled the endpoints of $\ell$ as $a$ and $b$, and $\Omega(h_1,h_2)$ acts as the identity on all other sites in $\partial \sigma$.  $\Omega(h_1,h_2)$ acts non-trivially only near the endpoints of $\ell$. Hence, we can split $\Omega(h_1,h_2)$ into an operator $\Omega_a(h_1,h_2)$ acting at $a$ and an operator $\Omega_b(h_1,h_2)$ acting at $b$. The decomposition is unique up to a sign,\footnote{The $\ZZ_2$ fusion rules of $1$-form gauge charges reduce the ambiguity from a general phase to a sign \cite{C15}.} which may depend on $h_1$ and $h_2$. For example, we may write:
\begin{align} \label{Omega at a}
    \Omega_a(h_1,h_2)=\sum_{g_a}(-1)^{\rho(1,h_1,h_1h_2,g_a)}|\Psi^{\partial \sigma}_b, g_a\rangle \langle \Psi^{\partial \sigma}_b, g_a |,
\end{align}
which acts as the identity away from the endpoint $a$. Alternatively, we could modify both $\Omega_a(h_1,h_2)$ and $\Omega_b(h_1,h_2)$ by a sign $(-1)^{\beta(1,h_1,h_1h_2)}$, where $\beta$ is an arbitrary group cochain in $C^2(G,\ZZ_2)$.

The symmetry fractionalization can be shown by analyzing the associativity of the restricted group action. The associativity of the $\mathcal{V}_\ell(h)$ operators implies (see Ref. \cite{EN15}):
\begin{align} \label{eq: Omega associativity}
    \Omega(h_1,h_2)&\Omega(h_1h_2,h_3)= \\ \nonumber
    &\mathcal{V}_\ell(h_1)\Omega(h_2,h_3){\mathcal{V}_{\ell}}^\dagger(h_1)\Omega(h_1,h_2h_3).
\end{align}
If the effective symmetry action can be written as onsite, then Eq.~\eqref{eq: Omega associativity} is satisfied independently at endpoints $a$ and $b$ for some choice of $\beta$. (This follows from Appendix B of Ref.~\cite{EN15}.) On the other hand, if Eq.~\eqref{eq: Omega associativity} is not satisfied independently at the endpoints for any choice of $\beta$, then there is an obstruction to realizing the effective symmetry action onsite.

For $\Omega_a$ in Eq.~\eqref{Omega at a}, Eq.~\eqref{eq: Omega associativity} only holds up to a $G$-dependent sign at the endpoint $a$:
\begin{multline} \label{eq: Omega a associativity}
    \Omega_a(h_1,h_2)\Omega_a(h_1h_2,h_3)=(-1)^{\rho(1,h_1,h_1h_2,h_1h_2h_3)}\\\times \mathcal{V}_\ell(h_1)\Omega_a(h_2,h_3){\mathcal{V}_\ell}^\dagger(h_1)\Omega_a(h_1,h_2h_3).
\end{multline}
The sign $(-1)^{\rho(1,h_1,h_1h_2,h_1h_2h_3)}$ in Eq.~\eqref{eq: Omega a associativity} captures the symmetry fractionalization on the $1$-form gauge charge. Taking into account the ambiguity in defining $\Omega_a$, we see that $\rho$ is well defined up to a group coboundary $\delta\beta$, with $\beta$ an arbitrary element of $C^2(G,\ZZ_2)$. In other words, the symmetry fractionalization on the loop-like excitation is described by an element of the group cohomology $H^3(G,\ZZ_2)$. Therefore, when $\rho$ represents a nontrivial class in $H^3 (G, \ZZ_2)$, there is a nontrivial symmetry fractionalization on the $1$-form gauge charges of the shadow model.

The group cohomology class represented by $\rho$ defines a quantized invariant of the symmetry-enriched twisted toric code phase. To make this explicit, we note that any state belonging to the same symmetry-enriched phase can be constructed (approximately) from $\ket{\Psi_s}$ by applying a FDQC built of symmetric local unitaries. If we modify $\Us$ by a FDQC built of symmetric local unitaries, the calculation above is unchanged. Thus, the symmetry fractionalization on a gauge charge is given by the same group cohomology class for any state in the symmetry-enriched phase. 

In the next section, we see that this quantized invariant may be pushed forward to characterize the corresponding fSPT obtained from condensing the emergent fermion in the shadow model.  

\subsection{Fermionic SPT  } \label{sec: supercohomology model}

Finally, we condense the emergent fermion in the shadow model to construct a fSPT Hamiltonian corresponding to the supercohomology data $(\rho, \nu)$. In the process, we find a FDQC $\Uf$ that prepares the fSPT ground state from a product state. We argue that our models exhibit the expected responses to probing with fermion parity defects, by referring to the properties of the shadow models. Lastly, we interpret the equivalence relation on supercohomology data in Ref.~\cite{GW14} using the stacking rules. 
% Lastly, using the stacking rules for supercohomology phases, we verify the equivalence relation on supercohomology data in Ref.~\cite{GW14}. 

Fermion condensation, described in Section~\ref{sec: atomic insulator}, can be readily applied to any Hamiltonian expressed in terms of the operators $\bar{G}_e$, $\bar{U}_f$, and $W_t$. The fermion is then condensed by mapping the operators according to: 
\begin{align} \label{fermion condense map}
    \bar{G}_e \rightarrow 1, \quad \bar{U}_f \rightarrow S_f, \quad W_t \rightarrow P_t.
\end{align}
The result is a model defined on a Hilbert space with a single spinless complex fermion at each tetrahedron.
We emphasize again that this mapping requires the spatial manifold to admit a spin structure.

To apply the fermion condensation duality in Eq.~\eqref{fermion condense map} to the shadow model:
\begin{align}
    {H_s = \Us H_\text{ttc}^G \Us^\dagger},
\end{align}
we write $H_\text{ttc}^G$ and $\Us$ in terms of $\bar{G}_e$, $\bar{U}_f$, and $W_t$ operators. By definition, $H_\text{ttc}^G$ can be written using $\bar{G}_e$ and $W_t$. As for $\Us$, the Pauli X and Pauli Z operators can be commuted to write (see Appendix~\ref{app: fermionization shadow circuit} for a derivation):
\begin{align} \label{Us nice form}
    \Us = \prod_t e^{2 \pi i O_t \nuhat(t)} \xi_{\bar{\rho}}(M){\prod_{f}} \bar{U}_f^{\rhohat(f)} \prod_{t} W_t^{\int\rhohat \cup_2 \boldsymbol{t}}.
\end{align}
% \begin{align} \label{Us nice form}
%     \Us = \prod_t e^{2 \pi i\nuhat(t)} \overline{\prod_{f}} U_f^{\rhohat(f)} \prod_{t= \langle 1234 \rangle} W_t^{\rhohat(123)+\rhohat(134)},
% \end{align} 
% Here, $\nuhat(t)$ is the operator:
% \begin{align}
%     \nuhat(t) = \sum_{\{g_v\},\ba_f} \conu(t) \ket{\{g_v\},\ba_f}\bra{\{g_v\},\ba_f},
% \end{align}
Here, we have defined an arbitrary ordering of faces on $M$:
\begin{align}
    \{f_1,\ldots,f_i,\ldots \}, \quad (f_1 < \cdots < f_i < \cdots),
\end{align}
and the product of $\bar{U}_f$ operators is determined by the order of faces in $M$:
\begin{align}
    {\prod_{f}} \bar{U}_f^{\rhohat(f)} = \left( \cdots \bar{U}_{f_i}^{\rhohat(f_i)}\cdots \bar{U}_{f_1}^{\rhohat(f_1)} \right).
\end{align}
$\xi_{\bar{\rho}}(M)$ in Eq.~\eqref{Us nice form} is a sign that compensates for the order dependence of the product of $\bar{U}_f$ operators, so that $\Us$ is {independent} of the choice of ordering. Specifically, $\xi_{\bar{\rho}}(M)$ is given by (Appendix~\ref{app: fermionization shadow circuit}):
\begin{align} \label{xi rhobar def}
    \xi_{\bar{\rho}}(M) \equiv  \prod_{i,i'|i'<i} (-1)^{\rhohat(f_{i'})\rhohat(f_{i}) \int \bface_{i'} \cup_1 \bface_{i} } .
\end{align}
We emphasize that, although $\xi_{\bar{\rho}}(M)$ and the product of $\bar{U}_f$ operators depend on a choice of ordering of the faces in $M$, the FDQC $\Us$ does \emph{not} depend on the choice of ordering.\footnote{The independence on the ordering can be derived from the commutation relations: $ \bar{U}_f \bar{U}_{f'} = (-1)^{\int (\bface^\prime \cup_1 \bface+\bface \cup_1 \bface^\prime)} \bar{U}_{f^\prime} \bar{U}_f$, as described in Appendix~\ref{app: fermionization shadow circuit}.}
% \begin{align}
%     \xi_{\bar{\rho}}(F) \equiv \prod_{i} \left[ \prod_{i'>i} (-1)^{\rhohat(f_i) \cdot \rhohat(f_{i'}) \cdot \sum_t \bface_{i} \cup_1 \bface_{i'}(t) } \right].
% \end{align}

% \begin{figure}[t]
% \centering
% \includegraphics[width=0.21\textwidth, trim={760 500 1000 300},clip]{Gai_dof.png}
% \caption{Our model for a fSPT phase is defined on a Hilbert space with $G$ {d.o.f.} on the vertices of a triangulation of $M$ and a single spinless complex fermion at the center of each tetrahedron. }
% \label{fig: Gai dof}
% \end{figure}

With this, we apply the mapping of operators in Eq.~\eqref{fermion condense map} to $H_s$ to condense the emergent fermion. First, $H^G_\text{ttc}$ is mapped to an atomic insulator and a decoupled $G$ paramagnet (Appendix \ref{fermioncondensationHttc}):
\begin{align} \label{GAI H def}
    H_\text{AI}^G \equiv -\sum_t P_t - \sum_v \mathcal{P}_v.
\end{align}
Second, $\Us$ is mapped to a FDQC $\Uf$:
\begin{align} \label{eq: uf def}
  \Uf \equiv \prod_t e^{2 \pi i O_t \nuhat(t)} \xi_{\bar{\rho}}(M){\prod_{f}} S_f^{\rhohat(f)} \prod_{t} P_t^{\int \rhohat \cup_2 \boldsymbol{t}}. 
\end{align}
Thus, fermion condensation leads to the exactly-solvable fermionic Hamiltonian:
\begin{align} \label{eq: fSPT H}
    H_f \equiv \Uf H_\text{AI}^G \Uf^\dagger.
\end{align}
This is precisely the Hamiltonian described in Section~\ref{sec: supercohomology models 2}. 

We note that $H_f$ is symmetric due to the symmetry of both $H_\text{AI}^G$ and $\Uf$. To see that $\Uf$ is symmetric, we consider the bosonic FDQC $\Us$. $\Us$ commutes with the symmetry up to factors of $\bar{G}_e$ (see Appendix~\ref{app: sym var Us}), and since $\bar{G}_e$ maps to the identity under fermion condensation, $\Uf$ must commute with the symmetry.

\vspace{1.5mm}
\noindent \begin{center}\emph{Equivalence relation on supercohomology data:}\end{center}
\vspace{1.5mm}

% \textcolor{orange}{We still need to show that this is indeed trivial supercohomology data.}
% \textcolor{orange}{This needs to be revised to weaken the claim of reproducing the equivalence relation. Instead, what we can show is that the models have the expected response to probing with symmetries.}
For each set of supercohomology data, we can now construct a fSPT Hamiltonian $H_f$. However, the Hamiltonians constructed from two \textit{a priori} different sets of supercohomology data, say $(\rho,\nu)$ and $(\rho',\nu')$, may be within the same phase. This motivates imposing an equivalence relation on the supercohomology data, so that two sets of supercohomology data are equivalent if and only if they describe the same characteristic response of the SPT phase. Ref.~\cite{GW14} used spacetime methods to argue that the appropriate equivalence relation $\sim$ on the supercohomology data is:
\begin{multline} \label{supercohomology equivalence relation}
    (\rho,\nu) \sim \\ (\rho+\delta\beta,\nu+\delta \eta + \frac{1}{2}\beta \cup \beta + \frac{1}{2}\beta \cup_{1} \delta \beta + \frac{1}{2}\rho \cup_2 \delta \beta).
\end{multline}
% which is 
% in agreement with the equivalence relation identified using spacetime methods in Ref.~\cite{GW14}.
% as introduced using spacetime methods in Ref.~\cite{GW14}.
% {\color{red} With the stacking law derived in Appendix \ref{app:composition laws}: 
% \begin{equation}
%     (\rho_1,\nu_1) + (\rho_2,\nu_2)  = (\rho_1 + \rho_2, \nu_1 + \nu_2 + \frac{1}{2} \rho_1 \cup_2 \rho_2),
% \end{equation}
% the equivalence relation on the supercohomology data is simply:
% \begin{eqs}
%   (0,0) \sim ( \delta \beta, \delta \eta + \frac{1}{2}\beta \cup_{1} \delta \beta + \frac{1}{2}\beta \cup \beta).
% \end{eqs}}
Here, $\beta$ and $\eta$ are arbitrary group cochains:
% \footnote{More generally, we use $C^m(G,A)$ to denote the set of homogeneous functions from $G^{m+1}$ to $A$.}
\begin{align}
    \beta \in C^2(G,\ZZ_2), \quad \eta \in C^3(G,\mathbb{R}/\ZZ),
\end{align}
and the group cup products can be written explicitly using the general formulas in Appendix~\ref{app:groupcoho}.
% \begin{align} \nonumber
% \beta \cup_{1} \delta \beta (g_1,g_2,g_3,g_4,g_5) &\equiv \beta(g_1,g_4,g_5)\delta\beta(g_1,g_2,g_3,g_4) \\ 
% &+\beta(g_1,g_2,g_5) \delta\beta(g_2,g_3,g_4,g_5)\\
%     \beta \cup \beta (g_1,g_2,g_3,g_4,g_5) &\equiv \beta(g_1,g_2,g_3)\beta(g_3,g_4,g_5).
% \end{align}
% To justify the equivalence relation in Eq.~\eqref{supercohomology equivalence relation}, 
% one must show that (i) equivalent sets of supercohomology data lead to Hamiltonians in the same phase and (ii) inequivalent sets of supercohomology data give rise to Hamiltonians in distinct phases. In what follows, we confirm (i) and verify (ii) up to bosonic SPT phases.
% we must show that both (i) equivalent sets of supercohomology data lead to Hamiltonians in the same phase and (ii) inequivalent sets of supercohomology data give rise to Hamiltonians in distinct phases.

We point out that the equivalence relation can be phrased in terms of the operation $\boxtimes$ induced by stacking. 
As discussed in Section~\ref{sec: supercohomology models 2}, the stacking operation can be determined by composing FDQCs. The stacking operation applied to $(\rho,\nu)$ and $(\rho',\nu')$ gives:
\begin{align} \label{stacking by composition2}
    (\rho,\nu) \boxtimes (\rho',\nu')  = (\rho + \rho',\nu + \nu' + \frac{1}{2} \rho \cup_2 \rho').
\end{align}
With this, we see that any two equivalent sets of supercohomology data can be related by stacking a ``trivial'' set of supercohomology data:
\begin{align} \label{supercohomology trivial}
    (\rho_0,\nu_0) \equiv (\delta \beta, \delta \eta + \frac{1}{2}\beta \cup \beta + \frac{1}{2}\beta \cup_1 \delta \beta),
\end{align}
for some choice of $\beta$ in $C^2(G,\ZZ_2)$ and $\eta$ belonging to $C^3(G,\mathbb{R}/\ZZ)$.
Explicitly, stacking $(\rho_0,\nu_0)$ with an arbitrary set of supercohomology data $(\rho,\nu)$ yields:
\begin{multline} \label{stacking trivial}
    (\rho,\nu)\boxtimes (\delta \beta, \delta \eta + \frac{1}{2}\beta \cup \beta + \frac{1}{2}\beta \cup_1 \delta \beta)= \\(\rho+\delta\beta,\nu+\delta \eta + \frac{1}{2}\beta \cup \beta + \frac{1}{2}\beta \cup_{1} \delta \beta + \frac{1}{2}\rho \cup_2 \delta \beta),
\end{multline}
which is equivalent to $(\rho,\nu)$ according to Eq.~\eqref{supercohomology equivalence relation}.

The equivalence relation given in Eq.~\eqref{supercohomology equivalence relation} can be motivated in terms of our supercohomology models. We show this by arguing that (i) equivalent sets of supercohomology data lead to Hamiltonians in the same phase and (ii) inequivalent sets of supercohomology data give rise to Hamiltonians in distinct phases (up to stacking bosonic SPT phases). 

To see that supercohomology models built from equivalent sets of supercohomology data belong to the same phase, we consider the ground state $\ket{\Psi_f^{\rho_0\nu_0}}$ of the Hamiltonian corresponding to a set of trivial data $(\rho_0,\nu_0)$. In Appendix \ref{app: trivial circuit}, we show that $\ket{\Psi_f^{\rho_0\nu_0}}$ can be written as: 
 %In the case that the bosonic SPT phase is non-trivial, this is only possible by the use of the fermionic hopping operators.
 \begin{eqs}
    \ket{\Psi^{\rho_0\nu_0}_f} = \prod_t e^{2 \pi i O_t \hat{\boldsymbol \eta}(t)}
    \xi_\beta(M){\prod_{f}} S_f^{\hat{\boldsymbol \beta}(f)} \sum_{\{g_v\}}
     \ket{\{g_v\},\text{vac}},
\end{eqs}
where $\xi_\beta(M)$ is defined as in Eq.~\eqref{xi rhobar def}.
% In this form, we see that, $\ket{\Psi_f^{\rho_0\nu_0}}$ can be constructed from a product state by a FDQC composed of local symmetric unitaries.
Due to the homogeneity of $\eta$ and $\beta$, we see that $\ket{\Psi_f^{\rho_0\nu_0}}$ can be prepared from a product state by a FDQC composed of symmetric local unitaries.
% In this form, we see that, due to the homogeneity of $\eta$ and $\beta$, $\ket{\Psi_f^{\rho_0\nu_0}}$ can be constructed from a product state by a FDQC composed of local symmetric unitaries.
This implies that $\ket{\Psi_f^{\rho_0\nu_0}}$ belongs to a trivial fSPT phase. Therefore, the ground state of the Hamiltonian constructed from $(\rho,\nu)$ belongs to the same phase as the ground state of the Hamiltonian built from the equivalent set of data ${(\rho,\nu)\boxtimes(\rho_0,\nu_0)}$ (see the discussion of the stacking rule in Section~\ref{sec: supercohomology models 2}).

%  This can be seen by considering the ground state $\ket{\Psi_f^{\rho_0\nu_0}}$ $\Uf^{\rho_0\nu_0}$ corresponding to the set of trivial supercohomology data $(\rho_0,\nu_0)$. In Appendix~[APPENDIX], we show that $\Uf^{\rho_0\nu_0}$ can be expressed as a FDQC composed of symmetric local unitaries, for any choice of $(\rho_0,\nu_0)$ in Eq.~\eqref{supercohomology trivial}. This implies that the Hamiltonians corresponding to $(\rho,\nu)$ and ${(\rho,\nu)\boxtimes(\rho_0,\nu_0)}$ are in the same phase, given

We note that in the special case that $\beta$ and $\eta$ in Eq.~\eqref{supercohomology trivial} are closed, the trivial data $(\rho_0,\nu_0)$ is equal to $(0,\frac{1}{2}\beta \cup \beta)$. This corresponds to the data of a bosonic SPT phase, which according to the equivalence relation, must be trivial when considered as a fermionic SPT phase.\footnote{An example (though for an anti-unitary symmetry) was given in  Ref.~\cite{GW14}, where it was observed that the in-cohomology ($3+1$)D bosonic SPT phase protected by time-reversal symmetry is trivial in the presence of fermions.} Although $\frac{1}{2}\beta \cup \beta$ may be a nontrivial cocycle, the bosonic SPT model can nonetheless be disentangled by a FDQC composed of local symmetric unitaries if fermionic hopping operators are used.

% \begin{eqs}
%     \ket{\Psi^{\rho_0\nu_0}_f} = \prod_t e^{2 \pi i O_t \hat{\boldsymbol \eta}(t)}
%     S_{\beta} \sum_{\{g_v\}}
%      \ket{\{g_v\},\text{vac}},
% \end{eqs}
% where $S_{\beta}$ is the FDQC:
% \begin{eqs}
%     S_{\beta} &\equiv \prod_{i,i'|i'<i} (-1)^{\hat{\boldsymbol \beta}(f_{i'})\hat{\boldsymbol \beta}(f_{i}) \int \bface_{i'} \cup_1 \bface_{i} } {\prod_{f}} S_f^{\hat{\boldsymbol \beta}(f)}.
% \end{eqs}
%with $\xi_\beta(M)$ given by:
%\begin{align}
%    \xi_\beta(M) &\equiv \prod_{i,i'|i'<i} (-1)^{\hat{\boldsymbol \beta}(f_{i'})\hat{\boldsymbol \beta}(f_{i}) \int \bface_{i'} \cup_1 \bface_{i} }.
%\end{align}

%The algebra of the fermionic hopping operators ensures that the bosonic SPT state can be disentangled using a FDQC built from local symmetric unitaries (see Appendix~\ref{app: trivial circuit}). %In Ref.~\cite{GW14}, it was observed that, as a consequence, the ($3+1$)D bosonic SPT phase protected by time reversal symmetry is trivial in the presence of fermions.

Next, we argue that supercohomology models constructed using inequivalent sets of supercohomology data belong to distinct fSPT phases. To make this precise, suppose $(\rho,\nu)$ and $(\rho',\nu')$ are inequivalent, or in other words, the stack: 
\begin{align} \label{inequivalent supercohomology}
   (\tilde{\rho},\tilde{\nu}) \equiv (\rho,\nu)\boxtimes(\rho',\nu')^{-1}
\end{align} 
is not equal to $(\rho_0,\nu_0)$, for any choice of $(\rho_0,\nu_0)$ in Eq.~\eqref{supercohomology trivial}. 
There are then two possibilities for $(\tilde{\rho},\tilde{\nu})$ to be nontrivial. The first possibility, which is the focus of our discussion, is that $\tilde{\rho}$ is a nontrivial element of $H^3(G,\ZZ_2)$, so it cannot be written as $\delta \beta$ for any choice of group cochain $\beta \in C^2(G,\ZZ_2)$. The second possibility is that $\tilde{\rho}$ is trivial (i.e., $\tilde{\rho}=\delta \beta$), but $\tilde{\nu}$ is nontrivial [i.e., $\tilde{\nu}$ is not equal to ${\delta \eta + \frac{1}{2}\beta \cup \beta + \frac{1}{2} \beta \cup_1 \delta \beta}$ for any choice of $\eta \in C^3(G,\mathbb{R}/\ZZ)$]. In this case, $(\tilde{\rho},\tilde{\nu})$ is equivalent to a set of supercohomology data of the form $(0,\tilde{\nu}')$, for some $\tilde{\nu}'$. Hence, $(\rho,\nu)$ and $(\rho',\nu')$ differ by a bosonic group cohomology SPT phase with a $G$ symmetry. To show that $(\rho,\nu)$ and $(\rho',\nu')$ belong to different fSPT phases, one needs to determine whether the bosonic SPT phase is trivialized in the presence of fermions.
Below, we address only the first possibility and leave the question of the trivialization of bosonic SPT phases in the presence of fermions for future studies.

% There are two possibilities in which $(\tilde{\rho},\tilde{\nu})$ is not equivalent to any $(\rho_0,\nu_0)$, and we address them in turn. The first possibility is that $\tilde{\rho}$ is a nontrivial element of $H^3(G,\ZZ_2)$, so it cannot be written as $\delta \beta$ for any choice of group cochain $\beta \in C^2(G,\ZZ_2)$. The second possibility is that $\tilde{\rho}$ is trivial, i.e., it can be written as $\delta \beta$, but $\tilde{\nu}$ is not equal to ${\delta \eta + \frac{1}{2}\beta \cup \beta + \frac{1}{2} \beta \cup_1 \delta \beta}$ for any choice of $\eta \in C^3(G,\mathbb{R}/\ZZ)$. 

Considering the first possibility, if $\tilde{\rho}$ represents a nontrivial class in $H^3(G,\ZZ_2)$, then we can use the symmetry fractionalization properties of the shadow model to argue that $(\tilde{\rho},\tilde{\nu})$ must correspond to a nontrivial fSPT phase. To derive a contradiction, suppose that the Hamiltonian $H_f^{\tilde{\rho}\tilde{\nu}}$ built using $(\tilde{\rho},\tilde{\nu})$ is in a trivial fSPT phase. Then, we can find a path of symmetric gapped Hamiltonians connecting $H_f^{\tilde{\rho}\tilde{\nu}}$ to the atomic insulator Hamiltonian $H^G_\text{AI}$. For each Hamiltonian in this path, we can gauge the fermion parity symmetry -- or ``ungauge'' the anomalous $2$-form symmetry. More precisely, we can map the fermionic operators to bosonic operators according to the duality in Table \ref{table: gauging fermion parity}. To avoid ambiguity in this ungauging process, we conjugate the bosonic operators by local projectors onto the $\bar{G}_e=1$ subspace and add a term $-\sum_e \bar{G}_e$ to enforce the $\bar{G}_e=1$ constraint. The result of ungauging the anomalous $2$-form symmetry is a symmetric gapped path of Hamiltonians connecting the shadow model corresponding to $(\tilde{\rho},\tilde{\nu})$ to the twisted toric code Hamiltonian $H^G_\text{ttc}$. This is a contradiction, because the shadow model describes a nontrivial symmetry-enriched twisted toric code when $\tilde{\rho}$ is nontrivial in $H^3(G,\ZZ_2)$, while $H^G_\text{ttc}$ is in a trivial symmetry-enriched phase. Therefore, $H_f^{\tilde{\rho}\tilde{\nu}}$ is in a nontrivial SPT phase, and $(\rho,\nu)$ and $(\rho',\nu')$ must correspond to distinct fSPT phases.

We note that the procedure for ungauging an anomalous $2$-form symmetry, briefly described here, can be applied to any fermionic model. We emphasize that it is equivalent to gauging the fermion parity symmetry -- the point-like $1$-form fluxes in the shadow model correspond to fermion parity gauge charges and the loop-like $1$-form gauge charges correspond to fermion parity gauge fluxes. 
% We also point out that it was important to analyze the stack of $(\rho,\nu)$ and $(\rho',\nu')^{-1}$, since gauging the fermion parity of supercohomology Hamiltonians in distinct phases can yield the same symmetry-enriched twisted toric code \cite{EF19}. 
%This phenomenon is exhibited by the example discussed in the next section. 

\section{Gapped boundaries through symmetry extension} \label{sec: sym extension boundary}

The supercohomology models, described in Sections~\ref{sec: bulk Z2f} and \ref{sec:bulkG}, characterize the bulk of the SPT phase, i.e., the Hamiltonians are defined on manifolds without boundary. In this section, we consider models on manifolds with boundary. In particular, we describe supercohomology models that feature gapped, topologically ordered boundaries. To get started, we review gapped boundaries for group cohomology models, constructed via a symmetry extension \cite{WWW19}. We then generalize the symmetry extension construction to supercohomology models. Our generalization relies on the connection between $2$-group SPT phases and supercohomology phases. We provide the details of the construction, starting from a $2$-group SPT model, in Appendix~\ref{app: super from 2gauge}.

\subsection{Review of gapped boundary construction of group cohomology models} \label{sec: group coho boundary}

As shown in Ref.~\cite{WWW19}, symmetric gapped boundaries for group cohomology models can be constructed by first enlarging the symmetry at the boundary. One can then partially gauge the extended symmetry to produce a symmetric, topologically ordered boundary. To illustrate the construction, we consider a $(3+1)$D group cohomology model corresponding to the group cocycle $\nu \in H^4(G,\RR/\ZZ)$ with a unitary $G$ symmetry.

We motivate the gapped boundary construction by reviewing the proof that the ground state of the group cohomology model
is symmetric on a manifold without boundary. Recall that the ground state of the group cohomology model is:
\begin{align} \label{groupcoho gs2}
    \ket{\Psi_b} = \sum_{\{g_v\}} \Psi_b(\{g_v\}) \ket{\{g_v\}},
\end{align}
with the amplitude $\Psi_b(\{g_v\})$ given by:
\begin{align}
    \Psi_b(\{g_v\}) \equiv \prod_{t = \langle 1234 \rangle}e^{2 \pi i O_t \nu(1,g_1,g_2,g_3,g_4)}.
\end{align}
Applying the symmetry action $V(h)$, for $h \in G$, to $\ket{\Psi_b}$ yields:
\begin{eqs} \label{sym on group coho gs}
       V(h) \ket{\Psi_b} =  \sum_{\{g_v\}} \tilde{\Psi}_b(\{g_v\})\ket{\{g_v\}}.
\end{eqs}
Here, we have shifted the indices and used the homogeneity of $\nu$ to define:
\begin{align} \label{tilde group coho amplitude}
    \tilde{\Psi}_b(\{g_v\}) \equiv \prod_{t = \langle 1234 \rangle}e^{2 \pi i O_t \nu(h,g_1,g_2,g_3,g_4)}.
\end{align}
We evaluate Eq.~\eqref{tilde group coho amplitude} further by using the cocycle property of $\nu$, which tells us:
\begin{eqs} \label{nu coboundary}
    \nu (h,g_1, g_2, g_3,g_4) = \nu (1,g_1, g_2, g_3,g_4) - \nu (1,h, g_2, g_3,g_4) \\
    +\nu (1,h,g_1, g_3,g_4) - \nu (1,h,g_1, g_2,g_4)+\nu (1,h,g_1, g_2, g_3). 
\end{eqs}
The last four terms in Eq.~\eqref{nu coboundary} each correspond to a face of the tetrahedron $\langle 1234 \rangle$. 
Therefore, in substituting Eq.~\eqref{nu coboundary} into Eq.~\eqref{tilde group coho amplitude}, the terms associated to faces cancel in pairs. We are left with: $\tilde{\Psi}_b(\{g_v\})=\Psi_b(\{g_v\})$, which implies that $\ket{\Psi_b}$ is symmetric.

The cancellation of the face terms in the calculation above relies crucially on the fact that the manifold has no boundary. 
The wave function in Eq.~\eqref{groupcoho gs2} is not guaranteed to be symmetric on a manifold $M$ with boundary $\partial M$, since the terms associated with faces fail to cancel at the boundary. In this case, the symmetry action $V(h)$ on $\ket{\Psi_b}$ leaves a residual phase factor $\mathcal{V}_h({{\{g_v\}}})$ on the boundary of $M$, i.e.:
\begin{align} \label{residual phases from sym}
    \tilde{\Psi}_b(\{g_v\}) = \mathcal{V}_h({{\{g_v\}}}) \Psi_b(\{g_v\}),
\end{align}
where $\mathcal{V}_h({{\{g_v\}}})$ is the phase:
\begin{align} \label{residual phase def}
    \mathcal{V}_h({{\{g_v\}}}) \equiv  \prod_{f_\partial = \langle 123 \rangle} e^{2 \pi i O_{f_\partial}\nu(1,h,g_1,g_2,g_3)}.
\end{align}
Here, the product is over faces $f_\partial$ in the boundary of $M$ and $O_{f_\partial} \in \{-1,+1\}$ is $-1$ if the orientation of $f_\partial$ points out of $M$ and $+1$ otherwise.
The residual phase factor is indicative of the anomalous symmetry action at the boundary of the SPT phase \cite{EN15}. To find a gapped boundary, we search for a modification of $\ket{\Psi_b}$ near the boundary to ``saturate'' the anomaly.

The key observation, made in Refs.~\cite{WWW19} and \cite{PWW18}, is that the anomaly can be saturated by enlarging the symmetry at the boundary. To make this precise, we define $L$ to be a central extension of $G$ by a group $K$,\footnote{\unexpanded{By a central extension, we mean that $K$ belongs in the center of $L$.}} giving the short exact sequence:
\begin{equation}
    1 \rightarrow K \rightarrow L \xrightarrow[]{\pi} G \rightarrow 1.
    \label{equ:bosonicgroupextension}
\end{equation}
% We note that an element of $H$ can then be specified by a pair $(g,k)$ with $g \in G$ and $k \in K$. 
The group cocycle $\nu$ can then be pulled back by $\pi$ to form a cocycle $\nu^* \in H^4(L,\RR/\ZZ)$:
\begin{eqs}
    \nu^*(\ell_0,\ell_1,\ell_2,&\ell_3,\ell_4) \equiv \\ &\nu\boldsymbol{(}\pi(\ell_0),\pi(\ell_1),\pi(\ell_2),\pi(\ell_3),\pi(\ell_4)\boldsymbol{)},
\end{eqs}
with $\ell_0,\ell_1,\ell_2,\ell_3$, and $\ell_4$ in $L$. According to Refs.~\cite{WWW19} and \cite{T17-2}, one can always find an extension $L$ such that $\nu^*$ is a coboundary, i.e.:
\begin{align} \label{coboundary mu}
    \nu^* = \delta \eta,
\end{align}
for some $\eta$ in $C^4(L,\RR/\ZZ)$. As described below, the cochain $\eta$ can be used to absorb the residual phase factor in Eq.~\eqref{residual phases from sym}. 

\begin{figure}[t]
\centering
\includegraphics[width=0.4\textwidth, trim={800 450 650 300},clip]{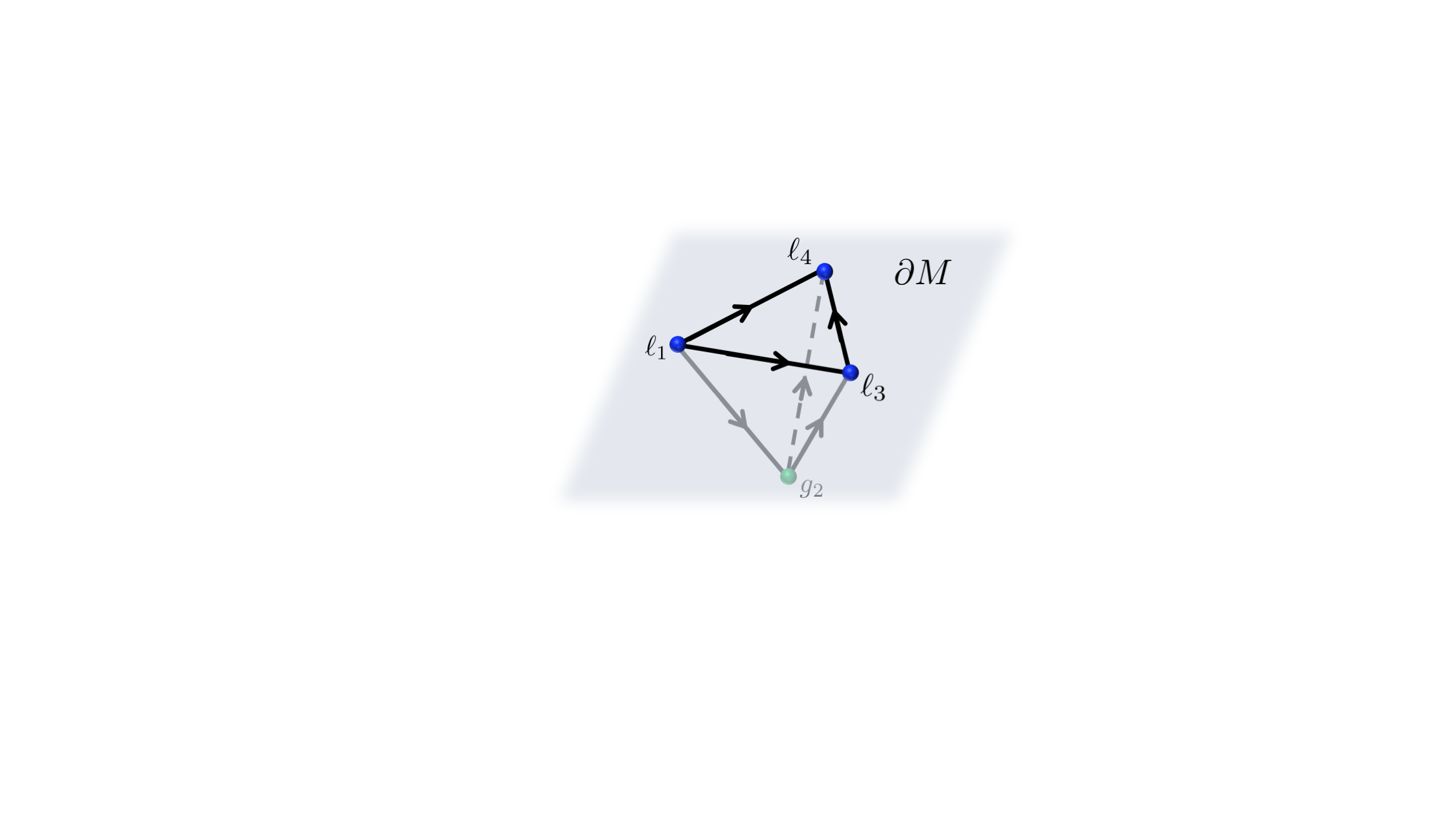}
\caption{The $L$ symmetric state $\ket{\Psi^L_b}$ is defined on a Hilbert space with $L$ d.o.f. (blue) on the boundary of $M$. The vertices on the interior of $M$ host $G$ d.o.f. (green), as in Fig.~\ref{fig: groupcohodof}.}
\label{fig: bdrydof}
\end{figure}

To build a symmetric wave function using $\eta$, we extend the global symmetry of the group cohomology model to $L$ by replacing each $G$ d.o.f. on the boundary with an $L$ d.o.f. (see Fig.~\ref{fig: bdrydof}).  
We denote a configuration of the $L$ and $G$ d.o.f. by $\{\ell_{{\bdryv}},g_v\}$, where $\ell_{{\bdryv}}$ is an element of $L$ labeling the boundary vertex ${\bdryv}\in \partial M$ and $g_v$ belongs to $G$ and labels a vertex $v$ in the bulk $v \in {M \setminus \partial M}$.  
% We use $\{h_v\}_{\partial M}$ to denote a configuration of the $H$ d.o.f. at the boundary of $M$ and specify a configuration of the $G$ d.o.f. in the bulk of $M$ by $\{g_v\}_{\bar{M}}$, where $\bar{M} = M \setminus \partial M$.
For $h \in L$, the global $L$ symmetry is then represented by:
\begin{eqs} \label{H symmetry}
   V(\ell) \equiv \sum_{\{\ell_{{\bdryv}},g_v \}} \ket{\{\ell \ell_{{\bdryv}},\pi(\ell)g_v\}}\bra{\{\ell_{{\bdryv}},g_v\}}.
\end{eqs}
% \begin{eqs} \label{H symmetry}
%   &V(h) \equiv \\ &\sum_{\{h_v\}_{\partial M}} \ket{\{hh_v\}_{\partial M}}\bra{\{h_v\}_{\partial M}}  \otimes \sum_{\{g_v\}_{\bar{M}}} \ket{\{\pi(h)g_v\}_{\bar{M}}}\bra{\{g_v\}_{\bar{M}}}.
% \end{eqs}
% To build a symmetric wave function using $\mu$, we extend the global symmetry of the group cohomology model to $H$ by adding a $K$ d.o.f. to each vertex in the boundary.  Each boundary vertex thus hosts both a $G$ d.o.f. and a $K$ d.o.f., which can be interpreted as a single $H$ d.o.f. via a (non-unique) identification of each $h\in H$ with a pair $(\pi(h),k) \in G \times K$.\footnote{\unexpanded{We note that the identification is a homomorphism only if $H$ splits into $G \times K$.}}  For $h \in H$, we represent the global $H$ symmetry on the extended Hilbert space by:
% \begin{eqs} \label{H symmetry}
%   &V(h) \equiv \\ &\sum_{\{h_v\}_{\partial M}} \ket{\{hh_v\}_{\partial M}}\bra{\{h_v\}_{\partial M}}  \otimes \sum_{\{g_v\}_{\bar{M}}} \ket{\{\pi(h)g_v\}_{\bar{M}}}\bra{\{g_v\}_{\bar{M}}}.
% \end{eqs}
% Here, $\{h_v\}_{\partial M}$ is a configuration of the $H$ d.o.f. at the boundary of $M$, and $\{g_v\}_{\bar{M}}$ is a configuration of the $G$ d.o.f. in $\bar{M} = M \setminus \partial M$.
% \begin{eqs}
%   V(h) \equiv V_{\partial M}(h) \otimes V_{\bar{M}}\boldsymbol{(}\pi(h)\boldsymbol{)},
% \end{eqs}
% where $V_{\partial M}(h)$ acts on the sites in $\partial M$ and $V_{\bar{M}}\boldsymbol{(}\pi(h)\boldsymbol{)}$ acts on the sites in $\bar{M} = M \setminus \partial M$.
Heuristically, the symmetry acts as $L$ on the boundary sites and $G$ on the bulk sites.

Now, we consider a modified state built using both $\nu$ and $\eta$ and show that it is invariant under the $L$ symmetry in Eq.~\eqref{H symmetry}. In particular, we consider the state $\ket{\Psi_b^L}$ defined as:
\begin{align} \label{groupcoho gs3}
    \ket{\Psi^L_b} \equiv \sum_{\{\ell_{{\bdryv}},g_v\}} \Psi_\eta(\{\ell_{{\bdryv}}\}) \Psi_b(\{\pi( \ell_{{\bdryv}}),g_v\}) \ket{\{ \ell_{{\bdryv}},g_v\}}.
\end{align}
% Here, $\Psi_\nu(\{h_{v_\partial},g_v\})$ is analogous to the amplitude ${\Psi_b}(\{g_v\})$:
% \begin{align}
%     \Psi_\nu(\{h_{v_\partial},g_v\}) \equiv \prod_{t = \langle 1234 \rangle}e^{2 \pi i O_t \nu(1,g_1,g_2,g_3,g_4)}.
% \end{align}
% However, in the formula above, $\nu$ is implicitly a function of $\pi(h_{v_\partial})$, for any vertices $v_\partial$ in $\partial M$. 
Here, $\Psi_\eta(\{ \ell_{{\bdryv}}\})$ is a product of $\eta$ dependent phase factors corresponding to faces in $\partial M$:
\begin{align}
    \Psi_\eta(\{ \ell_{{\bdryv}}\}) \equiv \prod_{f_\partial = \langle 123 \rangle} e^{-2 \pi i O_{f_{\partial}}\eta(1, \ell_1, \ell_2, \ell_3)}.
\end{align}
% \begin{eqs} \label{groupcoho gs3}
%     \ket{\Psi^H_b} \equiv \sum_{\{h_{v_\partial},g_v\}} \Bigg[ &\prod_{f_\partial = \langle 123 \rangle} e^{-2 \pi i O_{f_{\partial}}\mu(1,h_1,h_2,h_3)} \\ \times &\prod_{t = \langle 1234 \rangle}e^{2 \pi i O_t \nu(1,g_1,g_2,g_3,g_4)}\Bigg]\ket{\{h_{v_\partial},g_v\}}.
% \end{eqs}
% \begin{eqs} \label{groupcoho gs3}
%     \ket{\Psi^H_b} \equiv \sum_{\{h_{v_\partial},g_v\}} \Psi_b^H(\{h_{v_\partial},g_v\}) \ket{\{h_{v_\partial},g_v\}}.
% \end{eqs}
% \begin{eqs}
%   \Psi_b^H(\{h_{v_\partial},g_v\}) = \prod_{f_\partial = \langle 123 \rangle} &e^{-2 \pi i O_{f_{\partial}}\mu(1,h_1,h_2,h_3)} \\ \times &\prod_{t = \langle 1234 \rangle}e^{2 \pi i O_t \nu(1,g_1,g_2,g_3,g_4)}
% \end{eqs}
% Note that, in the expression above, $\nu$ is implicitly a function of $\pi(h_{v_\partial})$, for vertices $v_\partial$ in $\partial M$. 
% Note that, $\nu$ is implicitly composed with $\pi$ for any argument corresponding to a vertex on the boundary of $M$.  

To see that $\ket{\Psi_b^L}$ is invariant under the $L$ symmetry, we act on the state with $V(\ell)$ in Eq.~\eqref{H symmetry}. After shifting the indices and using the homogeneity of $\nu$ and $\eta$, we find:
\begin{align}
    V(\ell)\ket{\Psi^L_b} = \sum_{\{ \ell_{{\bdryv}},g_v\}} \tilde{\Psi}_\eta(\{ \ell_{{\bdryv}}\}) \tilde{\Psi}_b(\{\pi( \ell_{{\bdryv}}),g_v\}) \ket{\{ \ell_{{\bdryv}},g_v\}},
\end{align}
where $\tilde{\Psi}_\eta(\{ \ell_{{\bdryv}}\})$ is the phase:
\begin{align}
    \tilde{\Psi}_\eta(\{ \ell_{{\bdryv}}\}) \equiv \prod_{f_\partial = \langle 123 \rangle} e^{-2 \pi i O_{f_{\partial}}\eta(\ell, \ell_1, \ell_2, \ell_3)}. \label{tilde mu amplitude}
\end{align}
% \begin{align}
%     \tilde{\Psi}_\nu(\{h_{v_\partial},g_v\}) &\equiv \prod_{t = \langle 1234 \rangle}e^{2 \pi i O_t \nu(\pi(h),g_1,g_2,g_3,g_4)}, \\ 
%     \tilde{\Psi}_\mu(\{h_{v_\partial}\}) &\equiv \prod_{f_\partial = \langle 123 \rangle} e^{-2 \pi i O_{f_{\partial}}\mu(h,h_1,h_2,h_3)}. \label{tilde mu amplitude}
% \end{align}
% \begin{eqs} \label{groupcoho gs3}
%     V(h) \ket{\Psi^H_b} =  &\sum_{\{h_{v_\partial},g_v\}} \Bigg[ \prod_{f_\partial = \langle 123 \rangle} e^{-2 \pi i O_{f_{\partial}}\mu(h,h_1,h_2,h_3)} \\ &\times \prod_{t = \langle 1234 \rangle}e^{2 \pi i O_t \nu(\pi(h),g_1,g_2,g_3,g_4)}\Bigg]\ket{\{h_{v_\partial},g_v\}}.
% \end{eqs}
The phase factors $\tilde{\Psi}_\eta(\{ \ell_{{\bdryv}}\})$ and $\tilde{\Psi}_b(\{\pi( \ell_{{\bdryv}}),g_v\})$ can be simplified by using the coboundary relations for $\nu$ and $\eta$. Similar to Eq.~\eqref{residual phases from sym}, the coboundary relation for $\nu$ leads to a residual phase factor: 
\begin{align} \label{nu amplitude variation}
    \tilde{\Psi}_b(\{\pi( \ell_{{\bdryv}}),g_v\}) = \mathcal{V}_\ell({{\{ \ell_{{\bdryv}},g_v\}}}) \Psi_b(\{\pi( \ell_{{\bdryv}}),g_v\}),
\end{align}
with $\mathcal{V}_\ell({{\{ \ell_{{\bdryv}},g_v\}}})$ given by:
\begin{align} \label{phase factor pull back}
    \mathcal{V}_\ell({{\{ \ell_{{\bdryv}},g_v\}}}) \equiv  \prod_{f_\partial = \langle 123 \rangle} e^{2 \pi i O_{f_\partial}\nu^*(1,\ell, \ell_1, \ell_2, \ell_3)}.
\end{align}
% \begin{eqs} \label{groupcoho gs4}
%     V(h) \ket{\Psi^H_b} =  &\sum_{\{h_{v_\partial},g_v\}} \Bigg[ \prod_{f_\partial = \langle 123 \rangle} e^{-2 \pi i O_{f_{\partial}}\mu(h,h_1,h_2,h_3)} \\ \times \mathcal{V}_{\scriptscriptstyle{\{h_{v_\partial},g_v\}}}(h)  &\prod_{t = \langle 1234 \rangle}e^{2 \pi i O_t \nu(1,g_1,g_2,g_3,g_4)}\Bigg]\ket{\{h_{v_\partial},g_v\}},
% \end{eqs}
% We substitute the expression in Eq.~\eqref{nu coboundary} for the factors of $\nu$, which leaves the residual phase factor in Eq.~\eqref{residual phase factor} on the boundary. 
As for $\eta$, the coboundary relation in Eq.~\eqref{coboundary mu} tells us:
\begin{eqs}
   &\eta(\ell, \ell_1, \ell_2, \ell_3)=\nu^*(1,\ell, \ell_1, \ell_2, \ell_3) + \eta(1, \ell_1, \ell_2, \ell_3) \\
   &- \eta(1,\ell, \ell_2, \ell_3) +\eta(1,\ell, \ell_1, \ell_3)-\eta(1,\ell, \ell_1, \ell_2).
\end{eqs}
The last three terms correspond to edges in $\partial M$ and cancel pairwise, when substituted into Eq.~\eqref{tilde mu amplitude}. The $\nu^*$ term in the coboundary relation of $\eta$ produces a phase that precisely cancels the excess phase factor in Eq.~\eqref{phase factor pull back}:
\begin{align} \label{mu amplitude variation}
    \tilde{\Psi}_\eta(\{ \ell_{{\bdryv}}\}) = \mathcal{V}^{-1}_\ell({{\{ \ell_{{\bdryv}},g_v\}}}) {\Psi}_\eta(\{ \ell_{{\bdryv}}\}).
\end{align}
Inserting Eqs.~\eqref{nu amplitude variation} and \eqref{mu amplitude variation} into the expression for $V(\ell) \ket{\Psi_b^L}$, we see that $\ket{\Psi_b^L}$ is symmetric under the $L$ symmetry. The $\eta$-dependent phase factors at the boundary compensates for the failure of the bulk wave function to be symmetric. (We note the similarity with anomalous SPT states introduced in Ref.~\cite{WQG19}.)

The state $\ket{\Psi_b^L}$ can be prepared from the product state:
\begin{align}
    \ket{\Psi^L} \equiv \sum_{\{ \ell_{{\bdryv}},g_v\}} \ket{\{ \ell_{{\bdryv}},g_v\}}
\end{align}
by the FDQC $\mathcal{U}_b^L$:
\begin{align} \label{group coho boundary circuit}
    \mathcal{U}_b^L \equiv \prod_{f_\partial} e^{-2 \pi i O_{f_\partial} \etahat(f_\partial)} \prod_t e^{2 \pi i O_t \nuhat(t)}.
\end{align}
Here, $\etahat(f_\partial)$ is the operator given by:
\begin{align}
    \etahat\boldsymbol{(}\langle 123 \rangle \boldsymbol{)} \equiv \sum_{\{ \ell_{{\bdryv}},g_v\}} \eta(1, \ell_1, \ell_2, \ell_3) \ket{\{ \ell_{{\bdryv}},g_v\}} \bra{\{ \ell_{{\bdryv}},g_v\}},
\end{align}
and $\nuhat(t)$ is explicitly:
\begin{eqs}
    \nuhat\boldsymbol{(} t \boldsymbol{)} = \sum_{\{ \ell_{{\bdryv}},g_v\}} {{\overline{\boldsymbol{\nu}}}_{\scriptscriptstyle{\{\pi( \ell_{\bdryv}),g_v\}}}} \ket{\{ \ell_{{\bdryv}},g_v\}} \bra{\{ \ell_{{\bdryv}},g_v\}},
\end{eqs}
with ${{\overline{\boldsymbol{\nu}}}_{\scriptscriptstyle{\{\pi( \ell_{\bdryv}),g_v\}}}}$ defined in Eq.~\eqref{bold bar rho}. $\mathcal{U}_b^L$ can be used to create a gapped parent Hamiltonian for $\ket{\Psi_b^L}$ by conjugating a certain paramagnet Hamiltonian whose ground state is $\ket{\Psi^L}$.

To recover the $G$ symmetry, we can gauge the $K$ subgroup of the $L$ symmetry. This results in a $G$ symmetric system with a $K$ gauge theory at the boundary. The $K$ gauge theory lives only on the boundary d.o.f., because the $K$ subgroup acts as the identity on the bulk sites. We have thus constructed a group cohomology model with a symmetric, topologically ordered gapped boundary.

\subsection{Gapped boundary construction for supercohomology models} \label{sec: fermionic gapped boundary construction}

We now generalize the construction of gapped boundaries for group cohomology models to build gapped boundaries for ($3+1$)D supercohomology models. The first step of the construction for group cohomology models is to find an extension of the $G$ symmetry to `trivialize' the cocycle $\nu$. That is, to find an extension $L$ such that the pull back of $\nu$ is a coboundary. Similarly, for supercohomology models, we first identify an extension of the $G$ symmetry to an $L$ symmetry that trivializes the supercohomology data $(\rho,\nu)$. Here, we say the supercohomology data is trivialized if the pull back $(\rho^*,\nu^*)$ is of the form:
\begin{align} \label{trivialization super data}
    (\rho^*,\nu^*) = (\delta \beta, \delta \eta + \frac{1}{2}\beta \cup \beta + \frac{1}{2}\beta \cup_1 \delta \beta),
\end{align}
for some $\beta \in C^2(L,\ZZ_2)$ and $\eta \in C^3(L,\RR/\ZZ)$. (The trivial supercohomology data in Eq.~\eqref{trivialization super data} was identified in Section~\ref{sec: supercohomology model}.) We then use the data $(\beta,\eta)$ to build a symmetric, topologically ordered gapped boundary for the supercohomology model. The detailed derivation from a $2$-group SPT model is given in Appendix~\ref{app: super from 2gauge}.
In Appendix~\ref{app:2Dgappedboundary}, we describe a similar construction of gapped, spontaneous symmetry breaking boundaries for ($2+1$)D supercohomology models. 

Before defining the supercohomology models on a manifold with boundary using $(\beta,\eta)$, we argue that there exists a central extension of $G$ for which the supercohomology data is trivialized, as in Eq.~\eqref{trivialization super data}. To show that such an extension exists, we make two consecutive extensions of $G$. The first is given by the short exact sequence:
\begin{equation} \label{first extension}
    1 \rightarrow K_1 \rightarrow L' \xrightarrow[]{\pi_1} G \rightarrow 1,
\end{equation}
and is chosen to trivialize $\rho$. We denote the pull backs of $\rho$ and $\nu$ to $L'$ by $\rho'$ and $\nu'$, respectively. By the definition of this extension, $\rho'$ can be written as:
$\rho' = \delta \beta'$,
for some $\beta' \in C^2(L',\ZZ_2)$.
Using $\nu'$ and $\beta'$, we then construct a cocycle:
\begin{align} \label{omega cocycle}
     \nu' + \frac{1}{2}\beta' \cup \beta' + \frac{1}{2}\beta' \cup_1 \delta \beta'.
\end{align}
The second extension is defined to trivialize the cocycle in Eq.~\eqref{omega cocycle} and corresponds to the short exact sequence:
\begin{equation} \label{second extension}
    1 \rightarrow K_2 \rightarrow L \xrightarrow[]{\pi_2} L' \rightarrow 1.
\end{equation}
Consequently, there exists $\eta \in C^3(L,\RR/\ZZ)$ such that:
\begin{align}
    \delta \eta = \nu^{*} + \frac{1}{2}\beta \cup \beta + \frac{1}{2}\beta \cup_1 \delta \beta,
    \label{equ:deltaeta}
\end{align}
where $\nu^{*}$ and $\beta$ are the pull backs of $\nu'$ and $\beta'$ by $\pi_2$. The extensions in Eqs.~\eqref{first extension} and \eqref{second extension} are guaranteed to exist by the arguments presented in Ref.~\cite{T17-2}. Since the composition of two central extensions is itself a central extension, there exists an extension:
\begin{equation}\label{super extension}
    1 \rightarrow K \rightarrow L \xrightarrow[]{\pi} G \rightarrow 1,
\end{equation}
which trivializes the supercohomology data $(\rho,\nu)$ such that the pull back $(\rho^*,\nu^*)$ is:
\begin{align} \label{super trivialization}
    (\rho^*,\nu^*) = (\delta \beta, \delta \eta + \frac{1}{2}\beta \cup \beta + \frac{1}{2}\beta \cup_1 \delta \beta),
\end{align}
for $\beta \in C^2(L,\ZZ_2)$ and $\eta \in C^3(L,\RR/\ZZ)$. In Appendix \ref{sec:example}, we give an example of the trivialization of supercohomology data by extending a $G=\ZZ_2 \times \ZZ_4$ symmetry. We also note that the two consecutive extensions above were used in Ref.~\cite{KOT19} to construct gapped boundaries for supercohomology models using a spacetime formalism. 

We now use an extension $L$ of $G$, given in Eq.~\eqref{super extension}, and the data $(\beta,\eta)$, satisfying Eq.~\eqref{super trivialization}, to define supercohomology models with a topologically ordered gapped boundary. 
% We take $H$ to be an extension of $G$ that trivializes the supercohomology data $(\rho,\nu)$, and we define $\beta \in C^2(H,\ZZ_2)$ and $\eta \in C^3(H,\RR/\ZZ)$ to be the cochains satisfying:
% \begin{align}
%     \delta \beta &= \rho^* \\
%     \delta \eta &= \nu^* + \frac{1}{2}\beta \cup \beta + \frac{1}{2}\beta \cup_1 \delta \beta,
% \end{align}
% with $\rho^*$ and $\nu^*$ denoting the pull backs of $\rho$ and $\nu$ to $H$.
To this end, we first build a gapped Hamiltonian $H^L_f$ with an $L$ symmetry and bulk terms that are equivalent to those of a supercohomology model with a $G$ symmetry. The $K$ subgroup of the $L$ symmetry can then be gauged to obtain a $G$ symmetric supercohomology model hosting a $K$ gauge theory at the boundary.  

\begin{figure}[t]
\centering
\includegraphics[width=0.4\textwidth, trim={800 450 650 300},clip]{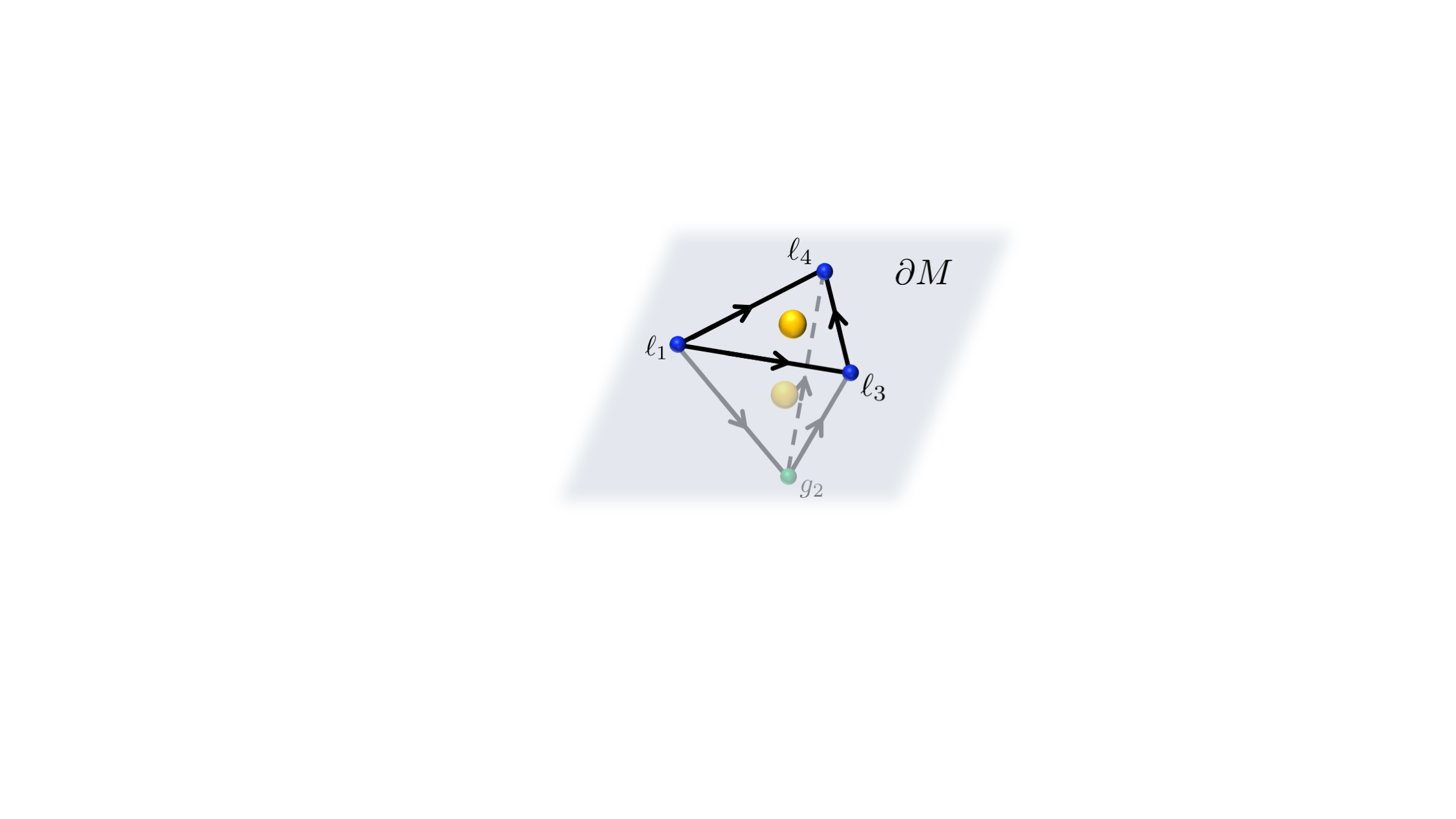}
\caption{The Hamiltonian $H^L_f$ acts on a Hilbert space with an $L$ d.o.f. (blue) on each boundary vertex and a spinless complex fermion (yellow) on every face of $\partial M$. The d.o.f. in the bulk, a $G$-valued spin (green) at every vertex and a spinless complex fermion on each tetrahedron, are the same as the d.o.f. in Fig.~\ref{fig: Gai dof2}.}
\label{fig: bdrydoff}
\end{figure}

\begin{figure}[t]
\centering
\includegraphics[width=0.4\textwidth, trim={900 550 620 200},clip]{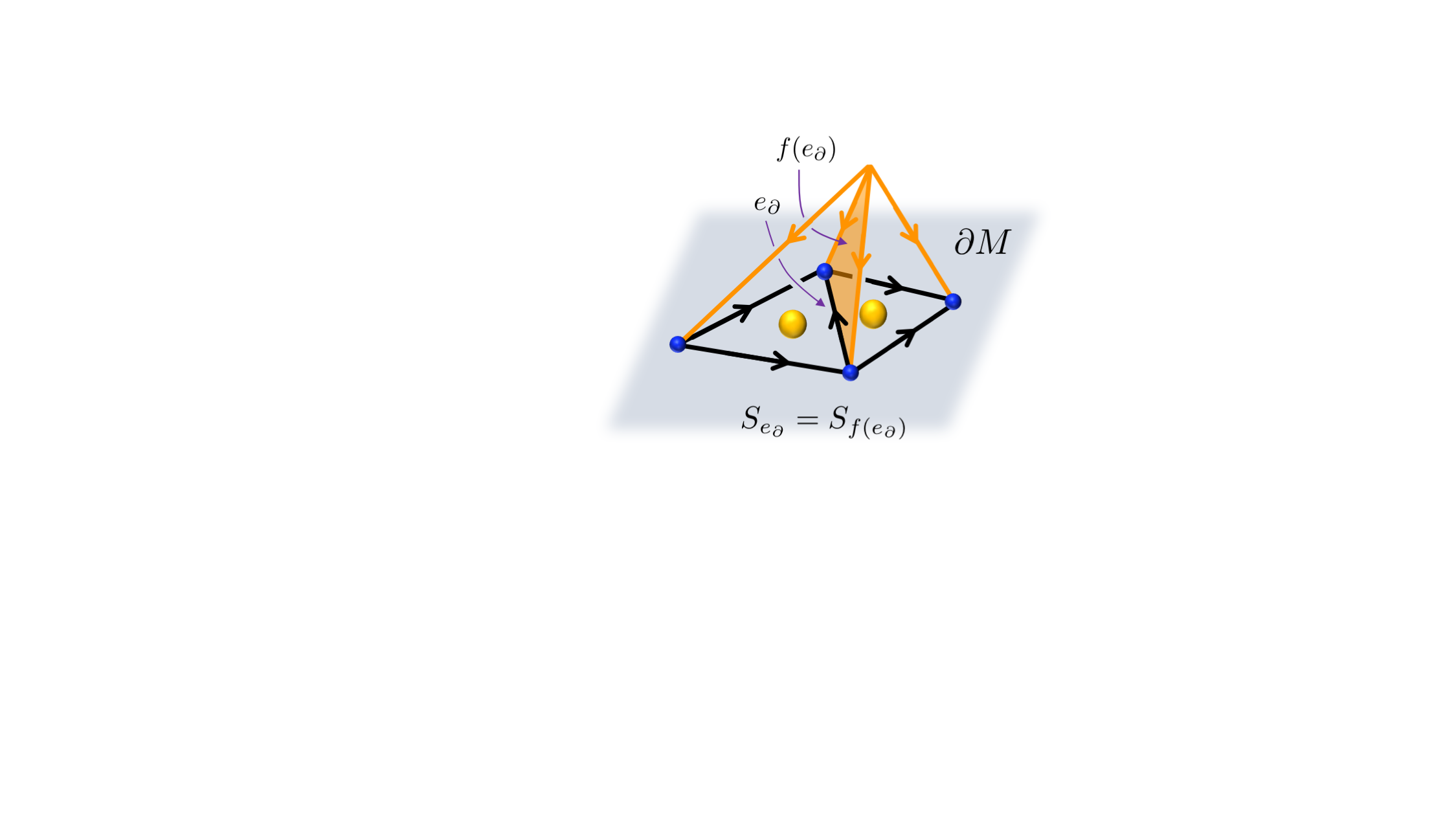}
\caption{The hopping operator $S_{e_\partial}$ on the boundary is defined by connecting the vertices in the boundary to an artificial vertex. The edges connecting to the artificial vertex are shown in orange and are oriented away from the additional point. The boundary edge $e_\partial$ is associated to the artificial face $f(e_\partial)$ (shaded orange). For clarity the bulk d.o.f. are not pictured.}
\label{fig: hoppingboundary}
\end{figure}

The Hilbert space for the $L$ symmetric Hamiltonian $H^L_f$ can be constructed from the Hilbert space of the bulk supercohomology models. We recall that the supercohomology models on a manifold without boundary are defined on a Hilbert space with a $G$ d.o.f. at each vertex and a single complex fermion on each tetrahedron, as in Fig.~\ref{fig: Gai dof2}. For $H^L_f$ on a manifold $M$ with boundary $\partial M$, we replace the $G$ d.o.f. on the boundary with $L$ d.o.f. and introduce a single spinless complex fermion on each face in $\partial M$ (see Fig.~\ref{fig: bdrydoff}). To define the hopping operators on this Hilbert space, we imagine extending the manifold to $\overline{M}$ by connecting all of the boundary vertices to an additional ``artificial'' vertex. Each fermionic d.o.f. on a boundary face can then be associated to a tetrahedron connected to the artificial vertex, and every boundary edge $e_\partial$ can be assigned a face $f(e_\partial)$ containing the artificial vertex, as pictured in Fig.~\ref{fig: hoppingboundary}. 
The trick of adding a vertex allows one to unambiguously determine the spin structure dependent sign in the definition of the hopping operator, discussed in Appendix~\ref{sec: review of boson-fermion duality}. We use $S_{e_\partial}$ to denote the hopping operator $S_{f(e_\partial)}$ between ferminionic d.o.f. at boundary faces:
\begin{align}
    S_{e_\partial} \equiv S_{f(e_\partial)}.
\end{align}
Similar to Eq.~\eqref{H symmetry}, the $L$ symmetry is represented by:
\begin{eqs} \label{H symmetry fermions}
   V(\ell) = \sum_{\{ \ell_{{\bdryv}},g_v \}} \ket{\{\ell \ell_{{\bdryv}},\pi(\ell)g_v\}}\bra{\{ \ell_{{\bdryv}},g_v\}},
\end{eqs}
tensored with the identity on the fermionic d.o.f..

The Hamiltonian $H^L_f$ is formed from a trivial Hamiltonian - with a product state ground state - by conjugation with a FDQC. The trivial Hamiltonian $H^L_\text{AI}$ is an atomic insulator with a decoupled $L$-paramagnet, defined as:
\begin{align}
    H^L_\text{AI} \equiv -\sum_{v \notin \partial M} \mathcal{P}^G_v - \sum_{w \in \partial M} \mathcal{P}_w^L - \sum_{f_\partial \in \partial M} P_{f_\partial} - \sum_{t \in M} P_t.
\end{align}
Here, $\mathcal{P}^G_v$ and $\mathcal{P}_w^L$ are projectors onto the symmetric state at the bulk vertex $v$ and boundary vertex $w$, respectively, and $P_{f_\partial}$ is the fermion parity operator at a face $f_\partial$ in $\partial M$.
% Here, $\mathcal{P}^G_v$ and $\mathcal{P}_w^H$ are projectors onto a symmetric state at a bulk vertex $v$ and boundary vertex $w$, respectively:
% \begin{align}
%     \mathcal{P}^G_v &= \frac{1}{|G|} \left(  \sum_{g_v} \ket{g_v} \right)\left( \sum_{g_v} \bra{g_v} \right) \\
%     \mathcal{P}_w^H &= \frac{1}{|H|} \left(  \sum_{h_w} \ket{h_w} \right)\left( \sum_{h_w} \bra{h_w} \right).
% \end{align}

We prepare $H^L_f$ from $H^L_\text{AI}$ by conjugation with the FDQC $\Uf^L$: 
% \begin{eqs} \label{eq: ufH def}
%   \Uf^H \equiv &\prod_{f_\partial} e^{-2 \pi i O_{f_\partial} \etahat(f_\partial)} \xi_\beta(\partial M) \prod_{e_\partial} S_{e_\partial}^{\betahat(e_\partial)} \\ 
%  \times &\prod_t e^{2 \pi i O_t \nuhat(t)} \xi_{\bar{\rho}}(M){\prod_{f}} S_f^{\rhohat(f)} \\
%  \times &\prod_{f_\partial} P_{f_\partial}^{\int_{\partial M} \bface \cup_1 \betahat} \prod_{t} P_t^{\int \rhohat \cup_2 \boldsymbol{t}} . 
% \end{eqs}
% \begin{eqs} \label{eq: ufH def}
%   \Uf^L \equiv &\prod_{f_\partial} P_{f_\partial}^{\int_{\partial M} \boldsymbol{f_\partial} \cup_1 \betahat}  \chi_\beta(\partial M) \prod_{e_\partial} S_{e_\partial}^{\betahat(e_\partial)} \prod_{f_\partial} e^{-2 \pi i O_{f_\partial} \etahat(f_\partial)} \\ 
%  \times &\prod_t e^{2 \pi i O_t \nuhat(t)} \xi_{\bar{\rho}}(M){\prod_{f}} S_f^{\rhohat(f)} \prod_{t} P_t^{\int_M \rhohat \cup_2 \boldsymbol{t}} . 
% \end{eqs}
\begin{eqs} \label{eq: ufH def}
  \Uf^L \equiv &\prod_{f_\partial} e^{-2 \pi i O_{f_\partial} \etahat(f_\partial)}  \chi_\beta(\partial M) \prod_{e_\partial} S_{e_\partial}^{\betahat(e_\partial)}  \prod_{f_\partial} P_{f_\partial}^{\int_{\partial M} \boldsymbol{f_\partial} \cup_1 \betahat} \\ 
 \times &\prod_t e^{2 \pi i O_t \nuhat(t)} \xi_{\bar{\rho}}(M){\prod_{f}} S_f^{\rhohat(f)} \prod_{t} P_t^{\int_M \rhohat \cup_2 \boldsymbol{t}} . 
\end{eqs}
% Let us break down the formula above. 
The first line of Eq.~\eqref{eq: ufH def} is a FDQC supported on the boundary sites, while the second line is the FDQC that prepares the bulk supercohomology model (composed with the projector $\pi$ on the boundary vertices). The $\etahat$ term in Eq.~\eqref{eq: ufH def} is the analog of the $\etahat$ term in Eq.~\eqref{group coho boundary circuit} for the group cohomology models. 
In the configuration basis, $\chi_\beta(\partial M)$ is a sign that depends on an ordering of the faces in $\partial M$ and makes up for the order dependence of the product of $S_{e_\partial}$ hopping operators; we give the explicit form of $\chi_\beta(\partial M)$ in Eq,~ \eqref{eq: chi beta def} in Appendix~\ref{app: super from 2gauge}. Furthermore, $\betahat(e_\partial)$ is the diagonal operator:
\begin{eqs}
    \betahat\boldsymbol{(}\langle 12 \rangle \boldsymbol{)} \equiv \sum_{\{ \ell_{{\bdryv}},g_v\}} \beta(1, \ell_1, \ell_2) \ket{\{ \ell_{{\bdryv}},g_v\}} \bra{\{ \ell_{{\bdryv}},g_v\}},
\end{eqs}
defined for an arbitrary edge $e_\partial$ in the boundary of $M$. 
% In particular, $\chi_\beta(\partial M)$ also depends on the ordering of the edges in $\partial M$ and is defined as:
% \begin{multline}
%     \chi_\beta(\partial M) \equiv (-1)^{\int_{\partial M}\left(\hat{\boldsymbol{\beta}} \cup_1 \betahat + \delta \betahat \cup_1 \betahat \right)} \\ \times \prod_{e_i,e_{i'} \in \partial M|i<i'} (-1)^{\betahat(e_{i})\betahat(e_{i'}) \int_{\partial M} \be_{i'} \cup \be_{i} },
% \end{multline}
% with $\hat{ \boldsymbol{\beta}}(f_\partial)$ the operator given by:
% \begin{align}
%     \hat{\boldsymbol{\beta}}\boldsymbol{(}\langle 123 \rangle \boldsymbol{)} \equiv \sum_{\{h_{{\bdryv}},g_v\}} \beta(h_1,h_2,h_3) \ket{\{h_{{\bdryv}},g_v\}} \bra{\{h_{{\bdryv}},g_v\}},
% \end{align}
% for any boundary face $f_\partial = \langle 123 \rangle$.
With this, $H_f^L$ is the Hamiltonian:
\begin{align}
    H_f^L \equiv \Uf^L H^L_\text{AI} \Uf^L{}^\dagger.
\end{align}

The derivation of the model above largely follows the construction of the bulk models, in that, we first build a model for a $2$-group SPT phase, then subsequently gauge the $1$-form symmetry and apply the fermionization duality. However, special care is needed at the boundary, and we give the full detail in Appendix~\ref{app: super from 2gauge}. As a consistency check, notice that when $\rho$ and $\beta$ are zero, the FDQC $\Uf^L$ reduces to the FDQC $\mathcal{U}_b^L$ for the group cohomology case in Section~\ref{sec: group coho boundary}. We also note that $H^L_f$ agrees with the $G$-symmetric bulk supercohomology model on the interior of $M$. This is because, away from the boundary of $M$, the action of $\Uf^L$ is equivalent to that of the bulk circuit $\Uf$.

$H^L_f$ has a global $L$ symmetry, so to recover a supercohomology model with a $G$ symmetry, one can gauge the $K$ subgroup of $L$. Due to the peculiar $L$ symmetry in Eq.~\eqref{H symmetry fermions}, the $K$ gauge fields live only on the boundary of $M$. Therefore, after gauging the $K$ symmetry, we obtain a $G$-symmetric supercohomology model with a gapped boundary, hosting a $K$ gauge theory.

\section{Discussion}\label{sec:conclusion}

We have constructed exactly solvable lattice Hamiltonians for supercohomology fermionic SPT (fSPT) phases in ($3+1$)D. Moreover, we have identified finite-depth quantum circuits (FDQCs) that prepare the SPT ground states from symmetric product states. The Hamiltonians are of the simple form:
\begin{align} \label{Hfform}
    H_f = \Uf H_\text{AI}^G \Uf^\dagger,
\end{align}
where $\Uf$ is the FDQC that prepares the ground state and $H_\text{AI}^G$ has a unique unentangled symmetric ground state. With our models, we are able to explicitly compute the supercohomology invariants by gauging the fermion parity symmetry and calculating the symmetry fractionalization on the flux loops. We also generalized the gapped boundary construction for group cohomology models in Ref.~\cite{WWW19} to construct gapped boundaries for the supercohomology models through a symmetry extension. 

Our construction is based on the correspondence between certain bosonic $2$-group SPT phases and the supercohomology fSPT phases, first recognized in Refs.~\cite{BGK17} and \cite{KT15}. By gauging the $\ZZ_2$ $1$-form symmetry of the $2$-group SPT phase, we obtain the shadow model -- a $\ZZ_2$ gauge theory with an emergent fermion. The emergent fermion can be interpreted as the gauge charge of an fSPT phase after gauging fermion parity. The supercohomology model results from condensing the emergent fermion, or equivalently, applying the fermionization duality of Ref.~\cite{CK18}. 

We would like to point out that by adding a disordered or quasi-periodic onsite potential to the Hamiltonian in Eq.~\eqref{Hfform}, the abelian supercohomology models can in principle exhibit many-body localization, or at least a long-lived pre-thermal regime \cite{PV16,1d_MBL_SPT_1,1d_MBL_SPT_2,Pal,Chandran1,Bauer1,EBN17}. This is because each eigenstate of the disordered $H_f$ is short-range entangled and can be viewed as a representative ground state of the supercohomology SPT phase. 

We also comment further on the relation between our models and the Lagrangian formulation of Ref.~\cite{LZW18}.
In Ref.~\cite{LZW18}, it was shown that the shadow theory can be thought of as a $G$-symmetry-enriched $\ZZ_2$ gauge theory with a Steenrod square topological term in the action, which transmutes the statistics of the point-like gauge charge. In our case, we built the $2$-group SPT model from a $2$-group cocycle $\alpha = \nu + \frac{1}{2} \rho \cup_1 \eps + \frac{1}{2} \eps \cup \eps$, described in Section~\ref{sec: 2-groupspt} and Appendix~\ref{sec: 2-group extension}. The cocycle $\alpha$ is cohomologous to $\alpha' \equiv \nu + \frac{1}{2} \eps \cup_1 \rho + \frac{1}{2} \eps \cup \eps$, which can be expressed in terms of the Steenrod square $Sq^2$ as $\alpha'=\nu + \frac{1}{2} Sq^2 \eps$. $\alpha'$ is in the same form as the action in Ref.~\cite{LZW18}, and accordingly, our shadow model Hamiltonians can be understood as the Hamiltonian formulation of the Lagrangians in Ref.~\cite{LZW18}, for a unitary $G_f=G\times \ZZ_2^f$.

There are many interesting potential generalizations of our models and avenues for future work. We briefly comment on them below.

% \vspace{.2mm}
% \noindent \begin{center}\emph{Connection to the Lagrangian formulation:}\end{center}
% \vspace{.2mm}

% We expect that their topological term can be recovered by constructing a Euclidean path integral from the $2$-group SPT Hamiltonian in this paper. 
% It would be interesting to work out the details explicitly.

% \textbf{Anomalous symmetry-enriched topological orders:} 

% \vspace{.2mm}
% \noindent \begin{center}\emph{Supercohomology models in higher dimensions:}\end{center}
% \vspace{.2mm}

\textbf{Supercohomology models in higher dimensions:} We conjecture that supercohomology models in $(n+1)$D can be constructed using a similar approach. In general, the supercohomology data $(\rho,\nu)$ belongs to $Z^n(G,\ZZ_2) \times C^{n+1}(G, \RR / \ZZ)$ and satisfies the constraints:
\begin{align}
    \delta \rho = 0, \quad \delta \nu = \frac{1}{2} \rho \cup_{n-2} \rho.
\end{align}
Using $\rho$ and $\nu$ one can first build an auxiliary bosonic SPT model with an $(n-1)$-group symmetry with the $(n-1)$-group cocycle:
\begin{equation}
    \alpha = \nu + \frac{1}{2} \rho \cup_{n-2} \eps_{n-1} + \frac{1}{2} \eps_{n-1} \cup_{n-3} \eps_{n-1}.
\end{equation}
Here, $\eps_{n-1}$ can be pulled back to $M$ to give a cochain $\boldsymbol{\eps}_{n-1} \in C^{n-1}(M,\ZZ_2)$ satisfying $\delta \boldsymbol{\eps}_{n-1} = \boldsymbol{\rho}$.\footnote{For simplicity, we have suppressed the configuration dependence in the subscript used in Section~\ref{sec: supercohomology data}.} In principle, one can gauge an $(n-2)$-form $\ZZ_2$ subgroup to build the shadow model and then apply the fermionization duality of Ref.~\cite{C19-2} to construct the fSPT model. We therefore, see no obstruction to finding symmetric FDQCs that prepare the ground states from a symmetric product state, unlike the beyond cohomology model in Ref.~\cite{HFH18} and the beyond supercohomology model in Appendix G of Ref.~\cite{EF19}.

% \vspace{.2mm}
% \noindent \begin{center}\emph{Time-reversal and nontrivial extensions by fermion parity:}\end{center}
% \vspace{.2mm}

\textbf{Time-reversal and nontrivial extensions by fermion parity:} Our supercohomology models are protected by unitary symmetries of the form $G_f = G \times \ZZ_2^f$. An important generalization is to symmetries which may include anti-unitary symmetries, such as time-reversal, and for which $G_f$ is a nontrivial central extension of $G$ by fermion parity. These cases are outside of the supercohomology framework, however, we expect some of our results to apply more broadly. 

To include time-reversal symmetries, we can modify $\nu$ as in Refs.~\cite{CGLW13} and \cite{GW14} so that the homogeneity of $\nu$ is replaced with:
\begin{eqs} 
        (-1)^{s(h)}\nu(g_0,g_1,g_2,g_3,g_4)&=\nu(hg_0,hg_1,hg_2,hg_3,hg_4),
\end{eqs}
where $s(h) \in \{0,1\}$ is $1$ if $h$ includes time-reversal. We expect that, with this modification, the FDQC $\Uf$ will prepare the ground state of the corresponding fSPT model -- the symmetry fractionalization on fermion parity flux loops can be generalized to anti-unitary symmetries as described in Appendix B of Ref.~\cite{C15}. However, the equivalence relation in Eq.~\eqref{supercohomology equivalence relation} relies on the assumption that $G$ is unitary. In fact, the models with time-reversal can be ``trivialized'' by accounting for the beyond supercohomology data as described in Refs.~\cite{WG18} and \cite{WQG19}.

More generally, $G_f$ can take the form $G_f = G \times_\psi \ZZ_2^f$, where $G_f$ is a nontrivial central extension of $G$ by $\ZZ_2^f$, specified by a 2-cocycle $\psi \in H^2(G,\ZZ_2)$. Each element of $G_f$ can be written as $g\Pi^m$, with $g \in G$, $m \in \ZZ_2$ and $\Pi$ denoting global fermion parity. The group laws in $G_f$ are defined by:
\begin{align}
    \left(g\Pi^m \right)\left( h\Pi^n \right)= (gh)\Pi^{m+n+\psi(1,g,gh)}.
\end{align}
The supercohomology data $(\rho,\nu)$ is modified to satisfy \cite{WG18}:
\begin{eqs} \label{modified guwen extension}
    \delta \rho = 0, \quad \delta \nu = \frac{1}{2} \rho \cup_1 \rho + \frac{1}{2} \psi \cup \rho.
\end{eqs}

We can define a model for the fSPT protected by $G_f = G \times_\psi \ZZ_2^f$ corresponding to $(\rho,\nu)$ in Eq.~\eqref{modified guwen extension}. To do so, we first describe the representation of the symmetry on a Hilbert space with $G$ d.o.f. on vertices and a spinless complex fermion on each tetrahedron. For any $g\Pi^m$ in $G_f$, the $G_f$ symmetry acts as:
\begin{align} \label{Gfsymextension}
    \mathbb{V}(g\Pi^m) \equiv \prod_v \mathbb{V}_v(g\Pi^m),
\end{align}
with the product over vertices and $\mathbb{V}_v(g\Pi^m)$ the symmetry action associated to $v$. Here, $\mathbb{V}_v(g\Pi^m)$ is defined as:
\begin{align}
    \mathbb{V}_v(g\Pi^m) \equiv V_v(g) \prod_{t=\langle 1234 \rangle | v=\langle 1\rangle}P_t^m,
\end{align}
where $V_v(g)$ is the regular representation of $g$ at $v$, and we have associated the fermionic d.o.f. at the tetrahedra $t=\langle 1234 \rangle$ to the vertex $\langle 1 \rangle$. 

It can be checked that $\Uf$ built with the modified $(\rho,\nu)$ is symmetric under the representation of $G_f$ in Eq.~\eqref{Gfsymextension}, so we can define a Hamiltonian $H_f$ as in Eq.~\eqref{Hfform}. Furthermore, it can be shown that the symmetry fractionalizes on the fermion parity flux loops according to $\rho$ as in Section~\ref{sec: shadow model}. Finally, using a similar argument as in Ref.~\cite{EF19}, it can be verified that the symmetry fractionalization on the fermion parity gauge charges is determined by $\psi$, in accordance with Ref.~\cite{CTW17}. However, in this case, more work is needed to understand both the equivalence relations and the corresponding auxiliary bosonic SPT phases.

% \vspace{.2mm}
% \noindent \begin{center}\emph{Beyond supercohomology models:}\end{center}
% \vspace{.2mm}

\textbf{Beyond supercohomology models:} We showed in Section~\ref{sec: supercohomology models 2} that the ground states of the supercohomology models have $(0+1)$D junctions of symmetry domain walls decorated by complex fermions. The domain wall decoration picture can be extended to the beyond supercohomology phases, where the fixed point wave functions feature $(1+1)$D and $(2+1)$D junctions of symmetry domains decorated by Majorana wires and $p+ip$ superconductors, respectively \cite{WG18}. 
Although exactly-solvable models for beyond supercohomology phases have been constructed in Refs.~\cite{FCV13,KT15,TF16,BGK17,KT17,WNC18,WG182,SA19,K20}, it would be interesting to search for models of the form in Eq.~\eqref{Hfform} -- related to a trivial SPT Hamiltonian by conjugation with a locality preserving unitary. Such a construction might have implications for the classification of fermionic quantum cellular automata, analogous to the beyond group cohomology models in Ref.~\cite{HFH18}. 
% we look forward to work relating the decorated domain wall models in $(3+1)$D with a physical picture for the responses due to probing with symmetry defects, as in Ref.~\cite{CBYG18}. 
It would also be interesting to study the boundaries of the beyond supercohomology models in $(3+1)$D using exactly-solvable models, similar to Ref.~\cite{M19}.

\vspace{0.1in}
\noindent{\it Acknowledgements -- }
YC thanks Anton Kapustin, Ryohei Kobayashi, Po-Shen Hsin, and Bowen Yang for many useful discussions. YC was supported in part by the US
Department of Energy, Office of Science, Office of High Energy Physics, under Award Number DE-SC0011632.
TDE would like to acknowledge Sujeet Shukla and Lukasz Fidkowski for their work on a preliminary construction of the ($3+1$)D fSPT models. TDE also thanks Lukasz Fidkowski for carefully explaining the results of Ref.~\cite{FVM17}, Davide Gaiotto for a useful discussion related to the model in Section~\ref{sec ttc}, and Theo Johnson-Freyd for valuable discussions regarding group supercohomology. TDE is grateful for the hospitality of Perimeter Institute, where much of his work was completed. Perimeter Institute is supported in part by the Government of Canada through the Department of Innovation, Science and Economic Development Canada and by the Province of Ontario through the Ministry of Economic Development, Job Creation and Trade. NT is grateful to Ashvin Vishwanath for helpful discussions and would also like to thank Zheng-Cheng Gu and the participants of the Croucher summer course "Quantum Entanglement and Topological Order" at CUHK for various discussions. NT is supported by NSERC. Part of this work was done during the Simons Collaboration on Ultra Quantum Matter Workshop, which was supported by a grant from the Simons Foundation (651440).

\appendix

\section{Terminology from cohomology} \label{app:term}

Here, we compile the cohomology notation used in the main text. This includes the group cohomology notation used to define the supercohomology data as well as the simplicial cohomology notation employed to describe the construction of the supercohomology models. For both, we define group cochains, the coboundary operator, and cup products.

 \subsection{Group cohomology}\label{app:groupcoho}
 
For our purposes, a $p$-cochain is a homogeneous function from $G^{p+1}$ to $A$, where $G$ is a finite group and $A$ is either ${\ZZ_2=\{0,1\}}$ or ${\mathbb{R}/\ZZ=[0,1)}$. By a homogeneous function, we mean that the $p$-cochain $c$ satisfies:
\begin{align}
    c(g_0,\ldots,g_{p}) = c(hg_0,\ldots,hg_{p}), \quad \forall h \in G.
\end{align}
The collection of $p$-cochains is denoted as $C^p(G,A)$.

The coboundary operator $\delta$ maps a $p$-cochain to a $(p+1)$-cochain. Explicitly, the coboundary operator maps $c$ to the $(p+1)$-cochain $\delta c$, defined as:
\begin{align} \label{groupcoboundarydef}
     \delta c (g_0,\ldots,g_{p+1}) \equiv \sum_{i=0}^{p+1} (-1)^{i}c(g_0,\ldots,\widehat{g_i},\ldots,g_{p+1}),
\end{align}
where $\widehat{g_i}$ indicates that $g_i$ has been omitted. When $A=\ZZ_2$, the sign in Eq.~\eqref{groupcoboundarydef} can be ignored.
% If $A=\ZZ_2$, the minus sign in Eq.~\eqref{groupcoboundarydef} can be ignored. 
A $p$-cochain $c$ satisfying $\delta c = 0$ is called a $p$-cocycle, and we use $Z^p(G,A)$ to denote the set of $p$-cocycles. 
 
% Throughout the main text, we use the homogeneous version of group cohomology.
% A $p$-cochain $\nu$ is a homogeneous function from $G^{p+1}$ to an abelian group $A$. The homogeneity condition is
%  \begin{align}
%      \nu(g_0,\ldots,g_p) = \nu(hg_0,\ldots,hg_p).
%  \end{align}
% and $A$ is either $\RR/\ZZ$ or $\ZZ_2$ with addition as the group multiplication.

% In parallel to simplicial cohomology, we can also define coboundary operators and cup products. The group coboundary operator is defined as
% \begin{align}
%      \delta \nu (g_0,\ldots,g_p) \equiv \sum_{i=0}^p (-1)^{i}\nu(g_0,\ldots,\widehat{g_i},\ldots,g_p).
% \end{align}
% where $\widehat{g_i}$ means omitting $g_i$. In the case of $\ZZ_2$ group cochains, the minus sign can be dropped.

We can impose an equivalence relation on $Z^p(G,A)$ to define the $p^\text{th}$ group cohomology. We call the $p$-cocycles $c$ and $c'$ equivalent if there exists a $(p-1)$-cochain $d\in C^{p-1}(G,A)$ such that:
\begin{align}
    c' = c + \delta d.
\end{align}
The set of equivalence classes under the equivalence relation above defines the $p^\text{th}$ group cohomology, denoted $H^p(G,A)$. 

Throughout our calculations, we assume that the cocycles are normalized. That is, the $p$-cocycle $c$ satisfies:
\begin{align}
    c(1,1,g_2,\ldots,g_p) = 0, \quad \forall g_i, \,\, i \in \{2,\ldots,p\},
\end{align}
where $1$ is the identity in $G$. This assumption is justified by the fact that every group cohomology equivalence class has a normalized representative. 

Lastly, for $A=\ZZ_2$, we define the cup-$n$ products $\cup_n$ with $n \in \{0,1,2\}$. The cup-$n$ product maps a $p$-cochain $c$ and a $q$-cochain $d$ to a ${(p+q-n)}$-cochain ${c \cup_n d}$. We note that the cup-$0$ product is referred to as simply the cup product and is denoted by $\cup$. The cup product of the homogeneous group cochains $c \in C^p(G,\ZZ_2)$ and $d \in C^q(G,\ZZ_2)$ is the $(p+q)$-cochain given by:
\begin{align}
    c \cup d (g_0,\ldots,g_{p+q}) \equiv c(g_0,\ldots,g_p)d(g_p,\ldots,g_{p+q}).
\end{align}
The cup-$1$ product of a $p$-cochain $c$ and a $q$-cochain $d$ is the $(p+q-1)$-cochain defined by:
\begin{align}
    &c \cup_1 d (g_0,\ldots,g_{p+q})= \\ \nonumber
    &\sum_{i=0}^{p-1}c( g_0, \ldots , g_i, g_{q+i}, \ldots ,g_{p+q-1})d( g_i,\ldots,g_{q+i}). 
\end{align}
We refer to Ref.~\cite{C19-2} for the general formula for the cup-$2$ product. In the main text, we only ever use the cup-$2$ product between two group $3$-cochains. Hence, we give the explicit cup-$2$ product of $c$ and $d$ with $c,d \in C^3(G,\ZZ_2)$:
\begin{align} \label{explicit cup2}
    c \cup_2 d (g_1,g_2,g_3,g_4,g_5) &\equiv  c(g_1,g_2,g_3,g_4) d(g_1,g_2,g_4,g_5) \nonumber \\ 
    &+ c(g_1,g_3,g_4,g_5) d(g_1,g_2,g_3,g_5) \nonumber \\ 
    &+ c(g_1,g_2,g_3,g_4) d(g_2,g_3,g_4,g_5) \nonumber \\ 
    &+ c(g_1,g_2,g_4,g_5) d(g_2,g_3,g_4,g_5).
\end{align}

\subsection{Simplicial cohomology} \label{app: terminology}
% \subsection{Simplicial cohomology on $M$ with coefficients in $\mathbb Z_2$} \label{app: terminology}

Simplicial cohomology on $M$ with coefficients in $\ZZ_2$ was introduced in Section~\ref{sec: 1formspt} in the context of the $1$-form SPT model. Here, we summarize the terminology from Section~\ref{sec: 1formspt} and give the cup product relations that are used in the appendices.

% The concepts were introduced to discuss the $1$-form SPT model in Section \ref{sec: 1formspt}. 

Given a triangulation of a manifold $M$, we denote the vertices, edges, faces, and tetrahedra by $v$, $e$, $f$, and $t$, respectively. We often denote a $p$-simplex by its $p+1$ vertices, i.e., $\langle 0 ,\ldots ,  p \rangle$. (Elsewhere in the text, we omit the commas between vertices for simplicity.) We define a $p$-chain as a formal sum ($\text{mod }2$) of $p$-simplices in the manifold $M$.

A $p$-cochain on $M$ is a linear, $\ZZ_2$-valued function of $p$-chains. The set of $p$-cochains on $M$ is denoted by $C^p(M,\ZZ_2)$. We use a bold font for cochains on $M$, e.g., $\boldsymbol{c} \in C^p(M,\ZZ_2)$.

A cochain labeled by a $p$-simplex is a $p$-cochain that evaluates to $1$ on the corresponding $p$-simplex and $0$ otherwise. For example, $\bv$ denotes the $0$-cochain dual to the vertex $v$, i.e.:
\begin{align} 
    \bv(v') = \begin{cases} 
      1 & v'=v \\
      0 & \text{otherwise}. 
   \end{cases}
\end{align} 
Likewise, for an edge $e$, we have:
\begin{align} 
    \be(e') = \begin{cases} 
      1 & e'=e \\
      0 & \text{otherwise},
   \end{cases}
\end{align} 
and for a face $f$:
\begin{align} 
    \bface(f') = \begin{cases} 
      1 & f'=f \\
      0 & \text{otherwise}. 
   \end{cases}
\end{align} 

The coboundary operator $\delta$ is a linear map from $p$-cochains to $(p+1)$-cochains:
\begin{align}
    \delta: C^p(M,\ZZ_2) \to C^{p+1}(M,\ZZ_2).
\end{align}
The coboundary of a $p$-cochain $\boldsymbol{c}$ is defined as the $(p+1)$-cochain $\delta \boldsymbol{c}$ such that:
\begin{align} \label{coboundarydef1}
    \delta \boldsymbol{c}(s) = \boldsymbol{c}(\partial s),
\end{align}
for an arbitrary $(p+1)$-simplex $s$ and $\partial s$ its boundary. Explicitly, $\partial s$ is an equally weighted sum of the $p$-simplices in $s$.

A $p$-cochain $\boldsymbol{c}$ is called closed, if $\delta \boldsymbol{c}=0$. We denote the collection of closed $p$-cochains on $M$ as $Z^p(M,\ZZ_2)$.
We also note that $\delta \delta = 0$, which follows from Eq.~\eqref{coboundarydef1} and the fact that $\partial \partial = 0$. Therefore, $\delta \boldsymbol{d}$ is a closed $p$-cochain for any $\boldsymbol{d}\in C^{p-1}(M,\ZZ_2)$.

The cup product $\cup$ maps a $p$-cochain and a $q$-cochain to a $(p+q)$-cochain:
\begin{align}
    \cup: C^p(M,\ZZ_2) \times C^q(M,\ZZ_2) \to C^{p+q}(M,\ZZ_2).
\end{align}
The cup product of $\boldsymbol{c} \in C^p(M,\ZZ_2)$ and $\boldsymbol{d} \in C^q(M,\ZZ_2)$ evaluated on a $(p+q)$-simplex $\langle 0 \ldots p+q \rangle$ is:
\begin{align}
    \boldsymbol{c} \cup \boldsymbol{d} \boldsymbol{(}\langle 0 \ldots p+q \rangle\boldsymbol{)} = \boldsymbol{c} \boldsymbol{(}\langle 0 \ldots p \rangle \boldsymbol{)} \boldsymbol{d}\boldsymbol{(}\langle p  \ldots p+q\rangle\boldsymbol{)}.
\end{align}
The coboundary operator is a derivation, meaning it satisfies:
\begin{align} \label{Liebnizrule}
    \delta \left(\boldsymbol{c} \cup \boldsymbol{d}\right) = \delta \boldsymbol{c} \cup \boldsymbol{d} +  \boldsymbol{c} \cup \delta \boldsymbol{d}.
\end{align}

The cup-$1$ product $\cup_1$ produces a $(p+q-1)$-cochain from a $p$-cochain and a $q$-cochain:
\begin{align}
    \cup_1 : C^p(M,\ZZ_2) \times C^q(M,\ZZ_2) \to C^{p+q-1}(M,\ZZ_2).
\end{align}
For $\boldsymbol{c} \in C^p(M,\ZZ_2)$ and $\boldsymbol{d} \in C^q(M,\ZZ_2)$ the cup-$1$ product $\boldsymbol{c} \cup_1 \boldsymbol{d}$ evaluated on the $(p+q-1)$-simplex $\langle 0, \ldots , p+q-1 \rangle$ is:
\begin{align}
    &\boldsymbol{c} \cup_1 \boldsymbol{d} \boldsymbol{(} \langle 0, \ldots, {p+q-1} \rangle \boldsymbol{)}= \\ \nonumber
    &\sum_{i=0}^{p-1}\boldsymbol{c}\boldsymbol{(} \langle 0, \ldots , i, {q+i}, \ldots ,{p+q-1} \rangle \boldsymbol{)} \boldsymbol{d} \boldsymbol{(} \langle i,\ldots,{q+i} \rangle \boldsymbol{)}. 
\end{align}
Furthermore, the cup-$1$ product satisfies \cite{S47}:
\begin{align} \label{cup1liebniz}
    \delta (\boldsymbol{c} \cup_1 \boldsymbol{d}) = \delta \boldsymbol{c} \cup_1 \boldsymbol{d} + \boldsymbol{c} \cup_1 \delta \boldsymbol{d} + \boldsymbol{c} \cup \boldsymbol{d} + \boldsymbol{d} \cup \boldsymbol{c}.
\end{align}
Finally, we introduce the cup-$2$ product
\begin{align}
    \cup_2 : C^p(M,\ZZ_2) \times C^q(M,\ZZ_2) \to C^{p+q-2}(M,\ZZ_2).
\end{align}
The general formula for the cup-$2$ product is given in Ref.~\cite{C19-2}. We provide the explicit formula for the cup-$2$ product of a $3$-cochains $\boldsymbol{c} \in C^3(M,\ZZ_2)$ and a $4$-cochain $\boldsymbol{d} \in C^4(M,\ZZ_2)$: 
\begin{eqs}
    \boldsymbol{c} \cup_2 d &\boldsymbol{(}\langle 1234 \rangle \boldsymbol{)} \equiv \\ &\boldsymbol{c}\boldsymbol{(} \langle 123 \rangle \boldsymbol{)}\boldsymbol{d}\boldsymbol{(} \langle 1234 \rangle \boldsymbol{)}+\boldsymbol{c}\boldsymbol{(} \langle 134 \rangle \boldsymbol{)}\boldsymbol{d}\boldsymbol{(} \langle 1234 \rangle \boldsymbol{)},
\end{eqs}
for an arbitrary tetrahedron $\langle 1234 \rangle$.
% \begin{align}
% \begin{split}
%     &\boldsymbol{c}_p \cup_2 \boldsymbol{d}_q \boldsymbol{(} \langle 0, \ldots, {p+q-2} \rangle \boldsymbol{)}= \\
%     &\sum_{i=0}^{p-1}\sum_{j=1}^{q-1}\boldsymbol{c}_p\boldsymbol{(} \langle 0, \ldots , i,i+j , \ldots ,p+j-1 \rangle \boldsymbol{)} \\& \times \boldsymbol{d}_q \boldsymbol{(} \langle i,\ldots i+j,p+j-1 ,\ldots p+q-2 \rangle \boldsymbol{)}. 
%     \end{split}
% \end{align}
The cup-$2$ product satisfies:
\begin{align} \label{cup2liebniz}
    \delta (\boldsymbol{c} \cup_2 \boldsymbol{d}) = \delta \boldsymbol{c} \cup_2 \boldsymbol{d} + \boldsymbol{c} \cup_2 \delta \boldsymbol{d} + \boldsymbol{c} \cup_1 \boldsymbol{d} + \boldsymbol{d} \cup_1 \boldsymbol{c}.
\end{align}

To simplify the expressions in the text, we also introduce the notation $\int_N \boldsymbol{c}$ as shorthand for the sum:
\begin{align}
    \int_N \boldsymbol{c} = \sum_{s} \boldsymbol{c}(s).
\end{align}
Here, $N$ is a $p$-dimensional manifold, $\boldsymbol{c}$ is a $p$-cochain, and the sum is over all $p$-simplices in $N$. If unspecified, it should be assumed that the integral is over the manifold $M$.

% The cup-$2$ product $\cup_2$ is a map of a $p$-cochain and a $q$-cochain to a $(p+q-2)$-cochain:
% \begin{align}
%     \cup_2 : C^{p}[M,\ZZ_2] \times C^{q}[M,\ZZ_2] \to C^{p+q-2}[M,\ZZ_2].
% \end{align}
% Let $\boldsymbol{c}_p$ be in $ C^{p}[M,\ZZ_2]$ and $\boldsymbol{c}'_q$ be in $C^{q}[M,\ZZ_2]$. Then the cup-$2$ product of $\boldsymbol{c}_p$ and $\boldsymbol{c}'_q$ evaluated on the $(p+q-2)$-simplex $\langle 0, \ldots , p+q-2 \rangle$ is \textcolor{green}{TE: I could use some help determining the bounds on the sums. Also, this formula is not quite right. There are too many arguments in $\boldsymbol{c}'_q$}:
 %\begin{widetext}
 %\begin{align}
  %   \boldsymbol{c}_p \cup_2 \boldsymbol{c}'_q \boldsymbol{(} &\langle 0, \ldots , p+q-2 \rangle \boldsymbol{)}= \\ \nonumber
  %   &\sum_{i_2} \sum_{i_1} \sum_{i_0} \boldsymbol{c}_p \boldsymbol{(} \langle 0 ,\ldots, i_0, i_1 ,\ldots, i_2, q+i_0-i_1+i_2 ,\ldots, p+q-2 \rangle \boldsymbol{)} 
  %   \boldsymbol{c}'_q \boldsymbol{(} \langle i_0, \ldots, i_1, i_2,\ldots,q+i_0-i_1+i_2 \rangle \boldsymbol{)}
 %\end{align}
 %\end{widetext}

\section{Explicit $1$-form SPT Hamiltonian} \label{App: 1form}

In this appendix, we derive the $1$-form SPT Hamiltonian $H_1$ in Eq.~\eqref{1formsptHsimplified}, i.e.:
\begin{align} \label{1formsptHsimplified2}
     H_1 = -\sum_e \left( X_e   \prod_{f}  W_f^{\int \delta \be \cup_1 \bface} \right),
\end{align}
and demonstrate that it is indeed symmetric under the $1$-form symmetry in Eq.~\eqref{1formgenerator}.
We begin with $H_1$ defined in Eq.~\eqref{1formsptH} as:
\begin{align} \label{1formHspt}
    H_1=-\sum_e {\Uspt} X_e \Uspt^\dagger,
\end{align}
with ${\Uspt}$ given by:
\begin{align}
     {\Uspt} = \sum_{\ba_e} \prod_{t} (-1)^{\ba_e \cup \delta \ba_e (t)} \ket{\ba_e}\bra{\ba_e}.
\end{align}

To compute ${\Uspt} X_e \Uspt^\dagger$ in Eq.~\eqref{1formHspt} explicitly, we first evaluate ${\Uspt} X_e$. We find:
\begin{align} \label{UXcalc1}
    {\Uspt} X_e &= \sum_{\ba_e} \prod_{t} (-1)^{\ba_e \cup \delta \ba_e (t)} \ket{\ba_e}\bra{\ba_e} \sum_{\ba'_e} \ket{\ba'_e+\be}\bra{\ba'_e} \\ \nonumber
    &=\sum_{\ba_e} \prod_{t} (-1)^{(\ba_e+\be) \cup \delta (\ba_e+\be) (t)} \ket{\ba_e+\be}\bra{\ba_e} \\ \nonumber
    &= X_e \sum_{\ba_e} \prod_{t} (-1)^{(\ba_e+\be) \cup \delta (\ba_e+\be) (t)} \ket{\ba_e}\bra{\ba_e}
\end{align}
Expanding the cup product in the last line of Eq.~\eqref{UXcalc1}, we obtain:
\begin{align} \label{UXcalc2}
    {\Uspt}& X_e = \\ \nonumber
    &X_e \sum_{\ba_e} \prod_{t} (-1)^{(\be\cup \delta \ba_e + \ba_e \cup \delta \be + \be \cup \delta \be )(t)} \ket{\ba_e}\bra{\ba_e} {\Uspt}
\end{align}
$\be \cup \delta \be(t)$, in the expression above [Eq.~\eqref{UXcalc2}], is zero for all tetrahedra $t$. This can be seen by evaluating $\be \cup \delta \be$ on an arbitrary tetrahedron $\langle 1234 \rangle$:
\begin{align}
    \be \cup \delta \be \boldsymbol{(} \langle 1234 \rangle \boldsymbol{)} = \be \boldsymbol{(} \langle 12 \rangle \boldsymbol{)} \left[ \be \boldsymbol{(} \langle 23 \rangle \boldsymbol{)} + \be \boldsymbol{(} \langle 34 \rangle \boldsymbol{)} + \be \boldsymbol{(} \langle 24 \rangle \boldsymbol{)} \right ].
\end{align}
The right hand side is zero if $e \neq \langle 12 \rangle$. However, if $e=\langle 12 \rangle$, then the term in square brackets must be zero. Hence, we see that $\be \cup \delta \be(t)=0$. We then have:
\begin{align} \label{UXcalc3}
    {\Uspt} X_e = 
    X_e \sum_{\ba_e} \prod_{t} (-1)^{(\be\cup \delta \ba_e + \ba_e \cup \delta \be )(t)} \ket{\ba_e}\bra{\ba_e} {\Uspt}.
\end{align}

We simplify the right hand side of Eq.~\eqref{UXcalc3} further by employing the identities in Eqs.~\eqref{Liebnizrule} and \eqref{cup1liebniz}. An application of Eq.~\eqref{Liebnizrule} gives us:
\begin{align} 
    {\Uspt} X_e = 
    X_e \sum_{\ba_e} \prod_{t} (-1)^{(\delta \be\cup  \ba_e + \ba_e \cup \delta \be )(t)} \ket{\ba_e}\bra{\ba_e} {\Uspt},
\end{align}
where we have also used that $M$ is closed. Then, using Eq.~\eqref{cup1liebniz}, we see:
\begin{align} \label{UXcalc4}
    {\Uspt} X_e = 
    X_e \sum_{\ba_e} \prod_{t} (-1)^{\delta \ba_e \cup_1 \delta \be (t)} \ket{\ba_e}\bra{\ba_e} {\Uspt},
\end{align}
where again we have used that $M$ is closed.

We can express the term:
\begin{align}\label{cup1term}
    \sum_{\ba_e} \prod_{t} (-1)^{\delta \ba_e \cup_1 \delta \be (t)} \ket{\ba_e}\bra{\ba_e}
\end{align}
in Eq.~\eqref{UXcalc4} using Pauli Z operators. To do so, we notice that, for any face $f=\langle 123 \rangle$:
\begin{align}
    \sum_{\ba_e} (-1)^{\delta \ba_e (f)}\ket{\ba_e}\bra{\ba_e} = 
    \sum_{\ba_e} (-1)^{a_{12}+a_{23}+a_{13}}\ket{\ba_e}\bra{\ba_e} 
    =W_f,
\end{align}
where in the last equality we have defined:
\begin{align}
    W_f \equiv \prod_{e \subset f} Z_e.
\end{align}
Therefore, Eq.~\eqref{cup1term} can be written as:
\begin{align} \label{cup1term2}
     \sum_{\ba_e} \prod_{t} (-1)^{\delta \ba_e \cup_1 \delta \be (t)} \ket{\ba_e}\bra{\ba_e} = \prod_{t} \prod_f W_f^{\bface \cup_1 \delta \be (t)},
\end{align}
with $\prod_f$ a product over all faces in $M$. Exchanging the product over tetrahedra for a sum, we see that:
\begin{align} \label{cup1term3}
    \sum_{\ba_e} \prod_{t} (-1)^{\delta \ba_e \cup_1 \delta \be (t)} \ket{\ba_e}\bra{\ba_e} = \prod_f W_f^{\int \bface \cup_1 \delta \be }.
\end{align}
Then, plugging Eq.~\eqref{cup1term3} into Eq.~\eqref{UXcalc4}, we arrive at:
\begin{align} \label{UXcalc5}
    {\Uspt} X_e = 
    X_e \prod_f W_f^{\int \bface \cup_1 \delta \be} {\Uspt}.
\end{align}

Finally, we can compute ${\Uspt} X_e {\Uspt}^\dagger$. From Eq.~\eqref{UXcalc5}, we have:
\begin{align}
    {\Uspt} X_e {\Uspt}^\dagger &= \left( X_e \prod_f W_f^{\int \bface \cup_1 \delta \be} {\Uspt} \right) {\Uspt}^\dagger \\ \nonumber
    &= X_e \prod_f W_f^{\int \bface \cup_1 \delta \be}.
\end{align}
The $1$-form SPT Hamiltonian is then:
\begin{align}
     H_1 = -\sum_e \left( X_e   \prod_{f}  W_f^{\int \bface \cup_1 \delta \be} \right),
\end{align}
as claimed. 

Next, we show that $H_1$ is symmetric under the $\ZZ_2$ $1$-form symmetry. Recall that the symmetry is generated by operators of the form [Eq.~\eqref{1formgenerator}]:
\begin{align}
    A_\Sigma \equiv \prod_{e \perp \Sigma}X_e,
\end{align}
for a closed surface $\Sigma$ of the dual lattice.
We prove that $H_1$ is symmetric by showing that ${\Uspt}$ is symmetric, i.e.:
\begin{align}
    A_\Sigma {\Uspt} A_\Sigma = {\Uspt},
\end{align}
for all choices of $\Sigma$.

Conjugation of ${\Uspt}$ by an arbitrary 
generator of the $1$-form symmetry $A_\Sigma$ gives:
\begin{align}
    A_\Sigma {\Uspt} A_\Sigma &= \sum_{\ba_e} \prod_{t} (-1)^{\ba_e \cup \delta \ba_e(t)} \ket{\ba_e + \bsig}\bra{\ba_e + \bsig}. 
\end{align}
Redefining the summation, we have:
\begin{align}
      A_\Sigma {\Uspt} A_\Sigma  = \sum_{\ba_e} \prod_{t} (-1)^{(\ba_e + \bsig) \cup \delta (\ba_e + \bsig)(t)} \ket{\ba_e}\bra{\ba_e}.
\end{align}
Then we expand the exponent and use that $\delta \bsig = 0$ to obtain:
\begin{align}
    A_\Sigma {\Uspt} A_\Sigma = \sum_{\ba_e} \prod_{t} (-1)^{\ba_e \cup \delta \ba_e(t) + \bsig \cup \delta \ba_e(t)} \ket{\ba_e}\bra{\ba_e}. 
\end{align}
Finally, we employ the identity in Eq.~\eqref{Liebnizrule} and use that $M$ is closed to arrive at:
\begin{align}
    A_\Sigma {\Uspt} A_\Sigma &= \sum_{\ba_e} \prod_{t} (-1)^{\ba_e \cup \delta \ba_e(t) + \delta (\Sigma \cup \ba_e)(t)} \ket{\ba_e}\bra{\ba_e} \\ \nonumber
    &= \sum_{\ba_e} \prod_{t} (-1)^{\ba_e \cup \delta \ba_e(t)} \ket{\ba_e}\bra{\ba_e} \\ \nonumber
    &= {\Uspt}.
\end{align}
Thus, ${\Uspt}$ is symmetric.

\section{Gauging a $\ZZ_2$ $1$-form symmetry at the level of quantum states} \label{app: 1form gauging quantum states}

In Section \ref{sec ttc}, we gave a physically motivated description of the gauging procedure and the construction of the twisted toric code. We use this appendix to give a more careful construction of the twisted toric code using a gauging procedure defined at the level of quantum states \cite{Y16}.  

\begin{figure}
\begin{center}
\includegraphics[width=0.45\textwidth, trim={500 440 400 360},clip]{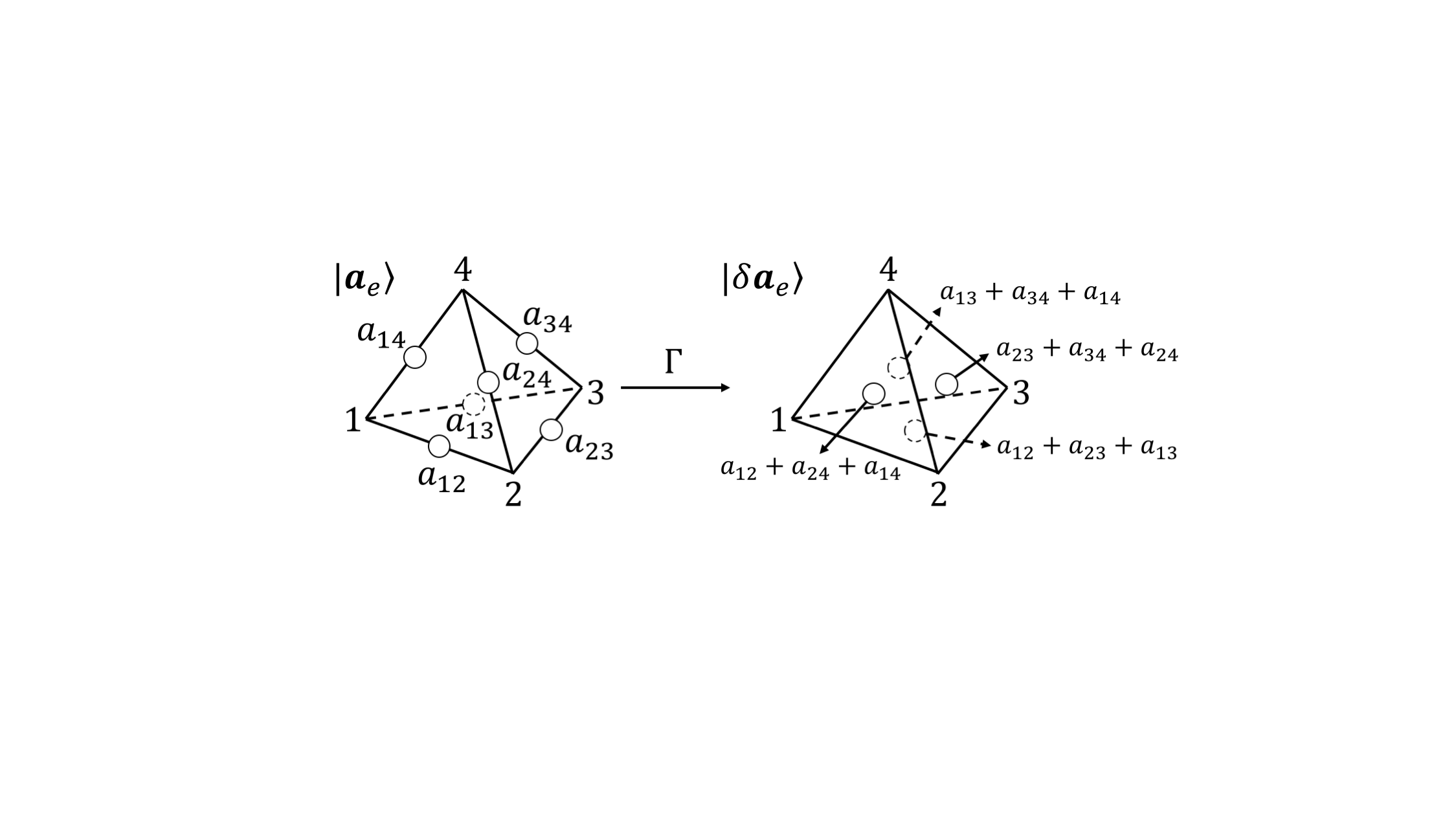}
\caption{$\Gamma$ maps the configuration state $\ket{\ba_e}$ to the state $\ket{\delta \ba_e}$. $\ket{\delta \ba_e}$ is an element of the Hilbert space formed from $\ZZ_2$ {d.o.f.} on faces (represented by circles).}
\label{fig: 1formgauging}
\end{center}
\end{figure}

We begin by denoting the Hilbert space of the $1$-form SPT model as $\mathcal{H}_1$ and the Hilbert space of the twisted toric code as $\mathcal{H}_\text{ttc}$. Then, in gauging the $\ZZ_2$ $1$-form symmetry, we map symmetric states from $\mathcal{H}_1$ to a certain constrained subspace of $\mathcal{H}_\text{ttc}$. For this purpose, we define the linear map $\Gamma$ by:
\begin{align} \label{Gammadef}
  \Gamma: \ket{\ba_e} \mapsto \ket{\delta \ba_e},
\end{align}
as shown schematically in Fig.~\ref{fig: 1formgauging}. We note that in Eq.~\eqref{Gammadef}, we have omitted a normalization factor for simplicity.\footnote{More precisely, $\Gamma$ maps:
\unexpanded{
$
\Gamma: \ket{\ba_e} \mapsto \frac{1}{\mathcal{A}}\ket{\delta \ba_e}.
$}
The normalization factor \unexpanded{$\mathcal{A}$} is given by 
\unexpanded{$
    \mathcal{A}={\sqrt{|Z^1(M,\ZZ_2)|}}
$},
where \unexpanded{$|Z^1(M,\ZZ_2)|$} is the cardinality of the set of closed $1$-cochains on $M$.}

$\Gamma$ is a many-to-one mapping, since configuration states differing by a $1$-form symmetry action are mapped to the same state. Explicitly, for an arbitrary closed $1$-cochain $\bsig$ ($\delta \bsig =0$), we have:
\begin{align} \label{gaugeredundant}
    \Gamma(A_\Sigma \ket{\ba_e}) = \Gamma(\ket{\ba_e + \bsig}) = \ket{\delta \ba_e + \delta \bsig} = \ket{\delta \ba_e}
\end{align}
Thus, $\ket{\ba_e}$ and $A_\Sigma \ket{\ba_e}$ map to the same state, namely $\ket{\delta \ba_e}$.

This observation suggests that a one-to-one mapping of states can be defined by restricting $\Gamma$ to the subspace of $\ZZ_2$ $1$-form symmetric states $\mathcal{H}_1^\text{sym}$, i.e.:
\begin{align}
    \mathcal{H}_{1}^\text{sym} \equiv \{ \ket{\Psi} \in \mathcal{H}_1 : A_\Sigma \ket{\Psi} = \ket{\Psi}, \, \forall \Sigma \text{ s.t. } \delta \bsig=0 \}.
\end{align}
$\Gamma$ then defines an isomorphism (a duality) between $\mathcal{H}_{1}^\text{sym}$ and its image, where the image of $\Gamma$ is spanned by 
% \begin{align}
%     \mathcal{H}_\text{ttc}^\text{con}\equiv \{ \ket{\Psi}\in \mathcal{H}_\text{ttc} : \ket{\psi}=\ket{\delta \ba_e}\text{ for some }\ba_e \},
% \end{align}
the set of configuration states $\ket{\delta \ba_e}$ for some $1$-cochain $\ba_e$. We denote the image of $\Gamma$ as $\mathcal{H}_\text{ttc}^\text{con}$:
\begin{align}
    \mathcal{H}_\text{ttc}^\text{con} \equiv \{ \ket{\Psi} \in \mathcal{H}_\text{ttc} : \ket{\Psi} = \sum_{\ba_e}C_{\ba_e}\ket{\delta \ba_e}, \, C_{\ba_e}\in \mathbb{C} \}.
\end{align}
% \begin{align}
%     \mathcal{H}_\text{ttc}^\text{con} \equiv \text{span}\left(\left\{ \ket{\delta \ba_e}_{\ba_e} \right\}\right).
% \end{align}

Alternatively, $\mathcal{H}_\text{ttc}^\text{con}$ can be expressed as a subspace of $\mathcal{H}_\text{ttc}$ with a particular $\ZZ_2$ $1$-form gauge constraint.
To motivate this, we notice that any basis state $\ket{\delta \ba_e} \in \mathcal{H}_\text{ttc}^\text{con}$ satisfies:
\begin{align}\label{1formgaugeconstraint}
    \prod_{f \subset \sigma}Z_f \ket{\delta \ba_e} = \ket{\delta \ba_e},
\end{align}
where $\sigma$ is an arbitrary closed surface ($\partial \sigma =0$) of the direct lattice, and the product is over faces contained in $\sigma$.
This follows from the definition of $Z_f$ [Eq.~\eqref{Paulionf}] and the definition of the coboundary [Eq.~\eqref{coboundarydef}]:
\begin{align}
    \prod_{f\subset \sigma} Z_f \ket{\delta \ba_e} = (-1)^{\delta \ba_e(\sigma)}\ket{\delta \ba_e}
    =(-1)^{\ba_e(\partial \sigma)}\ket{\delta \ba_e} 
    =\ket{\delta \ba_e}.   
\end{align}
Physically, Eq.~\eqref{1formgaugeconstraint} is a statement that $\ket{\delta \ba_e}$ contains no $1$-form gauge fluxes and has a trivial holonomy.
%\footnote{Note that $1$-form gauge fluxes are point-like while $1$-form gauge charges are loop-like.}
It can be checked
% \footnote{The definitions of \unexpanded{$\mathcal{H}_\text{ttc}^\text{con}$} are indeed equivalent. For example, consider \unexpanded{$\ket{\Psi}=\ket{\ba_f}$} such that \unexpanded{$\prod_{f\subset \sigma}Z_f\ket{\ba_f}=\ket{\ba_f}$} for all closed surfaces \unexpanded{$\sigma$}. It follows that \unexpanded{$\ba_f(\sigma)=0$} for any  \unexpanded{$\sigma$} with \unexpanded{$\partial \sigma =0$}. First, this implies that \unexpanded{$\ba_f$} is closed. For any \unexpanded{$2$}-simplex \unexpanded{$s$}, we have: \unexpanded{$\delta \ba_f (s) = \ba_f (\partial s) =0$}, since \unexpanded{$\partial s$} is closed. In addition, the contraints imply that the configuration \unexpanded{$\{a_f\}$} corresponding to \unexpanded{$\ba_f$} has trivial holonomy.  Together, the fact that $\ba_f$ is closed and the trivial holonomy mean that there exists \unexpanded{$\ba_e$} such that \unexpanded{$\delta \ba_e=\ba_f$}.} 
that an equivalent definition of $\mathcal{H}_\text{ttc}^\text{con}$ is then the set of constrained states:
\begin{align}
    \mathcal{H}_\text{ttc}^\text{con} = \{ \ket{\Psi} \in \mathcal{H}_\text{ttc} : \prod_{f \subset \sigma}Z_f \ket{\Psi} = \ket{\Psi}, \, \forall \sigma \text{ s.t. } \partial \sigma=0\}.
\end{align}
Here, the $\ZZ_2$ $1$-form gauge constraints are:
\begin{align}
    \prod_{f \subset \sigma}Z_f = 1,
\end{align}
for any closed surface $\sigma$. We note that the local constraints are generated by the set of operators:
\begin{align} \label{1formlocalgauge}
    W_t \equiv \prod_{f \subset t}Z_f =1
\end{align}
for all tetrahedra $t$.

$\Gamma$ restricted to $\mathcal{H}_1^\text{sym}$ defines a duality at the level of states. However, to apply the duality to $\onepara$ and $H_1$, we need to extend our understanding of $\Gamma$ to operators. In particular, we consider operators that preserve the subspace $\mathcal{H}_1^\text{sym}$, i.e., commute with $A_\Sigma$ for all $\Sigma$. The local, $1$-form symmetric operators can, in fact, be generated by:
$X_e$ and $W_f$. Accordingly, we focus on the image of $X_e$ and $W_f$ under the mapping $\Gamma$. For an arbitrary state $\ket{\Psi_\text{sym}}$ in $\mathcal{H}_1^\text{sym}$, we have:
\begin{align} 
    \Gamma(X_{e'} \ket{\Psi_\text{sym}}) &= \Gamma\left(\sum_{\ba_e} \ket{\ba_e+\be'}\langle \ba_e \ket{\Psi_\text{sym}}  \right) \\ \nonumber
    &= \sum_{\ba_e} \ket{\delta \ba_e+ \delta \be'}\langle \ba_e \ket{\Psi_\text{sym}} \\ \nonumber
    &= \prod_{f\supset e'}X_f \sum_{\ba_e} \ket{\delta \ba_e}\langle \ba_e \ket{\Psi_\text{sym}} \\ \nonumber
    &= \prod_{f\supset e'}X_f \Gamma( \ket{\Psi_\text{sym}} ),
\end{align}
where the product is over faces containing $e'$ in their boundary. 
For $W_f$, on the other hand, we find:
\begin{align}
    \Gamma(W_f \ket{\Psi_\text{sym}}) &= \Gamma\left(\sum_{\ba_e} (-1)^{\delta \ba_e(f)} \ket{\ba_e}\langle \ba_e \ket{\Psi_\text{sym}}  \right) \\ \nonumber
    &= \sum_{\ba_e}  (-1)^{\delta \ba_e(f)} \ket{\delta \ba_e}\langle \ba_e \ket{\Psi_\text{sym}} \\ \nonumber
    &= Z_f \sum_{\ba_e} \ket{\delta \ba_e}\langle \ba_e \ket{\Psi_\text{sym}} \\ \nonumber
    &= Z_f \Gamma( \ket{\Psi_\text{sym}} ).
\end{align}
Therefore, the operators $X_{e}$ and $W_f$ on $\mathcal{H}_1^\text{sym}$ are dual to the operators $\prod_{f\supset e}X_f$ and $Z_f$ on $\mathcal{H}_\text{ttc}^\text{con}$, respectively:
\begin{align} \label{Xeduality}
    X_e &\leftrightarrow \prod_{f\supset e}X_f \\ \label{Wfduality}
    W_f &\leftrightarrow Z_f.
\end{align}
We note that the commutation relations are preserved by the duality. In $\mathcal{H}_1$, $X_e$ anticommutes with $W_f$ if and only if $e$ is in the boundary of $f$. Likewise, in $\mathcal{H}_\text{ttc}$, $\prod_{f\supset e}X_f$ anticommutes with $Z_f$ if and only if $e$ is in the boundary of $f$.

% $\Gamma$ restricted to $\mathcal{H}_1^\text{sym}$ defines a duality at the level of states. However, to apply the duality to $\onepara$ and $H_1$, we need to extend our understanding of $\Gamma$ to operators. We do so as follows. Let $\mathcal{O}$ be an arbitrary operator that leaves the subspace $\mathcal{H}_1^\text{sym}$ invariant:
% \begin{align}
%     \mathcal{O} = \sum_{\ba_e,\ba'_e} C_{\ba_e,\ba'_e} \ket{\ba_e}\bra{\ba'_e}. 
% \end{align}
% Then we take $\Gamma(\mathcal{O})$ to be:
% \begin{align}
%     \Gamma(\mathcal{O}) = \sum_{\ba_e,\ba'_e} C_{\ba_e,\ba'_e} \ket{\delta \ba_e}\bra{\delta \ba'_e},
% \end{align}
% where, manifestly, the action of $\Gamma(\mathcal{O})$ is only defined in the constrained space $ \mathcal{H}_\text{ttc}^\text{con}$. Furthermore, in appendix [APPENDIX], we show that this is the unique choice of $\Gamma(\mathcal{O})$ that satisfies:
% \begin{align} \label{Gamma product condition} 
%     \Gamma(\mathcal{O})\Gamma(\ket{\psi_\text{sym}})=\Gamma(\mathcal{O}\ket{\psi_\text{sym}}),
% \end{align}
% for any $\ket{\psi_\text{sym}}\in \mathcal{H}_\text{ttc}^\text{con}$. 

% To demonstrate the gauging procedure, we first apply it to the $1$-form paramagnet and show that it is mapped to the usual $3$D toric code with a bosonic $1$-form gauge flux. 
We are now ready to apply the gauging duality to the models discussed in Section \ref{sec: 1formspt}. We begin with the $1$-form paramagnet. The ground state of the $1$-form paramagnet is:
\begin{align}
    \ket{\onetriv}= \sum_{\ba_e} \ket{\ba_e},
\end{align}
which certainly belongs to the subspace $\mathcal{H}_1^\text{sym}$. Applying $\Gamma$ to $\ket{\onetriv}$ then yields:
\begin{align}
     \ket{\Psi_\text{tc}} \equiv \Gamma\left( \ket{\onetriv} \right)= \sum_{\ba_e} \ket{\delta \ba_e}.
\end{align}
Using Eq.~\eqref{Xeduality}, the $1$-form paramagnet Hamiltonian:
\begin{align}
    \onepara=-\sum_{e}X_e,
\end{align}
is dual to the Hamiltonian $H_\text{tc}^\text{con}$:
\begin{align}
    H_\text{tc}^\text{con}=-\sum_{e} \prod_{f \supset e}X_f,
\end{align}
defined on $\mathcal{H}_\text{ttc}^\text{con}$. $ \ket{\Psi_\text{tc}}$ is the unique ground state of $H_\text{tc}^\text{con}$ in $\mathcal{H}_\text{ttc}^\text{con}$.

% We claim that this is a ground state of the $3$D toric code when defined on the unconstrained Hilbert space $\mathcal{H}_\text{ttc}$. 

% % To make this explicit, we first identify a parent Hamiltonian for $\ket{\onetriv}^\text{con}$ defined on $\mathcal{H}_\text{ttc}^\text{con}$. Since $\ket{\onetriv}$ is a $+1$ eigenstate of $X_e$
% To make this explicit, we apply $\Gamma$ to the $1$-form paramagnet:
% \begin{align}
%     \onepara=-\sum_{e}X_e.
% \end{align}
% The result is a Hamiltonian that acts within the constrained Hilbert space $ \mathcal{H}_\text{ttc}^\text{con}$.
% Since each term $X_e$ preserves the subspace $\mathcal{H}_1^\text{sym}$, we compute:
% \begin{align}\label{gammasymXemap}
%     \Gamma(X_e) &= \Gamma\left( \sum_{\ba_e} \ket{\ba_e+\be}\bra{\ba_e} \right) \\ \nonumber
%     &= \sum_{\ba_e} \ket{\delta \ba_e + \delta \be}\bra{\delta \ba_e} \\ \nonumber
%     &= \prod_{f \supset e}X_f \sum_{\ba_e}\ket{\delta \ba_e}\bra{\delta \ba_e}.
% \end{align}
% In the last line, the product is over faces containing $e$ in their boundary.
% From Eq.~\eqref{gammasymXemap}, we see that $\Gamma(\onepara)$ is:
% \begin{align}
%     \Gamma(\onepara)=\left(-\sum_{e}\prod_{f \supset e}X_f \right)  \sum_{\ba_e}\ket{\delta \ba_e}\bra{\delta \ba_e}.
% \end{align}
% Eq.~\eqref{Gamma product condition} implies that $\Gamma(\ket{\onetriv})$ is the ground state of $\Gamma(\onepara)$.

The final step of the gauging procedure is to impose the local constraints [Eq.~\eqref{1formlocalgauge}] of $ \mathcal{H}_\text{ttc}^\text{con}$ energetically. The resulting Hamiltonian is:\footnote{Strictly speaking, \unexpanded{$H_\text{tc}^\text{con}$} is ill-defined on the unconstrained space. However, we make the non-unique choice to map the operator \unexpanded{$\prod_{f \supset e}X_f$ on $\mathcal{H}_\text{ttc}^\text{con}$} to the operator \unexpanded{$\prod_{f \supset e}X_f$} on \unexpanded{$\mathcal{H}_\text{ttc}$}. Importantly, this choice preserves locality and \unexpanded{$\ket{\Psi_\text{tc}}$} remains a \unexpanded{$+1$} eigenstate regardless of the ambient Hilbert space.}
\begin{align}
     H_\text{tc} &\equiv H_\text{tc}^\text{con} - \sum_t W_t = -\sum_{e} \prod_{f \supset e}X_f - \sum_t W_t.
\end{align}
This is the Hamiltonian for the usual $3$D toric code (on the dual lattice). Bosonic point-like excitations ($1$-form gauge fluxes) can be created by the string operators:
\begin{align}
    \prod_{f\perp {p}}X_f,
\end{align}
where ${p}$ is a path in the dual lattice, and the product is over faces intersected by ${p}$.

Finally, we construct the twisted toric code from the nontrivial $\ZZ_2$ $1$-form SPT Hamiltonian $H_1$ by gauging the symmetry. The ground state of $H_1$ is given in Eq.~\eqref{1formsptwf} as:
\begin{align}
    \ket{\Psi_1}={\Uspt} \ket{\onetriv} = \sum_{\ba_e} \prod_{t} (-1)^{\ba_e \cup \delta \ba_e (t)} \ket{\ba_e}.
\end{align}
As shown in Appendix \ref{App: 1form}, ${\Uspt}$ is symmetric, so $\ket{\Psi_1}$ is in $\mathcal{H}_1^\text{sym}$. Therefore, we may apply $\Gamma$ to $\ket{\Psi_1}$:
\begin{align}
    \ket{\Psi_\text{ttc}}\equiv \Gamma(\ket{\Psi_1})=\sum_{\ba_e} \prod_{t} (-1)^{\ba_e \cup \delta \ba_e (t)}\ket{\delta \ba_e}.
\end{align}
When viewed as an element of the unconstrained space $\mathcal{H}_\text{ttc}$, $\ket{\Psi_\text{ttc}}$ is a ground state of the twisted toric code (also see Appendix \ref{app: ttc gs}).

The twisted toric code Hamiltonian can be derived by applying the operator duality in Eqs.~\eqref{Xeduality} and \eqref{Wfduality} to the $1$-form SPT  Hamiltonian:
\begin{align}
    H_1= -\sum_e \left( X_e   \prod_{f}  W_f^{\int \bface \cup_1 \delta \be} \right).
\end{align}
% However, to simplify the discussion in the main text, we take advantage of the ambiguity in the duality and first multiply $H_1$ by the identity:
% \begin{align}
%     1 = \prod_t \Big( \prod_{f\subset t} W_f \Big)^{\sum_{t'}\delta \be \cup_2 \boldsymbol{t}(t')} = \prod_f W_f^{\sum_t \delta \be \cup_2 \delta \bface}. 
% \end{align}
% Using the cup product relations in Appendix~\ref{app: terminology}, $H_1$ becomes:
% \begin{align}
%      H_1= -\sum_e \left( X_e   \prod_{f}  W_f^{\sum_t \bface \cup_1 \delta \be (t)} \right).
% \end{align}
This results in $H_\text{ttc}^\text{con}$ defined in the constrained space $\mathcal{H}_\text{ttc}^\text{con}$:
\begin{align}
    H_\text{ttc}^\text{con} = -\sum_e \bar{G}_e,
\end{align}
where $\bar{G}_e$ is defined as:
\begin{align}
    \bar{G}_e \equiv \prod_{f \supset e}X_f   \prod_{f}  Z_f^{\int \bface \cup_1 \delta \be}. 
\end{align}
Lastly, we impose the local constraints of $\mathcal{H}_\text{ttc}^\text{con}$ energetically:
\begin{align}
    H_\text{ttc} \equiv -\sum_{e} \bar{G}_e -\sum_t W_t.
\end{align}
$H_\text{ttc}$ is the Hamilitonian for the twisted toric code with a ground state given by $\ket{\Psi_\text{ttc}}$.

\section{A ground state of the twisted toric code} \label{app: ttc gs}

Here, we give a direct proof that the state $\ket{\Psi_\text{ttc}}$, defined in Eq.~\eqref{ttc gs} as:
\begin{align}
    \ket{\Psi_\text{ttc}}\equiv \sum_{\ba_e} \prod_{t} (-1)^{\ba_e \cup \delta \ba_e (t)}\ket{\delta \ba_e},
\end{align}
is a ground state of the twisted toric code Hamiltonian:
\begin{align}
    H_\text{ttc} = \sum_e \bar{G}_e -\sum_t W_t.
\end{align}
In particular, we show that $\ket{\Psi_\text{ttc}}$ is a $+1$ eigenstate of both $W_t$ and $\bar{G}_e$. This implies that $\ket{\Psi_\text{ttc}}$ is a ground state of $H_\text{ttc}$, since $W_t$ and $\bar{G}_e$ have eigenvalues $\pm 1$.\footnote{This follows from the observation that $W_t$ and $\bar{G}_e$ square to the identity}

Let us compute $W_t\ket{\Psi_\text{ttc}}$ and $\bar{G}_e\ket{\Psi_\text{ttc}}$ explicitly. For $W_t\ket{\Psi_\text{ttc}}$, we have: 
\begin{align}
    W_t \ket{\Psi_\text{ttc}} &= W_t \sum_{\ba_e} \prod_{t} (-1)^{\ba_e \cup \delta \ba_e (t)}\ket{\delta \ba_e} \\ \nonumber
    &=\sum_{\ba_e}(-1)^{\delta \ba_e(\partial t)}\prod_{t} (-1)^{\ba_e \cup \delta \ba_e (t)}\ket{\delta \ba_e} \\ \nonumber
    &=\sum_{\ba_e}(-1)^{\ba_e(\partial \partial t)}\prod_{t} (-1)^{\ba_e \cup \delta \ba_e (t)}\ket{\delta \ba_e} \\ \nonumber
    &=\ket{\Psi_\text{ttc}}.
\end{align}
While for $\bar{G}_e\ket{\Psi_\text{ttc}}$, we find:
\begin{align} \label{Geonttccalc1}
    \bar{G}_e\ket{\Psi_\text{ttc}} 
    &= \bar{G}_e  \sum_{\ba_e} \prod_{t} (-1)^{\ba_e \cup \delta \ba_e (t)}\ket{\delta \ba_e} \\ \nonumber
    &=  \sum_{\ba_e} \prod_{t} (-1)^{\ba_e \cup \delta \ba_e (t) + \delta \ba_e  \cup_1\delta \be(t)}\ket{\delta \ba_e + \delta \be}.
\end{align}
To evaluate this further, we focus on the sign:
\begin{align}
    \prod_{t} (-1)^{\ba_e \cup \delta \ba_e (t) + \delta \be \cup_1 \delta \ba_e (t)}.
\end{align}
Using Eqs.~\eqref{cup1liebniz}, \eqref{Liebnizrule} and that $M$ is closed, we can write the sign as:
\begin{align}
    \prod_t (-1)^{(\ba_e+\be) \cup \delta (\ba_e+\be) (t)}.
\end{align}
Now, we plug this sign into the expression for $\bar{G}_e\ket{\Psi_\text{ttc}}$ in Eq.~\eqref{Geonttccalc1}:
\begin{align}
    \bar{G}_e\ket{\Psi_\text{ttc}} &=  \sum_{\ba_e} \prod_t (-1)^{(\ba_e+\be) \cup \delta (\ba_e+\be) (t)}\ket{\delta \ba_e + \delta \be} \\ \nonumber
    &= \sum_{\ba_e} \prod_t (-1)^{\ba_e \cup \delta \ba_e (t)}\ket{\delta \ba_e} \\ \nonumber
    &= \ket{\Psi_\text{ttc}}.
\end{align}
Therefore, $\ket{\Psi_\text{ttc}}$ is a $+1$ eigenstate of $W_t$ and $\bar{G}_e$ for all $t$ and $e$, respectively.  

\section{Fermion condensation of the twisted toric code} \label{fermioncondensationHttc}

Here, we elaborate on the fermion condensation procedure for the twisted toric code:
\begin{align}
    H_\text{ttc}=-\sum_e \bar{G}_e - \sum_t W_t.
\end{align}
We begin by showing that the operators $\bar{G}_e$ are local generators of an anomalous $\ZZ_2$ $2$-form symmetry of the twisted toric code. Then, we apply the prescription for fermion condensation in Section~\ref{sec: atomic insulator} to the twisted toric code.

\vspace{1.5mm}
\noindent \begin{center}\emph{Anomalous $2$-form symmetry of the twisted toric code:}\end{center}
\vspace{1.5mm}

We prove the identity in Eq.~\eqref{Ge tildeS} -- for a path $p_e$ intersecting the faces meeting at the edge $e$:
\begin{align} \label{Ge tildeS2}
    \bar{G}_e = \tilde{\mathcal{S}}_{p_e},
\end{align}
where for reference, $\bar{G}_e$ is defined as:
\begin{align}
    \bar{G}_e = \prod_{f\supset e} X_f \prod_f Z_f^{\int \bface \cup_1  \delta \be}.
\end{align}
Eq.~\eqref{Ge tildeS2} says that $\bar{G}_e$ is equal to a small loop of emergent fermion string operator, i.e., it is a local generator of a $\ZZ_2$ anomalous $2$-form symmetry. We prove the equality in Eq.~\eqref{Ge tildeS2} as follows. First, by definition, $\tilde{\mathcal{S}}_{p_e}$ is:
\begin{align} \label{tildeSpe}
    \tilde{\mathcal{S}}_{p_e} \equiv  \overline{\prod_{f \in F_{\perp p_e}}} \bar{U}_f \prod_{f \in F_{\perp p_e}} W_{R(f)}.
\end{align}
We simplify the $W_t$ term and $\bar{U}_f$ term independently and then show that their product is $\bar{G}_e$.

To simplify the $W_t$ term, we note that the product $\prod_{f \in F_{\perp p_e}} W_{R(f)}$ can be rewritten as:
\begin{align} \label{eq: WR1}
    \prod_t W_t^{\boldsymbol{t} \cup_1 \be (t) + \be \cup_1 \boldsymbol{t}(t)}.
\end{align}
For $t=\langle 1234 \rangle$, the cochain ${\boldsymbol{t} \cup_1 \be (t) + \be \cup_1 \boldsymbol{t}(t)}$ evaluates to:
\begin{align}
    \be \boldsymbol{(}\langle 12 \rangle \boldsymbol{)}+\be \boldsymbol{(}\langle 23 \rangle \boldsymbol{)}+\be \boldsymbol{(}\langle 34 \rangle \boldsymbol{)}+\be \boldsymbol{(}\langle 14 \rangle \boldsymbol{)}.
\end{align}
The edges $\langle 12 \rangle$, $\langle 23 \rangle$, $\langle 34 \rangle$, and $\langle 14 \rangle$ above are precisely the edges $e$ of $t$ such that, for the two faces of $t$ meeting at $e$, the orientation of one is towards the center of $t$, while the orientation of the other is away from the center of $t$.
Therefore, for exactly one of these faces, we have $R(f)=t$. Using the higher cup product relations in Appendix~\ref{app: terminology}, we can re-express Eq.~\eqref{eq: WR1} in terms of a product over faces as:
\begin{align}
     \prod_t W_t^{\boldsymbol{t} \cup_1 \be (t) + \be \cup_1 \boldsymbol{t}(t)} =  \prod_f Z_f^{\int \left( \bface \cup_1 \delta \be + \delta \be \cup_1 \bface\right)}.
\end{align}

For the $\bar{U}_f$ term of Eq.~\eqref{tildeSpe}, we have:
\begin{align}
    \overline{\prod_{f \in F_{\perp p_e}}} \bar{U}_f = \prod_f Z_f^{\int \delta \be \cup_1  \bface} \prod_{f \supset e} X_f.
\end{align}
By commuting the Pauli Z operators to the right of the Pauli X operators, we find:
\begin{align}
    \overline{\prod_{f \in F_{\perp p_e}}} \bar{U}_f =(-1)^{\int \delta \be \cup_1 \delta \be} \prod_{f \supset e} X_f  \prod_f Z_f^{\delta \be \cup_1  \bface (t)} .
\end{align}
Using the cup product relations in Appendix~\ref{app: terminology}, it can be shown that $\int \delta \be \cup_1 \delta \be = 0$.
Therefore, the product in Eq.~\eqref{tildeSpe} is:
\begin{align}
    \tilde{\mathcal{S}}_{p_e} &= \prod_{f \supset e} X_f \prod_f Z_f^{\delta \be \cup_1 \bface  (t)} \prod_f Z_f^{\int \left( \bface \cup_1 \delta \be + \delta \be \cup_1 \bface\right)} \\ \nonumber
    &= \prod_{f\supset e} X_f \prod_f Z_f^{\int \bface \cup_1 \delta \be}.
\end{align}
This is exactly $\bar{G}_e$ in Eq.~\eqref{Ge tildeS2}.

\vspace{1.5mm}
\noindent \begin{center}\emph{Fermion condensation procedure:}\end{center}
\vspace{1.5mm}

Following the steps outlined in Section \ref{sec: atomic insulator}, we first introduce a fermionic {d.o.f.} at the center of each tetrahedron and impose the gauge constraint:
\begin{align} \label{gauge constraint condense}
    \tilde{U}_f\tilde{S}_f=1,
\end{align}
at each face $f$. As noted in the main text, the product of $\tilde{U}_f\tilde{S}_f$ over faces adjoined to the edge $e$ is precisely $\bar{G}_e$. Therefore, in the constrained space and after a shift of energy, $H_\text{ttc}$ becomes:
\begin{align}
    H'_\text{ttc}=- \sum_t W_t.
\end{align}
Next, we couple $H_\text{ttc}$ to the fermionic {d.o.f.} to make the Hamiltonian gauge invariant. We do so by replacing $W_t$ with $W_tP_t$:
\begin{align} \label{H''ttc}
     H''_\text{ttc} \equiv -\sum_t W_tP_t.
\end{align}
The last step is to fix a gauge in which the eigenvalue of each $Z_f$ is $+1$. Starting with $H''_\text{ttc}$, this gives us the atomic insulator Hamiltonian:
\begin{align}
    H_\text{AI} = - \sum_t P_t.
\end{align}
Thus, fermion condensation produces the atomic insulator from the twisted toric code Hamiltonian.

For completeness, let us show here that the gauge constraints in Eq.~\eqref{gauge constraint condense} are all mutually commuting. Using the definition $P_{L(f)}\tilde{S}_f=S_f$ and Eq.~\eqref{hopping commutation}, the commutation relations between $\tilde{S}_f$ and $\tilde{S}_{f'}$ are:
\begin{align}
\tilde{S}_f \tilde{S}_{f'} = (-1)^{\int \left( f' \cup_1 f + f \cup_1 f' \right)} \nonumber  (-1)^{\delta \bface'\boldsymbol{(}L(f)\boldsymbol{)}+\delta\bface \boldsymbol{(}L(f')\boldsymbol{)}}\tilde{S}_{f'}\tilde{S}_f.
\end{align} 
The expression $\delta \bface'\boldsymbol{(}L(f)\boldsymbol{)}+\delta\bface \boldsymbol{(}L(f')\boldsymbol{)}$ is $1$ when both $f$ and $f'$ share a tetrahedron $t$ and the orientation of only one of the faces points towards the center of $t$. Otherwise, the expression is $0$. This is also true of $\delta \bface'\boldsymbol{(}R(f)\boldsymbol{)}+\delta\bface \boldsymbol{(}R(f')\boldsymbol{)}$, so we may write: 
\begin{align}
\tilde{S}_f \tilde{S}_{f'} = (-1)^{\int \left( f' \cup_1 f + f \cup_1 f' \right)} \nonumber  (-1)^{\delta \bface'\boldsymbol{(}R(f)\boldsymbol{)}+\delta\bface \boldsymbol{(}R(f')\boldsymbol{)}}\tilde{S}_{f'}\tilde{S}_f.
\end{align} 
Given the commutation relations between $\bar{U}_f$ and $\bar{U}_{f'}$ in Eq.~\eqref{Ufcommutation}, for $\tilde{U}_f$ and $\tilde{U}_{f'}$, we have:
\begin{align}
    \tilde{U}_f \tilde{U}_{f'} = (-1)^{\int \left( f' \cup_1 f + f \cup_1 f' \right)} \nonumber  (-1)^{\delta \bface'\boldsymbol{(}R(f)\boldsymbol{)}+\delta\bface \boldsymbol{(}R(f')\boldsymbol{)}}\tilde{U}_{f'}\tilde{U}_f.
\end{align}
Therefore, for all faces $f$ and $f'$:
\begin{align} \nonumber
    \left(\tilde{U}_f\tilde{S}_f\right)\left(\tilde{U}_{f'}\tilde{S}_{f'}\right)=\left(\tilde{U}_{f'}\tilde{S}_{f'}\right)\left(\tilde{U}_f\tilde{S}_f\right).
\end{align}

\section{Bosonization duality in ($3+1$)D and spin structure}\label{sec: review of boson-fermion duality}

In this appendix, we review the operator-level duality between a fermionic theory and a $\ZZ_2$ lattice gauge theory in three spatial dimensions \cite{CK18}. The fermion condensation duality in Section~\ref{sec: atomic insulator} is functionally equivalent to applying the duality described here. We also elaborate on the spin structure dependent sign in the definition of the hopping operator. We note that the boson-fermion duality in ($2+1$)D is described in Ref.~\cite{CKR18}, and the duality in arbitrary dimensions is worked out in Ref.~\cite{C19-2}. 

The duality is defined for fermionic systems where each tetrahedron $t$ of the triangulated $3$-manifold $M$ hosts a single spinless complex fermion with an operator algebra generated by the Majorana operators $\gamma_t,\gamma'_t$. 
% The operator algebra for the fermionic d.o.f. at $t$ is equivalently generated by a pair of Majorana fermions:
% \begin{align}
%     \gamma_t \equiv c_t^\dagger + c_t, \quad \gamma'_t \equiv i(c^\dagger_t - c_t).
% \end{align} 
The fermion parity even algebra is generated by the site fermion parity:
\begin{equation}
    P_t=-i\gamma_t\gamma'_t,
\end{equation}
and the fermionic hopping operator:
\begin{equation} \label{hoppingdefapp}
    S_f= (-1)^{\bface(E)} i\gamma_{L(f)}\gamma'_{R(f)}.
\end{equation}
In the definition of $S_f$, $L(f)$ and $R(f)$ are the tetrahedra on either side of $f$, with the orientation of $f$ pointing out of $L(f)$ and into $R(f)$ (Fig. \ref{fig:Sf def2}). We discuss the spin structure dependent sign $(-1)^{\bface(E)}$ of $S_f$ in detail below.

The $2$-chain $E \in C_2(M,\ZZ_2)$ in Eq.~\eqref{hoppingdefapp}, is a formal sum of faces, corresponding to a choice of spin structure. $E$ is chosen such that the boundary of $E$ is equal to $w_2 \in C_1(M,\ZZ_2)$, a representative of the Poincar\'e dual of the second Stiefel-Whitney cohomology class $\boldsymbol w_2(TM)$. For a triangulated $3$D manifold, a representative $w_2$ is given by \cite{GT76,C19-2,WG18,T17}:
\begin{align}
    w_2 = \sum_e [1+N^+_{13}(e)+N^-_{02}(e)]_{\boldsymbol{2}} \cdot e,
\end{align}
where $[\cdots]_{\boldsymbol{2}}$ denotes that the coefficient of $e$ is taken modulo $2$, $N^+_{13}(e)$ is the number of positively oriented tetrahedra $\langle 0123 \rangle$ such that $\langle 13 \rangle = e$, and $N^-_{02}(e)$ is the number of negatively oriented tetrahedra $\langle 0123 \rangle$ such that $\langle 02 \rangle = e$.
% For a graphical interpretation of the spin structure in $2$D, we refer to Refs.~\cite{EF19} and \cite{GK16}.
% the sum of all vertices and the vertex $2$ of every negatively-oriented face $\langle 123 \rangle$ \cite{C19-2}.

\begin{figure}[t]
\centering
\includegraphics[width=0.59\textwidth, trim={140 200 900 100},clip]{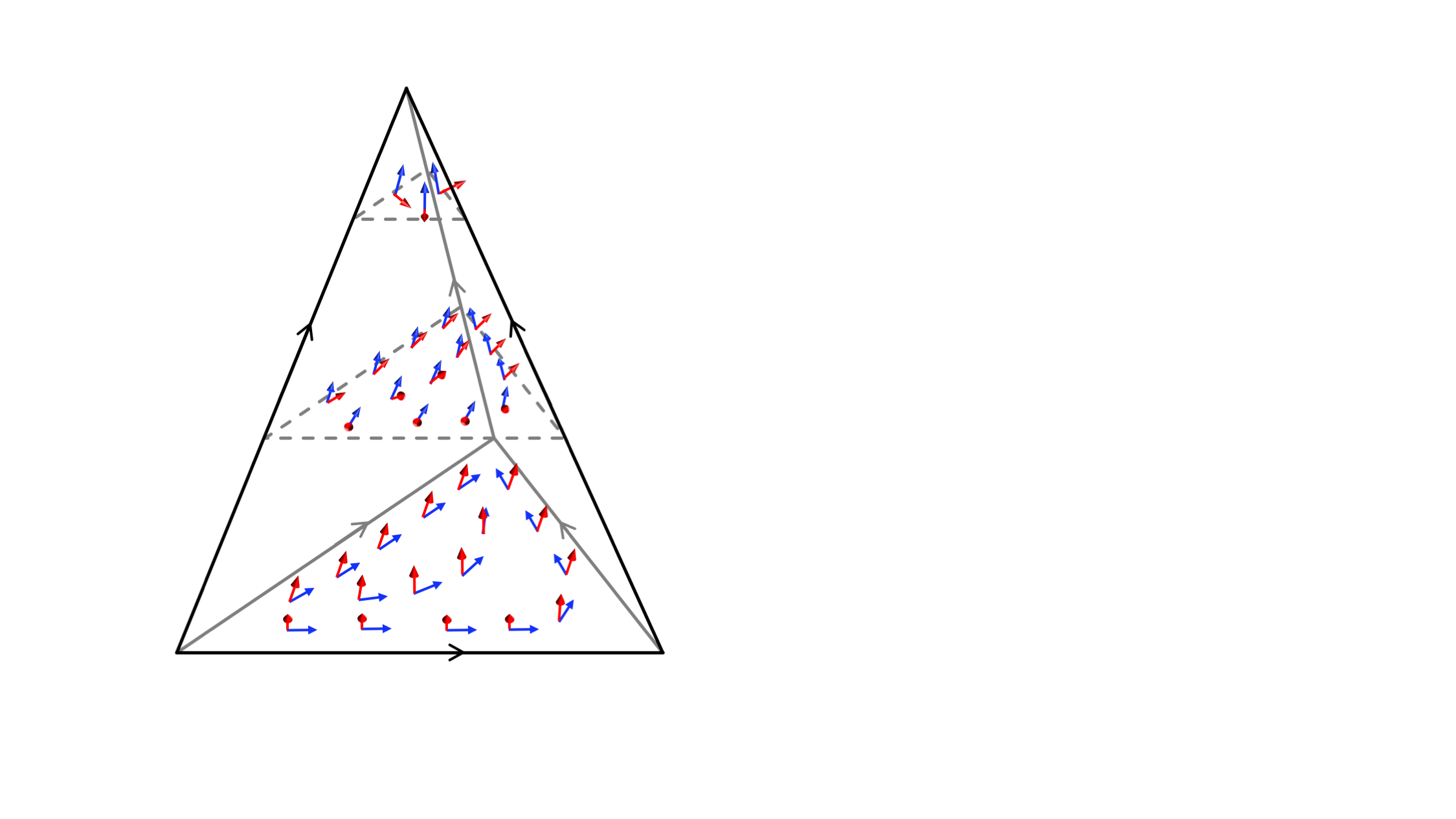}
\caption{A coordinate frame field can be constructed from the branching structure. We show cross sections of the frame field above. The $x$-axes (blue vectors) are given by an interpolation of the edge orientations into the interior of the tetrahedron. The $y$-axes (red vectors) are an interpolation of the face orientations into the interior of the tetrahedron. The $z$-axes (not pictured) are chosen with respect to the orientation of the manifold.}
\label{fig: branchingvector}
\end{figure}

\begin{figure}[t]
\centering
\includegraphics[width=0.5\textwidth, trim={100 300 750 310},clip]{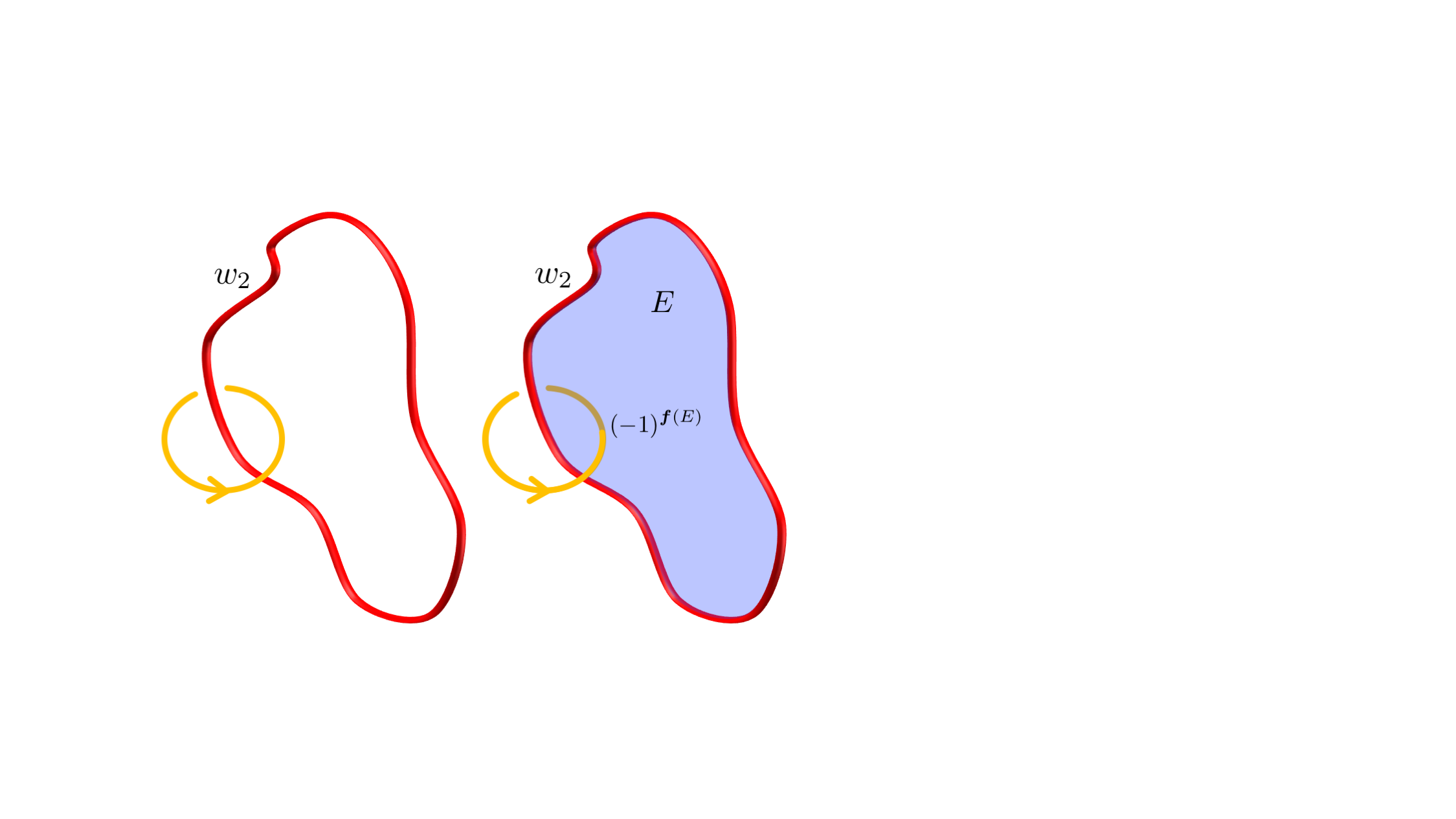}
\caption{The frames along a path (yellow) rotate by an odd multiple of $2 \pi$ when linked with the $1$-chain $w_2$ (red). The $2$-chain $E$ (blue) satisfies $\partial E = w_2$. The hopping operators are modified with the sign $(-1)^{\bface(E)}$ to account for the twisting of the framed path by an odd multiple of $2 \pi$.}
\label{fig: spinstatistics}
\end{figure}

The set of edges with coefficient $1$ in $w_2$ admits a graphical interpretation, which generalizes the graphical interpretation in Refs.~\cite{EF19} and \cite{GK16} for a spin structure in $2$D. To see this, we use the branching structure of $M$ to define a section of the frame bundle on $M$ - an assignment of a coordinate frame to each point in $M$. Similar to the $2$D case, we first interpolate the branching structure to a vector field on the interior of a tetrahedron of $M$, as depicted in Fig.~\ref{fig: branchingvector}. These vectors form the $x$-axes of the coordinate frames. The $y$-axes are then formed by interpolating the orientations of the faces to a vector field on the interior of the tetrahedron. Finally, the $z$-axes are determined by the orientation of $M$. The edges forming $w_2$ give precisely the singular edges of the frame field, i.e., for any path that encloses an odd number of edges in $w_2$, the frames along the path rotate by an odd multiple of $2 \pi$. Heuristically, a fermion that is moved around a singular edge is rotated by an odd multiple of $2 \pi$, and the sign in Eq.~\eqref{hoppingdefapp} compensates for the rotation via the spin-statistics of fermions (see Fig.~\ref{fig: spinstatistics}). 

To describe the bosonization duality, it is convenient to define a product of hopping operators corresponding to a $2$-cochain. For a $2$-cochain $\boldsymbol{\lambda}$, we define $S_{{\lambda}}$ to be: 
\begin{equation}
    \begin{split}
        S_{{\lambda}} &= \prod_{i,i'|i<i'} (-1)^{ \boldsymbol{\lambda}(f_i) \boldsymbol{\lambda}(f_{i'}) \int \boldsymbol f_i \cup_1 \boldsymbol f_{i'}} \prod_{f} S_f^{\boldsymbol{\lambda}(f)},
    \end{split}
\label{eq: S lambda}
\end{equation}
where the faces are arbitrarily ordered $\{ f_1, f_2, f_3, \cdots \}$ and the product is $\prod_{f \in \{ f_1, f_2,  \cdots, f_n \}} S_f = S_{f_n} \cdots  S_{f_2} S_{f_1}$. The order dependent sign ensures that the definition of $S_\lambda$ is independent of the choice of ordering. This follows from the commutation relations of the hopping operators:
\begin{equation}
    \begin{split}
        S_f S_{f'} &= (-1)^{\int (\boldsymbol{f} \cup_1 \boldsymbol{f}' + \boldsymbol{f}' \cup_1 \boldsymbol{f})}S_f S_{f'}.
    \end{split}
\label{eq: 2d fermionic hopping commutation relation}
\end{equation}
Given a pair of 2-cochains $\boldsymbol{\lambda}$ and $\boldsymbol{\lambda}'$, we have the identity:
\begin{equation}
S_{ \lambda + \lambda'} \equiv (-1)^{\int  \boldsymbol{\lambda} \cup_1  \boldsymbol{\lambda}'} S_{\lambda'} S_{\lambda}.
\label{eq: S on 1-cochain}
\end{equation}
% The explicit formula for $S_\lambda$ can be written as
% \begin{equation}
%     \begin{split}
%         S_\lambda &= \prod_{i,i'|i<i'} (-1)^{ \lambda(e_i) \lambda(e_{i'}) \int \boldsymbol e_i \cup \boldsymbol e_{i'}} \prod_{e} \left( (-1)^{\be(E)} i \gamma_{L(e)} \gamma^\prime_{R(e)} \right)^{\lambda(e)} \\
%         & \equiv \xi_\lambda(M_2) \prod_e S_e^{\lambda(e)},
%     \end{split}
% \label{eq: S lambda}
% \end{equation}
 
The two generators $P_t$ and $S_f$ satisfy the following constraint at each edge $e$ \cite{C19-2}:
\begin{equation} \label{3d hopping identity}
     S_{\delta \boldsymbol e} \prod_{t} P_t^{\int \boldsymbol e \cup_1 \boldsymbol t + \boldsymbol t \cup_1 \boldsymbol e} = 1
\end{equation}
The physical meaning of this identity is that moving a fermion along a small loop around a edge $e$ is an identity operator. We note that the spin structure dependent sign in the definition of the hopping operator guarantees that the product in Eq.~\eqref{3d hopping identity} is $1$.

The bosonic dual of this system has $\ZZ_2$-valued spins on the faces of the triangulation, with an operator algebra generated by $X$, $Z$ Pauli operators. For every tetrahedron $t$, we define the flux operator:
\begin{equation}
    W_t = \prod_{f \subset t} Z_f,
\end{equation}
and for every face $f$, we define a bosonic hopping operator:
\begin{eqs}
        \bar{U}_f = \prod_{f^\prime} Z_{f^\prime}^{\int  \boldsymbol f \cup_1 \boldsymbol f^\prime} X_f.
\end{eqs}
Similar to the fermionic hopping operator, we define a product of $\bar{U}_f$ for any $2$-cochain $\boldsymbol{\lambda}$. In this case, $\bar U_\lambda$ is defined as:
\begin{eqs}
        \bar{U}_\lambda = \prod_{f^\prime} Z_{f^\prime}^{\int  \boldsymbol{\lambda} \cup_1 \boldsymbol f^\prime} \prod_f X_f^{\boldsymbol{\lambda}(f)}.
\label{eq: U lambda}
\end{eqs}
To define a consistent duality between the even fermionic operator algebra and the algebra generated by $W_t$ and $\bar{U}_f$ operators, we also define $\bar{G}_e$, given by:
\begin{align}
    \bar{G}_e  = \prod_{f \supset e} X_f  \prod_{f^\prime} Z_{f^\prime}^{\int \boldsymbol f' \cup_1 \delta \boldsymbol e  }.
\end{align}

The duality in Ref.~\cite{CKR18}, is an isomorphism of the $C^*$ algebras $\mathcal{F}$ and $\mathcal{B}$, where $\mathcal{F}$ is the algebra of even fermion parity operators and $\mathcal{B}$ is the algebra generated by $W_t$ and $\bar U_f$ with the constraint $\bar G_e=1$. The mapping of operators is:
\begin{eqs} \label{3d duality}
    W_t \leftrightarrow P_t, \quad \bar{U}_f \leftrightarrow S_f.
\end{eqs}
We note the correspondence above is well-defined since the nontrivial relations map to constraints on the algebra, i.e.:
\begin{eqs}
    \bar G_e  &\leftrightarrow S_{\delta \boldsymbol e} \prod_{t} P_t^{\int \boldsymbol e \cup_1 \boldsymbol t + \boldsymbol t \cup_1 \boldsymbol e} = 1 \\
    \prod_t W_t = 1 &\leftrightarrow \prod_t P_t.
\end{eqs}
% \begin{widetext}
% \begin{equation}
% \begin{split}
% W_t = \prod_{f \subset t} Z_f&\leftrightarrow P_t = -i\gamma_t\gamma'_t, \\
% \bar{U}_f =  (\prod_{f^\prime} Z_{f^\prime}^{\int \boldsymbol f \cup_1 \boldsymbol f'})
% X_f &\leftrightarrow S_f = (-1)^{\boldsymbol f (E)} i\gamma_{L(f)}\gamma'_{R(f)},\\
% \bar{G}_e  = \prod_{f \supset e} X_f  (\prod_{f^\prime} Z_{f^\prime}^{\int \boldsymbol f' \cup_1 \delta \boldsymbol e  }) &\leftrightarrow S_{\delta \boldsymbol e} \prod_{t} P_t^{\int \boldsymbol e \cup_1 \boldsymbol t + \boldsymbol t \cup_1 \boldsymbol e} = 1, \\
% \prod_t W_t = 1 &\leftrightarrow \prod_t P_t.
% \end{split}
% \label{eq: 3d boson-fermion duality 1}
% \end{equation}
% \end{widetext}
Also, the operators $\bar U_\lambda$ and $S_\lambda$ are defined so that the duality in Eq.~\eqref{3d duality} maps:
\begin{align}
    \bar U_\lambda \leftrightarrow S_\lambda,
\end{align}
for any $2$-cochain $\boldsymbol \lambda$. This is because the Pauli X and Pauli Z operators in $\bar U_\lambda$ can be commuted past one another to obtain (see Appendix~\ref{app: fermionization shadow circuit}):
\begin{equation}
    \begin{split}
        \bar U_{{\lambda}} = \prod_{i,i'|i<i'} (-1)^{ \boldsymbol{\lambda}(f_i) \boldsymbol{\lambda}(f_{i'}) \int \boldsymbol f_i \cup_1 \boldsymbol f_{i'}} \prod_{f} \bar{U}_f^{\boldsymbol{\lambda}(f)}.
    \end{split}
\end{equation}

\section{Symmetry of the $2$-group SPT Hamiltonian}
\label{app: 2group sym}

We use this Appendix to prove that the $2$-group Hamiltonian $H_2=\Ugspt H^G_0 \Ugspt^\dagger$ in Section~\ref{sec: 2-groupspt} is symmetric. Given that $H^G_0$ is invariant under the $2$-group symmetry, it suffices to show that $\Ugspt$ is symmetric. For convenience, we re-write $\Ugspt$ here as:
\begin{align} \label{U2 def app}
    \Ugspt = \sum_{\{g_v\},\ba_e} \prod_{t} e^{2 \pi i O_t \coalpha(t)}\ket{\{g_v\},\ba_e}\bra{\{g_v\},\ba_e},
\end{align}
where the $3$-cochain $\coalpha$ is:
\begin{align}
    \coalpha = &\conu + \frac{1}{2}\corho \cup_1 \corho \\ \nonumber &+ \frac{1}{2}\corho \cup_1 \delta \ba_e  + \frac{1}{2}\ba_e\cup \delta \ba_e.
\end{align}
We note that we have expressed the $W_f$ term of $\Ugspt$ in terms of $\delta \ba_e$.

First of all, the FDQC $\Ugspt$ is symmetric under the $1$-form symmetry in Eq.~\eqref{2group 1formpart}. The $1$-form symmetry only affects the last two terms of $\coalpha$, those with $\ba_e$. The $\frac{1}{2}\ba_e \cup \delta \ba_e$ term is invariant, as shown in Appendix~\ref{App: 1form}. The $\frac{1}{2}\corho \cup_1 \delta \ba_e$ term is also symmetric, since it can be written in terms of the $1$-form symmetric operators $W_f$, as in Eq.~\eqref{u2 def}.

Second, we show that $\Ugspt$ commutes with the $0$-form symmetry operator $V_\rho(h)$, for all $h \in G$:
\begin{align} \label{2group 0form part app}
     V_\rho(h) = V(h) \prod_e X_e^{\rhohath(e)}.
\end{align}
Specifically, we compute:
\begin{align}
    V_\rho(h) \Ugspt V^\dagger_\rho(h).
\end{align}
First, in moving the product of Pauli X operators in Eq.~\eqref{2group 0form part app} past $\Ugspt$, we find:
\begin{align} \label{u2 sym step1}
    &V_\rho(h) \Ugspt V^\dagger_\rho(h) = \\ \nonumber &V(h)\sum_{\{g_v\},\ba_e} \prod_{t} e^{2 \pi i O_t \coalpha'(t)}\ket{\{g_v\},\ba_e}\bra{\{g_v\},\ba_e}V^\dagger(h),
\end{align}
where $\coalpha'$ is:
\begin{align} \nonumber
    \coalpha' = &{\overline{\boldsymbol{\nu}}}_{\scriptscriptstyle{\{g_v\}}} + \frac{1}{2}{\overline{\boldsymbol{\rho}}_{\scriptscriptstyle{\{g_v\}}}} \cup_1 {\overline{\boldsymbol{\rho}}_{\scriptscriptstyle{\{g_v\}}}} \\  &+ \frac{1}{2}{\overline{\boldsymbol{\rho}}_{\scriptscriptstyle{\{g_v\}}}} \cup_1 \delta (\ba_e+{\overline{\boldsymbol{\rho}}^h_{\scriptscriptstyle{\{g_v\}}}}) \\ \nonumber &+ \frac{1}{2}(\ba_e+{\overline{\boldsymbol{\rho}}^h_{\scriptscriptstyle{\{g_v\}}}})\cup \delta (\ba_e+{\overline{\boldsymbol{\rho}}^h_{\scriptscriptstyle{\{g_v\}}}}).
\end{align}
Then, conjugation by $V(h)$ on the right hand side of Eq.~\eqref{u2 sym step1} produces:
\begin{align}
    V_\rho(h) &\Ugspt V^\dagger_\rho(h) = \\ \nonumber &\sum_{\{g_v\},\ba_e} \prod_{t} e^{2 \pi i O_t {\overline{\boldsymbol{\alpha}}}_{\scriptscriptstyle{\{h^{-1}g_v\}}}'(t)}\ket{\{g_v\},\ba_e}\bra{\{g_v\},\ba_e},
\end{align}
with ${\overline{\boldsymbol{\alpha}}}_{\scriptscriptstyle{\{h^{-1}g_v\}}}'$ given explicitly by:
\begin{align} \nonumber
    {\overline{\boldsymbol{\alpha}}}_{\scriptscriptstyle{\{h^{-1}g_v\}}}' = &{\overline{\boldsymbol{\nu}}}_{\scriptscriptstyle{\{h^{-1}g_v\}}} + \frac{1}{2}{\overline{\boldsymbol{\rho}}_{\scriptscriptstyle{\{h^{-1}g_v\}}}} \cup_1 {\overline{\boldsymbol{\rho}}_{\scriptscriptstyle{\{h^{-1}g_v\}}}} \\ \label{effect of 0form on U2} &+ \frac{1}{2}{\overline{\boldsymbol{\rho}}_{\scriptscriptstyle{\{h^{-1}g_v\}}}} \cup_1 \delta (\ba_e+{\overline{\boldsymbol{\rho}}^h_{\scriptscriptstyle{\{h^{-1}g_v\}}}}) \\ \nonumber &+ \frac{1}{2}(\ba_e+{\overline{\boldsymbol{\rho}}^h_{\scriptscriptstyle{\{h^{-1}g_v\}}}})\cup \delta (\ba_e+{\overline{\boldsymbol{\rho}}^h_{\scriptscriptstyle{\{h^{-1}g_v\}}}}).
\end{align}

We now simplify the expression for ${\overline{\boldsymbol{\alpha}}}_{\scriptscriptstyle{\{h^{-1}g_v\}}}'$ in Eq.~\eqref{effect of 0form on U2}. To do so, we make use of the cochain $\kappah$, defined on an arbitrary edge $\langle 12 \rangle$ as:
\begin{align}
    \kappah \boldsymbol{(} \langle 12 \rangle \boldsymbol{)} = \rho(h,1,g_1,g_2).
\end{align}
It is also useful to introduce ${\widetilde{\boldsymbol{\nu}}}^{\, h}_{\scriptscriptstyle{\{g_v\}}}$:
\begin{align}
    {\widetilde{\boldsymbol{\nu}}}^{\, h}_{\scriptscriptstyle{\{g_v\}}}\boldsymbol{(} \langle 123 \rangle \boldsymbol{)} \equiv \nu(h,1,g_1,g_2,g_3),
\end{align}
for an arbitrary face $\langle 123 \rangle$.
We then employ the following identities:
\begin{align} \label{nubar relation}
    {\overline{\boldsymbol{\nu}}}_{\scriptscriptstyle{\{h^{-1}g_v\}}} =& \conu + \delta {\widetilde{\boldsymbol{\nu}}}^{\, h}_{\scriptscriptstyle{\{g_v\}}} + \frac{1}{2}\delta \kappah \cup_1 \corho \\ \nonumber &+\frac{1}{2}\kappah \cup \delta \kappah + \delta(\kappah \cup_1 \corho ), \\[10pt] 
 \label{rhobar relation}
    {\overline{\boldsymbol{\rho}}_{\scriptscriptstyle{\{h^{-1}g_v\}}}} =& \corho + \delta \kappah,\\[10pt]
    \label{kappa relation}
    {\overline{\boldsymbol{\rho}}^h_{\scriptscriptstyle{\{h^{-1}g_v\}}}} =& \kappah.
\end{align}
%  \begin{align} \label{nubar relation}
%     {\overline{\boldsymbol{\nu}}}_{\scriptscriptstyle{\{h^{-1}g_v\}}} =& \conu + \frac{1}{2}\delta \kappah \cup_1 \corho \\ \nonumber &+\frac{1}{2}\kappah \cup \delta \kappah + \delta(\kappah \cup_1 \corho ) \\ \label{rhobar relation}
%     {\overline{\boldsymbol{\rho}}_{\scriptscriptstyle{\{h^{-1}g_v\}}}} =& \corho + \delta \kappah \\ \label{kappa relation}
%     {\overline{\boldsymbol{\rho}}^h_{\scriptscriptstyle{\{h^{-1}g_v\}}}} =& \kappah.
% \end{align}
The first relation, in Eq.~\eqref{nubar relation}, can be derived using the coboundary relation of $\nu$ [Eq.~\eqref{guweneqs}], the homogeneity of $\nu$ [Eq.~\eqref{rho homogeneous}], and the cup product relations in Appendix~\ref{app: terminology}. The second identity, [Eq.~\eqref{rhobar relation}], follows from the fact that $\rho$ is closed [Eq.~\eqref{guweneqs}] as well as the homogeneity of $\rho$ [Eq.~\eqref{rho homogeneous}]. The final relation, [Eq.~\eqref{kappa relation}], is a result of the homogeneity of $\rho$. Plugging Eqs.~\eqref{nubar relation}, \eqref{rhobar relation}, and \eqref{kappa relation} into the right hand side of Eq.~\eqref{effect of 0form on U2} and using the cup product relations, we find:
\begin{multline} \label{variation of coalpha}
    {\overline{\boldsymbol{\alpha}}}_{\scriptscriptstyle{\{h^{-1}g_v\}}}' = \coalpha \\ + \frac{1}{2} \delta \left[{\widetilde{\boldsymbol{\nu}}}^{\, h}_{\scriptscriptstyle{\{g_v\}}} + \kappah \cup_1 (\corho + \delta \ba_e) + \ba_e \cup \kappah \right].
\end{multline}

On a closed manifold, the coboundary term in Eq.~\eqref{variation of coalpha} vanishes after taking the product over all tetrahedron in Eq.~\eqref{U2 def app}, by Stokes' theorem. Therefore, for arbitrary $h\in G$:
\begin{align}
    V_\rho(h)\Ugspt V^\dagger_\rho(h) = \Ugspt.
\end{align}
Since $\Ugspt$ commutes with the symmetry, $H_2$ must be symmetric under the $2$-group symmetry.

\section{Properties of $\Us$}

The goal of this Appendix is to establish the properties of the FDQC $\Us$ employed in the main text. We first show that $\Us$ is symmetric up to factors of $\bar{G}_e$. This is used to compute the fractionalization of the $G$ symmetry on the loop-like excitations of the shadow model (see Section~\ref{sec: shadow model}). Next, we express $\Us$ in terms of $\bar{U}_f$ and $W_t$, as required in Section~\ref{sec: supercohomology model}. As such, $\Us$ can be straightforwardly fermionized using the fermion condensation duality in Section~\ref{sec ttc}. Lastly, we derive the algebraic composition laws of the FDQCs for different sets of supercohomology data. With this, we determine the stacking laws of the corresponding fSPT phases in Section~\ref{sec: supercohomology model}.

To start, we make a minor simplification to $\Us$:
\begin{multline} \label{Us with Wt}
    \Us = \prod_f X_f^{\rhohat(f)} \prod_{t} \Big[e^{2 \pi i O_t \left[\nuhat(t)+\frac{1}{2}\rhohat \cup_1 \rhohat(t)\right]}  \prod_{f\subset t}Z_{f}^{\rhohat \cup_1 \bface (t)}\Big] \\ \times \prod_t W_t^{\int \rhohat \cup_2 \boldsymbol{t}}.
\end{multline}
We commute the product of $X_f$ operators past the product of $Z_f$ operators. This cancels the sign from $\frac{1}{2}\rhohat \cup_1 \rhohat(t)$, and we are left with:
\begin{multline} \label{Us X moved}
        \Us = \prod_{t} e^{2 \pi i O_t \nuhat(t)} \prod_{f}Z_{f}^{\int\rhohat \cup_1 \bface } \prod_f X_f^{\rhohat(f)} \prod_t W_t^{\int \rhohat \cup_2 \boldsymbol{t}}.
\end{multline}
We have also changed the product of $Z_f$ operators to a product over all faces in $M$. Since the $\cup_1$ product vanishes if $f$ is not contained in the boundary of $t$, the change of bounds does not affect $\Us$. We use the form of $\Us$ in Eq.~\eqref{Us X moved} for the calculations below. 

\subsection{Symmetry variation} \label{app: sym var Us}

Our goal is to show the identity:
\begin{align} \label{Us sym identity}
    V(h) \Us  =  \Us \Big[\prod_e \bar{G}_e^{\rhohat^{h^{-1}}(e)} \Big] V(h),
\end{align}
for an arbitrary $h \in G$, used in Section~\ref{sec: shadow model} [Eq.~\eqref{Us commutativity}]. We compute the action of the symmetry on $\Us$, i.e., $V(h)\Us V^\dagger(h)$ by conjugating terms in $\Us$ one at a time. 

First, conjugation of the $\nuhat$ term by $V(h)$ produces:
\begin{align} \label{nu sym var}
    V(h)\prod_{t} e^{2 \pi i O_t \nuhat(t)}V^\dagger(h) = \phi(h,\{g_v\})\prod_{t} e^{2 \pi i O_t \nuhat(t)}.
\end{align}
$\phi(h,\{g_v\})$ is a phase factor that depends on $h$ and the $\{g_v\}$-configuration. We put constraints on the phase factors that appear during the calculation at the end by using the symmetry of the ground state(s) of $H_s$.

Next, we conjugate the Pauli X and Pauli Z terms:
\begin{align}
    \prod_{f}Z_{f}^{\int \rhohat \cup_1 \bface} \prod_f X_f^{\rhohat(f)}.
\end{align}
After conjugation by the symmetry, $\rhohat$ becomes:
\begin{align} \label{rhohat transformation}
    \rhohat \rightarrow \rhohat + \delta \rhohat^{h^{-1}}.
\end{align}
Therefore, up to a phase factor $\phi_1'(h,\{g_v\})$, the Pauli X and Z terms of $\Us$ are mapped by the symmetry action to:
\begin{multline}\label{sym var Pauli}
    \prod_{f}Z_{f}^{\int \rhohat \cup_1 \bface} \prod_f X_f^{\rhohat(f)} \\ \times \phi_1'(h,\{g_v\}) \prod_f X_f^{\delta \rhohat^{h^{-1}}(f)} \prod_{f}Z_{f}^{\int \delta \rhohat^{h^{-1}} \cup_1 \bface} 
\end{multline}
Rearranging the last two products in Eq.~\eqref{sym var Pauli}, we find:
\begin{multline} \label{sym var Pauli2prime}
    \prod_{f}Z_{f}^{\int\rhohat \cup_1 \bface } \prod_f X_f^{\rhohat(f)} \\ \times \phi_2'(h,\{g_v\}) \prod_e \left( \prod_{f \supset e} X_f \prod_{f} Z_{f}^{\int \delta \be \cup_1 \bface} \right)^ {\rhohat^{g^{-1}} (e)},
\end{multline}
for another phase factor $\phi_2'(h,\{g_v\})$. The term in parentheses in Eq.~\eqref{sym var Pauli2prime} is equal to:
\begin{align}
    \prod_{f \supset e} X_f \prod_{f} Z_{f}^{\int \delta \be \cup_1 \bface} = \bar{G}_e \prod_t W_t^{\int \delta \be \cup_2 \boldsymbol{t}},
\end{align}
after using a cup product relation from Appendix~\ref{app: terminology}. Thus, the symmetry action on the Pauli X and Pauli Z terms yields:
\begin{multline} \label{sym var Pauli2}
    \prod_{f}Z_{f}^{\int \rhohat \cup_1 \bface} \prod_f X_f^{\rhohat(f)} \\ \times \phi_2'(h,\{g_v\}) \prod_e \bar{G}_e^ {\rhohat^{g^{-1}} (e)} \prod_t W_t^{\int \delta \rhohat^{h^{-1}} \cup_2 \boldsymbol{t}}.
\end{multline}

Lastly, we compute the symmetry action on the $W_t$ term of $\Us$:
\begin{align}
    \prod_t W_t^{\int \rhohat \cup_2 \boldsymbol{t}}.
\end{align}
$\rhohat$ is transformed as in Eq.~\eqref{rhohat transformation}, so we obtain:
\begin{align} \label{Wt sym var}
    \prod_t W_t^{\int \rhohat \cup_2 \boldsymbol{t}}\prod_t W_t^{\int \delta \rhohat^{h^{-1}} \cup_2 \boldsymbol{t}}.
\end{align}

Putting Eqs.~\eqref{nu sym var}, \eqref{sym var Pauli2}, and \eqref{Wt sym var} together, we have:
\begin{align} \label{almost sym of Us}
    V(h)\Us V^\dagger(h) = \phi''(h,\{g_v\}) \Us \prod_e \bar{G}_e^{\rhohat^{h^{-1}}},
\end{align}
where $\phi''(h,\{g_v\})$ is some yet undetermined phase factor. We note that the equality above differs from Eq.~\eqref{Us sym identity} by precisely $\phi''(h,\{g_v\})$. In what follows, we use the symmetry of the ground state(s) of $H_s$ to argue that $\phi''(h,\{g_v\})$ is indeed $1$.

A ground state of the shadow model is given by applying $\Us$ to a ground state of $H^G_\text{ttc}$:
\begin{align}
    \ket{\Psi_s}\equiv \Us \sum_{\{g_v\},\{\ba_e\}} (-1)^{ \sum_t \ba_e \cup \delta \ba_e(t)}\ket{\{g_v\},\{ \delta \ba_e\}}.
\end{align}
Furthermore, the state $\ket{\Psi_s}$ is invariant under the $G$ symmetry, i.e.:
\begin{align} \label{Psis invariance}
    V(g)\ket{\Psi_s}=\ket{\Psi_s}.
\end{align}
The symmetry of $\ket{\Psi_s}$ follows from the symmetry of $H_s$ and the fact that it can be prepared from a ground state of $H_\text{ttc}^G$ by a FDQC.

With this, we argue that the phase factor $\phi''(h,\{g_v\})$ is $1$. We apply $V(h)$ to $\ket{\Psi_s}$ to find:
\begin{align} \label{Psis invariance?}
    &V(h)\ket{\Psi_s} \\ \nonumber &= V(h)\Us \sum_{\{g_v\},\{\ba_e\}} (-1)^{ \sum_t \ba_e \cup \delta \ba_e(t)}\ket{\{g_v\},\{ \delta \ba_e\}}. \\ \nonumber
    &= \Us \sum_{\{g_v\},\{\ba_e\}} \phi(h,\{g_v\}) (-1)^{ \sum_t \ba_e \cup \delta \ba_e(t)}\ket{\{g_v\},\{ \delta \ba_e\}},
\end{align}
where we have used both Eq.~\eqref{almost sym of Us} and that the ground state(s) of $H^G_\text{ttc}$ are $+1$ eigenstates of $\bar{G}_e$. Now, in comparing Eq.~\eqref{Psis invariance} and the last line of Eq.~\eqref{Psis invariance?}, we see that $\phi''(h,\{g_v\})$ must be $1$ for every $\{g_v\}$. This implies that:
\begin{align}
    V(h)\Us  = \Us \prod_e \bar{G}_e^{\rhohat^{h^{-1}}} V(h),
\end{align}
as claimed.

\subsection{Fermionizability} \label{app: fermionization shadow circuit}

Here, we show that the FDQC $\Us$ in Eq.~\eqref{Us X moved} can be expressed in terms of $\bar{U}_f$ and $W_t$ operators. The fermion condensation duality (Table \ref{table: gauging fermion parity}) can then be immediately applied to $\Us$ to construct the FDQC $\Uf$.

% First, we identify factors of $W_t$ in $\Us$. We notice that the product of Pauli Z operators:
% \begin{align}
%   \prod_{t} \prod_{f\subset t}Z_{f}^{\rhohat \cup_2 \delta \bface (t)},
% \end{align}
% in $\Us$ is equivalent to the product of $W_t$ operators:
% \begin{align}
%     \prod_t W_t^{\sum_{t'}\rhohat \cup_2 \boldsymbol{t}(t')}.
% \end{align}
% Therefore, $\Us$ can be written as:
% \begin{multline} 
%     \Us = \prod_f X_f^{\rhohat(f)} \prod_{t} \Big[e^{2 \pi i O_t \left[\nuhat(t)+\frac{1}{2}\rhohat \cup_1 \rhohat(t)\right]} \prod_{f\subset t}Z_{f}^{\rhohat \cup_1 \bface (t)}\Big] \\ \times \prod_t W_t^{\sum_{t'}\rhohat \cup_2 \boldsymbol{t}(t')}.
% \end{multline}

By re-arranging the Pauli X and Pauli Z operators in Eq.~\eqref{Us with Wt}:
\begin{multline} \label{Us for fermionizing app}
        \Us = \prod_{t} e^{2 \pi i O_t \nuhat(t)} \prod_{f}Z_{f}^{\int \rhohat \cup_1 \bface } \prod_f X_f^{\rhohat(f)} \prod_t W_t^{\int \rhohat \cup_2 \boldsymbol{t}},
\end{multline}
we can form $\bar{U}_f$ operators. To show this, we decompose the product of $Z_f$ operators into:
\begin{align}
    \prod_{f}Z_{f}^{\int \rhohat \cup_1 \bface} = \prod_f\prod_{f'}Z_{f'}^{\rhohat(f) \sum_t  \bface \cup_1 \bface' (t)}.
\end{align}
We then see that we can form a factor of $\bar{U}^{\rhohat(f)}_f$ for each face $f$:
\begin{align}
    \bar{U}^{\rhohat(f)}_f = X_f^{\rhohat(f)}\prod_{f'}Z_{f'}^{\rhohat(f)\sum_t \bface \cup_1 \bface' (t)}.
\end{align}
Given the commutation relations of $\bar{U}_f$ operators, the resulting product of $\bar{U}^{\rhohat(f)}_f$ operators will generically depend on a choice of ordering. Hence, we choose an arbitrary ordering of the faces ($f_1<\cdots<f_i<\cdots$) of the set $F$ of faces in $M$. We aim to form a product of $\bar{U}^{\rhohat(f)}_f$ according to the ordering on $F$, i.e.:
\begin{align}\label{Uf ordered prod}
    \prod_{f \in F} \bar{U}^{\rhohat(f)}_f = (\cdots \bar{U}^{\rhohat(f_i)}_{f_i} \cdots \bar{U}^{\rhohat(f_1)}_{f_1}).
\end{align}
We re-order the Pauli X and Pauli Z operators of $\Us$ into the product in Eq.~\eqref{Uf ordered prod} by first, ordering the product of $Z_f$ operators by the ordering on $F$:
\begin{multline}
    \prod_{f\in F}\prod_{f'}Z_{f'}^{\rhohat(f) \sum_t  \bface \cup_1 \bface' (t)} = \\  \Big[\cdots \prod_{f'}Z_{f'}^{\rhohat(f_i)\sum_t \bface \cup_1 \bface' (t)}\cdots \prod_{f'}Z_{f'}^{\rhohat(f_1)\sum_t \bface \cup_1 \bface' (t)} \Big].
\end{multline}
To form $\bar{U}^{\rhohat(f_i)}_{f_i}$, $X_{f_i}^{\rhohat(f_i)}$ in Eq.~\eqref{Us for fermionizing app} is commuted past the Pauli Z operators of $\bar{U}^{\rhohat(f_{i'})}_{f_{i'}}$ for each $i'<i$. This produces the sign:
\begin{align}
    \prod_{i'<i} (-1)^{\rhohat(f_{i'})\rhohat(f_{i})\sum_t \bface_{i'} \cup_1 \bface_{i}(t) }.
\end{align}
In creating the product in Eq~\eqref{Uf ordered prod}, we thus accrue the sign:
\begin{align}
    \xi_{\bar{\rho}}(F) = \prod_{i} \left[ \prod_{i'<i} (-1)^{\rhohat(f_{i'})\rhohat(f_{i}) \int \bface_{i'} \cup_1 \bface_{i} } \right].
\end{align}

In summary, we have shown that $\Us$ can be written as:
\begin{align} 
    \Us = \prod_t e^{2 \pi i O_t \nuhat(t)} \xi_{\bar{\rho}}(F){\prod_{f\in F}} \bar{U}_f^{\rhohat(f)} \prod_{t} W_t^{\int \rhohat \cup_2 \boldsymbol{t}}.
\end{align}
which matches the form of $\Us$ in Eq.~\eqref{Us nice form}. Fermion condensation is implemented by replacing $\bar{U}_f$ with $S_f$ and $W_t$ with $P_t$. This gives the circuit $\Uf$ in Eq.~\eqref{eq: uf def}:
\begin{align}
  \Uf \equiv \prod_t e^{2 \pi i O_t \nuhat(t)} \xi_{\bar{\rho}}(M){\prod_{f}} S_f^{\rhohat(f)} \prod_{t} P_t^{\int \rhohat \cup_2 \boldsymbol{t}}.
\end{align}

\subsection{Composition laws}\label{app:composition laws}

As argued in Section~\ref{sec: supercohomology models 2}, the stacking laws of supercohomology phases can be determined by composing the FDQCs $\Uf$ that prepare the supercohomology models. For convenience, we evaluate the composition of the FDQCs $\Us$ that prepare the shadow model. The composition of $\Uf$ operators follows from this by applying the fermionization duality to the $\Us$ circuits. For reference, the FDQC $\Us$ corresponding to the supercohomology data $(\rho,\nu)$ is:
% \begin{equation}
%     \begin{split}
%         \mathcal{U}_s (\nu,\rho) = & \prod_{t=\langle 1234\rangle} \left(
%         e^{ 2 \pi i \, \nu (1,g_1,g_2,g_3,g_4) O_t }  \right)  \\
%         & ~~~~ \prod_{f'} \left( Z_{f'}^{\int \bar{\rho} \cup_1 f'} \right) \prod_f \left( X_f^{\bar{\rho}(f)} \right) \prod_t \left(W_t^{\bar{\rho} \cup_2 t} \right)
%     \end{split}
% \end{equation}
\begin{multline}
        \Us^{\rho\nu} = \\ \prod_{t}
        e^{ 2 \pi i \, \nuhat(t) O_t }  
        \prod_{f'}  Z_{f'}^{\int \rhohat \cup_1 \boldsymbol f'}  \prod_f  X_f^{\rhohat(f)}  \prod_t W_t^{\int \rhohat \cup_2 \boldsymbol{t}} .
\end{multline}

% We want to study the stacking of two phases $(\nu_1,\rho_1)$ and $(\nu_2,\rho_2)$. Denote the new phase as $(\nu^\prime,\rho^\prime)$, which satisfies:
% \begin{equation}
%     \mathcal{U}_s (\nu^\prime,\rho^\prime)=\mathcal{U}_s (\nu_1,\rho_1) \mathcal{U}_s (\nu_2,\rho_2).
% \end{equation}
Given two sets of supercohomology data $(\rho,\nu)$ and $(\rho',\nu')$, we calculate the product of $\Us^{\rho\nu}$ and $\Us^{\rho'\nu'}$ directly:
% \begin{equation}
%     \begin{split}
%         &\mathcal{U}_s (\nu_1,\rho_1) \mathcal{U}_s (\nu_2,\rho_2)\\
%         =& \prod_{t=\langle 1234\rangle} \left(
%         e^{ 2 \pi i \, \nu_1 (1,g_1,g_2,g_3,g_4) O_t }  \right)  \\
%         & ~~~~ \prod_{f'} \left( Z_{f'}^{\int \bar{\rho}_1 \cup_1 f'} \right) \prod_f \left( X_f^{\bar{\rho}_1(f)} \right) \prod_t \left(W_t^{\bar{\rho}_1 \cup_2 t} \right)\\
%         & \prod_{t=\langle 1234\rangle} \left(
%         e^{ 2 \pi i \, \nu_2 (1,g_1,g_2,g_3,g_4) O_t }  \right)  \\
%         & ~~~~ \prod_{f'} \left( Z_{f'}^{\int \bar{\rho}_2 \cup_1 f'} \right) \prod_f \left( X_f^{\bar{\rho}_2(f)} \right) \prod_t \left(W_t^{\bar{\rho}_2 \cup_2 t} \right)\\
%         =& \prod_{t=\langle 1234\rangle} \left(
%         e^{ 2 \pi i \, \nu' (1,g_1,g_2,g_3,g_4) O_t }  \right)  \\
%         & ~~~~ \prod_{f'} \left( Z_{f'}^{\int \bar{\rho}' \cup_1 f'} \right) \prod_f \left( X_f^{\bar{\rho}'(f)} \right) \prod_t \left(W_t^{\bar{\rho}' \cup_2 t} \right)\\
%         =&\mathcal{U}_s (\nu',\rho')
%     \end{split}
% \end{equation}
\begin{equation} \label{composition of Us}
    \begin{split}
        &\Us^{\rho\nu} \Us^{\rho'\nu'} \\
        &= \prod_{t} e^{ 2 \pi i \, \nuhat(t) O_t }  
        \prod_{f'}  Z_{f'}^{\int \rhohat \cup_1 \boldsymbol f'}  \prod_f  X_f^{\rhohat(f)}  \prod_t W_t^{\int \rhohat \cup_2 \boldsymbol{t}} \\
        &\times  \prod_{t} e^{ 2 \pi i \, \nuhat'(t) O_t }  
        \prod_{f'}  Z_{f'}^{\int \rhohat' \cup_1 \boldsymbol f'}  \prod_f  X_f^{\rhohat'(f)}  \prod_t W_t^{\int \rhohat' \cup_2 \boldsymbol{t}} \\
        &= \prod_{t} e^{ 2 \pi i \, \nuhat''(t) O_t }  
        \prod_{f'}  Z_{f'}^{\int \rhohat'' \cup_1 \boldsymbol f'}  \prod_f  X_f^{\rhohat''(f)}  \prod_t W_t^{\int \rhohat'' \cup_2 \boldsymbol{t}}. 
    \end{split}
\end{equation}
% \begin{equation}
%     \begin{split}
%         &\Us^{\rho\nu} \Us^{\rho'\nu'} \\
%         =& \prod_{t=\langle 1234\rangle} \left(
%         e^{ 2 \pi i \, \nuhat (t) O_t }  \right)  \\
%         & ~~~~ \prod_{f'} \left( Z_{f'}^{\int \rhohat \cup_1 \bface'} \right) \prod_f \left( X_f^{\rhohat(f)} \right) \prod_t \left(W_t^{\int \rhohat \cup_2 \bt} \right)\\
%         & \prod_{t=\langle 1234\rangle} \left(
%         e^{ 2 \pi i \, \nuhat' (t) O_t }  \right)  \\
%         & ~~~~ \prod_{f'} \left( Z_{f'}^{\int \rhohat' \cup_1 \bface'} \right) \prod_f \left( X_f^{\rhohat'(f)} \right) \prod_t \left(W_t^{\int \rhohat' \cup_2 \bt} \right)\\
%         =& \prod_{t=\langle 1234\rangle} \left(
%         e^{ 2 \pi i \, \nuhat'' (t) O_t }  \right)  \\
%         & ~~~~ \prod_{f'} \left( Z_{f'}^{\int \rhohat'' \cup_1 \bface'} \right) \prod_f \left( X_f^{\rhohat''(f)} \right) \prod_t \left(W_t^{\int \rhohat'' \cup_2 \bt} \right)\\
%         =&\Us^{\rho''\nu''}
%     \end{split}
% \end{equation}
In the last line, we have combined the Pauli Z and Pauli X operators and defined $\rhohat''\equiv \rhohat+\rhohat'$. The operator $\nuhat''$ is a sum of $\nuhat$ and $\nuhat'$ and also incorporates the sign incurred from commuting the Pauli X and Pauli Z operators. In particular, to form the last line in Eq.~\eqref{composition of Us}, we commute the operators:
\begin{eqs} 
      \prod_{f^\prime} \left( {Z}_{f^\prime}^{\int \rhohat' \cup_1 f^\prime} \right) &\text{ past }  \prod_f \left( {X}^{\rhohat (f)}_f \right), \\
     \prod_f \left( {X}^{\rhohat' (f)}_f \right) &\text{ past }
     \prod_t \left(W_t^{\rhohat \cup_2 \boldsymbol{t}} \right).
\end{eqs}
This produces the sign:
\begin{align}
    (-1)^{\int \rhohat' \cup_1 \rhohat + \rhohat \cup_2 \delta \rhohat'},
\end{align}
so we define $\nuhat''$ as:
\begin{align}
    \nuhat'' \equiv \nuhat + \nuhat' + \frac{1}{2}\rhohat' \cup_1 \rhohat + \frac{1}{2}\rhohat \cup_2 \delta \rhohat'.
\end{align}

Now, we recover the supercohomology data corresponding to $\rhohat''$ and $\nuhat''$ from their diagonal matrix elements. For a face $\langle 123 \rangle$, the matrix element $\bra{\{g_v\}}\rhohat''\boldsymbol{(}\langle 123 \rangle\boldsymbol{)}\ket{\{g_v\}}$ is:
\begin{align}
    \bra{\{g_v\}}\rhohat''\boldsymbol{(}\langle 123 \rangle\boldsymbol{)}\ket{\{g_v\}} = \rho(1,g_1,g_2,g_3) + \rho'(1,g_1,g_2,g_3).
\end{align}
Therefore, $\rhohat''$ corresponds to the function $\rho'' = \rho + \rho'$. For $\nuhat''$, we compute the matrix element $\bra{\{g_v\}}\nuhat''\boldsymbol{(}\langle 1234 \rangle\boldsymbol{)}\ket{\{g_v\}}$, with an arbitrary tetrahedron $\langle 1234 \rangle$:
\begin{eqs} \label{nuprimeprimematrixelements}
    \bra{\{g_v\}}\nuhat''\boldsymbol{(}\langle 1234 \rangle\boldsymbol{)}&\ket{\{g_v\}} = \\
    &\nu(1,g_1,g_2,g_3,g_4) + \nu'(1,g_1,g_2,g_3,g_4) \\
    +&\frac{1}{2}\rho(1,g_1,g_2,g_3)\rho'(1,g_1,g_3,g_4) \\
         +& \frac{1}{2}\rho(1,g_2,g_3,g_4) \rho'(1,g_1,g_2,g_4) \\
         +& \frac{1}{2}\rho(1,g_1,g_2,g_3) \rho'(g_1,g_2,g_3,g_4)  \\
         +& \frac{1}{2}\rho(1,g_1,g_3,g_4) \rho'(g_1,g_2,g_3,g_4). \\
\end{eqs}
To obtain the right-hand side of Eq.~\eqref{nuprimeprimematrixelements}, we have used the explicit formulas for cup products in Appendix~\ref{app: terminology}. The sum of $\rho$ and $\rho'$ terms can be further simplified to ${\frac{1}{2} \rho \cup_2 \rho'(1,g_1,g_2,g_3,g_4)}$ with Eq.~\eqref{explicit cup2}. Thus, $\nuhat''$ corresponds to the function ${\nu'' \equiv \nu + \nu'' + \frac{1}{2} \rho \cup_2 \rho'}$.

In summary, we have found $\Us^{\rho\nu}\Us^{\rho'\nu'}=\Us^{\rho''\nu''}$, where $(\rho'',\nu'')$ is:
\begin{align}
    (\rho'',\nu'') = (\rho+\rho',\nu + \nu'' + \frac{1}{2} \rho \cup_2 \rho').
\end{align}
Since the composition rules are preserved under fermionization, we have that $\Uf^{\rho\nu}\Uf^{\rho'\nu'}=\Uf^{\rho''\nu''}$. Hence, the stacking rules for supercohomology phases is given by:
\begin{equation}
    (\rho,\nu) \boxtimes (\rho',\nu')  = (\rho + \rho', \nu + \nu' + \frac{1}{2} \rho \cup_2 \rho').
\end{equation}

\section{SPT state built from trivial supercohomology data} \label{app: trivial circuit}
% \section{Locally symmetric circuit for trivial supercohomology data}

In this appendix, we show that the fSPT state $\ket{\Psi_f^{\rho_0\nu_0}}$ corresponding to the trivial supercohomology data:
\begin{align}\label{trivial super data 2}
    (\rho_0, \nu_0) = (\delta \beta, \delta \eta + \frac{1}{2} \beta \cup \beta + \frac{1}{2} \beta \cup_1 \delta \beta),
\end{align}
can be constructed from a product state by a FDQC composed of local symmetric unitaries. According to the definition of fSPT phases, $\ket{\Psi_f^{\rho_0\nu_0}}$ must belong to the trivial fSPT phase. This verifies the claim made in the main text that the supercohomology data $(\rho_0,\nu_0)$ corresponds to the trivial phase. As a warmup, we consider the case in which $\beta$ and $\eta$ are closed. We then derive the more general statement for an arbitrary choice of $(\rho_0,\nu_0)$.
% From the definition of SPT phases, this wavefunction is belonged to a trivial phase. This verifies the claim in the main text that $(\delta \beta, \delta \eta + \frac{1}{2} \beta \cup \beta + \frac{1}{2} \beta \cup_1 \delta \beta)$ is supercohomology data corresponding to the trivial phase. As a warm up, we consider a simple case before deriving the general circuit.

% \subsection{$\nu= \frac{1}{2} \beta \cup \beta$ and $\rho=0$ for any 2-cocycle $\beta \in H^2(G,\ZZ_2)$}

\vspace{1.5mm}
\noindent \begin{center}\emph{Assuming $\beta$ and $\eta$ are closed:}\end{center}
\vspace{1.5mm}

In this case, $(\rho_0,\nu_0)$ is equal to $(0,\frac{1}{2}\beta \cup \beta)$. Note that $\nu_0$ is given by the group cocycle $\frac{1}{2}\beta \cup \beta$, which can be used to construct a group cohomology model following the discussion in Section~\ref{sec: group cohomology models}. If $\frac{1}{2}\beta\cup\beta$ is a nontrivial cocycle, then the model describes a nontrivial bosonic SPT phase.
However, after introducing fermions [either emergent (in the twisted toric code) or physical (in the fSPT model)], the phase factor associated to $\frac{1}{2}\beta \cup \beta$ can be produced by a FDQC comprised of local symmetric unitaries, as we show below. 

For convenience, we consider the shadow model. A ground state of the shadow model corresponding to the supercohomology data $(0,\frac{1}{2}\beta \cup \beta)$ is:
\begin{eqs}
    \ket{\Psi_s^{\rho_0\nu_0}} =& \sum_{\{g_v\}, \ba_e} (-1)^{\int \betahat \cup \hat{\boldsymbol \beta} + \ba_e \cup \delta \ba_e}\ket{\{g_v\},\delta \ba_e},
\label{eq: beta cup beta shadow wavefunction}
\end{eqs}
where we have introduced the notation:
\begin{eqs}
    & \hat{\boldsymbol \beta} \boldsymbol{(}\langle 123 \rangle \boldsymbol{)}
    \ket{\{g_v\}, \delta \ba_e} = 
    \beta(g_1,g_2,g_3) \ket{\{g_v\}, \delta \ba_e},\\
    &\betahat\boldsymbol{(}\langle 12 \rangle \boldsymbol{)} \ket{\{g_v\}, \delta \ba_e} = 
    \beta(1,g_1,g_2) \ket{\{g_v\}, \delta \ba_e}.
\end{eqs}
% In this special case, $\nu = \frac{1}{2} \beta \cup \beta$, for any 2-cocycle $\beta$, can give nontrivial bosonic SPT phases if $\frac{1}{2} \beta \cup \beta$ is a nontrivial 4-cocycle. However, after introducing the fermionic degrees of freedom, we can show that this phase factor can be written as a locally symmetric fermionic quantum circuit. We begin with our shadow wavefunction:
% \begin{eqs}
%     \ket{\Psi_s} =& ~\Us \ket{\{g_v\},\Psi_\text{ttc}}\\
%     =& \sum_{\{g_v\}, \ba_e} \prod_{t=\lr{1234}} (-1)^{\beta(1,g_1,g_2) \beta(g_2,g_3,g_4)} \\
%     & ~~~~ (-1)^{\int \ba_e \cup \delta \ba_e}\ket{\{g_v\},\delta \ba_e}
% \end{eqs}
% where we have used $\Us$ defined in \eqref{eq: Us def} and $\ket{\Psi_\text{ttc}}$ defined in \eqref{ttc gs}. By defining the shorthand notation:
% \begin{eqs}
%     & \hat{\boldsymbol \beta} \boldsymbol{(}\langle 123 \rangle \boldsymbol{)}
%     \ket{\{g_v\}, \delta \ba_e} = 
%     \beta(g_1,g_2,g_3) \ket{\{g_v\}, \delta \ba_e},\\
%     &\betahat\boldsymbol{(}\langle 12 \rangle \boldsymbol{)} \ket{\{g_v\}, \delta \ba_e} = 
%     \beta(1,g_1,g_2) \ket{\{g_v\}, \delta \ba_e},
% \end{eqs}
% the shadow wavefunction can be written as
% \begin{eqs}
%     \ket{\Psi_s} =& \sum_{\{g_v\}, \ba_e} (-1)^{\int \betahat \cup \hat{\boldsymbol \beta} + \ba_e \cup \delta \ba_e}\ket{\{g_v\},\delta \ba_e}.
% \label{eq: beta cup beta shadow wavefunction}
% \end{eqs}
Since $\beta$ is a cocycle, we have $\delta \betahat = \hat{\boldsymbol \beta}$. Using the cup product relations in Appendix~\ref{app: terminology}, the exponent in Eq.~\eqref{eq: beta cup beta shadow wavefunction} can be written as:
\begin{multline}
    \int \betahat \cup \delta \betahat + \ba_e \cup \delta \ba_e = \\ \int (\ba_e + \betahat) \cup \delta (\ba_e + \betahat) +
    \hat{\boldsymbol \beta} \cup_1 \delta \ba_e.
\end{multline}
By defining $\ba_e' \equiv \ba_e + \betahat$, the ground state of the shadow model in Eq.~\eqref{eq: beta cup beta shadow wavefunction} becomes:
\begin{eqs}
    &~\ket{\Psi^{\rho_0\nu_0}_s} \\ &= \sum_{\{g_v\}, \ba_e} (-1)^{\int \hat{\boldsymbol \beta} \cup_1 (\delta \ba_e'+ \hat{\boldsymbol \beta}) + \ba_e' \cup \delta \ba_e'} \prod_f X_f^{\hat{\boldsymbol \beta} (f)}\ket{\{g_v\},\delta \ba_e'} \\
    &= \sum_{\{g_v\}, \ba_e'} \prod_{f'} Z_{f'}^{\int \hat{\boldsymbol \beta} \cup_1 \bface'}
    \prod_f X_f^{\hat{\boldsymbol \beta} (f)}
    (-1)^{\int \ba_e' \cup \delta \ba_e'}\ket{\{g_v\},\delta \ba_e'} \\
    &=  \sum_{\{g_v\}, \ba_e'} \bar{U}_{\beta} (-1)^{\int \ba_e' \cup \delta \ba_e'}\ket{\{g_v\},\delta \ba_e'}. \\
\end{eqs}
According to the fermionization duality reviewed in Appendix~\ref{sec: review of boson-fermion duality} $\ket{\Psi^{\rho_0\nu_0}_s}$ maps to the fermionic state:
\begin{eqs}
    \ket{\Psi^{\rho_0\nu_0}_f} = \sum_{\{g_v\}} S_{\beta} \ket{\{g_v\},\text{vac}}.
\end{eqs}
Here, $S_{\beta}$ is the FDQC built from the product of hopping operators:
\begin{eqs}
    S_{\beta} &\equiv \xi_\beta(M){\prod_{f}} S_f^{\hat{\boldsymbol \beta}(f)},
\end{eqs}
with $\xi_\beta(M)$ given by:
\begin{align}
    \xi_\beta(M) &\equiv \prod_{i,i'|i'<i} (-1)^{\hat{\boldsymbol \beta}(f_{i'})\hat{\boldsymbol \beta}(f_{i}) \int \bface_{i'} \cup_1 \bface_{i} }.
\end{align}
The FDQC $S_\beta$ is composed of local symmetric unitaries due to the homogeneity of $\beta$. Therefore, $\ket{\Psi^{\rho_0\nu_0}_f}$ belongs to the trivial fSPT phase.

\vspace{1.5mm}
\noindent \begin{center}\emph{General trivial supercohomology data:}\end{center}
\vspace{1.5mm}

We now consider the more general case where $(\rho_0,\nu_0)$ takes the form in Eq.~\eqref{trivial super data 2} for any $\eta \in C^3(G,\RR/\ZZ)$ and $\beta \in C^2(G,\ZZ_2)$. A ground state of the shadow model (after the basis transformation introduced in Eq.~\eqref{eq: bulk basis transform}) is:
\begin{eqs}
    \ket{\Psi^{\rho_0\nu_0}_s} =& \sum_{\{g_v\}, \ba_e}  \prod_t e^{2 \pi i O_t \nuhat(t)}(-1)^{\int \rhohat \cup_1 (\delta \ba_e + \rhohat)} \\
    & ~~~~ (-1)^{\int \ba_e \cup \delta \ba_e} \prod_f X_f^{\rhohat(f)}\ket{\{ g_v\},\delta \ba_e}.
\end{eqs}
The $\nuhat$ term can be decomposed into:
\begin{eqs} \label{nubar decomposition}
    \prod_t e^{2\pi i O_t \hat{\boldsymbol \eta} (t)} (-1)^{\int \betahat \cup \hat{\boldsymbol \beta} + \rhohat \cup \betahat + \betahat \cup_1 \hat{\boldsymbol \rho}},
\end{eqs}
where $\hat{\boldsymbol \rho}$ and $\rhohat$ are defined by:
\begin{eqs}
    & \hat{\boldsymbol \rho} \boldsymbol{(}\langle 1234 \rangle \boldsymbol{)}
    \ket{\{g_v\}, \delta \ba_e} = 
    \rho(g_1,g_2,g_3,g_4) \ket{\{g_v\}, \delta \ba_e},\\
    &\rhohat\boldsymbol{(}\langle 123 \rangle \boldsymbol{)} \ket{\{g_v\}, \delta \ba_e} = 
    \rho(1,g_1,g_2,g_3) \ket{\{g_v\}, \delta \ba_e}.
\end{eqs}
To obtain Eq.~\eqref{nubar decomposition}, we have used $\hat{\boldsymbol \rho} = \delta \hat{\boldsymbol \beta}$ and the explicit $\cup_1$ product:
\begin{eqs}
    &\hat{\boldsymbol \beta} \cup_1 \delta \hat{ \boldsymbol \beta} \boldsymbol{(}\lr{01234}\boldsymbol{)} \\
    =& \hat{ \boldsymbol \beta} \boldsymbol{(}\lr{034}\boldsymbol{)} \delta \hat{ \boldsymbol \beta} \boldsymbol{(}\lr{0123}\boldsymbol{)} + \hat{ \boldsymbol \beta} \boldsymbol{(}\lr{014}\boldsymbol{)} \delta \hat{ \boldsymbol \beta} \boldsymbol{(}\lr{1234}\boldsymbol{)}  \\
    =& \rhohat\boldsymbol{(}\lr{123}\boldsymbol{)} \betahat\boldsymbol{(}\lr{34}\boldsymbol{)} + \betahat\boldsymbol{(}\lr{14}\boldsymbol{)} \hat{\boldsymbol \rho} \boldsymbol{(}\lr{1234}\boldsymbol{)}\\
    =& [\rhohat \cup \betahat + \betahat \cup_1 \hat{\boldsymbol \rho}] \boldsymbol{(}\lr{1234}\boldsymbol{)}.
\end{eqs}
Now, using the identity $\rhohat = \hat{\boldsymbol \beta} + \delta \betahat$ and defining the new variables $\ba_e' \equiv \ba_e + \betahat$, the ground state of the shadow model can be organized into:
\begin{eqs}
    \ket{\Psi^{\rho_0\nu_0}_s} =& \sum_{\{g_v\}, \ba_e'}  \prod_t e^{2 \pi i O_t \hat{\boldsymbol \eta}(t)}
    (-1)^{\int \hat{\boldsymbol \beta} \cup_1 (\delta \ba_e' + \hat{\boldsymbol \beta})} \\
    & ~~~~ (-1)^{\int \ba_e' \cup \delta \ba_e'} \prod_f X_f^{\rhohat(f)} X_f^{\delta \betahat(f)}\ket{\{ g_v\},\delta \ba_e'} \\
    =& \sum_{\{g_v\}, \ba_e'} 
    \prod_t e^{2 \pi i O_t \hat{\boldsymbol \eta}(t)}
    \prod_{f'} Z_{f'}^{\int \hat{\boldsymbol \beta} \cup_1 \bface'}
    \prod_f X_f^{\hat{\boldsymbol \beta} (f)} \\
    & ~~~~ (-1)^{\int \ba_e' \cup \delta \ba_e'}\ket{\{g_v\},\delta \ba_e'} \\
    =& \sum_{\{g_v\}, \ba_e'} 
    \prod_t e^{2 \pi i O_t \hat{\boldsymbol \eta}(t)}
    \bar{U}_\beta
    (-1)^{\int \ba_e' \cup \delta \ba_e'}\ket{\{g_v\},\delta \ba_e'}
\end{eqs}
This fermionizes to the fermionic state:
\begin{eqs}
    \ket{\Psi^{\rho_0\nu_0}_f} &= \sum_{\{g_v\}}
    \prod_t e^{2 \pi i O_t \hat{\boldsymbol \eta}(t)}
    S_{\beta} \ket{\{g_v\},\text{vac}} \\
    &= \prod_t e^{2 \pi i O_t \hat{\boldsymbol \eta}(t)}
    S_{\beta} \sum_{\{g_v\}}
     \ket{\{g_v\},\text{vac}},
\end{eqs}
which is constructed from a product state by a FDQC composed of local symmetric unitaries.

\section{Review of (2+1)D supercohomology models} \label{sec: review of (2+1)D supercohology fSPT}

In this appendix, we review the (2+1)D supercohomology fSPT construction with symmetry $G \times \ZZ^f_2$ in \cite{EF19}. This is a warm-up for (3+1)D construction described in Appendix~\ref{app: super from 2gauge}.

\subsection{Bulk construction}
We start with the bulk construction, i.e., SPT states on a closed manifold. The $(2+1)$D supercohomology data is $(n, \nu)$ with $n \in H^2(G,\ZZ_2)$ and $\nu \in C^3(G,\RR/\ZZ)$, satisfying the equation \cite{GW14}:
\begin{equation}
    \delta \nu=\frac{1}{2} n \cup n.
\end{equation}
Similar to the $(3+1)$D case described in the main text, we first construct an auxiliary bSPT model from the supercohomology data. The bosonic model is in a $0$-form SPT phase protected by a symmetry $G'$ with a normal $\ZZ_2$ subgroup. By gauging the $\ZZ_2$ subgroup, we obtain the shadow model, and by subsequently applying the fermionization duality of Ref.~\cite{CKR18}, we arrive at a fermionic model for the supercohomology phase.

The 2-cocycle $n \in H^2(G,\ZZ_2)$ corresponds to a central extension of $G$ by $\ZZ_2$, which we denote as $G' \equiv  \ZZ_2 \times_n G$ given by the short exact sequence: 
\begin{equation}
    0 \ra \ZZ_2 \ra G' \ra G \ra 1.
\end{equation}
The group elements in $G'$ are $g^{(a)}$ with label $g\in G$ and $a \in \ZZ_2 = \{0,1\}$, and the group law is
\begin{equation}
    g_1^{(a_1)} g_2^{(a_2)} = (g_1 g_2)^{\boldsymbol{(}a_1 + a_2 + n(1,g_1,g_1 g_2)\boldsymbol{)}}.
\label{eq: G' group law}
\end{equation}
We can define a cocycle in $H^3(G',\RR/\ZZ)$ by:
\begin{equation}
    \alpha_3 = \nu + \frac{1}{2} n \cup \eps_1
\label{eq: alpha3}
\end{equation}
where $\eps_1 (g_1^{(a_1)}, g_2^{(a_2)}) \equiv a_1+a_2+n(1,g_1,g_2)$ is a $1$-cochain in $G'$ satisfying $\delta \eps_1 = n$ (where $n$ and $\nu$ are implicitly pulled back to $G'$). One can check that $\eps_1$ is homogeneous, i.e. $\eps_1(h^\prime g_i^\prime, h^\prime g_j^\prime) = \eps(g_i^\prime, g_j^\prime)$.
% In inhomogeneous form of $\eps_1$, the above definition for $\eps_1$ is:
% \begin{equation}
%     \begin{split}
%         \eps_1 ((g_i^{(a_i)})^{-1} g_j^{(a_j)}) 
%         &= \eps_1 ( (g_i^{-1}g_j)^{(a_i+a_j+n(1,g_i,g_j))} )\\
%         & \equiv a_i + a_j + n(1, g_i, g_j),
%     \end{split}
% \end{equation}

% which is equivalent to $\eps_1(g^{(a)}) = a$ defined in \cite{EF19} (we have assumed $n$ is normal, i.e. $n(1,1,g) = 0$).
Following the familiar group cohomology construction, we build an auxiliary bSPT state with symmetry $G'$ on a spatial manifold $M_2$ corresponding to the $3$-cocycle $\alpha_3$: 
%\begin{widetext}
\begin{equation}
\begin{split}
    &\ket{\Psi_\text{b}}\\
    &= \sum_{\{g_v\},\{a_v\}} \prod_{f=\langle 123 \rangle} e^ {2\pi i \alpha_3 (1,g_1^{(a_1)},g_2^{(a_2)},g_3^{(a_3)})O_{f}} \ket{\{g_v\},\{a_v\}} \\
    &=\sum_{\{g_v\},\{a_v\}}  \prod_{f=\langle 123 \rangle} \left[ e^{2 \pi i\nu(1,g_1,g_2,g_3) O_{f}} \times \right. \\
    & ~~~~ ~~~~ ~~~~ \left.  (-1)^{n(1,g_1,g_2) \boldsymbol{(}a_2+a_3 + n(1,g_2,g_3)\boldsymbol{)}} \right] \ket{\{g_v\},\{a_v\}}.
\end{split}
\label{eq:2d auxiliary SPT state}
\end{equation}
%\end{widetext}
where $O_{f}$ denotes the orientation of the face $f$. This state is invariant under multiplication of constant $h^{(b)}$ on all vertices, i.e $g_v^{(a_v)} \ra (h g_v)^{\boldsymbol{(}a_v + b + n(1,h,h g_v)\boldsymbol{)}}$. In other words, $\ket{\Psi_\text{b}}$ is invariant under the symmetry action:
\begin{equation}
    \begin{split}
     \ket{\{ g_v \} , \{ a_v \} } \ra \ket{\{ hg_v \} , \{ a_v + b + n(1,h,h g_v)\} },
    \end{split}
\label{eq: symmetry action aux bSPT}
\end{equation}
for all $h \in G$ and $b \in \ZZ_2$. By introducing the operators
\begin{eqs}
    \nuhat\boldsymbol{(} \langle 123 \rangle \boldsymbol{)} \ket{\{g_v\},\{a_v\}} &= \nu(1,g_1,g_2,g_3) \ket{\{g_v\},\{a_v\}} \\
    \nhat\boldsymbol{(} \langle 12 \rangle \boldsymbol{)} \ket{\{g_v\},\{a_v\}} &= \nu(1,g_1,g_2) \ket{\{g_v\},\{a_v\}},
\end{eqs}
and representing $\{a_v\}$ by a 0-cochain $\ba_v \in C^0(M_2,\ZZ_2)$, the SPT state can alternatively be written as:
\begin{eqs}
    \ket{\Psi_\text{b}}
    =\sum_{\{g_v\},\ba_v}  \prod_{f}  e^{2 \pi i\nuhat(f) O_{f}}     (-1)^{\int_{M_2} \nhat \cup (\delta \ba_v + \nhat) }\ket{\{g_v\},\ba_v}.
\end{eqs}

\begin{figure}[t]
\centering
\includegraphics[width=0.45\textwidth]{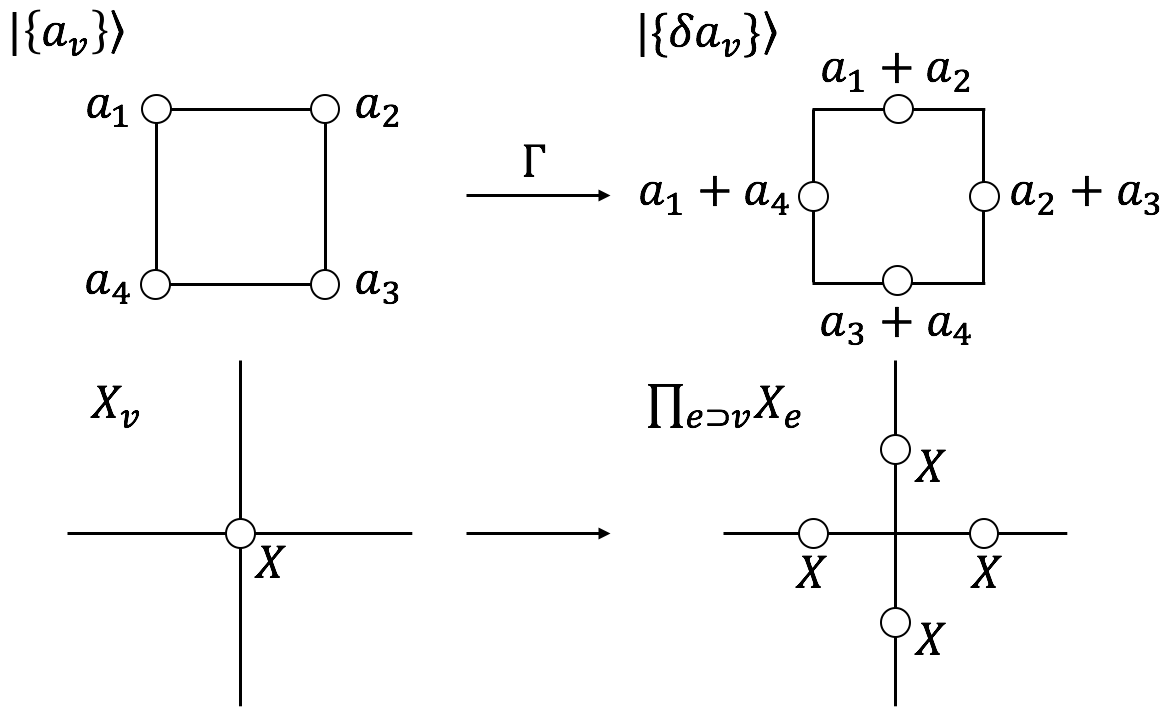}
\caption{The gauging map $\Gamma$ for 0-form $\ZZ_2$ symmetry. The Hilbert space $\mathcal{H}_1$ on the left contains $\ZZ_2$ spins at all vertices, with the symmetry constraint $\prod_v X_v =1$. The Hilbert space $\mathcal{H}_2$ on the right contains $\ZZ_2$ spins at all edges, with the gauge constraint $\prod_{e \subset f} Z_e = 1$ for each face $f$. For non simply-connected $M_2$, there are extra constraints that the product of $Z$ around nontrivial cycles are equal to $1$. The gauging map is an isomorphism between $\mathcal{H}_1$ and $\mathcal{H}_2$.}
\label{fig: 2d_gauging}
\end{figure}

The next step is to gauge the 0-form $\ZZ_2$ subgroup. This is a duality mapping $\{a_v\}$ d.o.f. on vertices to $\{a_e\}$ d.o.f. on the edges, where each $\{a_e\}$ configuration can be labeled by a $1$-cochain $\ba_e \in C^1(M_2,\ZZ_2)$. Since each configuration $\{a_v \}$ can be represented by the cochain $\boldsymbol a_v$,
the gauging map is defined as:
\begin{equation}
    \Gamma (\ket{ \boldsymbol a_v }) = \ket{ \delta \boldsymbol a_v  }
\end{equation}
where $\delta \boldsymbol a_v$ are $\ZZ_2$ fields living on edges, i.e. $\delta \boldsymbol a_v(e_{ij}) = a_i + a_j$, shown in Fig. \ref{fig: 2d_gauging}.
The ground state of the bosonic shadow model is:
\begin{equation}
    \begin{split}
        &\ket{\Psi_s} = \Gamma (\ket{\Psi_\text{b}}) \\
        &=\sum_{\{g_v\},\ba_v}  \prod_{f} \left[ e^{2 \pi i\nuhat(f) O_{f}}  \right]    (-1)^{\int_{M_2} \nhat \cup (\delta \ba_v + \nhat)} \ket{\{g_v\}, \delta \ba_v}.
    \end{split}
\label{eq: 2d bosonic shadow}
\end{equation}
% %\begin{widetext}
% \begin{equation}
%     \begin{split}
%         &\ket{\Psi_s} = \Gamma (\ket{\Psi_{b}}) \\
%         =&\sum_{\{g_v\},\{a_v\}} \prod_{f=\langle 123 \rangle} \left( e^{2 \pi i\nu(1,g_1,g_2,g_3) O_f} \times \right. \\
%         & ~~~~ ~~~~ ~~~~ \left.  (-1)^{n(1,g_1,g_2) (a_2+a_3 + n(1,g_2,g_3)}) \right) \ket{\{g_v\},\delta \ba_v}\\
%         =&\sum_{\{g_v\},\{a_v\}} \prod_{f=\langle 123 \rangle} \left( e^{2 \pi i\nu(1,g_1,g_2,g_3) O_f}  (-1)^{\bar{n} (12) (a_2+a_3 + \bar{n}(23))} \right) \\
%         & ~~~~ ~~~~ ~~~~ ~~~~ \ket{\{g_v\}, \delta \ba_v + \bar n }' \\
%     \end{split}
% \label{eq: 2d bosonic shadow}
% \end{equation}
% %\end{widetext}
Now, we define a new basis of states by a unitary transformation $\mathcal{R} \equiv \prod_e X_e^{\nhat(e)}$:
\begin{eqs}
    \ket{\{g_v\}, \ba_e}^\prime &\equiv \mathcal{R }\ket{\{g_v\},\ba_e } \\
    &=\ket{\{g_v\},\ba_e + {\bar{\boldsymbol{n}}}_{\scriptscriptstyle{\{g_v\}}}},
\label{eq: 2d bulk basis transformation}
\end{eqs}
where ${\bar{\boldsymbol{n}}}_{\scriptscriptstyle{\{g_v\}}}$ is the $1$-cochain given by:
\begin{align}
    {\bar{\boldsymbol{n}}}_{\scriptscriptstyle{\{g_v\}}}\boldsymbol{(} \langle 12 \rangle \boldsymbol{)} \equiv n(1, g_1, g_2).
\end{align}
% We can introduce the variables $ b_e \equiv a_i + a_j + n(1, g_i, g_j)$ ($\ZZ_2$ fields living on edges) and Pauli operators $Z_e \equiv (-1)^{b_e}$ which measures the $b_e$ variables.
One can check that the symmetry action \eqref{eq: symmetry action aux bSPT} acting on this new basis becomes
\begin{equation}
    \begin{split}
        V(h^{(b)}):~\ket{\{ g_v \}, \ba_e }^\prime \ra \ket{\{ hg_v \}, \ba_e }^\prime,
    \end{split}
\label{eq: new symmetry of 2d bosonic shadow}
\end{equation}
which is an onsite symmetry acting only on the vertex variables $\{g_v\}$.
Pauli matrices $X_{e'}$ and $Z_{e'}$ are defined to be acting on the second entry of states in the new basis, i.e.,
\begin{equation}
    \begin{split}
    X_{e'} \ket{\{ g_v \} , \boldsymbol a_e }^\prime &= \ket{\{ g_v \} , \boldsymbol a_e + \boldsymbol e' }^\prime \\
    Z_{e'} \ket{\{ g_v \} , \boldsymbol a_e}^\prime &= (-1)^{\boldsymbol a_e(e')} \ket{\{ g_v \} , \boldsymbol a_e }^\prime.
    \end{split}
\end{equation}
The bosonic shadow state with an onsite $G$ symmetry is then:
\begin{equation}
    \begin{split}
        &\ket{\Psi_s} =\sum_{\{g_v\},\ba_v}  \prod_{f}  e^{2 \pi i\nuhat(f) O_{f}}  \prod_{e'}  Z_{e'}^{\int_{M_2} \nhat \cup \be' } 
        \ket{\{g_v\}, \delta \ba_v + \nhat}' \\
        &=\sum_{\{g_v\},\ba_v}  \prod_{f}  e^{2 \pi i\nuhat(f) O_{f}}    \prod_{e'}  Z_{e'}^{\int_{M_2} \nhat \cup \be' } \prod_{e} X_{e}^{\nhat(e)}
        \ket{\{g_v\}, \delta \ba_v}'. 
    \end{split}
\end{equation}

Finally, we apply the $(2+1)$D fermionization duality of Refs.~\cite{CK18,CKR18} to $\ket{\Psi_\text{b}}$ to find the fSPT state. To apply the duality to $\ket{\Psi_\text{b}}$, we introduce the notation [similar to Eq.~\eqref{eq: U lambda}]:
\begin{eqs}
        \bar{U}_{\bar n} = \prod_{e^\prime} Z_{e^\prime}^{\int  \nhat \cup \boldsymbol e^\prime} \prod_e X_e^{\nhat (e)}.
\end{eqs}
With this, the shadow state is:
\begin{eqs}
\ket{\Psi_\text{s}} = \sum_{\{g_v\},\{a_v\}} \prod_{f\in M_2}  e^{2 \pi i\nuhat(f) O_{f}} \bar U_{\bar n} \ket{\{g_v\}, \delta \ba_v }',
\end{eqs}
The operator $\bar U_{\bar n}$ is dual to the fermionic operator $S_{\bar n}$, defined as: 
\begin{align} \label{Sbarn}
    S_{\bar n} = \xi_{\bar{n}}(M_2) \prod_e S_e^{\nhat(e)},
\end{align}
where $\xi_n(M_2)$ is a sign that depends on the order of edges in $M_2$:
\begin{eqs}
    \xi_{\bar{n}}(M_2) = \prod_{e_i, e_{i'} \in M_2| i<i'} (-1)^{\nhat(e) \nhat(e') \int \be_i \cup \be_{i'} },
\end{eqs}
and $S_e$ is the following fermionic hopping operator:
\begin{align}
    S_e = (-1)^{\be(E)} i \gamma_{L(e)} \gamma'_{R(e)}.
\end{align}
Here, $E \in C_1(M_2,\ZZ_2)$ is a choice of spin structure, and $L(e)$,$R(e)$ are the faces to the left and right of the edge $e$, respectively \cite{CK18,C19-2,EF19}.
The supercohomology state is thus:
\begin{equation}
    \begin{split}
        &\ket{\Psi_f} 
        = \sum_{\{g_v\}} \prod_{f} e^{2 \pi i \nuhat(f) O_f}  S_{\bar n} \ket{\{g_v\}, \text{vac}},
    \end{split}
\end{equation}
where $\ket{\text{vac}}$ is the fermionic state with trivial fermion occupancy.
% Furthermore, $S_{\bar n} = \xi_{n}(M_2) \prod_e S_e^{\nhat(e)} $ is defined by:
% \begin{eqs}
%     \xi_{n}(M_2) = \prod_{e_i, e_{i' \in M_2| i<i'}} (-1)^{\nhat(e) \nhat(e') \int \be_i \cup \be_{i'} },
% \end{eqs}
% with the hopping operator:
% \begin{align}
%     S_e = (-1)^{\be(E)} i \gamma_{L(e)} \gamma'_{R(e)}
% \end{align}
% and $E \in C_1(M_2,\ZZ_2)$ the spin structure \cite{CK18,C19-2,EF19}.
The toric code ground state is mapped to the fermionic vacuum state, since the toric code Hamiltonian is mapped to the atomic insulator $H_f = - \sum_f P_f$. The symmetry in the fermionic model acts as:
\begin{equation}
    V(h): \ket{\{ g_v \} , \cdots } \ra \ket{\{ h g_v \} , \cdots },
\end{equation}
because the symmetry in Eq.~\eqref{eq: new symmetry of 2d bosonic shadow} only involves the $G$ degrees of freedom.
% and doesn't affect the $\ZZ_2$ part.

\subsection{Boundary construction}\label{app:2Dgappedboundary}

In this section, we demonstrate how to construct SPT states for (2+1)D supercohomology phases on a spatial manifold $M_2$ with boundary $\partial M_2$. The construction is based on extending the symmetry at the boundary, similar to Section~\ref{sec: fermionic gapped boundary construction} and Ref.~\cite{WWW19}. 
% The construction is simpler than the (3+1)D construction in the main text. 
In fact, since $\alpha_3 \in H^3(G',\RR/\ZZ)$ in Eq.~\eqref{eq: alpha3} describes a bSPT phase, we can apply the methods of Ref.~\cite{WWW19} to first build a gapped boundary for the auxiliary bSPT. To this end, we identify an extension of $G'$ to $G''$ that trivializes $\alpha_3$, i.e., for some $\mu_2$ in $C^2(G'',\RR/\ZZ)$, we have $\delta \mu_2 = \alpha^*_3$, where $\alpha^*_3$ is the pullback of $\alpha_3$. The extension is such that we can then gauge a $\ZZ_2$ subgroup to construct the shadow model and then fermionize according to Ref.~\cite{CKR18} to build the supercohomology state on $M_2$.   

In analogy to Section~\ref{sec: fermionic gapped boundary construction}, we find such an extension by composing two central extensions of $G$. The first trivializes $n$, so that $n'=\delta \beta'$, where $n'$ is the pullback to an extended group $L'$,
and $\beta'$ belongs to $C^1(L',\ZZ_2)$. The second extension trivializes the cocycle ${\nu'+\frac{1}{2}\beta' \cup \delta \beta'}$, with $\nu'$ the pullback of $\nu$ to $L'$. The existence of these two extensions is guaranteed by the results of Ref.~\cite{T17-2}. Therefore, there is an extension of $G$ to $L$ by $K$:
\begin{align}
     1 \ra K \ra L \xrightarrow[]{} G \ra 1,
\end{align}
such that the pullback of the supercohomology data $(n^*,\nu^*)$ is trivialized:
\begin{align} \label{2dpullback}
    (n^*,\nu^*) = (\delta \beta, \delta \eta + \frac{1}{2}  \beta \cup \delta \beta),
\end{align}
for some $\beta \in C^1(L,\ZZ_2)$ and $\eta \in C^2(L,\RR/\ZZ)$.
% The auxiliary bSPT model is constructed using the cocycle $\alpha_3 \in Z^3(G',\RR/\ZZ)$ in Eq. \eqref{eq: alpha3}, where $G'$ is a $\mathbb Z_2$ extension of $G$ given by $n_2$. 
% From the result of Ref. \cite{WWW19}, there always exists an extension to trivialize $\alpha_3$. That is, there exists a group extension:
% \begin{equation}
%     1 \ra K \ra G'' \xrightarrow[]{t'} G \ra 1,
% \label{eq: second group extension}
% \end{equation}
% which trivializes the supercohomology data:
% \begin{equation} 
%     (n^*, \nu^*) = (\delta \beta, \delta \eta + \frac{1}{2}  \beta \cup \delta \beta),
% \end{equation}
% with $n^*$ and $\nu^*$ denoting the pullbacks of $n$ and $\nu$ to $G''$, and with $\eta \in C^2(G'',\RR/\ZZ)$ and $\beta \in C^1(G'',\ZZ_2)$. 
Note that the right-hand side of Eq.~\eqref{2dpullback} corresponds to a trivial set of supercohomology data \cite{EF19}. Elements of the group $L$ can be written as $\ell = g^{(k)}$, where $g$ is in $G$ and $k$ is in $K$. The group law in $L$ is defined by a $2$-cocycle $m \in H^2(G,K)$:
\begin{align}
    g_1^{(k_1)}g_2^{(k_2)}=(g_1g_2)^{\boldsymbol{(}k_1 + k_2 + m(1,g_1,g_1g_2)\boldsymbol{)}},
\end{align}
with $K$ taken to be an additive group.

We claim that the pullback of $\alpha_3$ to $G''$, with $G''$ defined as:
\begin{align} \label{G''defs}
    G'' \equiv \ZZ_2 \times_{n^*} L = K \times_{m^*} G',
\end{align}
is trivial in $H^3(G'',\RR/\ZZ)$. In Eq.~\eqref{G''defs}, we have used ${\ZZ_2 \times_{n^*} L}$ to mean the extension of $L$ by $\ZZ_2$ corresponding to the pullback of $n$ to $L$, and we have used ${K\times_{m^*} G'}$ to denote the extension of $G'$ by $K$ corresponding to $m$ pulled back to $G'$. 
Using this, we find that the pullback of $\epsilon_1$ is $\eps_1^*\in C^1(G'',\ZZ_2)$:
\begin{equation}
\epsilon_1^*({\ell_1}^{(a_1)},{\ell_2}^{(a_2)}) = a_1 + a_2 + \beta(\ell_1,\ell_2) + \delta \bar \beta(\ell_1,\ell_2) ,
\end{equation}
where ${\ell_1}^{(a_1)},{\ell_2}^{(a_2)}$ are elments of $G''$ labeled by ${\ell_1},{\ell_2}$ in $L$ and $a_1,a_2$ in $\ZZ_2$, $\bar \beta(\ell_1)$ is defined by $\bar \beta(\ell_1) \equiv \beta(1,\ell_1)$, and $\beta$ is implicitly pulled back to $G''$.
Inserting this into the pullback of $\alpha_3$, we obtain:
\begin{align}
    \alpha_3^* &= \delta \eta + \frac{1}{2}  \beta \cup \delta \beta + \frac{1}{2} \delta \beta \cup \eps_1^* \nonumber\\
    &= \delta (\eta + \frac{1}{2} \beta \cup \epsilon_1^* ).
\end{align}
Hence, we can define:
\begin{equation}
    \mu_2 \equiv \eta + \frac{1}{2} \beta \cup \epsilon_1^*,
\label{eq: mu_2}
\end{equation}
such that $\alpha_3^* = \delta \mu_2$.

Following the discussion for bosonic SPT phases in Section \ref{sec: group coho boundary}, 
we extend the symmetry on the boundary to $G''$ by adding extra degrees of freedom $k_{v_\partial} \in K$ 
to each boundary vertex $v_\partial \in \partial M_2$.
We see that the symmetry action for each $h^{(k)} \in L$ is
\begin{equation}
    \begin{split}
        V(h^{(k)}): 
        &g_v \ra h g_v, \quad \forall v \in M_2, \\
        &{k_v} \ra {k + k_v + m(h^{-1},1,g)}, \quad \forall v \in \partial M_2 ,  \\
        &a_v \ra a_v + n(h^{-1}, 1, g_i), \quad \forall v \in M_2.
    \end{split}
\end{equation}
There is also a $0$-form $\ZZ_2$ symmetry (analogous to the $1$-form symmetry in the $(3+1)$D construction):
\begin{equation}
    \begin{split}
        A: a_v &\ra a_v + 1, \quad \forall v \in M_2.
    \end{split}
\label{eq: 2d Z2 symmetry}
\end{equation}
The symmetric wave function on $M_2$ is then:
\begin{equation}
    \begin{split}
        \ket{ \Psi_\text{b} } = &\sum_{\{ g_v, a_v,k_{v_\partial} \}}~~ \prod_{e_{12} \in \partial M_2} e^{-2\pi i \mu_2 (1, g_1^{(k_1)}, g_2^{(k_2)})O_{e_{12}}} \\
        &\prod_{f=\lr{123} \in M_2} e^{2\pi i \alpha_3 (1,g_1', g_2', g_3') O_f} \ket{ \{ g_v \},\{ k_{v_\partial} \}, \ba_v },
    \end{split} 
    \label{equ:2Dwavefunctionwithboundary}
\end{equation}
where $O_{e_{12}}$ is the orientation of the boundary edge $e_{12}$. Using the operators introduced in the previous section, $\ket{\Psi_\text{b}}$ can be written as:
\begin{eqs}
       &\ket{ \Psi_\text{b} } = \\
       &\sum_{\{g_v\},\{k_{v_\partial}\}, \ba_v}~~ \prod_{e \in \partial M_2}
        \left[ e^{-2\pi i \etahat(e) O_e }  (-1)^{\int_{\partial M_2} \betahat \cup (\delta \ba_v + \nhat) }
        \right]  \\
        & \prod_{f\in M_2} \left[ e^{2 \pi i\nuhat(f) O_{f}}  \right]    (-1)^{\int_{M_2} \nhat \cup (\delta \ba_v + \nhat) } \ket{\{g_v\}, \{ k_{v_\partial}\}, \ba_v },
\end{eqs}
where $\etahat$ and $\betahat$ are defined as:
\begin{eqs}
&\etahat\boldsymbol{(}\langle 12\rangle \boldsymbol{)} |\{g_v\},\{k_{v_\partial}\}, \ba_v\rangle =  \\ &~~~~~~~~~~~~~~~~~~~~\,~~~~\,~~~~~~~~~~\eta(1,g^{(k_1)}_1,g^{(k_2)}_2)  \ket{\{g_v\},\{k_{v_\partial}\},\ba_v}, \\
        &\betahat\boldsymbol{(}\langle 1 \rangle \boldsymbol{)} \ket{\{g_v\},\{k_{v_\partial}\},\ba_v} = \beta(1,g^{(k_1)}_1) \ket{\{g_v\},\{k_{v_\partial}\},\ba_v}.
\end{eqs}

We now gauge the $\ZZ_2$ symmetry in Eq.~\eqref{eq: 2d Z2 symmetry} to obtain the bosonic shadow model. To do so, we consider the manifold $\overline{M}_2$, formed by connecting all boundary vertices of $M_2$ to an additional vertex $0$. In other words, $\overline{M}_2$ is $M_2 \coprod C(\partial M_2)$, where the cone of $\partial M_2$, $C(\partial M_2)$, is attached to $M_2$. We define $a_0=0$ for this extra vertex and apply the gauging map $\Gamma$ on the simply connected manifold $\overline{M}_2$. The bosonic shadow state is $\ket{\Psi_{s}} \equiv \Gamma(\ket{\Psi_\text{b}})$, defined on a Hilbert space with $\{a_e\}$-configurations on edges labeled by a $1$-cochain $\ba_e \in C^1(\overline{M}_2,\ZZ_2)$.
% \begin{equation}
% \begin{split}
%     &\ket{\Psi_{s}} \equiv \Gamma(\ket{\Psi_\text{b}}).
%     % \\
%     % =&\sum_{\{g_v\},\{a_v\},\{k_{v_\partial}\}} \prod_{e=\lr{12}\in \partial M_2}
%     % \left[ e^{-2\pi i \eta (1,g''_1,g''_2) O_e } \right.\\
%     % & ~~~~ ~~~~ ~~~~ ~~~~ ~~~~ ~~~~
%     % \left. \times (-1)^{\beta(1,g''_1) (\delta \ba_v (12) + \bar n (12))}
%     % \right]  \\
%     % & \prod_{\substack{f=\langle 123 \rangle \\ \in M_2}} \left[ e^{2 \pi i\nu(1,g_1,g_2,g_3) O_{f}}  (-1)^{n(1,g_1,g_2) (\delta \ba_v (23) + \bar n(23) ) }  \right] \\
%     % &\ket{\{g_v\}, \{ k_{v_\partial}\}, \delta \ba_v }.
% \label{eq: 2d bosonic shadow 1}
% \end{split}
% \end{equation}

As in the bulk construction, we perform a change of basis to make the symmetry onsite. Including the edges $e=\lr{0i}$, for all $\lr{i} \in \partial M_2$, the basis transformation is:
% \begin{equation}
%     \ket{ \{ g_v \}, \{  k_{v_\partial} \}, \boldsymbol c_1} = \ket{ \{ g_v \}, \{  k_{v_\partial} \}, \boldsymbol c_1 + \boldsymbol B_e}' ~ \forall \boldsymbol c_1 \in C^1(M_2,\ZZ_2),
% \end{equation}
% \textcolor{orange}{Are the $\cdots$'s just the $\ba_e$ configuration?}
% \Yuan{Any congfiguration on edges. There is no $a_e$ in 2d. We may use $c_e$ representing any 1-cochain}
% \textcolor{orange}{Why is there no $\ba_e$ in 2d? Maybe we can discuss tonight}
\begin{equation} \label{basis trans 2d bdry}
   \ket{ \{ g_v \}, \{  k_{v_\partial} \}, \ba_e}' \equiv \ket{ \{ g_v \}, \{  k_{v_\partial} \}, \ba_e+ \boldsymbol{B}_e},
\end{equation}
where $\boldsymbol B_e$ is defined as:
\begin{equation}
    \begin{split}
        \boldsymbol B_e (\lr{ij}) &= n (1, g_i,g_j)  \\
        \boldsymbol B_e (\lr{0i}) &= \beta (1, g''_i).
    \end{split}
\end{equation}
We define Pauli operators in the new basis by:
\begin{eqs}
    X_{e'} \ket{ \{ g_v \}, \{  k_{v_\partial} \}, \boldsymbol a_e}' &\equiv  \ket{ \{ g_v \}, \{  k_{v_\partial} \}, \boldsymbol a_e + \be'}' \\
    Z_{e'} \ket{ \{ g_v \}, \{  k_{v_\partial} \}, \boldsymbol a_e}' &\equiv (-1)^{\boldsymbol a_e(e')} \ket{ \{ g_v \}, \{  k_{v_\partial} \}, \boldsymbol a_e}'.
\end{eqs}

After applying the basis transformation in Eq.~\eqref{basis trans 2d bdry}, the bosonic shadow state is:
\begin{equation}
\begin{split}
    \ket{\Psi_{s}} = &\sum_{\{g_v\},\{k_{v_\partial}\}, \ba_v} \prod_{e \in \partial M_2}
    \left[ e^{-2\pi i \etahat(e) O_e}\right] \bar U_{\bar \beta}\\
    &\prod_{f \in M_2} \left[ e^{2 \pi i\nuhat(f) O_{f}}\right] \bar U_{\bar n}
    \ket{\{g_v\}, \{ k_{v_\partial}\}, \delta \ba_v }'.
\label{eq: 2d bosonic shadow 2}
\end{split}
\end{equation}
% \begin{equation}
% \begin{split}
%     &\ket{\Psi_{s}}  \\
%     =&\sum_{\{g_v\},\{k_{v_\partial}\}, \ba_v} \prod_{e = \lr{12} \in \partial M_2}
%     e^{-2\pi i \etahat (12) O_e}
%     Z_{12}^{\betahat (1)}
%      \\
%     & \prod_{f=\langle 123 \rangle \in M_2} \left[ e^{2 \pi i \nuhat(123) O_{f}}  Z_{23}^{\nhat(12)} \right] \\
%     &\prod_{{v_\partial} \in \partial M_2} X_{\lr{0{v_\partial}} }^{\betahat( {v_\partial})}
%     \prod_{e\in M_2} X_e^{\nhat(e)}\ket{\{g_v\}, \{ k_{v_\partial}\}, \delta \ba_v }'\\
%     =& \sum_{\{g_v\},\{k_{v_\partial}\}, \ba_v} \prod_{e=\lr{12}\in \partial M_2}
%     \left[ e^{-2\pi i \etahat(12) O_e}\right] \bar U_{\beta}\\
%     & \prod_{f=\langle 123 \rangle \in M_2} \left[ e^{2 \pi i\nuhat(123) O_{f}}\right] \bar U_{\bar n}
%     \ket{\{g_v\}, \{ k_{v_\partial}\}, \delta \ba_v }',
% \label{eq: 2d bosonic shadow 2}
% \end{split}
% \end{equation}
Above, $\bar{U}_{\bar n}$ is the product of Pauli Z and Pauli X operators:
\begin{eqs}
        \bar{U}_{\bar n} = \prod_{e^\prime} Z_{e^\prime}^{\int_{M_2}  \nhat \cup \boldsymbol e^\prime} \prod_{e \in M_2} X_e^{\nhat (e)},
\end{eqs}
and $\bar{U}_{\bar \beta}$ is the operator:
\begin{equation}
    \bar{U}_{\bar \beta} = \prod_{\lr{12} \in \partial M_2} Z_{12}^{\betahat \boldsymbol{(}\lr{1} \boldsymbol{)}} 
    \prod_{{v_\partial} \in \partial M_2} X_{0 {v_\partial}}^{\betahat({v_\partial})}.
\end{equation}

The operators $\bar{U}_{\bar n}$ and $\bar{U}_{\bar \beta}$ in the expression for the shadow state can be fermionized following Refs.~\cite{CK18,CKR18}. $\bar{U}_{\bar n}$ maps to the product of hopping operators $S_{\bar n}$ in Eq.~\eqref{Sbarn}:
\begin{eqs}
    S_{\bar n} = \xi_{\bar{n}} (M_2) \prod_{e\in M_2} S_e^{\nhat(e)},
\end{eqs} 
while $\bar{U}_{\bar \beta}$ fermionizes to:
\begin{eqs}
\prod_{{v_\partial} \in \partial M_2} S_{ {v_\partial}}^{\betahat( {v_\partial})}.
\end{eqs}
Here, $S_{v_\partial}$ is shorthand for the operator $S_{0 v_\partial}$, and we note that there is no order dependence of the product of $S_{v_\partial}$ operators, since the $S_{v_\partial}$ are all commuting.

The (2+1)D boson-fermion duality maps the bosonic shadow state to the supercohomology state:
\begin{equation}
\begin{split}
    \ket{\Psi_{f}} = \sum_{\{g_v\},\{k_{v_\partial}\}} &\prod_{\substack{e  \in \partial M_2}}
    \left[ e^{-2\pi i \etahat(e) O_e}\right] \prod_{{v_\partial} \in \partial M_2} S_{ {v_\partial}}^{\betahat( {v_\partial})}\\
    & \prod_{\substack{f \in M_2}} \left[ e^{2 \pi i\nuhat(f) O_{f}}\right] \xi_n (M_2) \prod_{e\in M_2} S_e^{\nhat(e)}\\
    &~~~~~~~~~~~~~~~~~~~~~~~~~~~~~~~~\ket{\{g_v\}, \{ k_{v_\partial}\}, \text{vac} }.
\label{eq: 2d boundaryfermionic SPT}
\end{split}
\end{equation}
$\ket{\Psi_{f}}$ is defined on $M_2$ with a spinless complex fermion at each face $f$ and at each edge $e \in \partial M_2$.

\section{2-groups and 2-gauge theories}\label{sec: 2-group extension}

{The goal of this appendix is to define $2$-groups and $2$-gauge theory \cite{KT15}. In Section~\ref{sec: 2-group}, we give a formal definition of strict 2-groups and argue that certain equivalence classes of strict $2$-groups, i.e., weak 2-groups, can be labeled by a $3$-cocycle. In Section~\ref{sec: 2-gauge theory}, we describe 2-gauge theories \cite{KT15}, built from 2-groups. The machinery of $2$-gauge theory is used to derive the $2$-group SPT model and symmetric gapped boundaries of supercohomology models in the subsequent appendix, Appendix~\ref{app: super from 2gauge}. }

% \textcolor{gray}{The goal of this appendix is to define $2$-groups and $2$-gauge theory \cite{KT15}, and the $2$-group SPT model in terms $2$-gauge theory. In Section~\ref{sec: 2-group}, we give a formal definition of strict 2-groups and argue that equivalence classes of strict $2$-groups, i.e., weak 2-groups, can be labeled by certain $3$-cocycles.
% In Section~\ref{sec: 2-gauge theory}, we describe 2-gauge theories \cite{KT15}, built from 2-groups. In Section~\ref{app: super from 2gauge}, we then study the symmetry action of the bosonic 2-group SPT states from the point of view of 2-gauge theories, for both closed manifolds and manifolds with boundaries. We have constructed the symmetric gapped boundary for supercohomology models on a manifold with boundary via a group extension.}

\subsection{Definition of $2$-groups}\label{sec: 2-group}
% \vspace{1mm}
% \noindent \begin{center}\emph{Definition of $2$-groups:}\end{center}
% \vspace{1mm}
% \subsection{2-group and 2-group extension}\label{sec: 2-group}

{We first define strict $2$-groups before describing the notion of weak $2$-groups, also simply referred to as $2$-groups, which are used directly in the definition of $2$-gauge theory in the next section. Following Ref.~\cite{KT15}, a strict 2-group is given by $\GG=(G',A',t',\alpha')$, where $G'$ and $A'$ are groups, $t':A'\to G'$ is a group homomorphism, and $\alpha': G' \to \text{Aut}(A')$ is an action of $G'$ on $A'$. Furthermore, for any $g'\in G'$ and $a',a'_1,a_2' \in A'$, the maps $t'$ and $\alpha'$ satisfy:
\begin{equation}
     t'\boldsymbol{(}\alpha'[g'](a')\boldsymbol{)}=g' t'(a') {g'}^{-1},\quad \alpha' [t'(a_1')](a_2^\prime)= a'_1 a_2^\prime {a'}_1^{-1}.
\end{equation}}

{Weak $2$-groups, on the other hand, correspond to certain equivalence classes of strict $2$-groups sharing the same $\ker{t'}$ and $\coker{t'}=G'/\Ima t'$. Given any strict $2$-group $(G',A',t',\alpha')$ with $\ker{t'}=A$ and $\coker{t'}=G$,
it can be organized into the short exact sequence:
\begin{equation}
    1 \ra A \xrightarrow[]{\iota} A^\prime \xrightarrow[]{t^\prime} G^\prime \xrightarrow[]{\pi} G \ra 1,
\label{eq:double extension}
\end{equation}
also known as a double extension of $G$ by $A$. The double extensions of $G$ by $A$ can be labeled by an action of $G$ on $A$ given by a map $\alpha:G \to \text{Aut}(A)$ (induced from $\alpha'$) and a cocycle $\varrho \in H^3(G,A)$. A weak $2$-group is specified by the quadruple $(G,A,\alpha,\varrho)$.}

{We find the prescription, described below, for determining the cocycle $\varrho$ associated to the double extension illuminating.
% We find it particularly illuminating to see how the group cocycle $\varrho$ can be determined from a double extension. 
For example, we take $A$ to be $\ZZ_2$, in which case, $\alpha$ is trivial, and consider the double extension of $G$ by $\ZZ_2:$
\begin{equation}
    1 \ra \ZZ_2 \xrightarrow[]{\iota} A^\prime \xrightarrow[]{t^\prime} G^\prime \xrightarrow[]{\pi} G \ra 1,
\label{eq:double extension2}
\end{equation}
with the action $\alpha':G' \to \text{Aut}(A')$.
First, any section $s:G \ra G^\prime$ satisfies the group law projectively:
\begin{equation}
    s(g)s(h)=f(g,h)s(gh)
\end{equation}
with $f: G \times G \ra \ker \pi$. Furthermore, by the associative property, $f$ must obey:
\begin{equation}
    [s(g)f(h,k)s(g)^{-1}]f(g,hk)=f(g,h)f(gh,k).
\label{eq:associative}
\end{equation}
Finally, since $\Ima t^\prime = \ker \pi$, we can lift $f$ to $F:G\times G \ra A^\prime$, where Eq.~\eqref{eq:associative} is satisfied projectively:
\begin{equation}
    \boldsymbol{(}\alpha^\prime[s(g)]F(h,k)\boldsymbol{)}F(g,hk)= \iota\boldsymbol{(}\varrho(g,h,k)\boldsymbol{)} F(g,h)F(gh,k).
\end{equation}
It can be shown that $\varrho$ is a 3-cocycle, whose cohomology class is independent on the choice of $s$ and $F$, and therefore the double extension can be labeled by $H^3(G,A)$.}
We point out that the calculation of $\varrho$ here is analogous to the calculation of the $G$-symmetry fractionalization on loop-like excitations of the shadow model in Section~\ref{sec: shadow model} \cite{EN15}.

\subsection{Review of $2$-gauge theory} \label{sec: 2-gauge theory}
% \vspace{1mm}
% \noindent \begin{center}\emph{Review of $2$-gauge theory}\end{center}
% \vspace{1mm}

% \subsection{2-gauge theory on a lattice} \label{sec: 2-gauge theory}
We now define a 2-gauge theory, using the data $(G,A,\alpha,\varrho)$ described in the previous section \cite{KT15}. For our purposes, we consider $A$ to be $\ZZ_2$, in which case, only a trivial $\alpha$ 
% action ($G$ on $\ZZ_2$) 
exists. {The $2$-gauge theory is defined on a discrete spacetime manifold $N$ with a branching structure,} and the field configurations are given by an assignment of an element $g_e \in G$ to each edge and an element $b_f\in \ZZ_2$ to each face. 
% A branching structure on the triangulation is chosen. 
We denote a configuration of $\{ g_e\}$ and $\{b_f\}$ fields by $(\bar{g},\bar{b})$, where $\bar{g}$ and $\bar{b}$ are shorthand for $ \{ g_e\}$ and $\{b_f\}$, respectively. The allowable field configurations must satisfy the following ``flatness condition''. First, for every  
% triangle $\Delta_{012}$
face $f=\langle 012 \rangle$, $\bar{g}$ must satisfy:
\begin{equation}
    g_{01} g_{12} = g_{02}.
\label{eq: constraint 1}
\end{equation}
Second, on each tetrahedron, $(\bar{g},\bar{b})$ has to satisfy:
\begin{equation}
    \delta b = b_{012}+b_{013}+b_{023}+b_{123}=\varrho(g_{01},g_{12},g_{23}).
\label{eq: constraint 2}
\end{equation}

The 0-form gauge transformations are parameterized by $\bar{h} = \{h_v\}$, i.e., a choice of $h_v \in G$ for every vertex $v$, and are defined by:
{\begin{equation}
    \begin{split}
        \bar{g}&\ra \bar{g}^h = \{g^h_{ij}\} = \{h^{-1}_i g_{ij} h_j\} \\
        \bar{b}&\ra \bar{b}^h = \{b^h_{ijk}\} = \{b_{ijk} + \zeta(g_{ij},g_{jk},h_i,h_j,h_k)\}
    \end{split}
\label{eq: 0-form gauge transformation}
\end{equation}}
% \begin{equation}
%     \begin{split}
%         \bar{g}\ra \bar{g}^h : \,& g^h_{ij} = h^{-1}_i g_{ij} h_j \\
%         \bar{b}\ra \bar{b}^h : \,& b^h_{ijk} = b_{ijk} + \zeta(g_{ij},g_{jk},h_i,h_j,h_k)
%     \end{split}
% \label{eq: 0-form gauge transformation}
% \end{equation}
where $\zeta$ is a $\ZZ_2$-valued $2$-cochain on $N$ with the property:
\begin{equation}
    \delta \zeta (\bar{g},\bar{h})=\rho(\bar{g}^h)-\rho(\bar{g}).
\label{eq: zeta equation}
\end{equation}
% where we take $\zeta$ as a 2-cochain.
Here and below, we use the homogeneous notation $\rho$ for the cocycle $\varrho$, i.e. $\rho(1,g,gh,ghk)=\varrho(g,h,k)$.
{Heuristically, one can show that a solution for $\zeta$ in Eq.~\eqref{eq: zeta equation} always exists by considering $\rho$ as a label for a ``Dijkgraaf-Witten-type'' theory (see Section~\ref{sec:zeta}). }
% \textcolor{gray}{A solution for $\zeta$ in Eq.~\eqref{eq: zeta equation} always exists since we can consider $\rho$ as the label for ``Dijkgraaf-Witten-type'' theory (gauge theories are classified by the cocycle $\rho$).}
Therefore, the $0$-form gauge transformation only changes the values of $\varrho$ in Eq.~\eqref{eq: constraint 2} by coboundary terms.
In what follows, we take $\zeta$ to be:
% Our choice of $\zeta$ is (derived in the following section \ref{sec:zeta})
\begin{equation}
    \begin{split}
    &\zeta(g_{12},g_{23},h_1,h_2,h_3) =\rho(1,g_{12},g_{12}g_{23},g_{12}g_{23}h_3) +\\ & \rho(1,g_{12},g_{12}h_2,g_{12}g_{23}h_3) + \rho(1,h_1,g_{12}h_2,g_{12}g_{23}h_3),
    \end{split}
\label{eq: zeta}
\end{equation}
derived in Section~\ref{sec:zeta}.
In particular, we note that for the trivial $G$ configuration $g_e=1$, $\forall e$, the gauge transformation on the $\bar{b}$ fields yields:
\begin{equation}
    \zeta(1,1,h_1,h_2,h_3) = \rho (1,h_1,h_2,h_3).
\end{equation}
The 1-form symmetry depends on a $\ZZ_2$-valued 1-cochain $\bar{\lambda}=\{ \lambda_e \}$. The transformation is as usual:
\begin{equation}
    \begin{split}
        g_e &\ra g_e \\
        b_f &\ra b_f + \delta \lambda_e. 
    \end{split}
\label{eq: 1-form gauge transformation}
\end{equation}
% The theory defined as such is called a 2-gauge theory. 
% We emphasize that the space configuration $d\Phi$ defined in Fig. \ref{fig: CM_dPhi} can be interpreted as applying a gauge transformation $\Phi$ on the trivial spacetime configuration.

The classifying space of a $2$-group $\GG$, denoted as $B\GG$ allows us to define a 2-gauge theory via the map $M \ra B\GG$. It can be described by a $\Delta$-complex structure. $B \GG$ contains one vertex and edges labeled by $g \in G$. Its $2$-simplices $\langle 012 \rangle$ are labeled by $(g_{01},g_{12},g_{02},b_{012})$ such that $g_{01} g_{12} = g_{02}$ and $b_{012} = 0,1$. Its $3$-simplices $\langle 0123 \rangle$ contains boundary $2$-simplices $\langle 012 \rangle$, $\langle 013 \rangle$, $\langle 023 \rangle$, and $\langle 123 \rangle$ such that
\begin{equation}
    \begin{split}
        \varrho(g_{01},g_{12},g_{23}) =& \rho(1, g_{01}, g_{01}g_{12}, g_{01} g_{12} g_{23})\\
        =& b_{012} + b_{013}+b_{023}+ b_{123}
    \end{split}
\label{eq: 2-group constraint}
\end{equation}
{For $n \geq 4$, we glue $n$-simplices to $(n-1)$-cycles to eliminate all higher homotopy groups.} 
% \textcolor{gray}{For $n \geq 4$, we glue the $n$-simplex to any $(n-1)$-cycle.}
% More explicitly, we glue the boundary of a $n$-simplex $\langle 01\dots n \rangle$ with $(n-1)$-simplices $\langle 0 \dots \hat{i} \dots n \rangle$ for $i= 0,1, \dots, n$, where $\hat{i}$ means that $i$ is omitted. 
According to Dijkgraaf and Witten \cite{DW90}, topological gauge theories with gauge group $G$ in $(n+1)$ spacetime dimensions are classified by $H^{d+1} (G, \RR/\ZZ) \equiv H^{n+1}(BG,\RR/\ZZ)$. This was generalized to 2-group gauge theories by Kapustin and Thorngren \cite{KT15}: 2-gauge theories with $2$-group $\GG$ in $(n+1)$D are classified by $H^{n+1} (B \GG, \RR/\ZZ)$.

\subsection{Derivation of $\zeta$} \label{sec:zeta}
Here, we derive the choice of $\zeta$ in \eqref{eq: zeta}.
% \begin{equation}
%     \begin{split}
%     &\zeta(g_{12},g_{23},h_1,h_2,h_3) =\rho(1,g_{12},g_{12}g_{23},g_{12}g_{23}h_3) +\\ & \rho(1,g_{12},g_{12}h_2,g_{12}g_{23}h_3) + \rho(1,h_1,g_{12}h_2,g_{12}g_{23}h_3),
%     \end{split}
% \end{equation}
Consider a tetrahedron $\langle 0123 \rangle$ with group elements $g,h,k$ on edges $\langle01 \rangle,\langle12\rangle,\langle23\rangle$. The value of cocycle $\rho$ on this tetrahedron is expressed as $\rho(1,g,gh,ghk)$. First, we perform a gauge transformation on vertex $0$ by group element $c_0$ (and identity element on all other vertices); $g,h,k$ becomes $c_0^{-1}g,h,k$ and the value of cocycle is (using the homogeneity of $\rho$):
% \begin{equation}
%     \begin{split}
%     &\rho(1,c_0^{-1}g,c_0^{-1}gh,c_0^{-1}ghk) = \rho(c_0,g,gh,ghk)\\
%     &=\rho(1,c_0,g,ghk) - \rho(1,c_0,g,gh) \\
%     & -\rho(1,c_0,gk,ghk) + \rho(1,g,gh,ghk)
%     \end{split}
% \end{equation}
\begin{eqs}
    \rho(c_0,g,gh,ghk)
    &=\rho(1,c_0,g,ghk) - \rho(1,c_0,g,gh) \\
     &-\rho(1,c_0,gk,ghk) + \rho(1,g,gh,ghk)
    \end{eqs}
To satisfy \eqref{eq: zeta equation}, we can define the gauge transformation of $b_{ijk}$ by $c_0$ at the vertex $i$ of a face $\langle ijk \rangle$ as $\rho(1,c_0,g_{ij},g_{ij}g_{jk})$, or explicitly { $\zeta(g_{ij},g_{jk},c_0,1,1)=\rho(1,c_0,g_{ij},g_{ij}g_{jk})$ }.

Secondly, we perform a gauge transformation on vertex $1$ by group element $c_1$. $g,h,k$ becomes $g c_1, c_1^{-1} h,k$ and the value of the cocycle is
\begin{equation}
    \begin{split}
    \rho(1,gc_1,gh,ghk)
    &= \rho(1,g,gc_1,gh)-\rho(1,g,gc_1,ghk) \\
    &+ \rho(g,gc_1,gh,ghk) +\rho(1,g,gh,ghk) \\
    &= \rho(1,g,gc_1,gh)-\rho(1,g,gc_1,ghk) \\
    &+ \rho(1,c_1,h,hk) +\rho(1,g,gh,ghk).
    \end{split}
\end{equation}
The first two terms after the last equality indicate that the gauge transformation on the vertex $j$ of a face $\langle ijk \rangle$ is $\rho(1,g_{ij},g_{ij}c_1,g_{jk})$, or { $\zeta(g_{ij},g_{jk},1,c_1,1)=\rho(1,g_{ij},g_{ij}c_1,g_{jk})$}, and the third term after the second equality is consistent with the previous case.

Similarly for the gauge transformation on the vertex $2$ by group element $c_2$, we have 
\begin{equation}
    \begin{split}
    \rho(1,g,gh c_2,ghk)
    &= \rho(g,gh,gh c_2,ghk)-\rho(1,g,gh,ghc_2) \\
    &+ \rho(1,g,gh c_2,ghk) +\rho(1,g,gh,ghk) \\
    &= \rho(1,h,h c_2,hk)-\rho(1,g,gh,ghc_2) \\
    &+ \rho(1,g,gh c_2,ghk) +\rho(1,g,gh,ghk).
    \end{split}
\end{equation}
The term $\rho(1,g,gh,ghc_2)$ after the last equality implies { $\zeta(g_{ij},g_{jk},1,1,c_2)=\rho(1,g_{ij},g_{ij}g_{jk},g_{ij}g_{jk} c_2)$}; the remaining terms are consistent with the previous cases.

We can check the gauge transformation on the $3$ vertex of the tetrahedron. However, it just gives us a consistency check. The previous three cases specify the gauge transformation. 

Now, we combine the three cases and apply the three gauge transformations $h_3$, $h_2$, $h_1$ on vertices $3,2,1$ of a face $\langle 123 \rangle$ in sequence. First, we add $\rho(1,g_{12},g_{12}g_{23},g_{12}g_{23} h_3)$ to $b_{123}$ and $g_{12},g_{23}$ is mapped to $g_{12},g_{23} h_3 $. Second, we gain a factor of $\rho(1,g_{12},g_{12} h_2,g_{12}g_{23} h_3)$ and $g_{12},g_{23} h_3$ becomes $g_{12} h_2  ,h_2^{-1} g_{23} h_3$. Finally, we add $\rho(1,h_1,g_{12} h_2 ,g_{12}g_{23} h_3)$. In total, the gauge transformation on $b_{123}$ is
\begin{equation}
    \begin{split}
    &\zeta(g_{12},g_{23},h_1,h_2,h_3)=\rho(1,g_{12},g_{12}g_{23},g_{12}g_{23}h_3) + \\  &\rho(1,g_{12},g_{12}h_2,g_{12}g_{23}h_3) + \rho(1,h_1,g_{12}h_2,g_{12}g_{23}h_3).
    \end{split}
\end{equation}

\section{Spacetime construction of $2$-group SPT and boundary of supercohomology models} \label{app: super from 2gauge}
% \subsection{Symmetry of 2-group SPT phases from 2-gauge theories }

We use the language of $2$-gauge theory, summarized in Appendix~\ref{sec: 2-group extension}, to describe a $2$-gauge theory dual to the $2$-group SPT models of Section~\ref{sec: 2-groupspt}. We then provide the details behind the construction of the topologically ordered symmetric gapped boundaries of the supercohomology models from Section~\ref{sec: fermionic gapped boundary construction}. We start by constructing the partition function for the $2$-group model on a manifold without boundary from the corresponding $2$-gauge theory in Section~\ref{app: bulk 2group}. Then, in Section~\ref{app: super with boundary 2gauge}, we describe the construction of the supercohomology models on a manifold with boundary starting with the spacetime description of the $2$-group model.

\begin{figure}[t]
\centering
\includegraphics[width=0.51\textwidth]{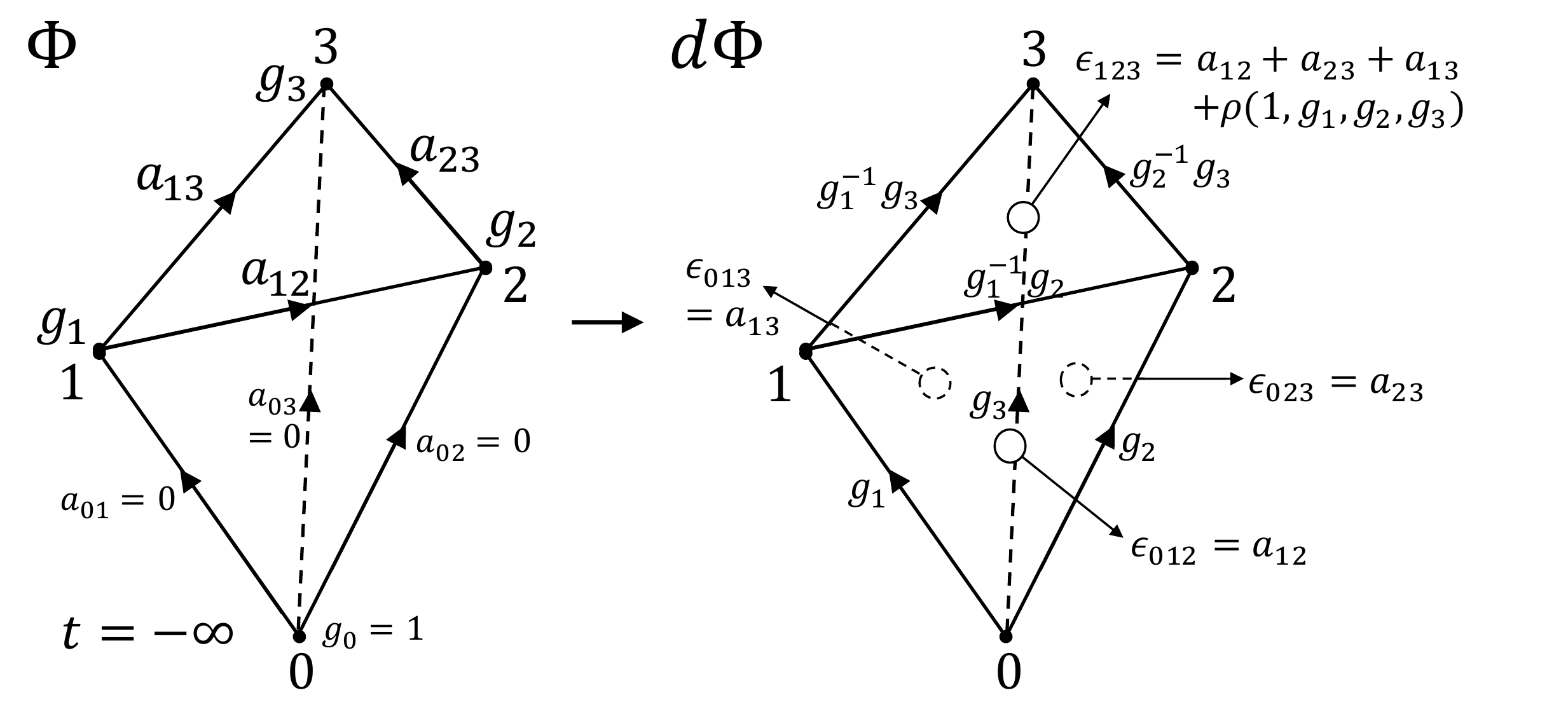}
\caption{{For a representative tetrahedron including the $0$ vertex, we show the mapping of the configuration $\Phi$ (left) to the configuration $d\Phi$ (right). In general, $\eps_{ijk}$ is $a_{ij}+a_{ik}+a_{jk}+{\rho}(1,g_i,g_j,g_k)$, with the labels determined by $\Phi$.  We have assumed that $\rho$ is normalized, i.e.,  $\rho(1,1,g,h)=0$. Notice that $d\Phi$ can be interpreted as applying gauge transformation specified by the configuration $\Phi$ on the trivial configuration of a 2-gauge theory.}
% The spacetime configuration where the 2-group cocycle $\alpha$ acts \cite{KT15}. Given a spatial configuration $\ket{\Phi} = \ket {\{g_v\},\ba_e}$, the spacetime configuration $\ket{d\Phi}$ is defined.
% We have used normalized cocycles $\rho$, i.e.  $\rho(1,1,g,h)=0$. Notice that this $d\Phi$ can be interpreted as applying gauge transformation specified by the configuration $\Phi$ on the trivial configuration of a 2-gauge theory.
}
\label{fig: CM_dPhi}
\end{figure}

% \subsection{Symmetry of 2-group SPT phases from 2-gauge theories }\label{sec: symmetry of 2-group SPT as 2-gauge theory}

\begin{figure*}[t]
\centering
\includegraphics[width=0.8\textwidth]{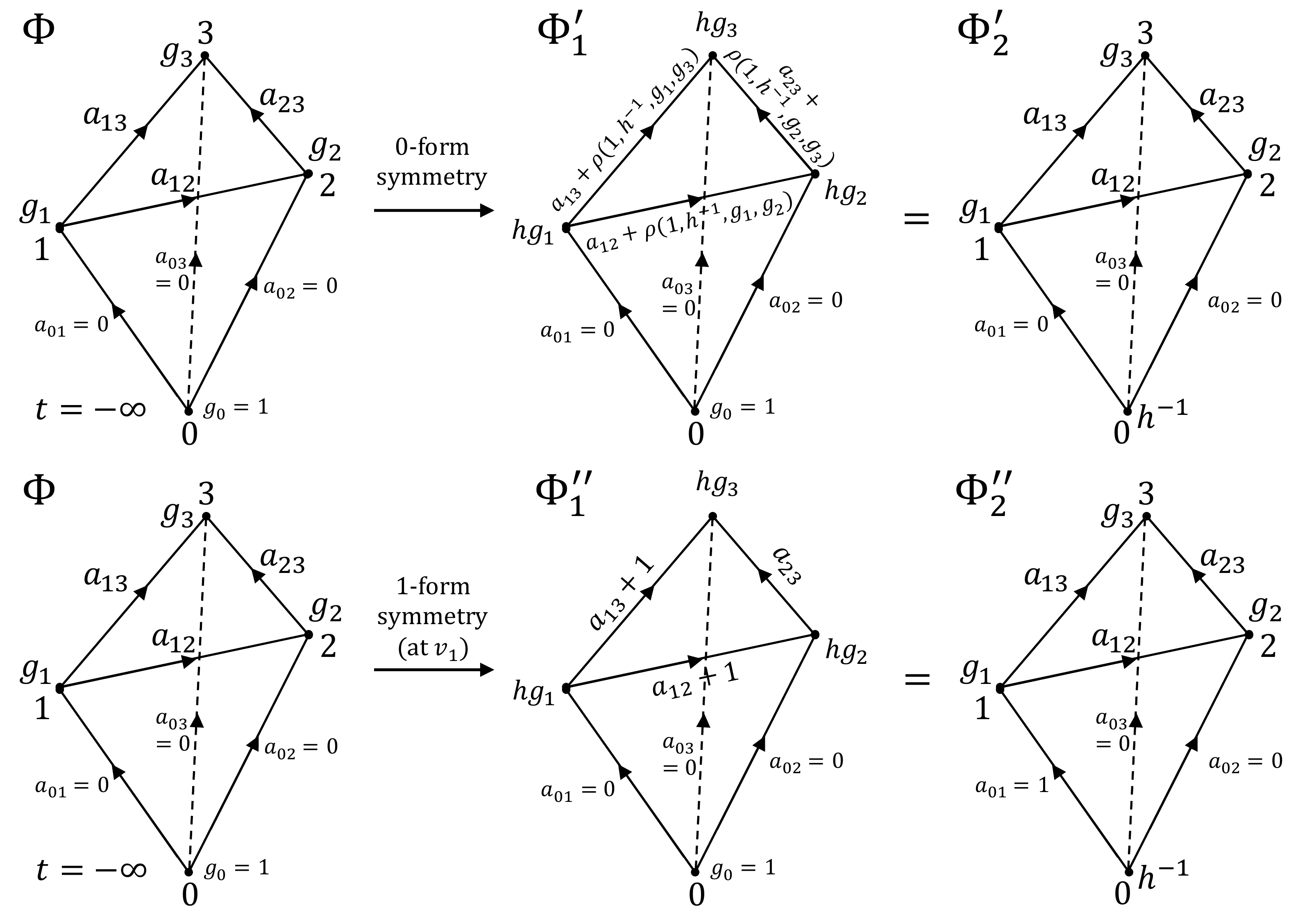}
\caption{The first row represents the $0$-form symmetry action [represented by $V_\rho(h)$]. Acting on the configuration $\Phi$, it multiplies $h$ on spatial vertices $v \in M$ and modifies spatial edges $e_{ij} \in M$ by $\rho (1,h^{-1},g_i, g_j)$. The resulting configuration is $\Phi'_1$. Given that $\alpha$ depends only on $d\Phi'_1$, we can consider this configuration to be equivalent to $\Phi'_2$, which is generated from $\Phi$ by changing $g_0 = 1$ to $h^{-1}$.  The second row represents the $1$-form symmetry action at the vertex $1$. The labels on the edges joined at $1$ change by $+1$. With respect to $\alpha$, this is equivalent to only changing the temporal edge $a_{01}$ by $+1$, i.e., $d\Phi''_1 = d\Phi''_2$.
}
\label{fig: symmetry_configuration_0+1}
\end{figure*}

% In this section, we will use spacetime configurations as a 2-gauge theory to analyze the properties of 2-group SPT states, including the case with the boundary construction.  The discussion here is similar to boundary constructions of bosonic SPT models in Section \ref{sec: group coho boundary}. Each corresponding step in the derivation for $2$-group SPT phases will be clearly elaborated.

\subsection{$2$-gauge theory interpretation of bulk $2$-group SPT} \label{app: bulk 2group}

{$2$-group SPT phases are dual to $2$-gauge theories, similar to the familiar duality between ordinary $0$-form SPT phases and conventional Dijkgraaf-Witten gauge theories \cite{LG12,WL15,GW14}. To see this, we consider a triangulated spatial $3$-manifold $M$ along with the spacetime manifold $CM$ formed by connecting each vertex of $M$ to a vertex $0$, sometimes referred to as the spacetime infinity vertex. We define a configuration $\Phi = \boldsymbol{\{}\{ g_v \}, \{ a_e\} \boldsymbol{\}}$ on $CM$ by assigning a $g_v \in G$ to each vertex $v \in M$ and an $a_e \in \ZZ_2$ to each edge $e \in M$; the temporal edges (connected to the $0$ vertex) are labeled with $0 \in \ZZ_2$ and the $0$ vertex is labeled with $1 \in G$.
%
% We define a spatial configuration $\Phi_M = \boldsymbol{\{}\{ g_v \}, \{ a_e\} \boldsymbol{\}}$ on $M$ by assigning a $g_v \in G$ to each vertex $v$ and an $a_e \in \ZZ_2$ to each edge $e$. 
% % We find it convenient to sometimes interpret the configuration $\{ a_e \}$ as  
% The configuration $\Phi_M$ is canonically associated to a configuration $\Phi$ on $CM$ with $0 \in \ZZ_2$ assigned to each temporal edge (connected to the $0$ vertex) and the identity $1 \in G$ assigned to the spacetime infinity vertex.
}
{Furthermore, we define a map $d$ from a configuration $\Phi$ to a configuration $d\Phi$ on $CM$. The configuration $d\Phi$ consists of a set of $G$ labels $g_{ij}$ for all edges in $CM$ and $\ZZ_2$ labels $a_{ijk}$ for all faces in $CM$. Specifically, for a configuration $\Phi = \boldsymbol{\{}\{ g_v \}, \{ a_e\} \boldsymbol{\}}$, the associated configuration $d\Phi$ is:
\begin{align}
    d\Phi = \boldsymbol{\{}\{  g_i^{-1} g_j \}, \{a_{ij}+a_{ik}+a_{jk} + \bar \rho(1,g_i,g_j,g_k)\} \boldsymbol{\}},
\end{align}
depicted in Fig.~\ref{fig: CM_dPhi}. 
% Here, the first element of $d\Phi$ is a $G$-configuration on edges $\langle ij \rangle$ and the second entry is a $\ZZ_2$-configuration on faces $\langle ijk \rangle$.
One can verify that this $d\Phi$ satisfies the constraint \eqref{eq: constraint 2} and therefore is an allowed configuration for a 2-gauge theory.}

% \textcolor{gray}{For simplicity, the artificial vertex is referred as the spacetime infinity vertex $0$ in this section. Connecting this vertex to the spatial manifold $M_3$ forms the spacetime manifold $CM_3$.
% Consider a spatial configuration $\Phi = \{ g_v \}, \{ a_e\}$ on $M_3$, where we put $g_v\in G$ on each vertex and $a_e \in \ZZ_2$ on each edge. Notice that we also have $0 \in \ZZ_2$ on the temporal edges and $1 \in G$ at the spacetime infinity vertex 0. Its spacetime configuration $d\Phi$ on $CM_3$ is
% \begin{eqs}
%     d\Phi = \{ g_{ij} =  g_i^{-1} g_j\},\{ \delta a_e + \bar \rho\},
% \label{eq: dPhi definition}
% \end{eqs}
% which shown explicitly in Fig. \ref{fig: CM_dPhi}. One can verify that this $d\Phi$ satisfies the constraint \eqref{eq: constraint 2} and therefore is an allowed configuration for a 2-gauge theory.} 

{Given a $2$-group cocycle $\alpha \in H^{4} (B \GG, \RR/\ZZ)$ we can now define the corresponding $2$-group SPT state. We first pull back $\alpha$ to a cocycle on $CM$. In a slight abuse of the notation introduced in the main text, for a configuration $d\Phi$, we denote the pullback of $\alpha$ to $CM$ as $\alpha_{d\Phi} \in Z^4(CM,\RR/\ZZ)$. When it is clear from context, we also omit the subscript $d\Phi$. A ground state in the associated $2$-group SPT phase can then be written as:
\begin{align}\label{2groupsptappstate}
    \ket{\Psi_\text{SPT}} \equiv \sum_{\Phi} \prod_{t = \langle 1234 \rangle \in M} e^{2 \pi i \alpha_{d\Phi}(01234)O_t}\ket{\Phi}
\end{align}
% \begin{align}
%     \ket{\Psi_\text{SPT}} \equiv \sum_{\Phi_M} e^{\sum_{t = \lr{1234} \in M} \alpha_{d\Phi}(01234) O_t}\ket{\Phi_M},
% \end{align}
% where the sum in the exponent is over all $4$-simplices in $CM$, $O_t$ is the orientation of $t$, and we have omitted the angled brackets around the $4$-simplex $\lr{01234}$.
% \begin{align}\label{2groupsptappstate}
%     \ket{\Psi_\text{SPT}} \equiv \sum_{\Phi_M} e^{2 \pi i \int_{CM} \alpha_{d\Phi}}\ket{\Phi_M}.
% \end{align}
Here, 
% the sum is over all configurations $\Phi_M$ with $0 \in \ZZ_2$ at each temporal edge (connected to the $0$ vertex) and the identity $1 \in G$ at the spacetime infinity vertex.
% Further, 
the product in Eq.~\eqref{2groupsptappstate} is over all $4$-simplices in $CM$, $O_t$ is the orientation of $t$, and we have omitted the angled brackets around the $4$-simplex $\lr{01234}$. In what follows, to simplify the notation, we often omit the angled brackets.
% Here, the sum is over all configurations $\Phi_M$ with $0 \in \ZZ_2$ at each temporal edge (connected to the $0$ vertex) and the identity $1 \in G$ at the spacetime infinity vertex.
% Further, the exponent in Eq.~\eqref{2groupsptappstate} is shorthand for the sum:
% \begin{align}
%     \int_{CM}\alpha_{d\Phi} \equiv \sum_{t = \lr{1234} \in M} \alpha_{d\Phi}(01234) O_t,
% \end{align}
% where the sum is over all $4$-simplices in $CM$, $O_t$ is the orientation of $t$, and we have omitted the angled brackets around the $4$-simplex $\lr{01234}$.
}

{We recover the $2$-group SPT states from Section~\ref{sec: 2-groupspt}, by considering the $2$-gauge theory corresponding to the weak $2$-group $(G,\ZZ_2,0,\rho)$\footnote{Here, $0$ denotes the trivial action of $G$ on $A$.} and the choice of $2$-group $4$-cocycle $\alpha$:
\begin{eqs} \label{alpha def app}
    \alpha = \nu + \frac{1}{2} \rho \cup_1 \eps + \frac{1}{2} \eps \cup \eps,
\end{eqs}
described in Section~\ref{sec: 2-groupspt}. We use the remainder of this section to verify that the SPT state associated to this choice of $\alpha$ is indeed symmetric. The calculation will prove useful in the construction of supercohomology models on a manifold with boundary in the next section.}

{The 2-group symmetry contains two types of symmetry actions: 0-form and 1-form. The 0-form part is:
\begin{equation}
\begin{aligned}
        V_\rho(h): 
        &g_v \ra h g_v,  &  &\forall {v} \in M \\ 
        &a_{ij} \ra a_{ij} + \rho(1,h^{-1},g_i, g_j), &   &\forall \lr{ij} \in M,
\label{eq: 0-form spatial symmetry}
\end{aligned}
\end{equation}
% \begin{equation}
%     \begin{split}
%         V(h): 
%         &\{g_v\} \ra \{h g_v\},  \\
%         &\{a_{ij}\} \ra \{a_{ij} + \rho(1,h^{-1},g_i, g_j)\},
%     \end{split}
% \label{eq: 0-form spatial symmetry}
% \end{equation}
and the local 1-form symmetry at the vertex $v \in M$ is:
\begin{equation}
\begin{aligned}
        A_{\delta \bv}: 
        ~a_{ij} \ra a_{ij} + \delta \bv \boldsymbol{(}\lr{ij}\boldsymbol{)}, \quad \forall \lr{ij} \in M. 
\label{eq: 1-form spatial symmetry}
\end{aligned}
\end{equation}
% \begin{equation}
%     \begin{split}
%         A_{\delta \bv}: 
%         &\{g_v\} \ra  \{g_v\}, \\
%         &\{a_{ij}\} \ra \{a_{ij} + \delta \bv \boldsymbol{(}\lr{ij}\boldsymbol{)}\}. 
%     \end{split}
% \label{eq: 1-form spatial symmetry}
% \end{equation}
These two symmetries are explicitly demonstrated in Fig.~\ref{fig: symmetry_configuration_0+1}.
Notice that the symmetry actions in Eq.~\eqref{eq: 0-form spatial symmetry} and Eq.~\eqref{eq: 1-form spatial symmetry} only change the spatial d.o.f., i.e., those belonging to $M$.} 

\begin{figure}[t]
\centering
\includegraphics[width=0.5\textwidth]{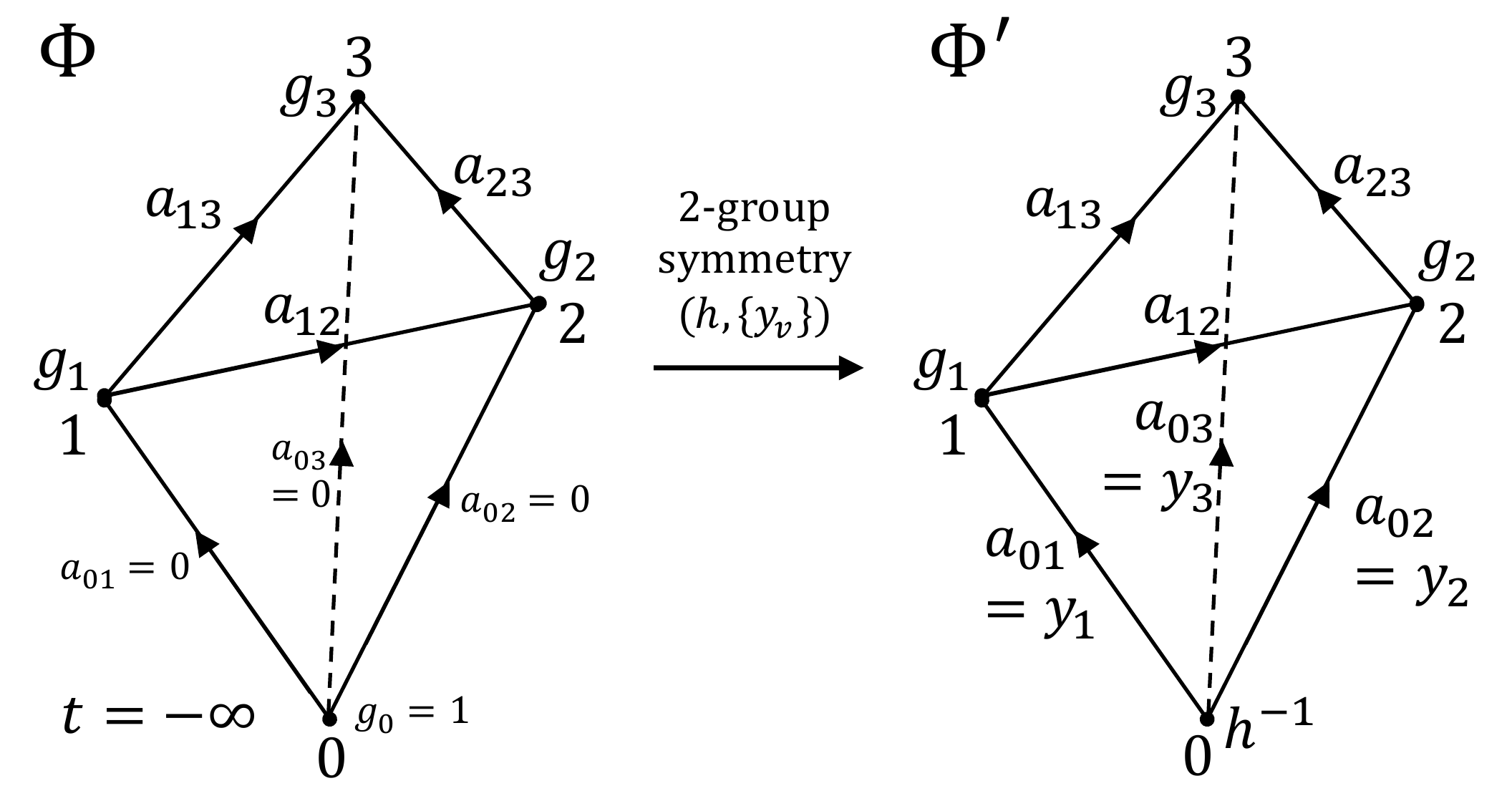}
\caption{{The (local) $2$-group symmetry is parameterized by $(h,\{y_v\})$. The configuration $\Phi$ is mapped by the symmetry to a configuration $\tilde{\Phi}$ that is equivalent to $\Phi'$, in the sense that $d\Phi_1'=d\Phi'$. $\Phi'$ differs from $\Phi$ on the temporal links, where $a_{0v}=y_v$, for $v \in M$, and at the $0$ vertex, where $g_0$ is $g_0=h^{-1}$.}
% The 2-group symmetry action is composed of 0-form symmetry $V_\rho(h)$ and 1-form symmetry at vertices labeled by $y_v = 0,1$. When $y_v=1$, we act 1-form symmetry at $v$. Combining the results in Fig. \ref{fig: symmetry_configuration_0+1}, the equivalent configuration after the 2-group symmetry action is simply changing $g_0 = h^{-1}$ and $a_{0v} = y_v$ for all $v \in M_3$.
}
\label{fig: symmetry_configuration_2}
\end{figure}

{To simplify the symmetry analysis, we use that the amplitude of the SPT state depends only on the configuration $d\Phi$ and not on the configuration $\Phi$. As such, we can consider two configurations $\Phi_1$ and $\Phi_2$ equivalent, with respect to the amplitude, if $d\Phi_1=d\Phi_2$.} 
{In Fig.~\ref{fig: symmetry_configuration_0+1}, the configuration $\Phi$ is mapped to the configuration $\Phi'_1$ by the $0$-form symmetry action. Furthermore, $\Phi_1'$ is equivalent to a configuration $\Phi_2'$ obtained from $\Phi$ by simply changing $g_0=1$ to $g_0=h^{-1}$, see Fig.~\ref{fig: symmetry_configuration_2}. Similarly, the $1$-form symmetry action by $A_{\delta \bv}$ is equivalent to changing $a_{0v} = 0$ to $a_{0v} = 1$ for all the temporal edges $\lr{0v}$. This is analogous to the homogeneous property of $0$-form group cocycles. More generally, the total 2-group symmetry can be labeled by $(h,\{y_v\})$ with $h\in G$ and $y_v=0,1$, $\forall v \in M$:
\begin{eqs}
    V_2 (h,\{y_v\}) \equiv V_\rho(h) \prod_{v} A_{\delta \bv}^{y_v},
\label{eq: combined 2-group symmetry action}
\end{eqs}
and with respect to the amplitude of $\ket{\Psi_\text{SPT}}$, this is effectively equivalent to mapping $g_0 = 1$ to $g_0 = h^{-1}$ at the $0$ vertex and $a_{0v} = 0$ to $a_{0v} = y_v$ on all temporal edges $\lr{0v}$.}

% \textcolor{gray}{However, the 0-form symmetry is gauge equivalent to changing $g_0 = 1$ to $g_0 = h^{-1}$ at the spacetime infinity $0$ vertex, and the 1-form symmetry is equivalent to changing $a_{0v} = 0$ to $a_{0v} = 1$ on the all of the temporal edges $\lr{0v}$, shown in Fig. \ref{fig: symmetry_configuration_0+1}. Here, the equivalence means that the two configurations satisfy: $d\Phi_1' = d\Phi_2'$ and $d\Phi_1''= d\Phi_2''$, see Fig. \ref{fig: symmetry_configuration_0+1}. The total 2-group symmetry can be labeled by $(h,\{y_v\})$ with $h\in G$ and $y_v=0,1$, $\forall v \in M$:
% \begin{eqs}
%     V_2 (h,\{y_v\}) \equiv V(h) \prod_{v} A_{\delta \bv}^{y_v},
% \label{eq: combined 2-group symmetry action}
% \end{eqs}
% which is equivalent to the composition of mapping $g_0 = 1$ to $g_0 = h^{-1}$ at $0$ and $a_{0v} = 0$ to $a_{0v} = y_v$ on all temporal edges $\lr{0v}$. 
% From now on, we will abuse the notation slightly by writing $\alpha$ acting on $\Phi$, $\alpha (\Phi)$, but the value of $\alpha$ actually depends only on $d\Phi$. }

{To show that the $2$-group state is invariant under the $2$-group symmetry, we consider the affect of the symmetry action on the amplitude of the state. The symmetry action maps a configuration $\Phi$ to a configuration equivalent to $\Phi'$, depicted in Fig.~\ref{fig: symmetry_configuration_2}. 
We thus need to compare $\alpha_{d\Phi'}$ to $\alpha_{d\Phi}$. To do so, it is useful to append a vertex $0^h$ to $CM$ by connecting each vertex of $CM$ to $0^h$. We assign a configuration on any $4$-simplices of the form $\lr{01234}$ with $\lr{1234}$ in $M$ according to $\Phi$, and we assign a configuration to any $4$-simplices $\lr{0^h1234}$ according to $\Phi'$. (We also fix $a_{00^h}=0$.)
We can then consider the difference:
\begin{eqs}\label{alpha difference}
     \alpha({0^h 1234})-\alpha({01234}) = &\alpha({00^h123}) - \alpha({00^h124})  \\
    & \alpha({00^h134}) - \alpha({00^h234}),
\end{eqs}
where the equality comes from the cocycle condition $\delta \alpha (00^h1234)=0$. The change in the amplitude depends on the sum over all tetrahedra in $M$:
\begin{eqs}
    \sum_{t=\lr{1234} \in M} [\alpha({0^h 1234})-\alpha({01234})] O_t.
\label{eq: sum of phases change}
\end{eqs}
Inserting the right-hand side of Eq.~\eqref{alpha difference} into the expression above, we find that each term appears exactly twice and with the opposite sign. This can be seen by noting that every term can be associated to a face, e.g., $\alpha({00^h123})$ corresponds to $f=\lr{123}$. Thus, the sum in Eq.~\eqref{eq: sum of phases change} vanishes, and the 2-group SPT state in Eq.~\eqref{2groupsptappstate} is invariant under 2-group symmetry action in Eq.~\eqref{eq: combined 2-group symmetry action}.} 

% \textcolor{gray}{Now, we denote $\lr{01234}$ as a 4-simplex with the $\Phi$ configuration and $\lr{0^h 1234}$ a 4-simplex with $\Phi'$ configuration in Fig. \ref{fig: symmetry_configuration_2}. We can glue them on $M_3$, forming a 5-simplex $\lr{0 0^h 1234}$, with $a_{{0 0^h}} =0$. When the 2-group symmetry action \eqref{eq: combined 2-group symmetry action} is performed on the SPT state \eqref{eq: spacetime 2-group SPT}, it produces a phase difference for each tetrahedron $t=\lr{1234} \in M_3$:
% \begin{eqs}
%     & [\alpha(\lr{0^h 1234})-\alpha(\lr{01234})] O_t \\
%     =& \left[ \alpha(\lr{00^h123}) - \alpha(\lr{00^h124})  \right. \\
%     & ~~~~ \left. \alpha(\lr{00^h134}) - \alpha(\lr{00^h234}) \right] O_t,
% \end{eqs}
% where the equality comes from the cocycle condition of $\alpha$, $\delta \alpha (\lr{0 0^h 1234})=0$. If the phases are summed over all tetrahedra, all terms cancel out since for each $f=\lr{123}$ $\alpha(\lr{00^h123})$ appears exactly twice with opposite signs in the adjacent tetrahedra:
% \begin{eqs}
%     \sum_{t=\lr{1234} \in M} [\alpha(\lr{0^h 1234})-\alpha(\lr{01234})] O_t = 0.
% \label{eq: sum of phases change}
% \end{eqs}
% This means that the 2-group SPT state \eqref{eq: spacetime 2-group SPT} is invariant under 2-group symmetry action \eqref{eq: combined 2-group symmetry action}.}

\subsection{Derivation of the supercohomology models with a boundary} \label{app: super with boundary 2gauge}

{We now describe the construction of the supercohomology models on a manifold with boundary, introduced in Section~\ref{sec: fermionic gapped boundary construction}, starting from a spacetime model for the $2$-group SPT phase. For a spatial manifold $M$ with boundary $\partial M$, the $2$-group SPT model is defined on the cone $CM$, and has a modified $2$-group symmetry action on the boundary of $CM$. With this spacetime $2$-group SPT model, we follow the prescription for the bulk construction of the supercohomology models, i.e., gauging the $1$-form symmetry followed by applying the fermionization duality.}

{As described in Section \ref{sec: fermionic gapped boundary construction}, the symmetry $G$ can be extended by $K$ to $L$ such that the supercohomology data is trivialized:
\begin{align}
    (\rho^*,\nu^*) = ( \delta \beta,\delta \eta + \frac{1}{2} \beta \cup_1 \delta \beta + \frac{1}{2} \beta \cup \beta),
\end{align}
for some $\beta \in C^2(L, \mathbb{Z}_2)$ and $\eta \in C^3(L,\mathbb{R}/ \mathbb{Z})$. The elements of $L$ can be written as $\ell = g^{(k)}$ with $g \in G$ and $k \in K$. If we let $m$ be the $2$-cocycle $m \in H^2(G,K)$ corresponding to the extension of $G$ by $K$, then the group law in $L$ is specified by:
\begin{align}
    g_1^{(k_1)}g_2^{(k_2)}=(g_1g_2)^{\boldsymbol{(}k_1 + k_2 + m(1,g_1,g_1g_2)\boldsymbol{)}}.
\end{align}}

{To define the spacetime $2$-group SPT models, we first describe the configurations $\Phi$ on $CM$. A configuration $\Phi$ is specified by a $G$ label on each vertex, a $K$ label on the vertices $v_\partial$ in $\partial M$, and a $\ZZ_2$ label on every edge in the boundary of $CM$, which we denote as $\overline{M}$. Unless otherwise specified, we fix the label at $0$ to be $1 \in L$ and the labels at the temporal edges $e \notin \overline{M}$ to be $0 \in \ZZ_2$.  We write such a configuration $\Phi$ as ${\Phi = \boldsymbol{\{} \{ g_v \}, \{ k_{v_\partial} \}, \{ a_e \} \boldsymbol{\}}}$. We note that, since each vertex in the boundary of $M$ is labeled by an element $g\in G$ and a $k \in K$, one can consider the boundary vertices to be labeled by an element of $L$.}

{The SPT model has a $0$-form symmetry parameterized by an element $h^{(k)}\in L$ as well as a $\ZZ_2$ $1$-form symmetry. The 0-form symmetry action is:
\begin{equation} \label{bdry spacetime 0form}
\begin{aligned}
        V_B(h^{(k)}):~ 
        &g_v \ra h g_v & &\forall v \in M \\
        &k_{v} \ra k + k_{v} + m(1,h,hg), & &\forall v \in \partial M \\
        &a_{ij} \ra a_{ij} + B^{h^{(k)}}_{ij}, & &\forall \lr{ij} \in \overline{M},
        \end{aligned}
\end{equation}
% \begin{eqs}
%         V(h^{(k)}):~ 
%         &g_0 \ra g_0, \\
%         &\{g_v\}_{v \neq 0} \ra \{h g_v\}_{v \neq 0}, \\
%         &\{ k_{v_\partial} \} \ra \{ k + k_{v_\partial} + m(1,h,hg) \}, \\
%         &\{a_{ij}\} \ra \{a_{ij} + B^{h^{(k)}}_{ij}\},
% \end{eqs}
where $B^{h^{(k)}}_{ij}$ is defined as:
\begin{eqs}
B^{h^{(k)}}_{ij} = \begin{cases} 
      \rho(1,h^{-1},g_i, g_j) & i\neq 0 \\
      \beta(1,{h^{(k)}}^{-1},g_j^{(k_j)}) & i = 0. 
   \end{cases}
\end{eqs}
% and $\overline{M}$ is the boundary of $CM$.
The local $1$-form symmetry action, for $v \in \overline{M}$, is given by:
\begin{eqs} \label{bdry spacetime 1form}
A_{\delta \bv}: ~ a_{ij} \ra a_{ij} + \delta \bv\boldsymbol{(} \lr{ij} \boldsymbol{)}, \quad \forall \lr{ij} \in \overline{M}.
\end{eqs}   
Note that, unlike the symmetry action in the previous section, the symmetry acts on some of the temporal links of $CM$ as well.
A more general $2$-group symmetry action is specified by a choice $(h^{(k)},\{y_v\})$ with $h \in G$, $k \in K$, and $y_v=0,1$, $\forall v \in \overline{M}$, and is represented by:
\begin{eqs}
    V_2 (h^{(k)},\{y_v\}) \equiv V_B(h^{(k)}) \prod_{v} A_{\delta \bv}^{y_v}.
\label{general 2group sym bdry}
\end{eqs}}

{We now (naively) follow the prescription for building a $2$-group SPT state - described in Section~\ref{app: bulk 2group} - using the $2$-group cocycle $\alpha^*$, the pullback of $\alpha$ to $H^4(B\mathbb{L},\RR/\ZZ)$. The amplitude of the state in Eq.~\eqref{2groupsptappstate} depends on the configuration $d\Phi$ as opposed to $\Phi$.
% , where $\Phi$, here, is any configuration with temporal edges $e \notin \overline{M}$ labeled by $0 \in \ZZ_2$ and the infinity vertex labeled by $1 \in L$. 
We use this fact to analyze the symmetry (or lack thereof) of the state constructed using $\alpha^*$.} 

\begin{figure}[t]
\centering
\includegraphics[width=0.5\textwidth]{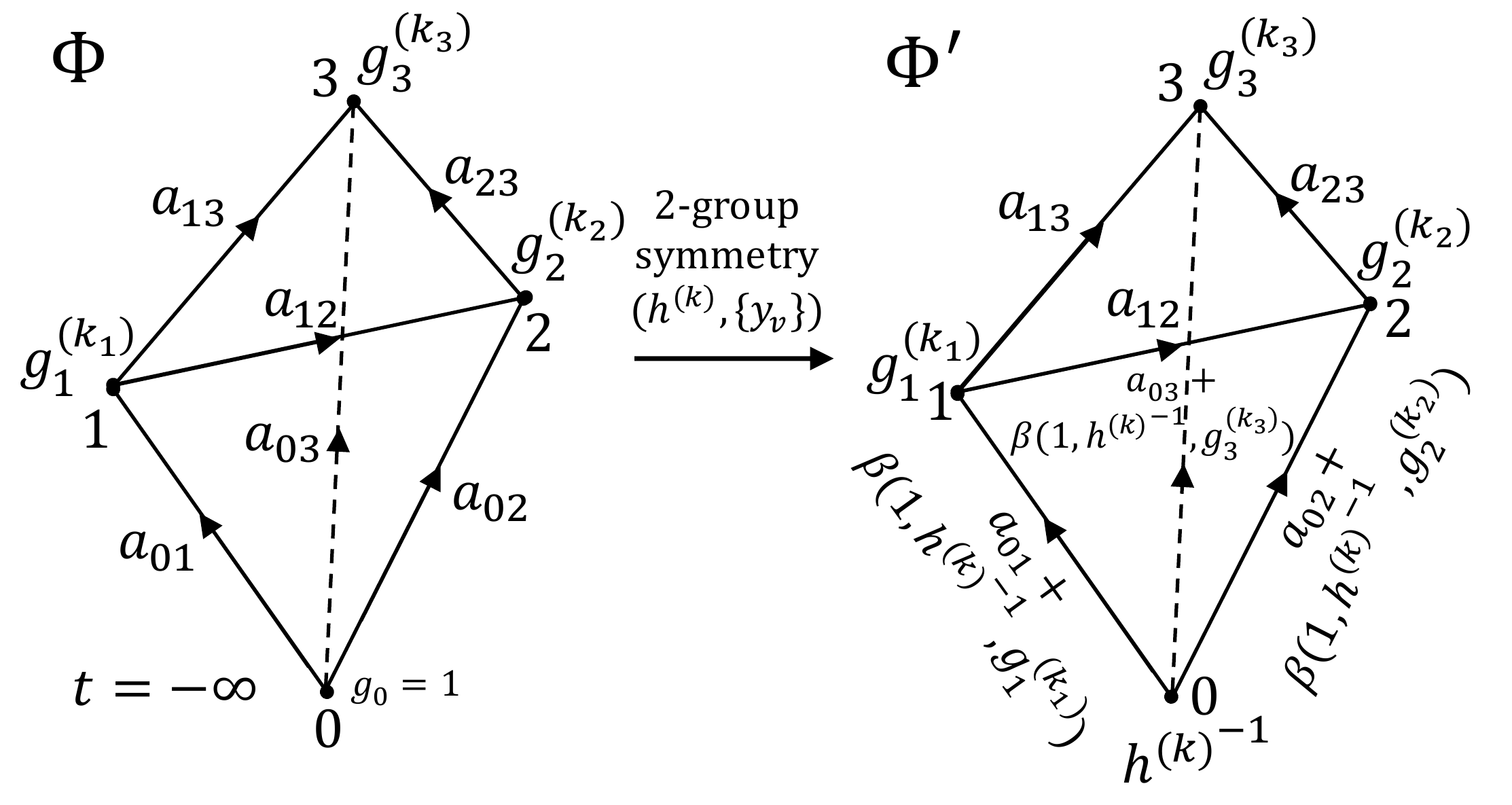}
\caption{{When $M$ has a boundary, the $2$-group symmetry action is modified to act on the temporal links connected to vertices $v_\partial \in \partial M$. In the figure above, $\lr{123}$ is a face in $\partial M$. The (local) $2$-group symmetry, parameterized by $(h^{(k)},\{y_v\})$, maps $\Phi$ to a configuration $\tilde{\Phi}$, which satisfies: $d\tilde{\Phi}=d\Phi'$. $\Phi'$ is equivalent to changing $\Phi$ so that $g_0=1$ goes to $g_0={h^{(k)}}^{-1}$ and $a_{0v_\partial}$ goes to $a_{0v_\partial} +\beta(1,{h^{(k)}}^{-1},g_{v_\partial}^{k_{v_\partial}} )$. Acting on $\Phi$ with the local $1$-form symmetry at a vertex in $\partial M$ does not affect $\Phi'$.} 
% The face $\lr{123}$ is on the boundary $\partial M_3$. Given a boundary vertex $v_\partial \in \partial M_3$, the 2-group symmetry acts differently on boundary temporal edges $\lr{0v_\partial} $. The 1-form symmetry at $v_\partial$ is equivalent to the identity, while the 0-form symmetry is equivalent changing $g_0 = 1 \ra {h^{(k)}}^{-1}$ and $a_{0v_\partial} \ra a_{0v_\partial} +\beta(1,{h^{(k)}}^{-1},g_{v_\partial}^{k_{v_\partial}} )$.
}
\label{fig: symmetry_configuration_2_boundary}
\end{figure}

{In particular, the symmetry action corresponding to $(h^{(k)},\{y_v\})$ on a configuration $\Phi$ maps it to a configuration $\tilde{\Phi}$ according to the maps in Eqs.~\eqref{bdry spacetime 0form} and \eqref{bdry spacetime 1form}. $d\tilde{\Phi}$ is equivalent to $d\Phi'$, where $\Phi'$ (pictured in Fig.~\ref{fig: symmetry_configuration_2_boundary}) is obtained from $\Phi$ by changing the label on the $0$ vertex to $(h^{(k)})^{-1}$ and adding $\beta(1,{h^{(k)}}^{-1},g_{v}^{(k_{v})})$ to any temporal link connected to a boundary vertex $v\in \partial M$.} 

{Similar to the argument in the previous section, we can introduce a vertex $0^h$ and use $\delta \alpha = 0$ to evaluate the difference:
\begin{align} \label{alpha pullback difference}
     \sum_{t=\lr{1234} \in M} [\alpha^*({0^h 1234})-\alpha^*({01234})] O_t.
\end{align}
Here, we have assigned the configuration $\Phi$ ($\Phi'$) to $4$-simplices formed from a tetrahedron in $M$ and the $0$ ($0^h$) vertex. Unlike the case in Section~\ref{app: bulk 2group}, where $M$ had no boundary, the sum in Eq.~\eqref{alpha pullback difference} does not vanish. Instead, we have:
\begin{multline} \label{alpha pullback difference2}
     \sum_{t=\lr{1234} \in M} [\alpha^*({0^h 1234})-\alpha^*({01234})] O_t \\ = \sum_{f=\lr{123} \in \partial M} \alpha^*({00^h 123}) O_f,
\end{multline}
where $O_f\in \{-1,+1\}$ is the orientation of the face $f$ with respect to the orientation of the boundary. Thus, the state constructed using $\alpha^*$ [Eq.~\eqref{2groupsptappstate}] is not invariant under the symmetry in Eq.~\eqref{general 2group sym bdry}.}

To remedy this, we first observe that the cocycle $\alpha^* \in H^4(B\mathbb{L},\RR/\ZZ)$ can be written as:
\begin{equation}
\begin{split}
    \alpha^* &= \nu^* + \frac{1}{2} \rho^* \cup_1 \eps^* + \frac{1}{2} \eps^* \cup \eps^* \\
    &{ = \delta \eta + \frac{1}{2} \beta \cup_1 \delta \beta + \frac{1}{2} \beta \cup \beta +\frac{1}{2} \delta \beta \cup_1 \eps^* +\frac{1}{2} \eps^* \cup \eps^* }\\
    &= \delta \mu + \frac{1}{2}(\eps^*+\beta) \cup (\eps^*+\beta),
\end{split}
\label{eq: pullback of alpha 4 (appendix)}
\end{equation}
where $\mu$ is defined as $\mu \equiv \eta + \frac{1}{2} \beta \cup_1 \eps$, and we have used the property $\delta \eps^* = \rho^*$ {as well as the cup-$1$ relation in Eq.~\eqref{cup1liebniz}}.
%: $\delta (a \cup_1 b) = \delta a \cup_1 b + a \cup_1 \delta b + a\cup b + b \cup a$.
The leftover factor $\alpha^*({00^h 123})$ in Eq.~\eqref{alpha pullback difference2} is equivalent to $\delta \mu({00^h 123})$. 
\begin{figure}[t!]
\centering
\includegraphics[width=0.25\textwidth]{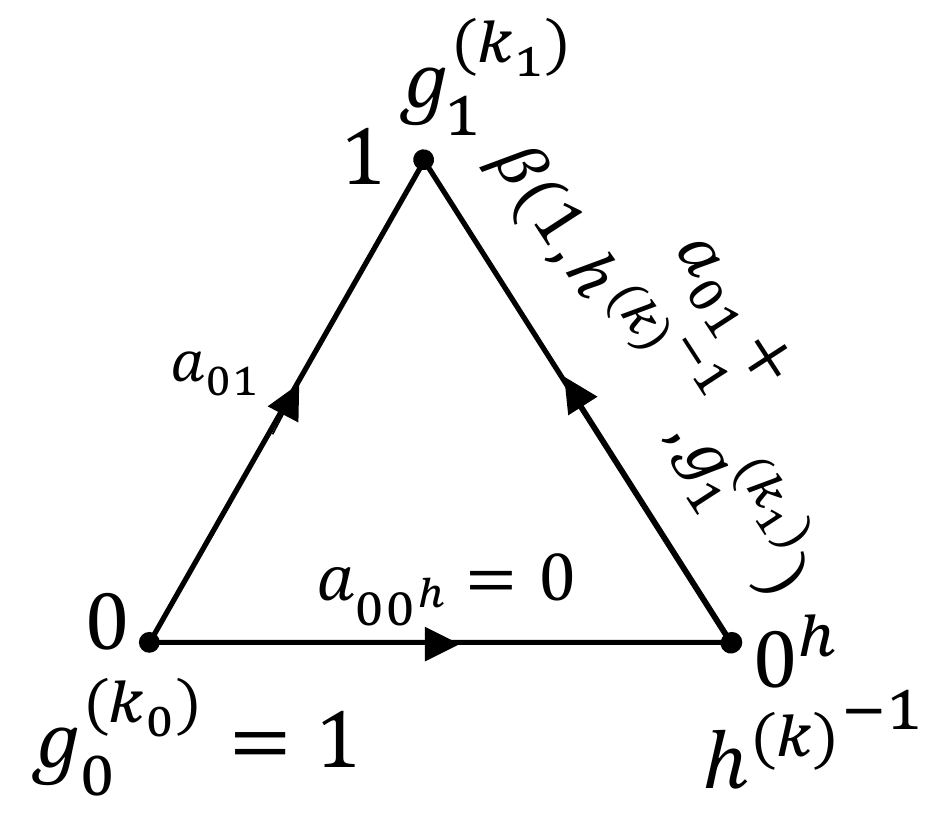}
\caption{{To determine the effect of the $0$-form symmetry action $V_B(h^{(k)})$ on $\alpha$, we append a vertex $0^h$, labeled by ${h^{(k)}}^{-1}$, to $CM$. The edges connected to $0^h$ and a vertex $v \in \partial M$ are labeled by $a_{0v}+\beta(1,{h^{(k)}}^{-1},g_v^{(k_v)})$. The vertex $0$ and $0^h$ are connected by the edge $\lr{00^h}$, labeled by $a_{00^h}=0$. When pulled back to this manifold, $(\eps+\beta)$ evaluates to zero on the face $\lr{00^h1}$, as shown in Eq.~\eqref{eq: eps + beta = 0}.}
% The spatial parts $M_3$ of $\Phi$ and $\Phi'$ in Fig. \ref{fig: symmetry_configuration_2_boundary} are glued and form higher dimensional simplices. The triangle formed by vertices $0$ (spacetime infinity vertex in $\Phi$), $0^h$ (spacetime infinity vertex in $\Phi'$), $1$ is shown here. We choose $a_{0 0^h}$ as the definition of gluing. $(\eps+\beta)$ acting on this triangle $\lr{00^h1}$ gives zero, calculated in \eqref{eq: eps + beta = 0}.
} 
\label{fig: 00h1}
\end{figure}
This is because $(\eps + \beta)(00^h1)$ is trivial (Fig.~\ref{fig: 00h1}): 
\begin{eqs}
    (\eps + \beta) ({0 0^h 1}) &= (\delta \ba + \bar \rho + \beta) ({0 0^h 1}) \\
    &= 2 \times \beta(1,{h^{(k)}}^{-1},g_1^{(k_1)}) + \rho(1,1,h^{-1},g_1) \\ &= 0,
\label{eq: eps + beta = 0}
\end{eqs}
where we have implicitly pulled back functions to cochains on the manifold with $0^h$, and we have used that $\rho$ is normalized. Therefore, the phase difference in Eq.~\eqref{alpha pullback difference2} is:
\begin{multline} \label{variation of mu gives alpha}
   \sum_{f=\lr{123} \in \partial M} \alpha^*({00^h 123)} O_f \\
   = \sum_{f=\lr{123} \in \partial M} [\mu({0^h 123}) - \mu({0 123}) ] O_f.
\end{multline}

To form a symmetric wave function, we can include a factor of $-\mu(0ijk)$ for each face $f=\lr{ijk}$ in the boundary of $M$. According to Eq.~\eqref{variation of mu gives alpha}, the change in this factor under the symmetry action will precisely cancel the change from the $\alpha^*$ terms. Consequently, the following state is invariant under the symmetry action:
\begin{multline}
        \ket{ \Psi_\text{b} } 
        = \sum_{\Phi} \prod_{f_\partial=\langle 123 \rangle \in \partial M}
        e^{- 2\pi i {\mu ({0123})}O_f} \\ 
         \prod_{t=\langle 1234 \rangle \in M} e^{2\pi i \alpha^* ({01234}) O_t} \ket{\Phi},
\end{multline}
where the sum is over all configurations $\Phi$ with the label for the $0$ vertex fixed as $1 \in L$ and the labels for the temporal edges $e \notin \overline{M}$ set to $0 \in \ZZ_2$. 
By replacing the configurations $\{a_e\}$ on the edges in $\overline{M}$ with a $1$-cochain $\ba_e$ and using the notation introduced in the main text, we can write the state $\ket{\Psi_\text{b}}$ more explicitly as:
\begin{widetext}
\begin{equation}
\begin{split}
        \ket{ \Psi_\text{b} }
        =&\sum_{\{ g_v, k_{v_\partial}\}, \, \ba_e} ~~ \prod_{f_\partial=\langle 123 \rangle \in \partial M}
        {
        e^{-2\pi i O_{f_\partial} \etahat(123)} (-1)^{\betahat(23) \delta \ba (0 12)} (-1)^{\betahat(13) [\delta \ba (123)+ \rhohat (123)]}
        }\\
        &\prod_{t=\langle 1234\rangle} e^{ 2 \pi i O_t \nuhat (1234)  }  (-1)^{ \delta \ba(012) \delta \ba(234)}
        (-1)^{ \rhohat (134) [\delta \ba(123)+ \rhohat (123)]+ \rhohat (124) [\delta \ba(234)+ \rhohat (234)] } \ket{ \{ g_v \}, \{  k_{v_\partial} \}, \ba_e}.
\end{split}
\label{eq: boundary state 3-2}
\end{equation}
\end{widetext}

{The next steps in the construction of the supercohomology models on a manifold with boundary are to gauge the $1$-form symmetry of $\ket{\Psi_\text{b}}$ and to apply the fermionization duality. The $1$-form symmetry can be gauged by applying the linear map $\Gamma$ defined in Appendix~\ref{app: 1form gauging quantum states}, which, up to a normalization factor, acts on a configuration state as:
\begin{equation}
    \Gamma\big( \ket{ \{ g_v \}, \{  k_{v_\partial} \}, \ba_e} \big)= \ket{ \{ g_v \}, \{  k_{v_\partial} \}, \delta \ba_e}.
\end{equation}
A ground state of the bosonic shadow model is then simply $\ket{\Psi_\text{s}}=\Gamma \big( \ket{\Psi_\text{b}} \big)$.}

% \textcolor{gray}{The bosonic shadow state is simply
% \begin{equation}
%     \ket{\Psi_s} = \Gamma(\ket{\Psi_\text{b}}),
% \end{equation}
% where $\Gamma$ is the gauging map
% \begin{equation}
%     \Gamma( \ket{ \{ g_v \}, \{  k_{v_\partial} \}, \ba_e} ) = \ket{ \{ g_v \}, \{  k_{v_\partial} \}, \delta \ba_e}.
% \end{equation}}

{Similar to the bulk construction in Section~\ref{sec:bulkG}, we make a change of basis to ensure that the $L$ symmetry is onsite. Here, the temporal edges in $\overline{M}$ are also included in the basis transformation. The basis change is implemented by the operator $\mathcal{R}$:
\begin{eqs}
    \mathcal{R} \equiv \prod_{e=\lr{ij} \in \partial M} X_{\lr{0ij}}^{\betahat(e)} \prod_{f \in M} X_f^{\rhohat(f)}.
\end{eqs}
Letting $\ba_f$ denote a $2$-cochain on $\overline{M}$, the action of $\mathcal{R}$  on a configuration state is:
\begin{equation}
    \mathcal{R} \ket{ \{ g_v \}, \{  k_{v_\partial} \}, \ba_f} = \ket{ \{ g_v \}, \{  k_{v_\partial} \}, \ba_f + \boldsymbol B_f},
\end{equation}
where $\boldsymbol B_f$ is the $2$-cochain satisfying:
\begin{equation}
    \begin{aligned}
        \boldsymbol B_f \boldsymbol{(}\lr{ijk}\boldsymbol{)} &= \rho (1, g_i,g_j,g_k), &  &\forall \lr{ijk} \in M \\
        \boldsymbol B_f \boldsymbol(\lr{0ij}\boldsymbol) &= \beta (1, g^{(k_i)}_i,g^{(k_j)}_j), & &\forall \lr{ij} \in \partial M.
    \end{aligned}
\label{eq: Bf def}
\end{equation}}
{Further, we define a new basis by the identification:
\begin{eqs}
          \ket{ \{ g_v \}, \{  k_{v_\partial} \}, \ba_f}' &\equiv \mathcal{R} \ket{ \{ g_v \}, \{  k_{v_\partial} \}, \ba_f}  \\
            &=\ket{ \{ g_v \}, \{  k_{v_\partial} \}, \ba_f + \boldsymbol B_f}.
\end{eqs}
In this basis, the $0$-form $L$ symmetry is onsite:
\begin{multline}
    V(h^{(k)}) \ket{\{ g_v \}, \{  k_{v_\partial} \}, \ba_f}' \\ 
    = \ket{\{ h g_v \}, \{  k+ k_{v_\partial} + m(1,h,hg_{v_\partial}) \}, \ba_f}'.
\end{multline}}
{We also re-define the Pauli operators in this basis so that:
\begin{eqs}
    X_{f^\prime} \ket{\{ g_v \}, \{ k_{v_\partial} \}, \boldsymbol a_f}^\prime &= \ket{\{ g_v \}, \{  k_{v_\partial} \}, \boldsymbol a_f + \boldsymbol{f}^\prime}^\prime \\
    Z_{f^\prime} \ket{\{ g_v \}, \{ k_{v_\partial} \}, \boldsymbol a_f}^\prime &= (-1)^{\boldsymbol a_f(f^\prime)} \ket{\{ g_v \}, \{  k_{v_\partial} \}, \boldsymbol a_f}^\prime.
\end{eqs}
Finally, the shadow model ground state $\ket{\Psi_\text{s}}$ can be written as:
\begin{widetext}
\begin{equation}
\begin{split}
        \ket{ \Psi_s }
        =&\sum_{\{ g_v, k_{v_\partial}\},\,\ba_e} ~~ \prod_{f_\partial=\langle 123 \rangle \in \partial M}
        e^{-2\pi i [\etahat(123) + \frac{1}{2} \betahat(12) \betahat(23)] O_{f_\partial} }
        Z_{012}^{\betahat(23) }  Z_{123}^{\betahat(13) } \\
        &\prod_{t=\langle 1234\rangle} e^{ 2 \pi i \, \nuhat (1234) O_t }  (-1)^{ \delta \ba(012) \delta \ba(234)}
        Z_{123}^{ \rhohat(134)} Z_{234}^{ \rhohat(124)  } \ket{ \{ g_v \}, \{  k_{v_\partial} \}, \delta \ba_e + \boldsymbol B_f}'. \\
\end{split}
\label{eq: boundary state 2}
\end{equation} 
\indent Next, we write $\ket{\Psi_\text{s}}$ in a form that can be more readily fermionized using the duality in Appendix~\ref{sec: review of boson-fermion duality} or equivalently Ref.~\cite{CK18}. First, the $\boldsymbol B_f$ factor in Eq.~\eqref{eq: boundary state 2} can be expressed using $X_f$ operators: 
\begin{equation}
\begin{split}
        \ket{ \Psi_s }=&\sum_{\{ g_v, k_{v_\partial}\}, \ba_e} ~~ \prod_{f_\partial=\langle 123 \rangle \in \partial M}
        {
        e^{-2\pi i [\etahat(123) + \frac{1}{2} \betahat(12) \betahat(23)] O_{f_\partial}}
        Z_{012}^{\betahat(23)}  Z_{123}^{\betahat(13) }
        \prod_{\lr{ij} \in \partial M} X_{0ij}^{\betahat (ij)} 
        }\\
        &\prod_{t=\langle 1234\rangle \in M} e^{ 2 \pi i \, \nuhat (1234) O_t }  (-1)^{ \delta \ba(012) \delta \ba(234)}
        Z_{123}^{ \rhohat(134)} Z_{234}^{ \rhohat(124)  } \prod_{f \in M} X_f^{\rhohat (f)} \ket{ \{ g_v \}, \{  k_{v_\partial} \}, \delta \ba_e }' 
\end{split}
\label{eq: boundary state 3}
\end{equation}
Then, the Pauli X and Pauli Z operators can be commuted to form the fermionizable operators $\bar{U}_{\bar \beta}$ and $\bar{U}_{\bar \rho}$, defined as:
% \begin{eqs}
% \bar{U}_{\bar \beta} \equiv \prod_{\lr{ij}\in \partial M}\prod_{f'\in \overline{M}} Z_{f'}^{\betahat(ij) \boldsymbol{\langle 0ij \rangle}\cup_1 \boldsymbol f'} \prod_{\lr{ij}\in \partial M}X_{0ij}^{\betahat(ij)}, ~~~~~~~~\text{ and }~~~~~~~~
% \bar{U}_{\bar \rho} \equiv \prod_{f' \in \overline{M}}Z_{f'}^{ \int_M \rhohat \cup_1 \boldsymbol f'}\prod_{f\in M}X_f^{\rhohat(f)}
% \end{eqs}
\begin{eqs}
\bar{U}_{\bar \beta} \equiv \prod_{f'\in \overline{M}} Z_{f'}^{\int_{\overline{M}} \hat{\boldsymbol \beta} \cup_1 \boldsymbol f'} \prod_{\lr{ij}\in \partial M}X_{0ij}^{\betahat(ij)}, ~~~~~~~~\text{ and }~~~~~~~~
\bar{U}_{\bar \rho} \equiv \prod_{f' \in \overline{M}}Z_{f'}^{ \int_M \rhohat \cup_1 \boldsymbol f'}\prod_{f\in M}X_f^{\rhohat(f)},
\end{eqs}
where $\hat{\boldsymbol{\beta}}\boldsymbol{(}\lr{ijk}\boldsymbol{)} \ket{\{g_v\},\{k_{v_\partial}\}, \ba_e} = \beta(g^{(k_i)}_i,g^{(k_j)}_j,g^{(k_k)}_k) \ket{\{g_v\},\{k_{v_\partial}\}, \ba_e}$ for $i,j,k \in \partial M \cup \{0\}$.
With this, the shadow state $\ket{\Psi_\text{s}}$ can be written as:
\begin{eqs} \label{shadow state fermionizable app}
 \ket{ \Psi_s }= &\prod_{f_\partial=\langle 123 \rangle \in \partial M}
        e^{-2\pi i [\etahat(123) + \frac{1}{2} \betahat(12) \betahat(23)] O_{f_\partial} }  \bar{U}_{\bar \beta} \prod_{t=\langle 1234\rangle}
         e^{ 2 \pi i \, \nuhat (1234) O_t } 
        \bar{U}_{\bar \rho}  \\
        &\sum_{\{ g_v, k_{v_\partial}\}, \ba_e}~~\prod_{f_\partial = \lr{123} \in \partial M} (-1)^{\ba(01) \delta \ba(123)}
        \prod_{t=\langle 1234\rangle} (-1)^{ \ba(12) \delta \ba (234)}
        \ket{ \{ g_v \}, \{  k_{v_\partial} \}, \delta \ba_e }'.
\end{eqs}
\end{widetext}
To write Eq.~\eqref{shadow state fermionizable app}, we have also used Stokes' theorem:
\begin{eqs}
    \int_{CM} \delta \ba \cup \delta \ba = \int_{\overline{M}} \ba \cup \delta \ba.
\end{eqs}
}

{The state in Eq.~\eqref{shadow state fermionizable app} can be fermionized straightforwardly. First of all, the second line in Eq.~\eqref{shadow state fermionizable app} is precisely a ground state of the twisted toric code on $\overline{M}$. Therefore, it is dual to the ground state of an atomic insulator. Furthermore, according to Appendix~\ref{sec: review of boson-fermion duality}, the operators $\bar{U}_{\bar \beta}$ and $\bar{U}_{\bar \rho}$ can be fermionized to the product of hopping operators:
\begin{eqs}
\bar{U}_{\bar \beta} \to  \xi_\beta(\partial M)  \prod_{\lr{0ij}} S_{0ij}^{\betahat(ij)}, ~\text{ and }~~
\bar{U}_{\bar \rho} \to \xi_{\bar{\rho}}(M) \prod_{f \in M} S_f^{\rhohat(f)}.
\end{eqs}
Here, $\xi_\beta(\partial M)$ is a sign that depends on an ordering of the faces $\lr{0ij}$, where $\lr{ij}$ is some edge in $\partial M$. It compensates for the order dependence of the product of hopping operators and is explicitly:
\begin{align}
    \xi_{\bar \beta}(\partial M) \equiv \prod_{e_i, e_{i'} \in \partial M |i<i'} (-1)^{\betahat{(e_{i})} \betahat{(e_{i'})} \int_{\partial M} \be_{i'} \cup \be_{i} }.
\end{align}
$\xi_{\bar{\rho}}(M)$ is defined in Section~\ref{sec: supercohomology model} and Appendix~\ref{app: fermionization shadow circuit}. Fermionization thus results in the supercohomology state:
\begin{eqs} \label{supercohomology bdry state app}
  &\ket{\Psi_f} = \prod_{f_\partial} e^{-2 \pi i O_{f_\partial} [\etahat(f_\partial)+\frac{1}{2}\betahat \cup \betahat(f_\partial)]}  \xi_{\bar \beta}(\partial M) \prod_{e_\partial} S_{e_\partial}^{\betahat(e_\partial)} \\ 
 &\prod_t e^{2 \pi i O_t \nuhat(t)} \xi_{\bar{\rho}}(M){\prod_{f}} S_f^{\rhohat(f)} \sum_{\{ g_v \}, \{  k_{v_\partial} \}} \ket{ \{ g_v \}, \{  k_{v_\partial} \}, \text{vac} }, 
\end{eqs}
where the second product is over edges $e_\partial$ in $\partial M$, $S_{e_\partial}$ is the hopping operator associated to the face formed by $e_\partial$ and the $0$ vertex, and $\ket{\text{vac}}$ is the fermionic state with zero fermion occupancy at each site. The fermionic d.o.f. on the tetrahedra in $\overline{M}\setminus M$ can be associated to faces in $\partial M$ as in Section~\ref{sec: fermionic gapped boundary construction}.}

{From Eq.~\eqref{supercohomology bdry state app}, we see that the supercohomology state $\ket{\Psi_f}$ can be prepared from a symmetric product state by the FDQC:
\begin{eqs}
\tilde{\mathcal{U}}_f^L \equiv &\prod_{f_\partial} e^{-2 \pi i O_{f_\partial} [\etahat(f_\partial)+\frac{1}{2}\betahat \cup \betahat(f_\partial)]}  \xi_{\bar \beta}(\partial M) \prod_{e_\partial} S_{e_\partial}^{\betahat(e_\partial)} \\ 
 &\prod_t e^{2 \pi i O_t \nuhat(t)} \xi_{\bar{\rho}}(M){\prod_{f}} S_f^{\rhohat(f)}
\end{eqs}
However, $\tilde{\mathcal{U}}_f^L$ is not symmetric. A symmetric FDQC can be formed by multiplying on the right by a product of parity operators:
\begin{eqs} \label{parity to make fdqc sym}
    \prod_{t} P_t^{\int_{\overline{M}} \boldsymbol B_f \cup_2 \boldsymbol t}.
\end{eqs}
This is analogous to the product of parity operators in the definition of $\Uf$ on a manifold without boundary in Eq.~\eqref{eq: uf def}. The product of parity operators does not change the state produced by $\tilde{\mathcal{U}}_f^L$, since $\ket{\text{vac}}$ is a $+1$ eigenstate of the parity operators. Using the definition of $\boldsymbol B_f$ in Eq.~\eqref{eq: Bf def} and the explicit formulas for the cup products, the product of parity operators in Eq.~\eqref{parity to make fdqc sym} can be decomposed into:
\begin{eqs}
    \prod_{t} P_t^{\int_{\overline{M}} \boldsymbol B_f \cup_2 \boldsymbol t} = \prod_t P_t^{\int_{M} \rhohat \cup_2 \boldsymbol t} \prod_{f \in \partial M} P_{f}^{\int_{\partial M} \bface \cup_1 \betahat}.
\end{eqs}
Thus, the FDQC $\Uf^L$ in the main text is produced by composing $\tilde{\mathcal{U}}_f^L$ with the product of parity operators above:
\begin{eqs}
  \Uf^L = &\prod_{f_\partial} e^{-2 \pi i O_{f_\partial} \etahat(f_\partial)}  \chi_{\bar \beta}(\partial M) \prod_{e_\partial} S_{e_\partial}^{\betahat(e_\partial)}  \prod_{f_\partial} P_{f_\partial}^{\int_{\partial M} \boldsymbol{f}_\partial \cup_1 \betahat} \\ 
 \times &\prod_t e^{2 \pi i O_t \nuhat(t)} \xi_{\bar{\rho}}(M){\prod_{f}} S_f^{\rhohat(f)} \prod_{t} P_t^{\int_M \rhohat \cup_2 \boldsymbol{t}}.
\end{eqs}
Here, we have absorbed the sign from commuting parity operators and the $\frac{1}{2}\betahat \cup \betahat$ factor into $\chi_\beta(\partial M)$, defined as:
\begin{eqs} \label{eq: chi beta def}
    \chi_{\bar \beta}(\partial M) \equiv & (-1)^{\int_{\partial M}\left(\hat{\boldsymbol{\beta}} \cup_1 \betahat + \delta \betahat \cup_1 \betahat +  \betahat \cup \betahat \right)} \xi_{\bar \beta}(\partial M).
\end{eqs}}

\section{Trivialization of supercohomology data: $G_f = \ZZ_2 \times \ZZ_4 \times \ZZ_2^f$}\label{sec:example}

In this appendix, we give an example of using a symmetry extension to trivialize the supercohomology data, as described in the construction of the gapped boundaries of supercohomology models in Section~\ref{sec: sym extension boundary}. We consider a fSPT phase protected by a $G_f = \ZZ_2 \times \ZZ_4 \times \ZZ_2^f$ symmetry, and we identify two possible symmetry extensions that trivialize the supercohomology data. We note that a gapped boundary for the particular $G_f = \ZZ_2 \times \ZZ_4 \times \ZZ_2^f$ phase has been found in Refs.~\cite{FVM17} and \cite{C19}.

% In this appendix, we give an example of our construction of a fermionic SPT protected by $G_f=\ZZ_2 \times \ZZ_4 \times \ZZ_2^f$, which cannot be realized by free fermions or stacking with bosonic SPTs\cite{CTW17}. We will describe both the construction on a closed manifold, and the case with a symmetric gapped boundary exhibiting topological order.

To get started, we describe the supercohomology data $(\rho, \nu)$ corresponding to our phase of interest. We represent a generic group element as $(g,h) \in \mathbb Z_2 \times \mathbb Z_4$ where $g=0,1$ and $h=0,1,2,3$, and use the notation $[\cdots]_{\boldsymbol{N}}$ to denote modulo $N$.
The cocycle $\rho$ can be obtained from a decorated domain-wall picture. Namely, a ($2+1$)D $\mathcal N=2$ supercohomology SPT with $\mathbb Z_2$ symmetry is decorated on the $\mathbb Z_4$ domain walls. The supercohomology data for the ($2+1$)D phase (in homogeneous variables) is given by\cite{GW14}:
\begin{align}
    n_2(g_0,g_1,g_2) &=  [g_0-g_1]_{\boldsymbol{2}}[g_1-g_2]_{\boldsymbol{2}},\\
    \nu_3(g_0,g_1,g_2,g_3) &= \frac{1}{4} [g_0-g_1]_{\boldsymbol{2}}[g_1-g_2]_{\boldsymbol{2}}[g_2-g_3]_{\boldsymbol{2}}. 
\end{align}
Therefore, the cocycle $\rho \in Z^3(\mathbb Z_2 \times \mathbb Z_4,\mathbb Z_2)$ takes the form:
\begin{equation}
    \rho = n_2 \cup \phi_1
\end{equation}
where the class of $\phi_1$ is the generator of $H^1(\mathbb Z_4,\mathbb Z_2)$ given by $\phi_1(h_0,h_1) = [h_0-h_1]_{\boldsymbol{2}}$, and both cocycles are implicitly pulled back via the projection map to each subgroup.
Explicitly, the supercohomology data can be chosen to be:
\begin{align}
    &\rho\boldsymbol{(}(g_0,h_0),(g_1,h_1),(g_2,h_2),(g_3,h_3)\boldsymbol{)}\\\nonumber &=  [g_0-g_1]_{\boldsymbol{2}}[g_1-g_2]_{\boldsymbol{2}}[h_2-h_3]_{\boldsymbol{2}},\\
    &\nu\boldsymbol{(}(g_0,h_0),(g_1,h_1),(g_2,h_2),(g_3,h_3),(g_4,h_4)\boldsymbol{)} \nonumber\\
     &= \frac{1}{4} [g_0-g_1]_{\boldsymbol{2}} [g_1-g_2]_{\boldsymbol{2}} [g_2-g_3]_{\boldsymbol{2}} [h_3-h_4]_{\boldsymbol{4}}
\end{align}
One can verify that they indeed satisfy the supercohomology equations in Eq.~\eqref{guweneqs}.

%\subsection{Symmetry extension for the gapped boundary}
%We show that a symmetry extension of at least $\mathbb Z_4$ is required to trivialize the supercohomology data of the intrinsically interacting Fermionic SPT in (3+1)D with symmetry $G=\mathbb Z_2 \times \mathbb Z_4$. The group is extended to $\mathbb Z_8 \times \mathbb Z_4$. By gauging the $\ZZ_4$ subgroup, the boundary becomes a $\mathbb Z_4$ topological order with anomalous $\ZZ_2 \times \ZZ_4$ symmetry. Ref. \cite{FVM17} also constructed a $\mathbb Z_4$ topological order as a gapped boundary of this phase.

To construct a symmetric gapped boundary, we perform two symmetry extensions to trivialize the supercohomology data. First, we extend by $\mathbb Z_2$ to trivialize $\rho_3$. The extension is given by:
\begin{equation} \label{example extension 1}
1 \rightarrow \mathbb Z_2  \xrightarrow[]{}\mathbb Z_4 \times \mathbb Z_4   \xrightarrow[]{\pi_1} \mathbb Z_2 \times \mathbb Z_4 \rightarrow 1
\end{equation}
where $\pi_1$ is the projector $\pi_1\boldsymbol{(}(g,h)\boldsymbol{)} = ([g]_{\boldsymbol 2},h)$.
The pulled-back cocycle $\rho^*$ is a coboundary of the cochain: 
\begin{align}
\beta_2\boldsymbol{(}(g_0,g_1),(g_1,h_1),(g_2,h_2)\boldsymbol{)} = \left \lfloor  \frac{g_0-g_1}{2} \right \rfloor [h_1-h_2]_{\boldsymbol 2}.
\end{align}
%i.e. $\delta \beta_2(\{t((g',h'))\}) = \rho(\{(g,h)\})$

The next step is to trivialize the following cocycle:
\begin{equation}
     \omega \equiv \nu^* +\frac{1}{2} \beta \cup \beta+ \frac{1}{2}  \beta \cup_1 \delta \beta 
     \label{equ:omegainBG'}
\end{equation}
which was defined in Eq. \eqref{equ:deltaeta}. However, because $\omega$ has a very complicated closed form, we instead identify the cohomology class of $\omega$ in $H^4(\mathbb Z_4 \times \mathbb Z_4,\mathbb R/\mathbb Z) \cong \ZZ_4^2$ by using topological invariants that completely distinguish the elements of the above cohomology group \cite{WL15,T17}. In particular, for inhomogeneous cocycles, the invariants of the cohomology group $H^4(\ZZ_{N_i} \times \ZZ_{N_j}, \RR/\ZZ) \cong \ZZ_{\text{gcd}(N_i,N_j)}^2$ are given by:
\begin{eqs}
    \mathcal I_1 =&\sum_{n=0}^{N_j-1} i_{(1,0)} \omega \boldsymbol{(}(0,1),(0,n),(0,1)\boldsymbol{)}, \nonumber\\
    \mathcal I_2 =& \sum_{n=0}^{N_i-1} i_{(0,1)} \omega \boldsymbol{(}(1,0),(n,0),(1,0)\boldsymbol{)}. \nonumber
\end{eqs}
where $i_{g} \omega$ is called the slant product of $g$ and $\omega$ and is defined as
\begin{eqs}
  i_{g}\omega(a,b,c)&=\omega(g,a,b,c)-\omega(a,g,b,c) \\ &+\omega(a,b,g,c)-\omega(a,b,c,g).
\end{eqs}
We stress that the quantities above are invariant under adding a coboundary to $\omega$.

In our case, the topological invariants are valued in $\RR/\ZZ$ and are multiples of $\frac{1}{4}$:
\begin{align}
    \mathcal I_1 =&\sum_{n=0}^{3} i_{(1,0)} \omega \boldsymbol{(}(0,n+2),(0,n+1),(0,1),(0,0)\boldsymbol{)},\\
    \mathcal I_2 =& \sum_{n=0}^{3} i_{(0,1)} \omega \boldsymbol{(}(n+2,0),(n+1,0),(1,0),(0,0)\boldsymbol{)}.
\end{align}
where the equations above are now using homogeneous cocycles.
% where $i_h \omega$ is the 3-cocycle:
% \begin{align}
%   i_{h}\omega(g_0,g_1,g_2,g_3)=&\omega(g_0,g_0+h,g_1+h,g_2+h,g_3+h)\nonumber\\
%   &-\omega(g_0,g_1,g_1+h,g_2+h,g_3+h) \nonumber\\
%   &+\omega(g_0,g_1,g_2,g_2+h,g_3+h)\nonumber\\
%   &-\omega_4(g_0,g_1,g_2,g_3,g_3+h).
%   \label{equ:slantproduct}
% \end{align}
Computing the invariants for the cocycle $\omega$ given in Eq. \eqref{equ:omegainBG'}, we find
$\mathcal I_1 = 0$ and $\mathcal I_2 =\frac{1}{2}$. 

One can also check that the following ``canonical'' cocycle
\begin{eqs}
&\omega^\text{can}\boldsymbol{(}(g_0,h_0),(g_1,h_1),(g_2,h_2),(g_3,h_3),(g_4,h_4)\boldsymbol{)} \\
&=\frac{1}{8} [g_0-g_1]_{\boldsymbol 4}[h_3-h_4]_{\boldsymbol 4} \\
&\times \left([g_1-g_2]_{\boldsymbol 4}+[g_2-g_3]_{\boldsymbol 4}-[g_1-g_3]_{\boldsymbol 4}\right)
%\frac{1}{8} g_1(g_2+g_3-[g_2+g_3]_4)h_4
\label{equ:canonicalcocycle}
\end{eqs}
% \begin{align}
% &\omega^\text{can}((g_0,h_0),(g_1,h_1),(g_2,h_2),(g_3,h_3),(g_4,h_4)) \\
% &=\frac{1}{8} [g_0-g_1]_{\boldsymbol 4}\left([g_1-g_2]_{\boldsymbol 4}+[g_2-g_3]_{\boldsymbol 4}-[g_1-g_3]_{\boldsymbol 4}\right)[h_3-h_4]_{\boldsymbol 4} \nonumber
% %\frac{1}{8} g_1(g_2+g_3-[g_2+g_3]_4)h_4
% \label{equ:canonicalcocycle}
% \end{align}
has the same topological invariants, and therefore is in the same cohomology class as the previous cocycle. It follows that the two must differ by a coboundary
\begin{align}
\omega= \omega^\text{can} + \delta \lambda
\end{align}
For some group 3-cochain $\lambda$. Although we do not have a closed form for $\lambda$, it can in principle be obtained by numerically solving linear equations $\delta \lambda = \omega - \omega^\text{can}$ using the Smith decomposition. We refer to Appendix G of Ref.~\cite{T17} for further details.

We can now perform a second symmetry extension. Here, we present two possible extensions. The first is given by
\begin{equation}
1 \rightarrow \mathbb Z_2  \xrightarrow[]{}\mathbb Z_8 \times \mathbb Z_4   \xrightarrow[]{\pi_{2}} \mathbb Z_4 \times \mathbb Z_4 \rightarrow 1,
\label{equ:extension1}
\end{equation}
where, $\pi_2\boldsymbol{(}(g,h)\boldsymbol{)} = ([g]_{\boldsymbol 4},h)$, and the second is given by
\begin{equation}
1 \rightarrow \mathbb Z_2  \xrightarrow[]{}\mathbb Z_4 \times \mathbb Z_8   \xrightarrow[]{\pi_{3}} \mathbb Z_4 \times \mathbb Z_4 \rightarrow 1,
\label{equ:extension2}
\end{equation}
where $\pi_3\boldsymbol{(}(g,h)\boldsymbol{)} = (g,[h]_{\boldsymbol 4})$. One can check that the pulled back canonical cocycles are trivialized in both cases by computing similar topological invariants of cocycles in the extended groups. This means that $\pi^*\omega^\text{can} = \delta \eta^\text{can}$ for some choice of $\eta$, where $\pi = \pi_2 \circ\pi_1$ or $\pi= \pi_3 \circ\pi_1 $ depending on the choice of extension. It follows that we can choose
\begin{align}
    \eta = \eta^\text{can} + \lambda,
\end{align}
so that $\delta \eta = \pi^*\omega$, as desired.

We have two extensions that trivialize the supercohomology data, one is an extension by $K=\ZZ_2 \times \ZZ_2$ and the other is by $K=\ZZ_4$. To build the gapped boundary, we gauge the $K$ symmetry, these give a $\ZZ_2 \times \ZZ_2$ gauge theory and a $\ZZ_4$ gauge theory, respectively. It would be interesting to compare the $\ZZ_4$ anomalous topological order here with that found in Refs.~\cite{FVM17} and \cite{C19}. It would also be interesting to study the anomalous $\ZZ_2 \times \ZZ_2$ topological order.

\onecolumngrid
\clearpage
\twocolumngrid

\bibliography{bibliography.bib}
\onecolumngrid

\end{document}